\documentclass[preprintnumbers,nofootinbib,showpacs,eqsecnum,pre]{revtex4}

\usepackage{graphicx}
\usepackage{amsmath,amsthm,amssymb}
\usepackage{mathrsfs,bbm}
\usepackage{color}
\usepackage{array} 			% define new table columns and allow for column type m 
\usepackage[sort&compress]{natbib}	% prsent citations in compressed mode, i.e. [10-19]
\usepackage{hyperref}   
\usepackage{hypernat}			% Allowing for use of natbib in compress mode together with hyperref
\definecolor{lgray}{gray}{0.9}          %background of the algorithm boxes
\graphicspath{ {Fig/} }

\makeatletter       % Macros for arabic section numbering

\renewcommand{\p@subsection}{}

\makeatother

\newtheorem*{definition}{Definition}
\newtheorem*{algorithm}{Algorithm}

\newcommand*{\smgroup}{\mbox{$SU(3)_C \times SU(2)_L \times U(1)_Y$} }
\newcommand*{\eweakgroup}{\mbox{$SU(2)_L \times U(1)_Y$} }
\newcommand*{\emgroup}{\mbox{$U(1)_{em}$} }

\newcommand*{\unitmatrix}{\mathbbm{1}}

\newcommand*{\Zthree}{\mathbbm{Z}_3}

\newcommand*{\abs}[1]{\left\lvert {#1} \right\rvert} % abs. value or norm
\newcommand*{\twomat}[1]{\underline{#1}}             % 2x2 matrix
\newcommand*{\tvec}[1]{\boldsymbol{#1}}              % 3 vector
\newcommand*{\trans}{\mathrm{T}}                     % transposed
\newcommand*{\by}{\!\times\!}                        % for nxn matrices
\newcommand*{\im}{\text{Im}}                        
\newcommand*{\re}{\text{Re}}                        
\newcommand*{\dcp}{\delta_{\kappa}'}
\newcommand*{\dedm}{\delta_{\text{EDM}}}
\newcommand*{\tl}[0]{\succ}                          % total ordering of term
\newcommand*{\tlex}[0]{\succ_{\text{lex}}}           % total ordering of term, lexicographic
\newcommand*{\tdeg}[0]{\succ_{\text{deg}}}           % total ordering of term, degree
\newcommand*{\kx}[0]{K[\mathbf{x}]}
\newcommand*{\di}{\qquad}
\renewcommand*{\sb}{s_\beta}
\newcommand*{\cb}{c_\beta}
\newcommand*{\tb}{t_\beta}
\newcommand*{\ct}{cot}
\newcommand*{\Rl}{R_\lambda}
\newcommand*{\Rk}{R_\kappa}
\newcommand*{\dlr}[1]{\stackrel{\!\!\leftrightarrow}{\partial^#1}}
\newcommand*{\opw}{{\mathbf I_W}^{\!\!\!\!\!\!a}\;}

\DeclareMathOperator{\trace}{tr}

\DeclareMathOperator{\sign}{sign}
\DeclareMathOperator{\normf}{normf}
\DeclareMathOperator{\lcm}{lcm} 
\DeclareMathOperator{\spol}{spol}               %S-polynomial
\DeclareMathOperator{\lp}{LP}                   %Leading Term
\DeclareMathOperator{\lc}{LC}                   %Leading Coefficient
\DeclareMathOperator{\diag}{diag}               %Diagonal matrix elements
\DeclareMathOperator{\BR}{BR}       		%Branching Ratio
\DeclareMathOperator{\D}{d}       		%derivative

\begin{document}

\preprint{HD-THEP-09-9}
\title{The Next-to-Minimal Supersymmetric extension of the Standard Model reviewed}
% PACS
% 12.60.Jv 	Supersymmetric models
% 14.80.Cp 	Non-standard-model Higgs bosons
% 14.80.Ly 	Supersymmetric partners of known particles

%\pacs{}

 \author{M. Maniatis}
 \email[E-mail: ]{M.Maniatis@thphys.uni-heidelberg.de}

 \affiliation{
 Institut f\"ur Theoretische Physik, Philosophenweg 16, 69120
 Heidelberg, Germany\\
 phone +49-6221-549-414, fax +49-6221-549-333
 }

 \begin{abstract}
 The next-to-minimal supersymmetric extension of the Standard Model (NMSSM) is
 one of the most favored supersymmetric models. After an introduction to
 the model, the Higgs sector and the neutralino sector are
 discussed in detail.  Theoretical, experimental, and
 cosmological constraints are studied. Eventually, the Higgs potential
 is investigated in the approach of bilinear functions. 
 Emphasis is placed on aspects which are different
 from the minimal supersymmetric extension.
 \end{abstract}

\maketitle

\tableofcontents

\newpage
\section{Introduction}
\label{intro}

Supersymmetry~\citep{Ramond:1971gb,Neveu:1971rx,Gervais:1971ji,Golfand:1971iw,Wess:1974tw,Volkov:1973ix,Fayet:1976et,Fayet:1977yc,Farrar:1978xj,Fayet:1979sa,Witten:1981nf,Dimopoulos:1981zb,Sakai:1981gr,Ohta:1982wn,Nilles:1983ge,Haber:1984rc,Gunion:1984yn,Gunion:1986nh,Lahanas:1986uc}
is one of the most appealing concepts of
physics beyond the Standard Model~(SM). 
Some of the main motivations for studying supersymmetry are:
\begin{itemize}
\item Supersymmetry is an extension of the Poincare algebra
which relates fermions to bosons.
\item As a local symmetry, supersymmetry is naturally connected to gravity.
\item Supersymmetry stabilizes the 
hierarchy~\citep{Witten:1981kv,Polchinski:1982an,Dimopoulos:1981zb,Sakai:1981gr}
between the
electroweak and the GUT or Planck scale. Thus, the quadratical divergences
occurring in the Higgs-boson mass loop corrections
are systematically canceled and fine tuning between the bare Higgs-boson
mass and the loop contributions is avoided.
\item Supersymmetry, realized below the TeV scale, unifies
the \smgroup couplings at the GUT scale.
\item Supersymmetry provides a cold dark matter candidate, supposed
matter-parity is conserved.
\end{itemize}

A minimal supersymmetric extension of the Standard Model (MSSM) was 
proposed already some time
ago. For a review of the MSSM see for instance~\citep{Martin:1997ns} 
and references therein. 
In the MSSM the minimal particle content is added to the SM
in order to arrive at a supersymmetric model.
The MSSM extents the SM by an additional Higgs doublet, 
necessary to give masses to
up- and down-type fermions and in order to keep the theory anomaly free.
Every field is promoted to 
a superfield, pairing fermionic and bosonic degrees of freedom.
However, none of the additional predicted particles has been observed up to now.
Thus the question arises, why should we study 
further extensions of the minimal supersymmetric model?
The reason is, that the MSSM does certainly {\em not} 
parameterize any supersymmetric extension of the SM.
This is in particular due to the fixed particle content we encounter in the MSSM.
For instance, although the Higgs sector in the MSSM consists
of two doublets in contrast to one in the SM, the Higgs sector turns out to
be highly restricted. Thus, at tree-level,
the lightest CP-even Higgs-boson is predicted to be
lighter than the $Z$-boson. Large quantum corrections
are necessary in order to comply with the experimental lower
bounds from LEP on the CP-even Higgs-boson mass, which in turn 
require a very large scalar--top mass. That is,
some new kind of fine-tuning is to be introduced in the MSSM.

Let us collect some reasons, making
it worthwhile to study extensions of the minimal
supersymmetric Standard Model:
\begin{itemize}
\item In the MSSM we encounter the so-called $\mu$-term 
in the superpotential, where $\mu$ is a dimensionful parameter.
This parameter has to be adjusted by hand to a value at the
electroweak scale, {\em before} spontaneous 
symmetry breaking occurs~\citep{Kim:1983dt}. This
is seen to be a problem of the model.
It is desirable to look for further extension, which do not
have this $\mu$-problem. In singlet extensions, for instance, 
an effective $\mu$-term may be generated dynamically.
\item As mentioned before, the Higgs sector is highly restricted in
the MSSM. The lower bounds on the Higgs-boson masses from
LEP measurements require large quantum corrections accompanied
by a large stop mass in this model. An extended Higgs sector may
relax this restrictions and thus circumvent
the lower experimental bounds.
\item The MSSM Higgs-boson sector is CP-conserving at tree level. Extending
the Higgs sector in an appropriate way, CP violating phases arise. 
Sufficient CP-violation would meet one of the necessary Sakharov 
criteria~\citep{Sakharov:1967dj}
in order to generate the baryon--antibaryon asymmetry in our Universe.
\item The baryon--antibaryon asymmetry may be generated by strong
electroweak phase transitions of 
first order~\citep{Kuzmin:1985mm,Shaposhnikov:1987tw,Rubakov:1996vz,Shaposhnikov:1996th}. 
The required cubic terms in the effective potential arise
in the SM and the MSSM only via generically small radiative corrections.
An explicit cubic term is possible in
extensions of the MSSM.
\end{itemize}
Here we will review the next-to-minimal supersymmetric extension of the Standard 
Model~(NMSSM)~\citep{Fayet:1974pd,Dine:1981rt,Nilles:1982dy,Derendinger:1983bz,Frere:1983ag,Veselov:1985gd,Ellis:1988er,Ellwanger:1993xa,King:1995vk,Ellwanger:1996gw,Franke:1995tc,Ananthanarayan:1996zv,Ellwanger:1999ji},
which has the capability to solve the mentioned limitations
of the MSSM. In the NMSSM an additional gauge singlet is introduced
which generates the $\mu$-term dynamically, that is, an effective
$\mu$-term arises spontaneously and the adjustment by hand drops out.
This is surely the main motivation for the NMSSM and may justify the price to pay,
that is, the introduction of an additional gauge-singlet superfield. The particle content
in the bosonic part of the singlet results in two additional Higgs bosons 
whereas in the fermionic part we have
one additional neutralino, called {\em singlino}. Altogether we
have seven Higgs bosons and five neutralinos in the NMSSM, compared to
five Higgs bosons and four neutralinos in the MSSM. The Higgs-boson sector
of the NMSSM is no longer CP-conserving at tree level, merely CP-conservation only arises if
the parameters of the Higgs-boson sector are chosen in an appropriate way.
Nevertheless, in most studies in the literature the special case
of a CP-conserving
Higgs sector is considered, in order to simplify matters. In case 
of a CP-conserving Higgs sector
we encounter altogether three CP-even Higgs-bosons, two CP-odd ones
and in addition two charged Higgs bosons in the next-to-minimal model. 
As we will see, the Higgs-boson sector is in deed much less restricted and
the lower mass bound prediction of a CP-even Higgs boson in the MSSM
is generally shifted substantially.
The Higgs-boson phenomenology can in general be very different from what 
to expect in the MSSM; in addition to supplement the total number of Higgs-bosons. 
For instance, in the NMSSM the possibility arises 
that a CP-even Higgs boson decays into two very light CP-odd ones 
which would have escaped detection at LEP and may even be difficult to
detect at the LHC. 
Also the Higgs potential is enriched, leading to interesting consequences.
Let us also address the last aforementioned advantage 
of the NMSSM over 
the MSSM. As we will see, the trilinear $A$-parameter soft supersymmetry breaking terms, 
corresponding to the superpotential, may account for the desired
strong first order electroweak phase transition without large fine-tuning.\\

The additional neutralino, that is, the singlino, in general mixes with
the other four neutralinos. Also in the neutralino sector there
may be a substantial change of phenomenology compared to the minimal 
supersymmetric model; in addition
to supplement the total number of neutralinos by a fifth neutralino. 
This is due
to the fact that the singlino is introduced as a gauge singlet. Only
through mixing with the other neutralinos this singlino has couplings to
the non-Higgs particles.
This opens the intriguing possibility to have a singlino-like neutralino
which moreover may become the lightest supersymmetric (partner-)particle~(LSP). 
Relying on matter-parity, every supersymmetric partner particle will in this
case eventually decay into this singlino-like LSP. In particular
the next-to-lightest supersymmetric (partner-)particle (NLSP) would,
due to the small couplings, decay very slowly into the LSP. These
large decay length' could be revealed by signatures of  {\em displaced vertices} in the detector.
But of course, in case the singlino-like neutralino is not the LSP, it
would be omitted or at least be suppressed in cascade decays.
Generally, in discussing the NMSSM phenomenology we draw the attention to differences 
to the MSSM. For the MSSM phenomenology we refer
to the extensive literature to this subject.\\

We start in Sect.~\ref{sec-nmssm} with the NMSSM superpotential. We review the main
motivations for the modifications compared to the MSSM and
also discuss the drawback which arises in context with the 
Peccei--Quinn symmetry, which is promoted to a
discrete $\Zthree$-symmetry by the introduction of
the singlet cubic selfcoupling in the superpotential.
We recall the arguments in order to circumvent the
occurrence of dangerous domain walls which emerge in context of
spontaneously broken discrete symmetries.\\

In Sect.~\ref{sub-Higgs} we present the mass matrices and
the parameters to describe the complete Higgs-boson sector at tree-level.
In this we treat the most general Higgs sector with
the possibility of CP-violation. The special case
of a CP-conserving Higgs sector may be easily inferred. 
We briefly mention the one-loop effective potential
involving the field-dependent Higgs-boson masses.
Then we discuss the Higgs-boson phenomenology, 
where we stress the prospects at the LHC. In particular
we review the discussion of a ``no-lose'' theorem,
that is, the interesting question, whether
it can be guaranteed that at least one Higgs boson
of the NMSSM will be detected at the LHC - supposed
the NMSSM is realized in Nature.\\

In Sect.~\ref{sec-nc} follows a recap of the neutralino sector,
where we present
the mixing matrix as well as neutralino phenomenology, which is
of special interest in case there is a singlino-like LSP.\\

In Sect.~\ref{sec-par} we consider parameter constraints
in the NMSSM, beginning with theoretical constraints. 
For instance a theoretical constraint comes from the requirement
to have a potential with a global minimum not
breaking electric or color charge.  
A further constraint restricts coupling parameters by
forcing them to be perturbative up to the GUT scale. Of course
the latter constraint relies on the fact that the
model is valid in the large range up to the GUT scale.
We will also discuss briefly an approach to quantify
fine-tuning in general extensions of the SM. 
On the experimental side we start with considering constraints
coming from colliders. Precision measurements of
the $Z$-boson width are discussed, followed by 
exclusion limits from searches for
neutralino and chargino pair production, 
as well as Higgs-boson production.
The possible contribution of the NMSSM to the anomalous
magnetic moment of the muon is discussed.
%, where
%we have more than 3--$\sigma$ deviations from what
%we expect from the SM. 
The constraints from
the $b \rightarrow s \gamma$ decay are presented,
which is a loop induced process and thus
highly sensitive to possible new particles
emerging in the loops. 
Turning to cosmology we start with the very
strong constraint origination from the new WMAP result
on the relic abundances of the LSP, supposed this is the candidate
for the observed  cold dark matter. The direct dark
matter detection experiments are recalled briefly.
Eventually we discuss the prospects of the NMSSM in terms of
strong first order electroweak phase transitions in order to explain
the baryon--antibaryon asymmetry.
Finally, we consider some of the newer parameter scans, which provide
some interesting insight for rather large
ranges of parameter space.\\

In section \ref{sec-global} we inspect the
Higgs potential with respect to stationary solutions.
A method is introduced employing a Groebner basis approach
in the framework of {\em gauge invariant functions}.
This method allows to certainly
determine the global minimum and
reveals a quite surprising stationarity structure.
However, the Groebner basis approach is restricted to the tree-level potential.\\

We close with a rather extended appendix.
We mention some of the frequently used computer tools, mainly used
in parameter scans, followed by a derivation of the
essential new Feynman rules in the NMSSM compared to the MSSM.
Then we present the formalism of
gauge-invariant functions as well as the Buchberger
algorithm which allows to compute a Groebner basis, 
employed in the determination of the global minimum.\\

In advance we apologize for any missing references in
the bibliography. But in view of the vast amount of publications with
view on the NMSSM, in particular
in the recent years, this seems to be hardly avoidable. 

\newpage
%%%%%%%%%%%%%%%%%%%%%%%%%%%%%%%%%%%%%%%%%%%%%%%%%%%%%%%%%%%%%%%%%
%%%%%%%%%%%%%%%%%%%%%%%%%%%%%%%%%%%%%%%%%%%%%%%%%%%%%%%%%%%%%%%%%
% The NMSSM superpotential
%%%%%%%%%%%%%%%%%%%%%%%%%%%%%%%%%%%%%%%%%%%%%%%%%%%%%%%%%%%%%%%%%
%%%%%%%%%%%%%%%%%%%%%%%%%%%%%%%%%%%%%%%%%%%%%%%%%%%%%%%%%%%%%%%%%
\section{The NMSSM superpotential}
\label{sec-nmssm}
In the NMSSM the superpotential is
\begin{equation}
\label{eq-W}
W=
 \hat{{u}} y_u (\hat{Q}^\trans \epsilon \hat{H}_u)
-\hat{{d}} y_d (\hat{Q}^\trans \epsilon \hat{H}_d)
-\hat{{e}} y_e (\hat{L}^\trans \epsilon \hat{H}_d)
+ \lambda \hat{S}  (\hat{H}_u^\trans \epsilon \hat{H}_d)
+ \frac{1}{3} \kappa \hat{S}^3
\end{equation}
with dimensionless couplings $y_u$, $y_d$, $y_e$, $\lambda$ and $\kappa$. Note
that the superpotential has cubic mass dimension.
The supermultiplets are denoted with a hat and are
given in~Tab.~\ref{NMSSM-cont} together with its bosonic and fermionic particle content and
its multiplicity respectively hypercharge with respect to 
\mbox{$SU(3)_C \times SU(2)_L \times U(1)_Y$}. 
Compared to the MSSM particle content there
is only one new supermultiplet ingredient, namely the gauge singlet $\hat{S}$,
that is, one complex spin-0 singlet $S$ and one spin-1/2 singlino $\tilde{S}$.
The weak isospin indices
are not written explicitly and $\epsilon$ is defined below in \eqref{eq-eps}.
Note that the convention for the signs of the
$\lambda$ and $\kappa$ terms in the superpotential differ in the literature.

\begin{table}[t]
\begin{tabular}{lc|cc|ccc}
\hline
\multicolumn{2}{c|}{chiral supermultiplets} & 
spin-$0$ & spin-$1/2$ & $SU_C(3)$ & $SU_L(2)$ & $U_Y(1)$\\
\hline
quark--squark \phantom{\Large{Q}}& $\hat{Q}$ & $\tilde{Q}=(\tilde{u}_L ,\;\tilde{d}_L)^\trans$ & $Q=({u}_L ,\;{d}_L)^\trans$
& $\mathbf{3}$ & $\mathbf{2}$ & $\phantom{+}1/6$\\
& $\hat{u}$ & $\tilde{u}_R^*$ & $u_R^\dagger$ & $\mathbf{3}$ & $\mathbf{1}$ & $-2/3$\\
& $\hat{d}$ & $\tilde{d}_R^*$ & $d_R^\dagger$ & $\mathbf{3}$ & $\mathbf{1}$ & $\phantom{+}1/3$\\
\hline
lepton--slepton & $\hat{L}$ & $\tilde{L}=(\tilde{\nu}_e,\; \tilde{e}_L)^\trans$ & $L=({\nu_e},\;{e}_L)^\trans$
& $\mathbf{1}$ & $\mathbf{2}$ & $-1/2$\\
& $\hat{e}$ & $\tilde{e}_R^*$ & $e_R^\dagger$ & $\mathbf{1}$ & $\mathbf{1}$ & $\phantom{+}1$\\
\hline
Higgs--Higgsino & $\hat{H}_u$ & ${H}_u=(H^+_u,\; H^0_u)^\trans$ & 
$\tilde{H}_u=(\tilde{H}^+_u ,\; \tilde{H}^0_u)^\trans$
& $\mathbf{1}$ & $\mathbf{2}$ & $\phantom{+}1/2$\\
& $\hat{H}_d$ & ${H}_d=(H^0_d,\; H^-_d)^\trans$ & 
$\tilde{H}_d=(\tilde{H}^0_d ,\; \tilde{H}^-_d)^\trans$
& $\mathbf{1}$ & $\mathbf{2}$ & $-1/2$\\
 & $\hat{S}$ & $S$ & $\tilde{S}$ & $\mathbf{1}$ & $\mathbf{1}$ & $\phantom{+}0$\\
\hline
\hline
\multicolumn{2}{c|}{gauge supermultiplets} & 
spin-$1/2$ & spin-$1$ & $SU_C(3)$ & $SU_L(2)$ &
$U_Y(1)$\\
\hline
gluon--gluino& & $\tilde{g}$ & $g$ & $\mathbf{8}$ & $\mathbf{1}$ & $\phantom{+}0$\\
W-boson--wino& & $\tilde{W}^\pm, \tilde{W}^0$ & ${W}^\pm, {W}^0$ & 
$\mathbf{1}$ & $\mathbf{3}$ & $\phantom{+}0$\\
B-boson--bino& & $\tilde{B}^0$ & $B^0$ & $\mathbf{1}$ & $\mathbf{1}$ & $\phantom{+}0$\\
\hline
\end{tabular}
\caption{\label{NMSSM-cont} Particle content of the NMSSM in terms of
decomposed supermultiplets together with the corresponding
multiplicity respectively hypercharge with respect to
\mbox{$SU(3)_C \times SU(2)_L \times U(1)_Y$}. 
Compared to the particle content of the 
minimal supersymmetric extension the
new ingredient in the next-to-minimal extension is the Higgs supermultiplet $\hat{S}$.
Only the first family is given; the other families are introduced analogously. The 
here used convention for the hypercharge is $Q=I_W^3+Y_W$ with $Q$ the electromagnetic 
charge in units of the positron charge, $I_W^3$ the weak isospin 3-component, and $Y_W$
the hypercharge.}
\end{table}

%%%%%%%%%%%%%%%%%%%%%%%%%%%%%%%%%%%%%%%%%%%%%%%%%%%%%%%%%%%%%%%%%%5
%%%%%%%%%%%%%%%%%%%%%%%%%%%%%%%%%%%%%%%%%%%%%%%%%%%%%%%%%%%%%%%%%%5
% Discussion of the superpotential
%%%%%%%%%%%%%%%%%%%%%%%%%%%%%%%%%%%%%%%%%%%%%%%%%%%%%%%%%%%%%%%%%%5
%%%%%%%%%%%%%%%%%%%%%%%%%%%%%%%%%%%%%%%%%%%%%%%%%%%%%%%%%%%%%%%%%%5
\subsection{Discussion of the superpotential}

We start from the MSSM potential in terms of scalar fields,
\begin{equation}
\label{eq-Wscalar3}
W_{\text{MSSM}}=
 \tilde{u}_R^* y_u (\tilde{Q}^\trans \epsilon H_u)
-\tilde{d}_R^* y_d (\tilde{Q}^\trans \epsilon H_d)
-{\tilde{e}_R^*} y_e (\tilde{L}^\trans \epsilon H_d)
+ \mu  (H_u^\trans \epsilon H_d) \,.
\end{equation}
This is the minimal superpotential, where we have instead of the
dimensionless terms
$\lambda {S} \left( H_u^\trans \epsilon H_d \right) + \frac{1}{3} \kappa S^3$
a so-called $\mu$-term, with $\mu$ a dimensionful parameter. In any case 
a term proportional to $H_u^\trans \epsilon H_d$ is
required to give masses to up- and down-type fermions. This follows immediately
from the F-terms derived from the superpotential; as given in \eqref{eq-VF2}. 
Now, we know from experiment that the Higgs vacuum-expectation-value
is $v \approx 246$~GeV, that is, of the order of the electroweak scale. 
With view on the F- and soft-breaking terms in the MSSM potential
we get potential terms $\left(|\mu|^2 + m_{H_u}^2\right)  |H_u^0|^2 +
\left(|\mu|^2 + m_{H_d}^2\right)  |H_d^0|^2$. 
(This can be seen in the following discussion of the Higgs potential
and can be read off from~\eqref{eq-VF} and \eqref{eq-vsoft} 
if we replace $|\lambda|^2 |S|^2$ by $|\mu|^2$). 
With the vacuum-expectation-values $\langle H_d^0 \rangle=v_d/\sqrt{2}$, 
$\langle H_u^0 \rangle= v_u/\sqrt{2}$ and $v^2=v_u^2+v_d^2$ we see that
without fine-tuning between the mass- and $\mu$-parameters,
the parameter
$\mu$ can not be much larger than $v$. 
This is, that we have to adjust the $\mu$-parameter by
hand to the electroweak scale which in the SM arises from
spontaneous symmetry breaking. 
This is the so-called {\em $\mu$-problem} of the MSSM.

%%%%%%%%%%%%%%%%%%%%%%%%%%%%%%%%%%%%%%%%%%%%%%%%%%%%%%%%%%%%%%%%%%5
% Peccei--Quinn symmetry
%%%%%%%%%%%%%%%%%%%%%%%%%%%%%%%%%%%%%%%%%%%%%%%%%%%%%%%%%%%%%%%%%%5

\subsubsection{Peccei--Quinn symmetry}
\label{sub-PQ}
The main idea is to circumvent this dimensionful parameter $\mu$ by introducing
a new scalar field $S$ which gets a vacuum expectation value to generate this
parameter effectively, that is, the $\mu$-term is replaced by
$\lambda {S} \left( H_u \epsilon H_d \right)$ and the required $\mu$-term
at the electroweak scale arises from spontaneous symmetry breaking
with
$\mu=\lambda \langle S \rangle$. In this way we have achieved a 
quasi-universal mechanism of spontaneous symmetry breaking.
This is considered to be the main motivation 
of the NMSSM.
For some other solutions to circumvent the $\mu$-problem we refer the
reader to App.~\ref{sec-varNMSSM}.
Now, since the bilinear $\mu$-term in the superpotential is replaced by 
a trilinear term, we can see that the superpotential is invariant
under a global phase transformation, the so-called Peccei--Quinn 
(PQ) symmetry
$U(1)_{PQ}$~\footnote{Note that this symmetry was first 
discussed by P.~Fayet.}~\citep{Fayet:1974pd,Peccei:1977ur,Peccei:1977hh}. 
If we assign PQ charges
$q_{\text{PQ}}$ to the chiral supermultiplets, denoted by $\hat{\phi}_i$ with $i$ running
over all chiral supermultiplets, 
in the following way~\citep{Fayet:1974pd,Panagiotakopoulos:1998yw},
\begin{equation}
\begin{array}{l|c|c|c|c|c|c|c|c}
\text{supermultiplet } \hat{\phi}_i & 
\hat{Q} & \hat{u} & \hat{d} & \hat{L} & \hat{e} & \hat{H}_u & \hat{H_d} & \hat{S}\\
\hline
\text{PQ-charge } q_{\text{PQ}} & -1 & 0 & 0 & -1 & 0 & +1 & +1 & -2
\end{array}\,,
\end{equation}
we see immediately that the superpotential is invariant under 
the global $U(1)_{PQ}$ 
transformations 
\begin{equation}
\hat{\phi}_i \rightarrow  e^{i q_{\text{PQ}} \theta} \hat{\phi}_i\;,
\end{equation}
with an arbitrary phase $\theta$.
Note that since we assign PQ-charges to supermultiplets also the
derived Lagrangian stays invariant under $U(1)_{PQ}$. 
In the minimal supersymmetric extension, MSSM, the PQ symmetry
is {\em explicitely} broken by the $\mu$-term. In the NMSSM this
PQ-symmetry
is {\em spontaneously} broken once the Higgs bosons
acquire vacuum-expectation-values. This spontaneous broken continuous
symmetry gives, following the Goldstone theorem, 
inevitable a massless so-called Peccei--Quinn mode.
Such axion has not been found
and exclusion limits were derived~\citep{Hagiwara:2002fs}. 
Experimentally there remains only a rather small
corresponding $\lambda$-parameter occurring in the 
superpotential term 
$ \lambda \hat{S}  (\hat{H}_u^\trans \epsilon \hat{H}_d)$
in the range $10^{-7} < |\lambda| < 10^{-10}$.
In order to get an effective $\mu$ term of the electroweak order, the 
vacuum-expectation-value (VEV) of $S$ would have
to be much larger compared to the electroweak scale - again a fine-tuning would be required. 

In the NMSSM thus an additional term is imposed in the superpotential in oder to
violate the Peccei-Quinn symmetry by means of a cubic selfcoupling term,
$\frac{1}{3} \kappa \hat{S}^3$, with $\kappa$ the PQ-symmetry breaking
parameter. In this way all parameters of the superpotential are still
dimensionless. Alternatively, one could impose a linear or quadratic term in 
order to break the Peccei-Quinn symmetry but then again a dimensionful
parameter would be necessary. Note that the superpotential has to have
mass dimension three.
Of course any term higher than
trilinear in the fields in the superpotential is forbidden by the
requirement of renormalizability.
Finally we arrive at the proposed NMSSM superpotential, written in its scalar form
\begin{equation}
\label{eq-Wscalar2}
W=
 \tilde{u}_R^* y_u (\tilde{Q}^\trans \epsilon H_u)
-\tilde{d}_R^* y_d (\tilde{Q}^\trans \epsilon H_d)
-{\tilde{e}_R^*} y_e (\tilde{L}^\trans \epsilon H_d)
+ \lambda {S}  (H_u^\trans \epsilon H_d) 
+ \frac{1}{3} \kappa S^3\;.
\end{equation}

%%%%%%%%%%%%%%%%%%%%%%%%%%%%%%%%%%%%%%%%%%%%%%%%%%%%%%%%%%%%%%%%%%5
% Discrete $\Zthree$ symmetry
%%%%%%%%%%%%%%%%%%%%%%%%%%%%%%%%%%%%%%%%%%%%%%%%%%%%%%%%%%%%%%%%%%5
\subsubsection{Discrete $\Zthree$ symmetry}

There is some debate, whether the NMSSM superpotential~(\ref{eq-Wscalar2}) is
viable. The main concern arises since the continuous Peccei-Quinn symmetry
is broken explicitly by the $\frac{1}{3} \kappa S^3$ term, but 
there remains a discrete $\Zthree$ symmetry of the superpotential~\citep{Zeldovich:1974uw}:
\begin{equation}
\hat{\phi}_i \rightarrow e^{i \frac{2 \pi}{3}} \hat{\phi_i}\,,
\end{equation}
where the index $i$ runs over all superfields occurring in the superpotential.
This discrete symmetry is spontaneously broken by the VEV of the complex scalar field~$S$.
Inevitably, this spontaneous breaking of a discrete symmetry leads to the {\em domain-wall} problem,
a special kind of a {\em topological defect}.
During electroweak phase transition of the early Universe, this spontaneously broken
discrete symmetry would cause a dramatic change of the evolution of the Universe and
would spoil the observed cosmic microwave background radiation. 

A study of the generation of domain walls in the NMSSM potential with numerical methods
can be found in~\citep{Abel:1995uc}.
Following~\citep{Zeldovich:1974uw,Vilenkin:1986hg} we want to sketch 
the mechanism of the formation of domain walls in a 
model with spontaneous breaking of a discrete symmetry
in the simple Goldstone model of a complex scalar field $\phi$ with Lagrangian
\begin{equation}
{\cal L} = \frac{1}{2} (\partial_\mu \phi^\dagger)(\partial^\mu \phi) - \frac{\lambda}{4} \left( \phi^2-\eta^2\right)^2,
\end{equation}
where $\eta$ is a constant.
The potential of this Lagrangian is the double well potential. The minima of the potential 
are at
\begin{equation}
\phi_\pm=\pm \eta,
\end{equation}
that is, we have in this case a discrete $\mathbbm{Z}_2$ symmetry of the Lagrangian which is spontaneously
broken by the vacuum.
Now, the domain wall is nothing but the intermediate region between these degenerate
minima $\phi_\pm$. Let us assume, without loss of generality, that this domain wall is situated in the $y$-$z$-direction.
From the Goldstone-model Lagrangian we determine the 
Euler--Lagrange equation~$\phi + \lambda \phi \left(\phi^2- \eta^2\right)=0$.
The solution of this equation is
\begin{equation}
\label{eq-solphi}
\phi = \eta \tanh ( \frac{x}{\Delta} ),
\end{equation}
with {\em thickness} of the wall $\Delta \equiv (\lambda/2)^{-1/2} \eta^{-1}$.
This solution describes the field in the intermediate domain-wall region between the two vacua. We
see that $\phi \rightarrow \pm\eta$ corresponds to $x\rightarrow \pm \infty$. 

In the transition region there is some additional energy proportional to the area of the wall.
The stress-energy tensor is defined as 
\begin{equation}
T_{\mu \nu} =(\partial_\mu \phi^\dagger)(\partial_\nu \phi^\dagger)
- g_{\mu \nu} {\cal L}
\end{equation}
 and inserting the solution~(\ref{eq-solphi}) we get
 \begin{equation}
(T_{\mu \nu}) = \frac{1}{2} \lambda \eta^4 \cosh^{-4} (\frac{x}{\Delta})
\diag{(1,0,-1,-1)}.
\end{equation}
Energy and stress of the domain wall are given by 
\begin{equation}
\begin{split}
\rho_W &= \langle T_{00} \rangle = \frac{1}{2} \lambda \eta^4 \cosh^{-4} (\frac{x}{\Delta})\,,\\
p_W &= \langle T_{ii} \rangle = - \frac{1}{3} \lambda \eta^4 \cosh^{-4} (\frac{x}{\Delta}).
\end{split}
\end{equation}
That is, the equation of state is
\begin{equation}
p_W = -\frac{2}{3} \rho_W
\end{equation}
The surface energy density of the domain wall follows as
\begin{equation}
\sigma= \int\limits_{-\infty}^{+\infty} T_{0 0} dx = \frac{2}{3} \sqrt{2 \lambda}\; \eta^3.
\end{equation} 
It is clear that the generation of domain walls with energy density $\sigma$ originates from the spontaneously broken
discrete symmetry.

Now we consider a macroscopic volume containing a large number of randomly oriented wall surfaces.
For simplicity we consider a flat Universe with Robertson-Walker metric
$ds^2=dt^2 - R(t)^2 (dx^2+dy^2+dz^2)$ with scale factor $R(t)$.
From the Einstein equation we can derive
\begin{equation}
\left( \frac{\dot{R}}{R} \right)^2 = \frac{8 \pi G}{3} \rho_W\;,
\end{equation}
with $G$ the gravitational constant.
Integrating this equation we arrive at
\begin{equation}
R(t) \sim t^2,
\qquad
\frac{E}{V} = \frac{3}{2} \pi G\; t^2.
\end{equation}
This means that the energy density of the walls will quickly overpower the radiation contribution, causing
a period of power-law inflation. An expansion period of this form would leave less time for galaxy formation.
Moreover the production rates during nucleosynthesis would change. In case this additional energy density
is observable in the present Universe, 
it would cause distortions in the cosmic microwave background, violating the limits
on homogeneity and isotropy. Quantitative studies show
that only surface energy densities
of a few MeV are allowed~\citep{Vilenkin:1984ib}.\\

If we trust this cosmological argument we see that the discrete $\Zthree$ symmetry 
we encounter in the NMSSM
must be broken explicitly
in order to avoid the formation of domain walls.
One idea is to introduce additional higher-order operators, that is, non-renormalizable $\Zthree$ breaking terms in 
the superpotential. These terms
are adjusted to be Planck-mass suppressed such that they do not have any effect at the low energy scale and 
renormalizability is of no concern. 
Nevertheless it can be shown that these additional terms generate
quadratically divergent tadpoles for the singlet~\citep{Abel:1995wk,Abel:1996cr}. 
The generic contribution of higher-order operators
to the effective potential with cutoff at the Planck-mass scale $M_P$
reads~\citep{Panagiotakopoulos:1998yw}
\begin{equation}
\label{eq-tad1}
\delta V =\xi M_P m_s^2 S + c.c.
\end{equation}
with soft supersymmetry breaking mass parameter $m_s$ and $\xi$ 
depending on the loop order at which the tadpole is generated.
This contribution to the potential changes the vacuum-expectation-value 
of the singlet
to values far above the electroweak scale, that is, destabilizes the
gauge hierarchy.

However one may impose new discrete symmetries in a way to forbid 
or at least loop suppress the dangerous tadpole contributions which 
arise from the additional Planck-suppressed terms in the
superpotential~\citep{Panagiotakopoulos:1998yw,Panagiotakopoulos:1999ah,Panagiotakopoulos:2000wp,Dedes:2000jp}. Imposing a $\mathbbm{Z}_2$ R-symmetry on
the non-renormalizable operators under which all superfields
flip sign, the dangerous operators are forbidden and there remains
only a harmless tadpole contribution to the potential,
\begin{equation}
\label{eq-tad2}
\delta V =\xi m_s^3 S + c.c.
\end{equation}
This term breaks the $\Zthree$ symmetry and thus
avoids the formation of domain walls. 
In the following we assume that the domain-wall problem
is circumvented in this way without any modifications except far beyond the
electroweak scale.

\newpage
%%%%%%%%%%%%%%%%%%%%%%%%%%%%%%%%%%%%%%%%%%%%%%%%%%%%%%%%%%%%%%%%%%5
%%%%%%%%%%%%%%%%%%%%%%%%%%%%%%%%%%%%%%%%%%%%%%%%%%%%%%%%%%%%%%%%%%5
% The Higgs-boson sector of the NMSSM
% revised 04.05.09
%%%%%%%%%%%%%%%%%%%%%%%%%%%%%%%%%%%%%%%%%%%%%%%%%%%%%%%%%%%%%%%%%%5
%%%%%%%%%%%%%%%%%%%%%%%%%%%%%%%%%%%%%%%%%%%%%%%%%%%%%%%%%%%%%%%%%%5
\section{The Higgs-boson sector of the NMSSM}
\label{sub-Higgs}

In this section we investigate the Higgs-boson sector of the NMSSM, starting
with the Higgs potential, electroweak symmetry breaking, followed by the
derivation of the Higgs-mass matrices and eventually discuss studies on 
Higgs-boson phenomenology. For studies of the Higgs potential, in particular
with respect to radiative corrections let us refer 
to~\citep{Pandita:1993tg,Miller:2003ay,Funakubo:2004ka}.

%%%%%%%%%%%%%%%%%%%%%%%%%%%%%%%%%%%%%%%%%%%%%%%%%%%%%%%%%%%%%%%%%%5
%%%%%%%%%%%%%%%%%%%%%%%%%%%%%%%%%%%%%%%%%%%%%%%%%%%%%%%%%%%%%%%%%%5
% The Higgs potential
%%%%%%%%%%%%%%%%%%%%%%%%%%%%%%%%%%%%%%%%%%%%%%%%%%%%%%%%%%%%%%%%%
%%%%%%%%%%%%%%%%%%%%%%%%%%%%%%%%%%%%%%%%%%%%%%%%%%%%%%%%%%%%%%%%%%5
\subsection{The Higgs potential}
\label{sec-Hpot}

The superpotential \eqref{eq-W} is a holomorphic function which 
determines all non-gauge interactions
of the chiral supermultiplets as is discussed in more detail in App.~\ref{app-A}.
Here we want to focus on the Higgs sector and first derive the
physical Higgs potential $V$ and in a second step
derive the Higgs mass matrices.
With view on Tab.~\ref{NMSSM-cont} we have the scalar fields in the
Higgs-boson sector
\begin{equation}
\label{eq-higgses}
H_u=
\begin{pmatrix}
H_u^+ \\ H_u^0
\end{pmatrix},
\qquad
H_d=
\begin{pmatrix}
H_d^0 \\ H_d^-
\end{pmatrix},
\qquad
S \; .
\end{equation}
The Higgs-boson potential gets contributions from
the chiral supermultiplets, the so-called F-terms encoded in the superpotential,
from so-called D-terms of the gauge multiplets as well
as from the soft supersymmetry breaking terms (see App.~\ref{app-A}).\\

We start with the scalar part of the superpotential, which reads
\begin{equation}
\label{eq-Wscalar}
W=
 \tilde{u}_R^* y_u (\tilde{Q}^\trans \epsilon H_u)
-\tilde{d}_R^* y_d (\tilde{Q}^\trans \epsilon H_d)
-{\tilde{e}_R^*} y_e (\tilde{L}^\trans \epsilon H_d)
+ \lambda {S}  (H_u^\trans \epsilon H_d) 
+ \frac{1}{3} \kappa S^3\,,
\end{equation}
where we just replaced each supermultiplet (denoted by a hat) in~(\ref{eq-W})
by its scalar component according to Tab.~\ref{NMSSM-cont}.
Here the term $(H_u^\trans \epsilon H_d)$ written
with the isoweak indices $\alpha, \beta=1,2$ reads 
$(H_u)_\alpha \epsilon^{\alpha \beta} (H_d)_\beta  = H_u^+ H_d^-  - H_u^0 H_d^0$,
where we defined
\begin{equation}
\label{eq-eps}
\epsilon =
\begin{pmatrix}
\phantom{+}0 & 1\\
-1 & 0
\end{pmatrix}.
\end{equation}

The so-called F-terms of the Higgs potential 
are derived from the superpotential in
the following way:
\begin{equation}
\label{eq-VF}
V_F= \sum_i \left| \frac{\delta W}{\delta \phi_i} \right|^2 = 
|\lambda|^2 |S|^2 \left( H_u^\dagger H_u + H_d^\dagger H_d \right) 
+ \left| \lambda (H_u^\trans \epsilon H_d)  + \kappa S^2\right|^2\,,
\end{equation}
where 
\begin{equation}
\phi=(H_u, H_d, S)
\end{equation}
denotes the tuple of
the three Higgs-boson fields.\\

The NMSSM Higgs singlet $S$ is supposed only to
couple through the superpotential $W$ to
the other Higgs-bosons and has no gauge couplings. This
restricts the D-terms to be exactly the same as in
the MSSM:
\begin{equation}
V_D=
%-\frac{1}{2} D^a D^a = 
\frac{1}{2} \sum\limits_{i,j} g_a^2 
( \phi_i^\dagger \mathbf{T}^a \phi_i)( \phi_j^\dagger \mathbf{T}^a \phi_j) =
\frac{1}{2} g_2^2 |H_u^\dagger H_d|^2 +\frac{1}{8} (g_1^2+g_2^2) \left( H_u^\dagger H_u - H_d^\dagger H_d \right)^2
\end{equation}
Here, $\mathbf{T}^a$ are the generators of the gauge groups and~$g_a$ the 
corresponding gauge couplings, that is
for the group~$U(1)_Y$ we have $g_a \mathbf{T}^a= g_1 \mathbf{Y}_W$ with 
$g_1$ the corresponding coupling and the 
hypercharge operator $\mathbf{Y}_W$ as well as 
for the group $SU(2)_L$ we have $g_a \mathbf{T}^a=g_2 \mathbf{I}^a_W= g_2 \sigma^a/2$, 
$a=1,2,3$ with $g_2$ the corresponding gauge coupling.
Note that we used the identity
$\sigma^a_{ij} \sigma^a_{kl} = 2 \delta_{il} \delta_{jk} - \delta_{ij} \delta_{kl}$.\\

Since there is no knowledge of the realization of the supersymmetry-breaking
mechanism (for a discussion of soft supersymmetry breaking scenarios we refer
to the general literature of supersymmetry and the MSSM), simply all possible additional terms are imposed which have couplings of mass dimension one or higher.
Only terms which violate {\em matter parity} are discarded; we briefly
recall the terminus of matter parity in App.~\ref{app-A}.

In the Higgs-boson sector the soft-breaking terms
corresponding to the superpotential $W$ are
\begin{equation}
\label{eq-vsoft}
\begin{split}
V_{\text{soft}} &= (m^2)^i_j \phi^{j\; \ast} \phi_i 
+\left( \frac{1}{2} b^{ij} \mu^{ij} \phi_i \phi_j + \frac{1}{6} a^{ijk} \lambda^{ijk} \phi_i \phi_j \phi_k + c.c.\right)\\
&= m_{H_u}^2 H_u^\dagger H_u + m_{H_d}^2 H_d^\dagger H_d + m_S^2 |S|^2
+\left( \lambda A_{\lambda} (H_u^\trans \epsilon H_d)  S
+\frac{1}{3} \kappa A_{\kappa} S^3 + c.c. \right)\; .
\end{split}
\end{equation}
The first line is the generic expression, where we get for the scalar fields
corresponding mass terms, a general $\mu$-term generates corresponding $b$-terms, and
a trilinear term in $W$ generates corresponding trilinear $A$-parameter terms.
Note that there is implicit summation over the indices $i,j$ respectively $i,j,k$.
The terms proportional to $b^{ij}$ in the first line of~\eqref{eq-vsoft} are absent in 
the NMSSM since we have no corresponding $\mu$-terms in the superpotential \eqref{eq-Wscalar}. 
However, we get the trilinear soft supersymmetry breaking $A$-parameter
terms, $A_\lambda$ and $A_\kappa$ in the NMSSM. 
The couplings $\lambda A_{\lambda}$ and $\kappa A_{\kappa}$ may be chosen to be real and positive since
any complex phase may be absorbed by a global redefinition of the scalar Higgs fields $\phi$. 
Since $\lambda$ and $\kappa$ are in general complex this means that also
$A_{\lambda}$ and $A_{\kappa}$ are complex in general.\\

Eventually we arrive at the scalar Higgs potential of the NMSSM
\begin{equation}
\begin{split}
\label{eq-V}
V =& V_F + V_D + V_{\text{soft}}\\
=& \phantom{+}
|\lambda|^2 |S|^2 \left( H_u^\dagger H_u + H_d^\dagger H_d \right) 
+ \left| \lambda (H_u^\trans \epsilon H_d) + \kappa S^2\right|^2 \\
& +
\frac{1}{2} g_2^2 |H_u^\dagger H_d|^2 +\frac{1}{8} (g_1^2+g_2^2) \left( H_u^\dagger H_u - H_d^\dagger H_d \right)^2\\
& +
m_{H_u}^2 H_u^\dagger H_u + m_{H_d}^2 H_d^\dagger H_d + m_S^2 |S|^2
+\left( \lambda A_{\lambda} (H_u^\trans \epsilon H_d) S
+\frac{1}{3} \kappa A_{\kappa} S^3 + c.c. \right)\; .
\end{split}
\end{equation}

%%%%%%%%%%%%%%%%%%%%%%%%%%%%%%%%%%%%%%%%%%%%%%%%%%%%%%%%%%
% Tadpole conditions 
%%%%%%%%%%%%%%%%%%%%%%%%%%%%%%%%%%%%%%%%%%%%%%%%%%%%%%%%%%
\subsection{Tadpole conditions}
\label{sec-tadpole}

We can always parameterize the complex entries of the fields $H_u$, $H_d$, and S in the
the form 
\begin{equation}
\label{eq-higgsespara}
H_d=
\begin{pmatrix}
\frac{1}{\sqrt{2}} \left( v_d+h_d +i a_d\right) \\ H_d^-
\end{pmatrix},
\qquad
H_u=
e^{i \phi_u}
\begin{pmatrix}
H_u^+ \\ \frac{1}{\sqrt{2}} \left( v_u+h_u +i a_u\right)
\end{pmatrix},
\qquad
S = \frac{1}{\sqrt{2}} e^{i \phi_s} \left( v_s+h_s +i a_s\right)\,,
\end{equation}
where we can choose the {\em vacuum-expectation values} $v_d$, $v_u$
and $v_s$ to be real and non-negative (any complex phase of $v_d$ can
be rotated away by gauge transformations and obviously the phases
$\phi_u$ and $\phi_s$ can be chosen to make $v_u$ and $v_s$ real and non-negative).
Note the convention with a factor $1/\sqrt{2}$ in front of the
vacuum-expectation-values. In the literature this is not handled consistently.
Of course the values $v_d$, $v_u$ and $v_s$ will take
the values for which the potential $V$ in~(\ref{eq-V}) has a global minimum, 
justifying the terminus vacuum-expectation values. 
Note that in the parameterization~(\ref{eq-higgsespara})
the vacuum denoted by $\langle \rangle$ precisely means 
\begin{equation}
\label{higgsvev}
\langle H_d \rangle=
\begin{pmatrix}
v_d/\sqrt{2} \\ 0
\end{pmatrix},
\qquad
\langle H_u \rangle=
e^{i \phi_u}
\begin{pmatrix}
0 \\ v_u/\sqrt{2} 
\end{pmatrix},
\qquad
\langle S \rangle= e^{i \phi_s} v_s/\sqrt{2} \; .
\end{equation}

The stationarity condition for the scalar Higgs potential $V$  
at the vacuum translates immediately 
into the following system of equations, so-called {\em tadpole conditions};
see for instance~\citep{Funakubo:2004ka}.
\begin{equation}
\label{eq-tadpole}
\begin{split}
\left\langle \frac{\partial{V}}{\partial h_d} \right\rangle&= v_d m_{H_d}^2 - R_\lambda v_u v_s 
+ \frac{g_1^2+g_2^2}{8} v_d (v_d^2- v_u^2)
+ \frac{|\lambda|^2}{2} v_d(v_u^2+v_s^2) - \frac{R}{2} v_u v_s^2 =0\;,\\
\left\langle \frac{\partial{V}}{\partial h_u} \right\rangle&= v_u m_{H_u}^2 - R_\lambda v_d v_s 
+ \frac{g_1^2+g_2^2}{8} v_u (v_d^2- v_u^2)
+ \frac{|\lambda|^2}{2} v_u(v_d^2+v_s^2) - \frac{R}{2} v_d v_s^2 =0\;,\\
\left\langle \frac{\partial{V}}{\partial h_s} \right\rangle&= v_s m_S^2 - R_\lambda v_d v_u 
+ R_{\kappa} v_s^2
+ \frac{|\lambda|^2}{2} v_s(v_d^2+v_u^2) + |\kappa|^2 v_s^3 - R v_d v_u v_s =0\;,\\
\left\langle \frac{\partial{V}}{\partial a_d} \right\rangle&=  
I_\lambda v_u v_s + \frac{I}{2} v_u v_s^2=0\;,\\
\left\langle \frac{\partial{V}}{\partial a_u} \right\rangle&=  
I_\lambda v_d v_s + \frac{I}{2} v_d v_s^2=0\;,\\
\left\langle \frac{\partial{V}}{\partial a_s} \right\rangle&=  
I_\lambda v_d v_u - I_\kappa v_s^2- I v_d v_u v_s=0\; ,
\end{split}
\end{equation}
where the following abbreviations are introduced
\begin{equation}
\begin{alignedat}{2}
\label{eq-phases}
&R= \re \left( \lambda \kappa^* e^{i(\phi_u-2\phi_s)} \right),
&\qquad
&I= \im \left( \lambda \kappa^* e^{i(\phi_u-2\phi_s)} \right),\\
&R_\lambda= \frac{1}{\sqrt{2}} \re \left( \lambda A_\lambda e^{i(\phi_u+\phi_s)} \right),&
&I_\lambda= \frac{1}{\sqrt{2}} \im \left( \lambda A_\lambda e^{i(\phi_u+\phi_s)} \right),\\
&R_\kappa= \frac{1}{\sqrt{2}} \re \left( \kappa A_\kappa e^{i 3 \phi_s} \right),&
&I_\kappa= \frac{1}{\sqrt{2}} \im \left( \kappa A_\kappa e^{i 3 \phi_s} \right).
\end{alignedat}
\end{equation}
Note that only these combinations of phases occur in the tadpole conditions.
Obviously the last three conditions of~(\ref{eq-tadpole}) can be written in the form
\begin{equation}
I_\lambda= -\frac{I}{2} v_s,
\qquad
I_\kappa=-\frac{3}{2} I \frac{v_d v_u}{v_s}\;
\end{equation}
and only one of the three imaginary parts $I$, $I_\lambda$, $I_\kappa$ is not fixed by the others through the vacuum conditions.\\

The MSSM may be obtained by taking real parameters $\lambda$, $\kappa$
in the limit $\lambda, \kappa \rightarrow 0$ with
the ratio $\kappa/\lambda$ kept constant and the product $\lambda v_s$ as well as the parameters $A_\lambda$
and $A_\kappa$ fixed. In particular this means $v_s \rightarrow \infty$ in this limit.  
This ensures that the singlet $S$ decouples completely and we end
up with the MSSM superpotential.

%%%%%%%%%%%%%%%%%%%%%%%%%%%%%%%%%%%%%%%%%%%%%%%%%%%%%%%%%%%%%%%%%%5
% Higgs-boson mass matrices in the NMSSM
%%%%%%%%%%%%%%%%%%%%%%%%%%%%%%%%%%%%%%%%%%%%%%%%%%%%%%%%%%%%%%%%%%5
\subsection{Higgs-boson mass matrices in the NMSSM}
\label{sub-massmatrices}

The mass squared matrix of the neutral Higgs scalars is obtained
by the second derivative of the Higgs potential with respect 
to the fields at the vacuum. The mass squared part of
the corresponding Lagrangian is
\begin{equation}
\label{eq-lagmass}
{\cal L}_{\text{neutr.}}^\text{Higgs mass} = 
-\frac{1}{2}
( \tvec{h}^\trans, \tvec{a}^\trans )
\begin{pmatrix}
M_S & M_{SP}\\
(M_{SP})^\trans & M_P
\end{pmatrix}
\begin{pmatrix}
\tvec{h}\\
\tvec{a}
\end{pmatrix}
\equiv
-\frac{1}{2}
( \tvec{h}^\trans, \tvec{a}^\trans )
M
\begin{pmatrix}
\tvec{h}\\
\tvec{a}
\end{pmatrix},
\end{equation}
in the basis $\tvec{h}^\trans=(h_d, h_u, h_s)$ and 
$\tvec{a}^\trans=(a_d, a_u, a_s)$.
The square block-matrices are explicitly
%checked 27.05.09
\begin{equation}
M_S=
\begin{pmatrix}
m_Z^2 \cb^2 + (\Rl + R v_s/2) v_s v_u/v_d &
|\lambda|^2 v_d v_u -m_Z^2 \sb \cb - \Rl v_s - R v_s^2/2 &
|\lambda|^2 v_d v_s -\Rl v_u - R v_u v_s \\
|\lambda|^2 v_d v_u -m_Z^2 \sb \cb - \Rl v_s - R v_s^2/2 &
m_Z^2 \sb^2 + (\Rl + R v_s/2) v_s v_d/v_u &
|\lambda|^2 v_u v_s -\Rl v_d - R v_d v_s \\
|\lambda|^2 v_d v_s -\Rl v_u - R v_u v_s &
|\lambda|^2 v_u v_s -\Rl v_d - R v_d v_s &
2 |\kappa|^2 v_s^2 +\Rl v_d v_u/v_s + \Rk v_s
\end{pmatrix},
\end{equation}
%
%checked 27.05.09
\begin{equation}
M_P=
\begin{pmatrix}
(\Rl+R v_s/2)v_s v_u/v_d &
(\Rl+R v_s/2)v_s &
(\Rl-R v_s)v_u \\
(\Rl+R v_s/2)v_s &
(\Rl+R v_s/2)v_s v_d/v_u &
(\Rl-R v_s)v_d \\
(\Rl-R v_s)v_u &
(\Rl-R v_s)v_d &
\Rl v_d v_u/v_s - 3 \Rk v_s + 2 R v_d v_u
\end{pmatrix},
\end{equation}
\begin{equation}
M_{SP}= I \cdot
\begin{pmatrix}
0 & 0 & - \frac{3}{2} v_u v_s\\
0 & 0 & - \frac{3}{2} v_d v_s\\
\frac{1}{2} v_u v_s &
\frac{1}{2} v_d v_s &
2 v_d v_u
\end{pmatrix}.
\end{equation}
Here the mass parameters $m_d^2$, $m_u^2$, $m_s^2$ as well
as $I_\lambda$ and $I_\kappa$ are replaced employing
the tadpole conditions~\eqref{eq-tadpole}.
From the mass squared matrix we see that in
general the neutral Higgs bosons mix. In the case
$I=0$ the mass squared matrix becomes block-diagonal and
the scalar fields $(h_d, h_u, h_s)$ do not mix with
the pseudo-scalar fields $(a_d, a_u, a_s)$, that
is there is no CP violation in this case in the Higgs-boson sector. 
Only in this case
the neutral Higgs bosons
can be assigned corresponding CP properties.

We may isolate the massless Goldstone field $G$ by
a change of basis with the $\beta$ rotation
($c_\beta \equiv \cos(\beta)$, $s_\beta \equiv \sin(\beta)$)
\begin{equation}
\label{eq-brot}
\begin{pmatrix}
a_d\\
a_u\\
a_s
\end{pmatrix}
=
\begin{pmatrix}
\cb & \sb & 0\\
-\sb & \cb & 0\\
0 & 0 & 1
\end{pmatrix}
\begin{pmatrix}
G\\
a\\
a_s
\end{pmatrix}
=
R(\beta) \; \tvec{a}'.
\end{equation}
With this rotation, the neutral scalar mass part of
the Lagrangian~(\ref{eq-lagmass}) reads
\begin{equation}
\label{eq-lagmassbeta}
{\cal L}_{\text{neutr.}}^\text{Higgs mass} = 
-\frac{1}{2}
( \tvec{h}^\trans, \tvec{a}'^\trans )
\begin{pmatrix}
M_S & M_{SP}'\\
(M_{SP}')^\trans & M_P'
\end{pmatrix}
\begin{pmatrix}
\tvec{h}\\
\tvec{a}'
\end{pmatrix},
\end{equation}
with changed matrices
\begin{equation}
\label{eq-Mp}
M_P'= R(\beta)^\trans M_P R(\beta) =
\begin{pmatrix}
0 & 0 & 0 \\
0 &
(2 \Rl+R v_s)v_s/ \sin (2 \beta)&
(\Rl-R v_s) v \\
0 &
(\Rl-R v_s) v &
\Rl v^2/v_s\cdot \sin (2 \beta)/2- 3 \Rk v_s + R v^2 \sin (2 \beta)
\end{pmatrix}
\end{equation}
and
\begin{equation}
\label{eq-mspnew}
M_{SP}'= M_{SP} R(\beta) =
\frac{I \; v}{2} \cdot
\begin{pmatrix}
0 & 0 & - 3 v_s \sb\\
0 & 0 & - 3 v_s \cb\\
0 & v_s & 4 v \sb \cb &
\end{pmatrix}.
\end{equation}
Here we introduced the convenient and usual abbreviations $\tb\equiv \tan(\beta)=v_u/v_d$ and
$v^2 \equiv v_d^2+v_u^2$. Since the massless Goldstone boson is now separated
(giving the longitudinal mode of the Z-boson) 
we can suppress the corresponding vanishing forth row and column
in the mass squared matrix in (\ref{eq-lagmassbeta}). We end up with five
physical neutral Higgs-boson fields.
The mass eigenstates of these five physical fields are obtained
by another orthogonal rotation $R$ in the basis $\phi=(h_d, h_u, h_s, a, a_s)^\trans$
yielding
\begin{equation}
\label{eq-Hrot}
H_i= R_{ij}\; \phi_j\;,
\qquad \text{ with } 
\diag(m_{H_1}^2, m_{H_2}^2,m_{H_3}^2,m_{H_4}^2,m_{H_5}^2)
= R M' R^\trans,
\end{equation}
that is, by diagonalizing the mass squared matrix 
\begin{equation}
\label{eq-finalmassmatrix}
M'=
\begin{pmatrix}
M_S & M_{SP}'\\
(M_{SP}')^\trans & M_P'
\end{pmatrix}\;.
\end{equation}
In the case of
no mixing of the scalar with the pseudoscalar Higgs bosons, the matrix
$M'$ is block diagonal and has non-vanishing entries only in the 
upper left $3 \times 3$ block and the lower right $2 \times 2$ block, which may 
in this case be diagonalized separately. Then we have three scalar mass
eigenstates, denoted by $H_1$, $H_2$, $H_3$ as well as two pseudoscalar
mass eigenstates, called $A_1$ and $A_2$. Without loss of generality
the Higgs bosons are finally put in ascending order, that is, $m_{H_1} \le m_{H_2} \le ...\le m_{H_5}$
or in case of a CP-conserving Higgs-boson sector $m_{H_1} \le m_{H_2} \le m_{H_3}$ and
$m_{A_1} \le m_{A_2}$.\\

Note that in case of $\kappa=0$ we get from~\eqref{eq-phases} $R=I=0$ as well as
$R_\kappa=I_\kappa=0$. In this case the mass squared matrix~\eqref{eq-finalmassmatrix}
becomes block diagonal, since $M_{SP}'$ in~\eqref{eq-mspnew} vanishes.
Moreover, in this CP conserving case the pseudoscalar mass squared matrix~\eqref{eq-Mp}
has an additional vanishing eigenvalue, giving the so-called {\em axion}.
This makes clear that a vanishing parameter $\kappa$ is
accompanied by an additional massless pseudoscalar state.\\

The mass squared matrix of the charged fields is
defined quite analogously from the 
Higgs potential with respect to the complex (charged) fields,
that is, in the basis $\left( (H_d^-)^*, H_u^+ \right)^\trans$ the mass
matrix reads
\begin{equation}
M_\pm= 
(R v_s^2/2 + R_\lambda v_s + s_\beta c_\beta m_W^2 - \frac{1}{2} v_u v_d |\lambda|^2)
\cdot
\begin{pmatrix}
\tan(\beta) & 1\\
1 & \cot (\beta)
\end{pmatrix}\;.
\end{equation}
Again, the $\beta$ rotation allows us to separate the Goldstone
mode
\begin{equation}
\label{eq-chb}
\begin{pmatrix}
(H_d^-)^*\\
H_u^+\\
\end{pmatrix}
=
\begin{pmatrix}
\cb &  \sb\\
-\sb & \phantom{+}\cb
\end{pmatrix}
\begin{pmatrix}
G^+\\
H^+
\end{pmatrix}
\end{equation}
with $G^- \equiv (G^+)^*$ and $H^- \equiv (H^+)^*$. 
From the diagonalization $R(\beta)^\trans M_\pm R(\beta)$ we can read the mass squares of
the charged Higgs bosons,
that is, we have two charged Goldstone modes $G^\pm$ as well as
the charged Higgs bosons with mass
\begin{equation}
\label{eq-repRl}
m_{H^\pm}^2= \trace (M_\pm) = m_W^2 -|\lambda|^2 v^2/2 +(2 \Rl + R\; v_s) v_s/\sin( 2\beta). 
\end{equation}
Note, that by means of this equation we may express the phase parameter
$\Rl$ in terms of the charged Higgs boson mass.
For instance, the pseudoscalar mass squared matrix 
\eqref{eq-Mp} reads, using \eqref{eq-repRl} 
\begin{equation}
\label{eq-MPMA}
M_P'=
\begin{pmatrix}
m_P^2 & 
\frac{1}{2} m_P^2 \frac{v}{v_s} \sin (2 \beta) - \frac{3}{2} R v v_s\\
\frac{1}{2} m_P^2 \frac{v}{v_s} \sin (2 \beta) - \frac{3}{2} R v v_s &
\frac{1}{4} m_P^2 \frac{v^2}{v_s^2} \sin^2 (2 \beta) + \frac{3}{4} R v^2 \sin (2\beta)-3R_\kappa v_s
\end{pmatrix}
\end{equation}
with the abbreviation for the upper left entry
\begin{equation}
\label{eq-mAparameter}
m_P^2 \equiv m_{H^\pm}^2 - m_W^2+ |\lambda|^2 v_s^2/2\,.
\end{equation}
In the MSSM limit of the model, we have no mixing with a singlet Higgs boson state $a_s$ and
the upper left mass squared entry becomes the single pseudoscalar mass squared 
$m_P^2=(m_A^{\text{MSSM}})^2$.  
But in general even in the CP-conserving Higgs case, the two pseudoscalar mass eigenstates in the
NMSSM, $m_{A_1}$ and $m_{A_2}$,
originate from the mixing in \eqref{eq-MPMA}.\\

The only modifications of the NMSSM compared
to the MSSM arise from the extended superpotential, that
is the generalized $\mu$-term, imposing a gauge singlet.
Nevertheless, the mixing of the corresponding
additional states with other states
may generate a coupling of the singlet
to the gauge bosons.
In table~\ref{NMSSM-mix} all new particles predicted by the NMSSM
are shown. The second column lists the gauge eigenstates whereas
the right column gives the corresponding mass eigenstates.

\begin{table}[t]
\begin{tabular}{l|l|l}
\hline
bosons & gauge eigenstates & mass eigenstates\\
\hline

         & $\tilde{e}_L$, $\tilde{e}_R$, $\tilde{\nu}_e$ &
$\tilde{e}_L$, $\tilde{e}_R$, $\tilde{\nu}_e$
\\
sleptons & $\tilde{\mu}_L$, $\tilde{\mu}_R$, $\tilde{\nu}_\mu$ &
$\tilde{\mu}_L$, $\tilde{\mu}_R$, $\tilde{\nu}_\mu$
\\
	& $\tilde{\tau}_L$, $\tilde{\tau}_R$, $\tilde{\nu}_\tau$ &
 $\tilde{\tau}_1$, $\tilde{\tau}_2$, $\tilde{\nu}_\tau$\\
& & \\
         & $\tilde{u}_L$, $\tilde{u}_R$, $\tilde{d}_L$, $\tilde{d}_R$ &
$\tilde{u}_L$, $\tilde{u}_R$, $\tilde{d}_L$, $\tilde{d}_R$\\
squarks & $\tilde{c}_L$, $\tilde{c}_R$, $\tilde{s}_L$, $\tilde{s}_R$ &
$\tilde{c}_L$, $\tilde{c}_R$, $\tilde{s}_L$, $\tilde{s}_R$\\
         & $\tilde{t}_L$, $\tilde{t}_R$, $\tilde{b}_L$, $\tilde{b}_R$ &
$\tilde{t}_1$, $\tilde{t}_2$, $\tilde{b}_1$, $\tilde{b}_2$\\
& & \\
Higgs bosons & $h_d$, $h_u$, $h_s$, $a$, $a_s$ \quad &
$H_1$, $H_2$, $H_3$, $H_4$, $H_5$ \\
& & ($H_1$, $H_2$, $H_3$, $A_1$, $A_2$) \\
 & $H_d^-$, $H_u^+$ & $H^\pm$ \\
\hline
fermions & & \\
\hline
neutralinos & $\tilde{B}^0$, $\tilde{W}^0$, $\tilde{H}_u^0$, $\tilde{H}_d^0$, $\tilde{S}$ &
$\tilde{\chi}^0_1$, $\tilde{\chi}^0_2$, $\tilde{\chi}^0_3$, $\tilde{\chi}^0_4$, $\tilde{\chi}^0_5$\\
& & \\
charginos & $\tilde{W}^\pm$, $\tilde{H}_d^-$, $\tilde{H}_u^+$ &
$\tilde{\chi}_1^\pm$, $\tilde{\chi}_2^\pm$\\
& & \\
gluino & $\tilde{g}$ & $\tilde{g}$\\
\hline
\end{tabular}
\caption{\label{NMSSM-mix} 
New particles in the NMSSM compared to the SM. 
The gauge eigenstates as well as the through mixing
generated mass eigenstates are shown. The Higgs-boson mass eigenstates
in brackets give the notation in the case of a CP-conserving
Higgs-boson sector, where we have three CP-even Higgs bosons $H_1$, $H_2$, $H_3$
as well as two CP-odd ones, $A_1$ and $A_2$ instead of
five mixed states $H_1,...,H_5$.}
\end{table}

%%%%%%%%%%%%%%%%%%%%%%%%%%%%%%%%%%%%%%%%%%%%%%%%%%%%%%%%%%%%%%%%%%5
% Stability and electroweak symmetry breaking of the global minimum
%%%%%%%%%%%%%%%%%%%%%%%%%%%%%%%%%%%%%%%%%%%%%%%%%%%%%%%%%%%%%%%%%%5
\subsection{Stability and electroweak symmetry breaking of the global minimum}

In this section we want to present conditions of the parameters in the NMSSM 
which follow from stability, as well as from the required symmetry breaking behavior at the
vacuum.
First of all, the Higgs potential has to be bounded from below 
in order to yield a stable vacuum
solution. From the F-terms as well as from the D-terms of the potential we get
quartic terms
in the $H_d$, $H_u$ and $S$ fields with positive coefficients 
for non-vanishing parameters $\lambda$ and $\kappa$.
These quartic terms dominate the potential value for
large field values. That is, stability is guaranteed in the NMSSM
Higgs potential if $\lambda \neq 0$ and $\kappa \neq 0$. 

Of course the potential has to have a minimum with the right electroweak symmetry breaking
behavior. In particular this means that the potential value at the vacuum has to be lower
than at the symmetric stationary point with $\langle H_d \rangle = \langle H_u \rangle =0$. 
Moreover a stationary point with $\langle S \rangle=0$ has to be avoided in order to
generate an effective $\mu$ term, necessary to give masses to up- and down-type fermions.
Since the Higgs potential is zero for vanishing scalar doublet and singlet fields we must 
have $\langle V \rangle <0$. Inserting the parameterization~(\ref{eq-higgsespara})
into the potential~(\ref{eq-V}), using the tadpole conditions and the replacement
of $\Rl$ by the charged Higgs mass~(\ref{eq-repRl}) we get an upper bound
\begin{equation}
\label{eq-mHpmup}
m_{H_\pm}^2 < 
2 \frac{|\lambda|^2 v_s^2}{\sin (2 \beta)} 
+ 2 \frac{|\kappa|^2 v_s^4}{v^2 \sin (2 \beta)} 
+ \frac{m_Z^2}{\tan^2 (2\beta)} 
- \frac{R v_s^2}{\sin (2 \beta)} + \frac{4 R_\kappa v_s^3}{3 v^2 \sin^2 (2 \beta)}
+ m_W^2\,.
\end{equation}
Note that this upper bound goes to infinity in the limit $v_s\rightarrow \infty$, that
is in the MSSM limit there is no upper bound.

A further constraint of a {\em minimum} of the potential is to have a positive definite Hessian
matrix in the scalar fields, that is, a positive definite scalar mass-squared matrix 
at the vacuum. Since the eigenvalues of the Hessian matrix are the mass eigenstates, they
have to be positive. 
In the simplified CP-conserving case where the scalar mass-squared matrix becomes block diagonal,
this condition requires $\det(M_P') >0$, that is
$(m_{H^\pm}^2 - m_W^2 +1/2 |\lambda|^2 v_s)(3/4 R v^2 \sin (2\beta)-R_\kappa v_s) > 3/4 R^2 v^2 v_s^2$, giving
\begin{equation}
\label{eq-mHpmlo}
m_{H_\pm}^2 > m_W^2 - \frac{1}{2} |\lambda|^2 v_s\,.
\end{equation}
that is, in the NMSSM we find for non-vanishing $\lambda$ 
a reduced lower bound for the charged Higgs-boson masses compared to the MSSM.
The conditions~(\ref{eq-mHpmup}), (\ref{eq-mHpmlo}) together with the tadpole equations are
necessary conditions for a viable global minimum, but not sufficient.
A stringent determination of the global minimum which has the required electroweak
symmetry breaking behavior can be found in~\citep{Maniatis:2006jd}
and will be discussed in more detail in Sect.~\ref{sec-global}.

%%%%%%%%%%%%%%%%%%%%%%%%%%%%%%%%%%%%%%%%%%%%%%%%%%%%%%%%%%%%%%%%%%5
% Parameters of the NMSSM Higgs potential
%%%%%%%%%%%%%%%%%%%%%%%%%%%%%%%%%%%%%%%%%%%%%%%%%%%%%%%%%%%%%%%%%%5
\subsection{Parameters of the NMSSM Higgs potential}
\label{sub-Higgspara}

The parameters of the NMSSM Higgs potential~(\ref{eq-V}) are
\begin{equation}
\label{eq-initpara}
\lambda, \kappa, A_\lambda, A_\kappa, m_{H_d}^2, m_{H_u}^2, m_S^2\,,
\end{equation}
in addition to the \eweakgroup~couplings $g_2$ and $g_1$. 
Since the potential is Hermitean we see that
$\lambda, \kappa, A_\lambda, A_\kappa$ are complex whereas
$m_{H_d}^2, m_{H_u}^2, m_S^2$ have to be real. 
In practice, it is often not useful to fix these initial parameters, 
but others like the vacuum expectation values of the Higgs fields,
$v_d$, $v_u$, $v_s$, or typically, written via
$\tan (\beta)=v_u/v_d$, $v^2=v_d^2+v_u^2$ for the two Higgs doublets in terms
of $v \approx 246$~GeV, $\tan (\beta)$, $v_s$. Moreover, we remark that the parameters
$m_{H_d}^2, m_{H_u}^2, m_S^2$ are {\em not} the physical parameters
of the Higgs bosons since the physical masses arise from the
diagonalization of the mixing matrices as discussed in Sect.~\ref{sub-massmatrices}.

The tadpole conditions~(\ref{eq-tadpole}) and 
relation~(\ref{eq-repRl}) can be employed to translate the new set of parameters,
\begin{equation}
\lambda, \kappa, |A_\kappa|, v, \tan (\beta), v_s, m_{H^\pm}^2, \sign(R_\kappa),
\delta_{\text{EDM}}, \delta_\kappa'
\end{equation}
into the initial parameter set~(\ref{eq-initpara}). Here the
phase combinations 
\begin{equation}
\delta_{\text{EDM}} \equiv \delta_\lambda+\phi_\kappa+\phi_s\,,
\qquad
\delta_\kappa' \equiv \delta_\kappa+ 3 \phi_s 
\end{equation}
are defined and the complex parameters are written as $\lambda = |\lambda| e^{i \delta_\lambda}$,
$\kappa = |\kappa| e^{i \delta_\kappa}$, $A_\lambda = |A_\lambda| e^{i \delta_{A_\lambda}}$,
$A_\kappa = |A_\kappa| e^{i \delta_{A_\kappa}}$.
We can easily see how to get the original parameters~(\ref{eq-initpara}) back from this new set:
via the phase combinations $\delta_{\text{EDM}}$ and $\delta_\kappa'$
we can determine $R$ and $I$ in~(\ref{eq-phases}).
The tadpole conditions fix then $I_\lambda$ and $I_\kappa$. Together
with the length $|A_\kappa|$ and $\sign(R_\kappa)$
the phase parameter $R_\kappa$ is determined. The physical charged
Higgs-boson mass $m_{H^\pm}^2$ gives via~(\ref{eq-repRl})
immediately $R_\lambda$, that is, all phase parameters~(\ref{eq-phases})
are known at this point. Together with $v, \tan (\beta), v_s $ we get
all mass parameters $m_{H_d}^2, m_{H_u}^2, m_S^2$
from the tadpole conditions. It only remains to fix the phases of
$A_\lambda$ and $A_\kappa$ to arrive at the initial set. This is done by using again
(\ref{eq-repRl}) since $|\lambda|$, $|\kappa|$, 
$|A_\lambda|$, $|A_\kappa|$ are known.
Note, that in the mass squared matrix of the Higgs bosons, the CP-violating entries 
in the matrix $M'_{SP}$ in~\eqref{eq-mspnew}
are proportional to
the imaginary part of $\exp [i (\dedm - \dcp) ]$.

%%%%%%%%%%%%%%%%%%%%%%%%%%%%%%%%%%%%%%%%%%%%%%%%%%%%%%%%%%%%%%%%%%5
% The one-loop effective potential
%%%%%%%%%%%%%%%%%%%%%%%%%%%%%%%%%%%%%%%%%%%%%%%%%%%%%%%%%%%%%%%%%%5
\subsection{The one-loop effective potential}

For later convenience, let us also introduce the contribution of
radiative corrections
of the Higgs-boson mass to the potential, which may be 
incorporated by studying the effective potential
\begin{equation}
V_{\text{eff}} = V + V_{\text{loop}}\,.
\end{equation}
The contributions of corrections from the light particles, that is
leptons and the quarks and squarks
of the first two families may be neglected due to the small Yukawa couplings
to the Higgs fields. 
At the one-loop order, the result for $V_{\text{loop}}$, first given by 
Coleman and Weinberg~\citep{Coleman:1973jx}, reads
\begin{equation}
\label{eq-colwei}
   V_{\text{loop}} = \frac{1}{64 \, \pi^2} \sum_{i}  C_i \:
   (-1)^{2 S_i} \: (2 S_i + 1) \: \bar{m}_i^4 \:
   \: \ln \left(\frac{\bar{m}_i^2}{Q^2}\right)\,, 
\end{equation}
where the sum is over all particles with field-dependent masses $\bar{m}_i$,
spin $S_i$ and color degrees of freedom $C_i$ and $Q$ is the
renormalization scale originating from the loop integral.
Of course, this corrections of the Higgs potential change the
tadpole conditions. The field-dependent masses, that is, the masses
{\em before} spontaneous symmetry breaking occurs, of the top, the stop, and
the gauge bosons are given in App.~\ref{ap-fielddependentmasses}.

\newpage
%%%%%%%%%%%%%%%%%%%%%%%%%%%%%%%%%%%%%%%%%%%%%%%%%%%%%%%%%%%%%%%%%%5
% Higgs-boson phenomenology
%%%%%%%%%%%%%%%%%%%%%%%%%%%%%%%%%%%%%%%%%%%%%%%%%%%%%%%%%%%%%%%%%%5
\subsection{Higgs-boson phenomenology}
\label{sub-Higgspheno}

In contrast to the minimal supersymmetric extension of the SM
we have in the next-to-minimal extension, NMSSM, due to the additional
singlet superfield $\hat{S}$ two additional Higgs bosons as well as one
additional singlino. In particular the phenomenology of the Higgs bosons
may be very different compared to the MSSM. This difference is
pointed out in this subsection.
We start with considering one of the advantages of the NMSSM over the
MSSM, namely the much less restricted Higgs-boson sector.
In the MSSM the lightest scalar Higgs boson is predicted to not exceed
the Z-boson mass,
\begin{equation}
\label{eqHtree}
(m_{H_1}^{\text{MSSM}})^2 < m_Z^2 \cos^2 (2 \beta) \,.
\end{equation}
This bound is softened through quantum corrections.
One-loop corrections to the lightest Higgs boson
where studied~\citep{Ellwanger:1993hn,Pandita:1993hx,Elliott:1993bs,Yeghian:1999kr}
as well as dominant two-loop corrections~\citep{Yeghian:1999kr,Ellwanger:1999ji}.
The largest
contribution comes from virtual top and stop loops, where
the one-loop corrections read 
\begin{equation}
\label{eq-shift1}
(\Delta m_{H_1}^{\text{MSSM}})^2 = c \frac{m_t^4}{v^2} 
\ln\left( \frac{m_{\tilde{t}_L} m_{\tilde{t}_R}}{m_t^2} \right)\,,
\end{equation}
with $c$ of the order 1.
Confronted with the experimental data, that is, LEP searches
for light Higgs bosons in the MSSM we have the lower bound~\citep{Schael:2006cr}
\begin{equation}
\label{eq-HboundM}
m_{H_1}^{\text{MSSM, exp}} > 92~\text{GeV} \,.
\end{equation}
That is, we need, considering \eqref{eqHtree} and \eqref{eq-shift1} rather large
stop masses, which enter only logarithmically in the radiative corrections, in order to increase the
mass of the lightest scalar Higgs-boson in the MSSM. On the other
hand, a large violation of the degeneracy of the superpartner
top and stop masses reintroduces a new kind of {\em unnatural}
fine-tuning.
This is in particular disturbing
since one of the main motivation for the introduction of supersymmetry
is to avoid this unnatural large quantum corrections to the scalar
Higgs-boson mass, which occur in the SM.
Let us note that there is some debate about this argument and refer
the reader to the discussion in Martin \citep{Martin:1997ns}.
We notice, that a large mass splitting of the superpartners
would reintroduce the initial naturalness problem we encounter in the SM.
Therfore, we expect the superpartner particle masses not
to exceed the scale of 1~TeV too much to be considered as natural.
Another argument for superpartner masses not to exceed 
the scale of 1~TeV too much is the unification of gauge couplings,
which is spoiled by very large mass splittings.\\

\begin{figure}[h!] 
\centering
\includegraphics[width=0.3\linewidth, angle=0,clip]{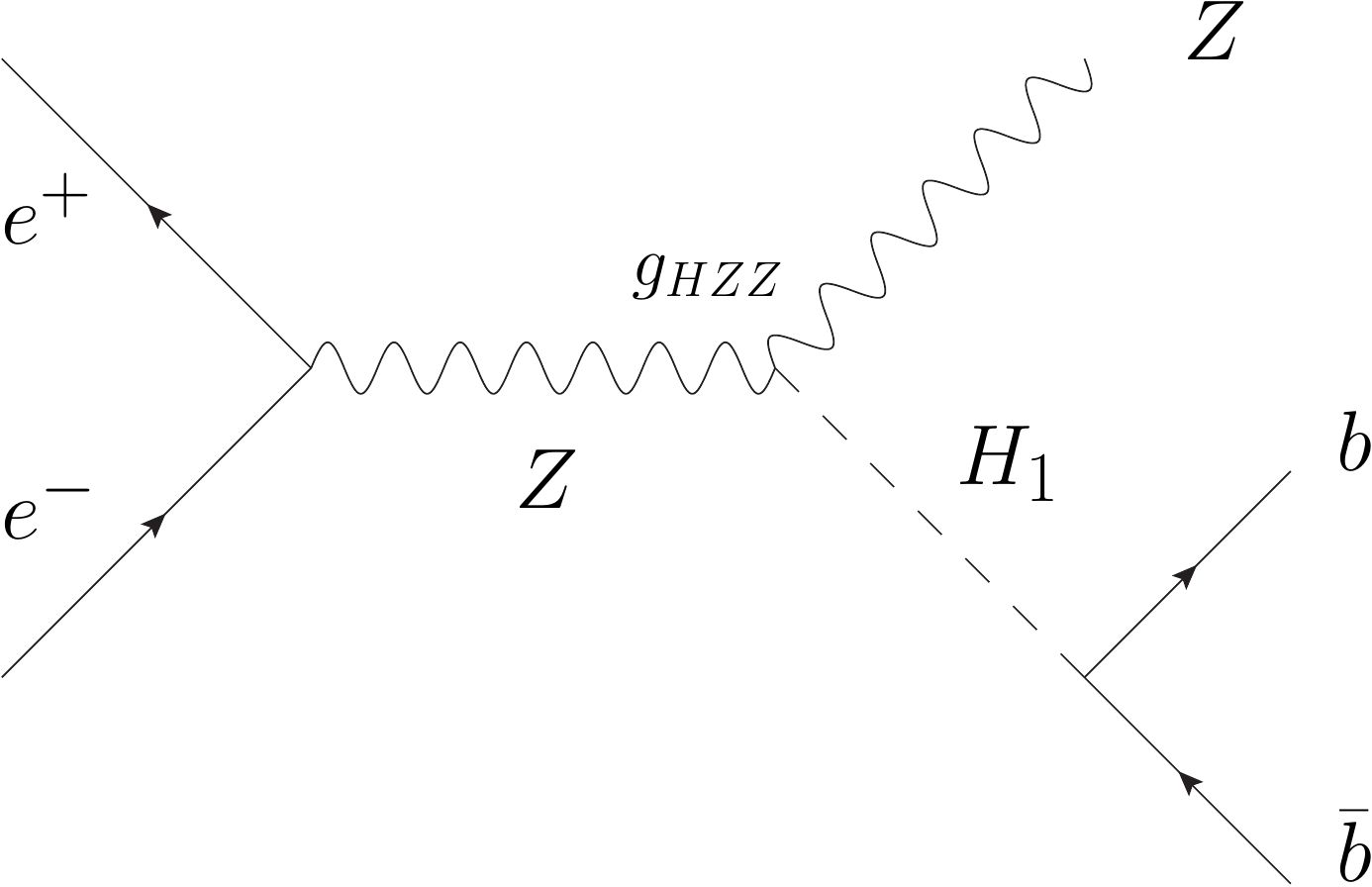}
\caption{\label{figHbb}
\it Higgs-boson production via Higgs-strahlung and subsequent decay into a
$b\bar{b}$ quark pair, as searched for at the LEP experiment \citep{Schael:2006cr}.
The coupling of the Higgs boson to the $Z$ bosons is denoted by $g_{HZZ}$ 
and given for the NMSSM in App.~\ref{app-feyn}.}
\end{figure}

The situation in the NMSSM is quite different. Firstly, the tree-level mass bound
\eqref{eqHtree} for the lightest Higgs boson is
no longer valid in the NMSSM. In the CP-conserving Higgs-boson sector 
case the lower bound is changed to~\citep{Drees:1988fc}
\begin{equation}
\label{eqHtreeN}
(m_{H_1}^{\text{NMSSM}})^2 < m_Z^2 \left( \cos^2 (2 \beta) +
\frac{2 |\lambda|^2 \sin^2 (2 \beta)}{g_1^2+g_2^2} \right)\,.
\end{equation}
Thus, the upper limit of the
lightest CP-even Higgs-boson mass is in general lifted compared to
the MSSM~\eqref{eqHtree}. 
Secondly, the experimental bound~\eqref{eq-HboundM} is
also much weaker in the NMSSM \citep{Dermisek:2005ar,Dermisek:2005gg,Dermisek:2008uu}.
This can be easily understood:
the main detection
strategy of the lightest scalar Higgs in the MSSM is
based on Higgs-strahlung off a $Z$-boson with subsequent decay
of the Higgs boson according to $H_1 \rightarrow b\bar{b}$ and $b$-tagging in the
detector; see Fig.~\ref{figHbb}.
But in the NMSSM the additional decay channel into
a pair of pseudoscalar Higgs bosons, $H_1 \rightarrow A_1 A_1$, is open, reducing
in general
the $b\bar{b}$ decay-channel branching fraction. 
In the case $m_{A_1} < 2 m_b$, the subsequent decays of
the pseudoscalar Higgs bosons can no longer proceed via b-quark pairs.
In this case, the pseudoscalar decays
into $\tau^+ \tau^-$, $c\bar{c}$ or $gg$ and may escape detection.
Then only the weaker decay mode independent LEP bound~\citep{Abbiendi:2002qp}
\begin{equation}
m_{H_1}^{\text{(N)MSSM, exp}} > 82~\text{GeV}
\end{equation}
applies. This experimental 
search is not based on the decay products of the Higgs boson,
but on the recoil mass spectrum of the Z boson in the
process $e^+e^- \rightarrow Z H_1$. Therefore, the actual
decay of the Higgs boson plays no role in this search and the given limit applies also
to the NMSSM.
To summarize, in the NMSSM we have a larger theoretical upper bound 
as well as a lower experimental bound 
and thus this model seems to be favored over the
MSSM with respect to the Higgs-boson sector restrictions.\\

\begin{figure}[h!] 
\centering
\includegraphics[width=0.4\linewidth, angle=0,clip]{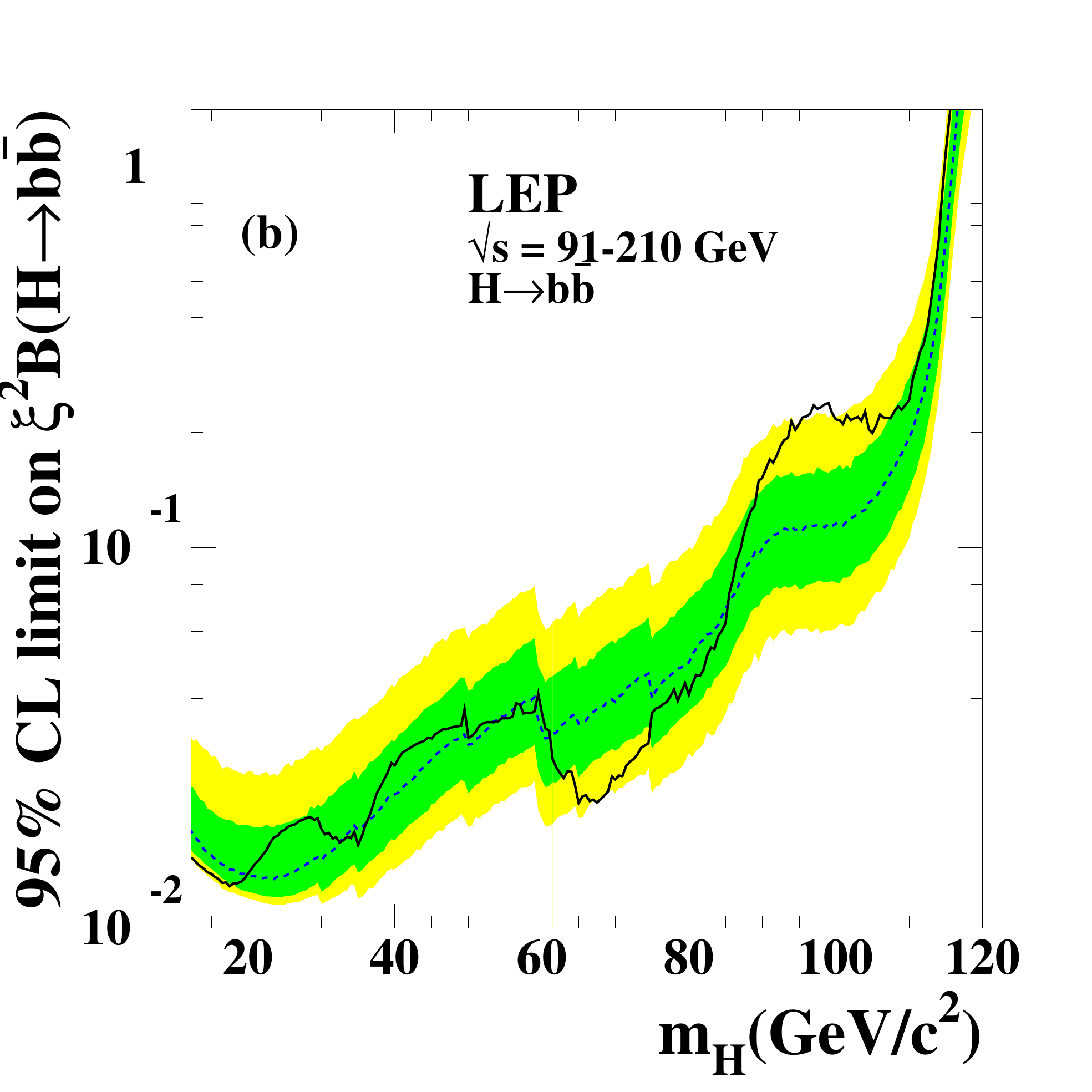}
\caption{\label{figLEPHexcess}
\it Upper limit on the ratio $\xi^2$ given in \eqref{eq-xi} times the
branching ratio of the Higgs boson into $b \bar{b}$ according to the SM.
The upper limit is gained from at LEP from LEP1 data at the $Z$ resonance
as well as from LEP2 data taken at energies between 161 and 209~GeV. 
The dark and bright band give the 1--$\sigma$ and 2--$\sigma$ deviations
from the central upper limit. The observation is shown (full line), exceeding the
upper limit for a Higgs-boson mass of about 98~GeV. The horizontal line
above represents the SM expectation on $\xi^2$. Figure taken 
from~\citep{Barate:2003sz}.
}
\end{figure}

With these remarks on the restrictions in the Higgs-boson sector, 
let us also briefly comment on the approach of Barate et al., \citep{Barate:2003sz},
on Higgs-boson phenomenology in extensions of the SM.
In Fig.~\ref{figLEPHexcess} the upper limit on the
ratio 
\begin{equation}
\label{eq-xi}
\xi^2= \left( \frac{g_{\text{HZZ}}}{g_{\text{HZZ}}^{\text{SM}}} \right)^2
\end{equation}
times the branching ratio $\BR(H\rightarrow b\bar{b})$ is shown, 
depending on the Higgs-boson mass. 
In this ratio $\xi^2$, the expression $g_{\text{HZZ}}^{\text{SM}}$ denotes
the SM Higgs--Z--Z coupling, whereas 
$g_{\text{HZZ}}$ denotes the same non-standard coupling.
The dark and bright bands give the 1--$\sigma$ and
2--$\sigma$ deviations from this limit. The limits
on $\xi^2$ are gained from LEP1 data at the $Z$-resonance 
as well as from LEP2 data taken at energies between
161 and 209~GeV. The Higgs boson is assumed to decay into
fermions and bosons according to the SM. The dominant
decay channel in the SM proceeds via the $b\bar{b}$ channel.
The Higgs-boson production
cross sections for the processes $e^+e^- \rightarrow H Z$,
$WW \rightarrow H$ and $ZZ \rightarrow H$ are scaled with 
the non-standard coupling squared $g_{\text{HZZ}}^2$.
Obviously, the observation exceeds the upper limit for an
assumed Higgs-boson mass of around 98~GeV.
As is argued in the works of Dermisek et al.~\citep{Dermisek:2005gg,Dermisek:2005ar}, the
observed excess is consistent with models which have a SM-like
$ZZH_1$ coupling but a reduced branching ratio $\BR (H_1 \rightarrow b \bar{b})$
due to new open decay channels. As aforementioned, in the NMSSM the
light scalar Higgs boson may decay into a pair
of pseudoscalars $H_1 \rightarrow A_1 A_1$. In case
$m_{A_1} < 2 m_b$ the pseudoscalar $A_1$ can not
subsequently decay into $b\bar{b}$ pairs,
but decays into $\tau$ pairs or jets.
As is argued, for a SM-like $ZZH_1$ coupling 
and $\BR (H_1 \rightarrow b \bar{b}) \approx 0.08$
and
$\BR (H_1 \rightarrow A_1 A_1) \approx 0.9$ the observed
excess at $m_{H_1}\approx 98$~GeV is nicely reproduced.\\

Let us turn our attention to the prospects of Higgs-bosons 
in the NMSSM at the
LHC, where we mention the approaches
\citep{Ellis:1988er,Gunion:1996fb,Rainwater:1998kj,Rainwater:1999sd,Yeghian:1999kr,
Plehn:1999xi,Ellwanger:1999ji,Zeppenfeld:2000td,Kauer:2000hi,Dobrescu:2000jt,
Ellwanger:2001iw,Ellwanger:2003jt,Ellwanger:2004gz,Dermisek:2005ar,Forshaw:2007ra}. 

In Ref.~\citep{Gunion:1996fb},
published in 1996, the 
Higgs detection capability of LEP2 and LHC was studied. It was stated, 
that there is large parameter space, where all Higgs bosons
in the NMSSM escape detection at both collider experiments, LEP2 and LHC. 
With respect to the LHC, an integrated luminosity as high as 600~fb$^{-1}$
was assumed.\\

However, in the meantime much improvements compared to the initial study
could be achieved.
In 2002 an investigation of Ellwanger, Gunion and Hugonie~\citep{Ellwanger:2001iw}
was titled rather promising: {\em Establishing a No-Lose Theorem
for NMSSM Higgs Boson Discovery at the LHC}. At this time there
were improvements at the theoretical side, like the two-loop corrections
to the effective potential \citep{Yeghian:1999kr,Ellwanger:1999ji} and 
the predictions of the $W$-fusion channel Higgs-boson production at the LHC 
\citep{Rainwater:1998kj,Rainwater:1999sd,Plehn:1999xi,Kauer:2000hi,Zeppenfeld:2000td}.
Moreover the improved LEP2 data were available, constraining the NMSSM parameter 
space further. 
Taking the newly predicted detection channels into account, that is
the associated production of Higgs bosons with a top pair and the two $W$-fusion
channel, 
altogether the following channel list is studied in \citep{Ellwanger:2001iw}:
\begin{equation}
\label{eq-Higgsprod}
\begin{split}
1)\quad & gg \rightarrow H \rightarrow \gamma \gamma\,,\\
2)\quad & WH \text{ or } t\bar{t}H \text{ production with } \gamma \gamma l^\pm \text{in the final state}\,,\\
3)\quad & t\bar{t}H \text{ with } H \rightarrow b\bar{b}\,,\\
4)\quad & gg \rightarrow H/A \text{ or } b\bar{b}H/A \text{ production with } H/A \rightarrow \tau \bar{\tau}\,,\\
5)\quad & gg \rightarrow H \rightarrow Z Z^* \rightarrow 4 \text{ leptons}\,,\\
6)\quad & gg \rightarrow H \rightarrow W W^* \rightarrow l^+ l^- \nu \bar{\nu}\,,\\
7)\quad & \text{at LEP2: } e^+e^- \rightarrow Z H \text{ and } e^+e^- \rightarrow H A\,,\\
8)\quad & WW \rightarrow H \rightarrow \tau \bar{\tau}\,,\\
9)\quad & WW \rightarrow H \rightarrow W W^*\,,
\end{split}
\end{equation}
where $H$ denotes any CP-even and $A$ any CP-odd Higgs-boson, that is
a CP-conserving Higgs-boson sector is considered.
In this study, the Higgs--to--Higgs as well as Higgs--to--top decays 
are not taken into account, 
that is, parameter sets leading to a particle spectrum which allows
kinematically for these decays are disregarded.
Explicitely, the disregarded decay channels are
$H\rightarrow HH$, $H\rightarrow AA$, $H\rightarrow H^+ H^-$, $H\rightarrow AZ$, $A\rightarrow H A$,
$A\rightarrow H Z$, $H/A \rightarrow H^\pm W^\mp$, $H/A \rightarrow t\bar{t}$, $t\rightarrow H^+b$.
The suppression of these decay channels is implemented by
invoking the following constraints:\\
\begin{equation}
\label{masskin}
\begin{split}
m_{H_3}\; &<\; 2 m_{H_1}, 2 m_{A_1}, 2 m_{H^\pm}, m_{A_1}+m_Z, m_{H^\pm}+m_W,\\
m_{A_2}\; &<\; m_{H_1}+m_{A_1}, m_{H_1}+m_Z, m_{H^\pm}+m_W,\\
m_{H^\pm} &>\; 155~GeV.
\end{split}
\end{equation}
Further assumptions are the absence of Landau singularities 
(see Sect.~\ref{sub-thconstraints})
and, as mentioned before, the LEP2 constraints on Higgs strahlung $e^+e^- \rightarrow Z H$~\citep{:2001xwa}
and $e^+e^-  \rightarrow H A$~\citep{:2001xx}. Moreover, 
in the parameter space it is required that 
$|\mu| \equiv \lambda v_s > 100$~GeV in order to avoid a light chargino, 
which is experimentally excluded. 
Neglecting kinematically the Higgs--to--Higgs, Higgs--to--top decays and Higgs--to--neutralino decays,
it is found that at least one Higgs boson will be detected at LHC for a luminosity of 300~fb$^{-1}$ for
arbitrary choices of remaining parameters, with a statistical discovery level of at least 5--$\sigma$.\\

In 2003 this investigation was followed by supplementing the missing $H \rightarrow AA $ decay channel via
$W$-fusion, that is, $WW \rightarrow H \rightarrow AA$~\citep{Ellwanger:2003jt}
({\em Towards a no-lose theorem for NMSSM Higgs discovery at the LHC}). There, 
special parameter sets are chosen, representing cases
which are rather difficult for detection at the LHC in this channel due
to difficulties with respect to the background.\\

In 2005 follows a study~\citep{Ellwanger:2005uu}
({\em Difficult scenarios for NMSSM Higgs discovery at the LHC})  with an
investigation aiming to
reveal parameter sets which yield a mass and coupling
spectrum such that no Higgs boson is detectable at the LHC.
A CP-conserving Higgs sector is considered.
Gaugino mass unification is assumed, where $M_2=1$~TeV is fixed at 
the electroweak scale, corresponding to $M_1 \approx 500$~GeV and
$M_{\tilde{g}} \approx 3$~TeV. 
For the soft supersymmetry breaking parameters it is set
$m_Q=m_{\tilde{u}}=m_{\tilde{d}}=m_L=m_{\tilde{e}}=1$~TeV for all three generations. 
Further, the $A$-parameters are fixed to $A_u=A_d=1.5$~TeV for all
generations. Eventually, a minimal charged Higgs-boson mass, $m_{H^\pm}>155$~GeV, 
is assumed in order 
to avoid detection of charged Higgs bosons in decays like $t \rightarrow H^\pm b$ for
moderate values of $\tan(\beta)$. 
This choice of parameters
suggests that the Higgs-boson detection might be the only
new signal at the LHC. Over the following ranges of
remaining parameters is scanned:
\begin{gather}
10^{-4} \le \lambda \le 0.75\,, \quad
-0.65 \le \kappa \le 0.65\,, \quad
1.6 \le \tan (\beta) \le 54\,, \quad
-1 \text{~TeV} \le \mu=\lambda v_s, A_{\lambda}, A_\kappa \le 1 \text{~TeV}\,.
\end{gather}
In this study the ``no-lose'' theorem is confirmed,
if the Higgs-to-Higgs decays are kinematically excluded via~\eqref{masskin}. As is
noted, this exclusion corresponds to large parts of available parameter space.
In the complementary region of parameter space in deed parameters are found, 
where no detection at LHC may be achievable, as is reported.
We repeat two example parameter sets, denoted by set 7 and 8, 
\begin{center}
\begin{tabular}{l|cccccc}
parameter set & $\lambda$ & $\kappa$ & $\tan(\beta)$ & $\mu$ & $A_\lambda$ & $A_\kappa$\\
\hline
7 &0.5 & -0.15 & 3.5 & 200  & 780 & 230\\
8 &0.27& 0.15  & 2.9 & -753 & 312 & 8.4
\end{tabular}
\end{center}
which might lead to a particle and coupling spectrum 
not detectable at the LHC. In the parameter set 7 the next-to-lightest
CP-even Higgs boson $H_2$ has SM-like couplings
and decays dominantly via $H_2 \rightarrow H_1 H_1$, with branching 
ratio $\BR(H_2 \rightarrow H_1 H_1)=0.93$.
Moreover, the $H_1$ has only small couplings to $b \bar{b}$ and $\tau^+ \tau^-$, following
from the fact that $H_1$ is mostly $H_u^0$-like in the mixing matrix.
This means that the signature $WW \rightarrow H_2 \rightarrow 2j\; \tau^+ \tau^-$
is suppressed and there remains only the $4 j$ final state which is of
course heavily plagued by QCD background.
In parameter set 8 the lightest CP-even Higgs boson $H_1$ is SM-like and
decays exclusively via $H_1 \rightarrow A_1 A_1$. In this scenario the pseudoscalar
$A_1$ Higgs boson is very light, that is, $m_{A_1}=1$~GeV, and decays only
into light jets. Thus, also this signature is overpowered by QCD background.
The authors note that at a complementary $e^+e^-$ collider, like the
proposed ILC, the scenarios, difficult to detect at the LHC, could
be observed in the decay-independent signature $e^+ e^- \rightarrow Z^* \rightarrow Z X$,
as discussed in the end of this section.\\

\begin{figure}[t] 
\centering
\includegraphics[width=0.4\linewidth, angle=0,clip]{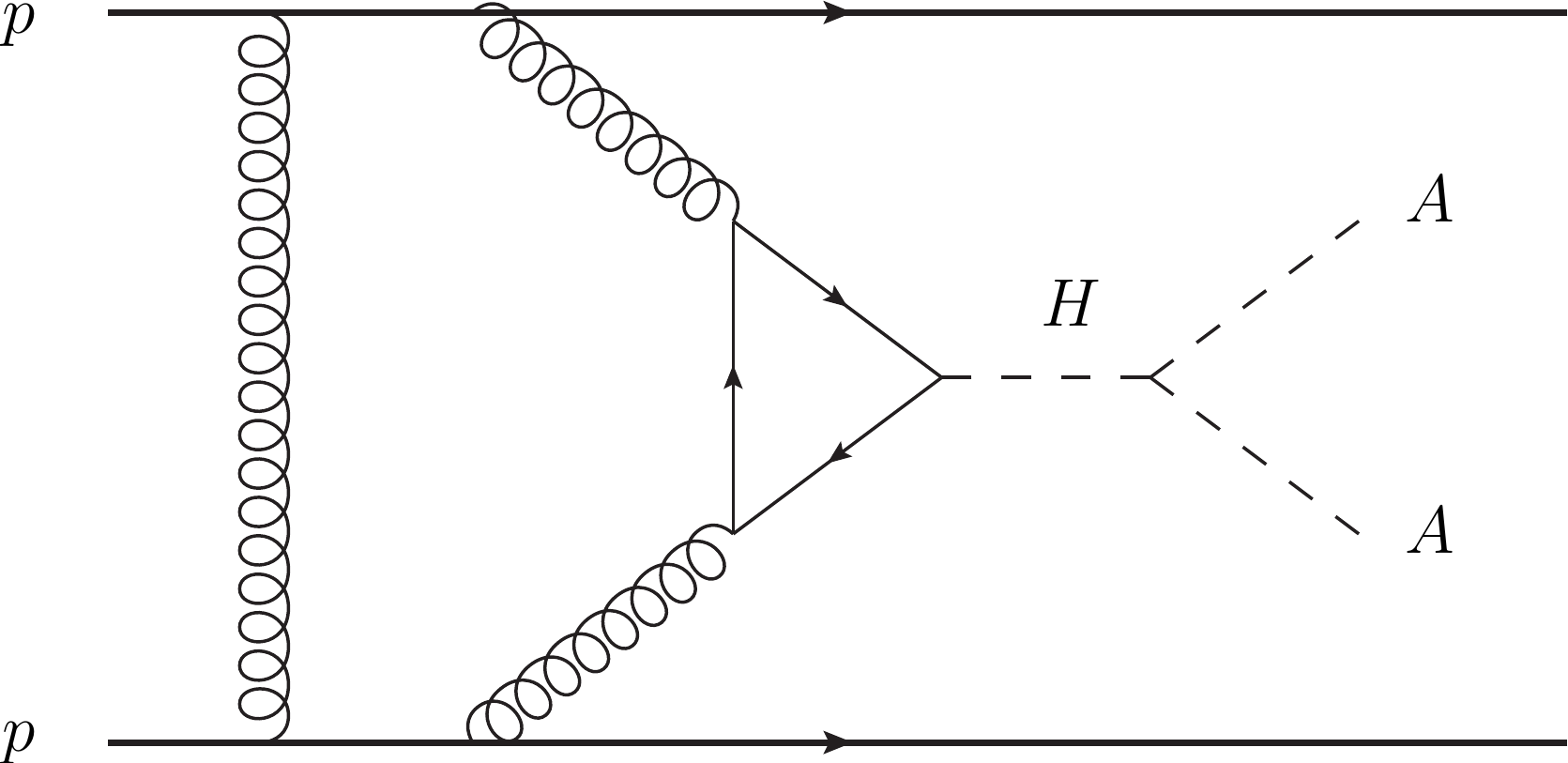}
\caption{\label{fig-pph}
\it Central exclusive production of a light CP-even Higgs boson~($H$)
in proton ($p$) collisions,
$pp \rightarrow ppH$, as suggested by~\citep{Forshaw:2007ra}. 
The Higgs boson $H$ decays into
a pair of pseudoscalar Higgs bosons $A$. The process proceeds via a fermion loop.}
\end{figure}

In the work~\citep{Forshaw:2007ra} 
({\em Reinstating the 'no-lose' theorem for NMSSM Higgs
discovery at the LHC})
a clear statement of a 5--$\sigma$ discovery
of at least one Higgs boson is given. 
This statement is achieved by employing an additional constraint:
it is argued that
absence of large fine-tuning in the NMSSM corresponds to parameter space
with a
CP-even Higgs boson in the mass range $90 \text{ GeV} < m_H < 100 \text{ GeV}$.
In this work fine-tuning is studied 
on a quantitative basis, introduced later in Sect.~\ref{sub-thconstraints}.
It is argued, that for a CP-even Higgs-boson with mass
around 98~GeV, the only remaining decay not excluded by
LEP may proceed via a light pair of pseudoscalars with
subsequent decay into $\tau$'s or jets, 
\begin{equation}
\label{eq-aa}
H \rightarrow A A \rightarrow \tau^+ \tau^- \tau^+ \tau^- \text{ or } 4j\,.
\end{equation}
The current experimental lower limits of this process with the decay
into $\tau$'s is $m_h^{\text{NMSSM, exp}} > 89$~GeV~\citep{Schael:2006cr} 
and with the decay into final state jets is
$m_h^{\text{NMSSM, exp}} > 82$~GeV
\citep{Abbiendi:2002qp} from the decay mode independent searches. This
means that Higgs production with subsequent decay  \eqref{eq-aa} 
is well beyond the constraints at LEP.
Further, it is noted that the requirement $m_A < 2 m_b$ is natural in the sense of
low fine-tuning. 
Nevertheless, it is found that this
decay channel is very challenging to detect at the LHC in
all so far studied Higgs production processes.
However as an additional search strategy, the authors suggest to
consider the well-known {\em central exclusive production} of a light Higgs boson $H$
via the process $pp \rightarrow ppH$.
This process with the subsequent decay into a pair of
pseudoscalar Higgs-bosons is shown in Fig.~\ref{fig-pph}. 
This detection mode requires the outgoing protons $p$ to be
detected in purpose-built low-angle detectors (FP420)
\citep{Cox:2004rv,Albrow:2005ig}. Via these
special detectors the four-momentum of the central
system could be reconstructed very accurately and
thus the masses of both the $H$ and the $A$
be determined on an event-by-event basis.
The investigation shows that this technique in deed
would allow for a discovery of the lightest Higgs bosons 
even when background is taken into account.\\

The decay chain \eqref{eq-aa} with 
$m_{A_1} < 2 m_b$  and large branching fractions
$BR(A_1 \rightarrow \tau^+ \tau^-)$
was also studied in
the publication of Belyaev et al.~\citep{Belyaev:2008gj} in order
to close the gap for a 'no-lose' theorem. 
The parameters of the Higgs-boson sector (see Sect.~\ref{sub-Higgspara})
were varied in large ranges and the remaining parameters,
which enter the Higgs-boson sector only at the suppressed loop level, were
fixed in this study.
The authors were looking for signatures, where two $\tau$'s 
decay into $\mu$'s and the other two $\tau$'s into jets. 
The Higgs-strahlung as well as the vector-boson-fusion
production channels were considered and, as is reported, thousands of
events are predicted, mostly originating from
vector-boson-fusion processes. An extended investigation,
taking also the background into account, is announced
to be in preparation.\\

The decay channel with $A_1$ decaying into a pair of
muons instead of taus has the advantage of much better detectability in the muon systems
of the detector of hadron colliders.
In~\cite{Lisanti:2009uy} it is reported that even for $2 m_\tau < m_{A_1}< 2 m_b$, the
subdominant decay into muons could be detected.
Constraints were deduced recently from negative searches for the decay
$H_1\rightarrow A_1 A_1$ with either both 
pseudoscalars $A_1$ decaying to $\mu$'s or one decaying to $\mu$'s
and the other to $\tau$'s 
by the D0 collaboration at Tevatron~\cite{Abazov:2009yi}.
For the negative search for the four $\mu$ final state 
the upper limit of 
$\sigma(p \bar{p} \rightarrow H + X) \cdot
\BR(H\rightarrow A_1 A_1) \cdot \BR(A_1 \rightarrow \mu^+ \mu^-)^2
\lesssim 10$~fb for $2 m_\mu < m_{A_1}< 2 m_\tau$ is given.
Note that the decay of even lighter pseudoscalars $m_{A_1}< 2 m_\mu$
is excluded by the model-independent search
in proton--copper collisions at Nikhef~\cite{Bergsma:1985qz}.\\

Eventually we note that at a future electron--positron collider like the 
{\em international linear collider} (ILC) with
center-of-mass energies up to 1~TeV, the study of the
Higgs-strahlung process $e^+e^- \rightarrow Z^* \rightarrow Z H$ would
be accessible for a large mass range of the Higgs bosons. 
In the ILC reference design report, referring to
an investigation in the constrained NMSSM \citep{Djouadi:2007ik},
it is stated that the measurement of the Higgs-boson masses with a resolution of the order of 100~MeV 
could be achieved, if the Higgs bosons are not too heavy.
A very direct approach available at an electron-positron collider is
the Higgs-boson detection, independent on the subsequent decay-mode. From the
recoil mass spectrum against the $Z$-boson in the process
$e^+ e^- \rightarrow Z^* \rightarrow Z X$ the
Higgs boson is detectable if the couplings to the Z-boson are
not too small. This decay-mode independent search is
in particular interesting in the case, where the Higgs-to-Higgs
decay is dominant and the LHC might fail to detect
the decay products.

\newpage
%%%%%%%%%%%%%%%%%%%%%%%%%%%%%%%%%%%%%%%%%%%%%%%%%%%%%%%%%%%%%%%%%%5
%%%%%%%%%%%%%%%%%%%%%%%%%%%%%%%%%%%%%%%%%%%%%%%%%%%%%%%%%%%%%%%%%%5
% Neutralinos 
% revised 26.05.09
%%%%%%%%%%%%%%%%%%%%%%%%%%%%%%%%%%%%%%%%%%%%%%%%%%%%%%%%%%%%%%%%%%5
%%%%%%%%%%%%%%%%%%%%%%%%%%%%%%%%%%%%%%%%%%%%%%%%%%%%%%%%%%%%%%%%%%5
\section{Neutralinos}
\label{sec-nc}

The modification in the NMSSM compared to the minimal supersymmetric
extension originates from
the superpotential terms
$\lambda \hat{S}  (\hat{H}_u^\trans \epsilon \hat{H}_d)
+ \frac{1}{3} \kappa \hat{S}^3$ accompanied by the introduction
of an additional singlet $\hat{S}$.
Here we focus on the
modification which arises from the fermionic part of the singlet $\hat{S}$, the
so-called {\em singlino} $\tilde{S}$. 
For studies of the MSSM
phenomenology we refer the reader to the review
of Martin~\citep{Martin:1997ns} including detailed references.
Note that we do not have additional
charginos compared to the MSSM.
The singlino component of the
superfield gives a fifth neutralino which in general mixes
with the bino $\tilde{B}^0$, wino $\tilde{W}^0$ and the Higgsinos 
$\tilde{H}_d^0$ and $\tilde{H}_u^0$; see Tab.~\ref{NMSSM-mix}. 
Collecting all quadratic terms
as performed in App.~\ref{app-feyn} we arrive in the basis
$\psi^0=(\tilde{B}^0, \tilde{W}^3, \tilde{H}_d^0, \tilde{H}_u^0, \tilde{S})^\trans$
at the symmetric neutralino mass matrix
\begin{equation}
\label{eq-neutralinomass}
M_{\tilde{\chi}^0}=
\begin{pmatrix}
M_1 & 0   & -\cb s_W m_Z &  \sb s_W m_Z  & 0\\
0   & M_2 &  \cb c_W m_Z & -\sb c_W m_Z & 0\\
-\cb s_W m_Z & \cb c_W m_Z & 0 & -\lambda v_s/\sqrt{2} & -\lambda v_u/\sqrt{2}\\
 \sb s_W m_Z &-\sb c_W m_Z & -\lambda v_s/\sqrt{2} & 0  & -\lambda v_d/\sqrt{2}\\
0 & 0 & -\lambda v_u/\sqrt{2} & -\lambda v_d/\sqrt{2} & \phantom{+}\sqrt{2}\kappa v_s
\end{pmatrix}.
\end{equation}
This neutralino mass matrix reflects the fact that
the singlino $\tilde{S}$ 
does not couple to the gauge bosons but only
to the Higgsino doublets $\tilde{H}_d^0$, $\tilde{H}_u^0$ in addition
to the selfcoupling in the lower diagonal entry. 
In case of a small mixing of the singlino with
the Higgsinos the singlino decouples. In this case the behavior of the four
neutralinos is MSSM-like and the mass of the singlino is at tree level
given by the lower diagonal entry $m_{\tilde{S}}^2 \approx 2 \kappa^2 v_s^2$. 
For large mass values $m_{\tilde{S}}$ the singlino may escape
detection and a distinction of the NMSSM from the MSSM is
very difficult~\citep{Choi:2004zx}.\\

After diagonalization with the unitary rotation $U$ 
of the mass matrix~\eqref{eq-neutralinomass} 
we end up with five neutralinos 
\begin{equation}
\label{eq-chi0mix}
\chi_i^0 = U_{j i}\, \psi_j^0\, \qquad \text{with } i,j=1,...,5
\end{equation}
which afterwards are 
arranged in ascending order, thus we end up in
Dirac notation with $\tilde{\chi}_i^0$, $i=1,...,5$,
with $\tilde{\chi}_1^0$ the lightest neutralino (more details are
given in~App.~\ref{app-feyn}).
The singlino components of
the mixed states do not couple to gauge bosons, gauginos,
leptons, sleptons, quarks and squarks. Thus, in addition to an increased 
total number of neutralinos, we expect in particular
a changed behavior compared to the MSSM for a
neutralino with a large singlino component, that is
for a {\em singlino-like} neutralino. 
It is worthwhile to note that the detection of a fifth neutralino would be a clear signal of an
extended supersymmetric model. However, the production cross section
of a singlino-like neutralino is rather small due to its small couplings.
Moreover it is evident from its small couplings that a singlino-like
neutralino, which is not the LSP, would be omitted
in cascade decays.\\

\begin{figure}[h!] 
\centering
\includegraphics[height=0.35\linewidth, angle=0,clip]{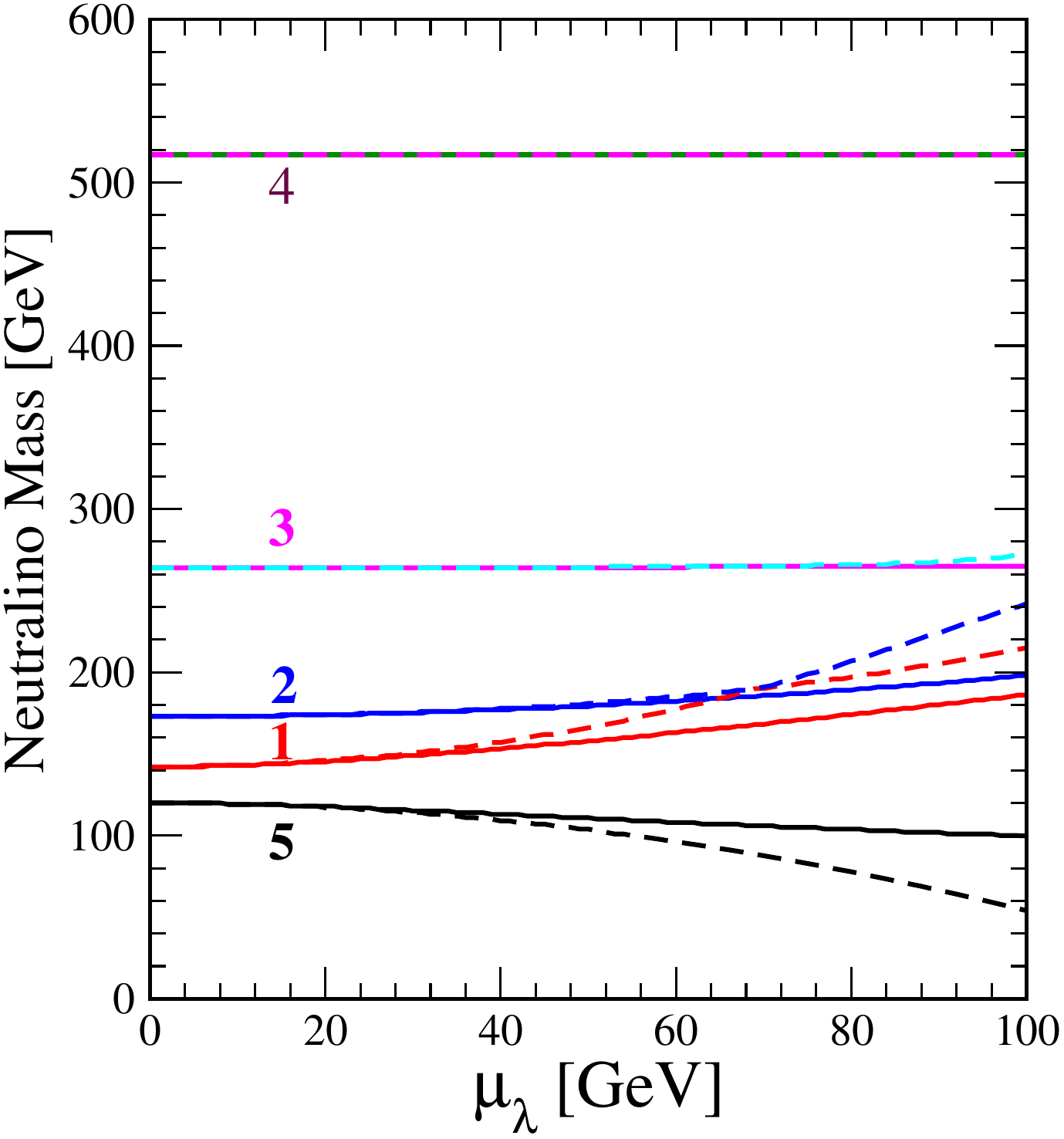}
\includegraphics[height=0.35\linewidth, angle=0,clip]{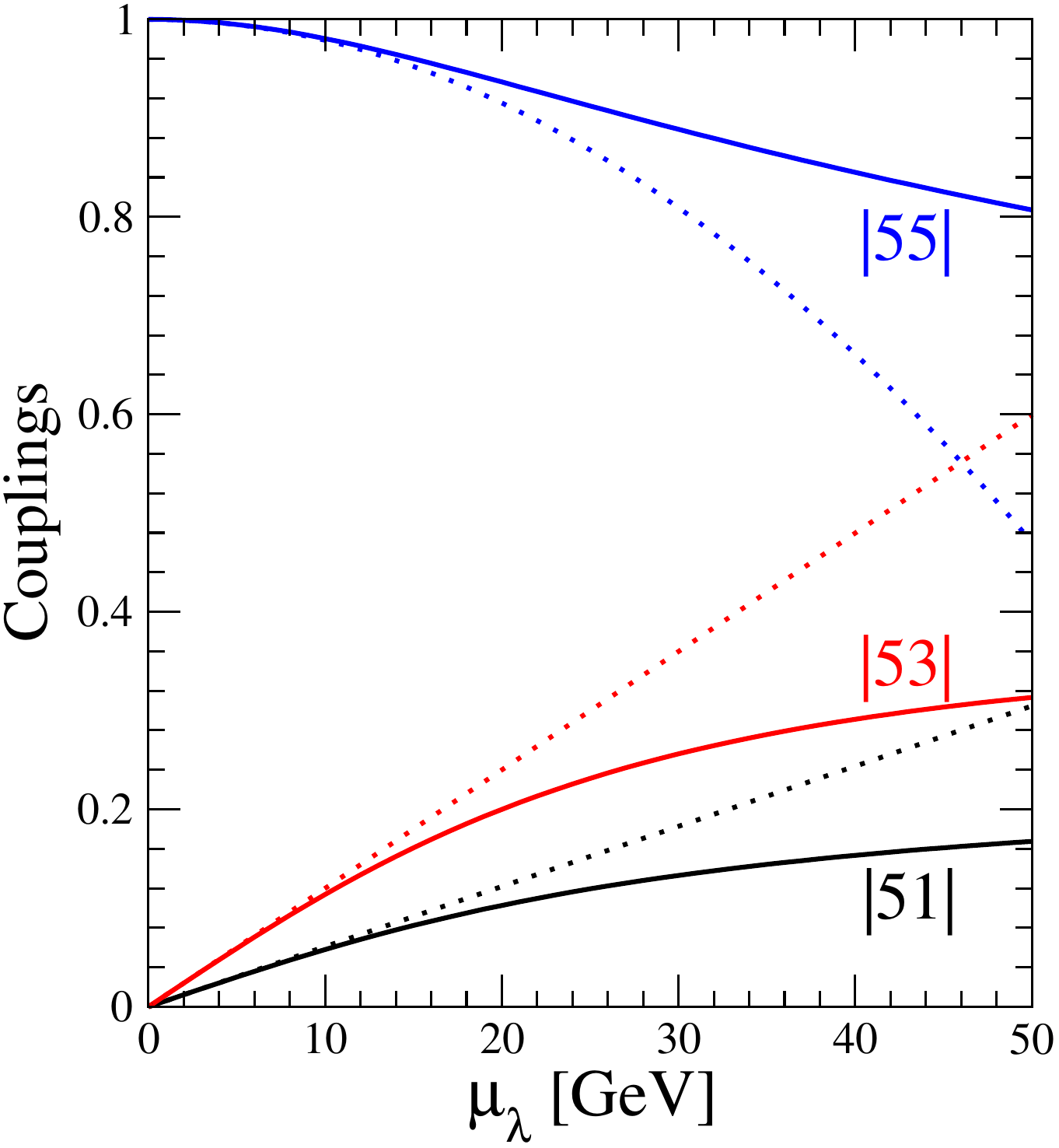}
\caption{\label{figZerwasneutralinos} \it
Neutralino masses (left) and mixings (right) depending on $\mu_\lambda \equiv \lambda v/\sqrt{2}$. 
The other parameters are fixed to $\sqrt{2} \kappa v_s= 120$~GeV (this is the lower diagonal
entry in the mixing matrix~\eqref{eq-neutralinomass}), $M_1=250$~GeV, $M_2=500$~GeV,
$\lambda v_s/\sqrt{2}=170$~GeV, $\tan(\beta)=3$. 
The numbering $i=1,...,5$ denotes the mass eigenstates in the mixing
$\chi_i^0 = U_{j i}\, \psi_j^0$~\eqref{eq-chi0mix}, before the arrangement in ascending order. 
The mixing entries in the matrix $U_{j i}$ are denoted in the right plot by
the indices, that is, for instance $|53|$ denotes $|U_{5 3}|$.  
The dashed respectively dotted curves show an approximation,
in case the singlino--Higgsino mixing is suppressed, that is, for $v \ll v_s$.
Figure taken from~\citep{Choi:2004zx}.
}
\end{figure}

In Fig.~\ref{figZerwasneutralinos} the mass spectrum of the five neutralinos as well
as the dominant mixings (couplings) of the mass eigenstates 
are shown depending on $\mu_\lambda \equiv \lambda v/\sqrt{2}$, as presented 
in~\citep{Choi:2004zx}. 
The parameters
chosen for this plot are given in the figure caption. 
The parameter $\mu_\lambda$
determines the mixing of the singlino with the Higgsinos, as is evident from
the mixing matrix~\eqref{eq-neutralinomass}. We see that in this
scenario the lightest neutralino is singlino-like, since the
mixing $U_{5 5}$ is dominant for small values of $\mu_\lambda$ 
as shown on the right hand side of Fig.~\ref{figZerwasneutralinos}. 
This is what is expected since
for small $\mu_\lambda$ the singlino decouples from the other neutralinos.
Let us note that in~\citep{Choi:2004zx} a study of
neutralino production cross sections in $e^+ e^-$ collisions and
decay rates of the neutralinos can also be found.\\

In the study by Hesselbach and Franke~\citep{Hesselbach:2002nv}
the associated singlino-like neutralino production is discussed
\begin{equation}
e^+e^- \rightarrow \tilde{\chi}_1^0 \tilde{\chi}_S^0\,,
\end{equation}
where $\tilde{\chi}_S^0$ denotes a singlino-like neutralino.
From the neutralino mixing matrix~\eqref{eq-neutralinomass} 
we see from the lower diagonal entry that we get a neutralino
with a large singlino component for a large vacuum-expectation-value $v_s$.
\begin{figure}[h!] 
\centering
\includegraphics[width=0.4\linewidth, angle=0,clip]{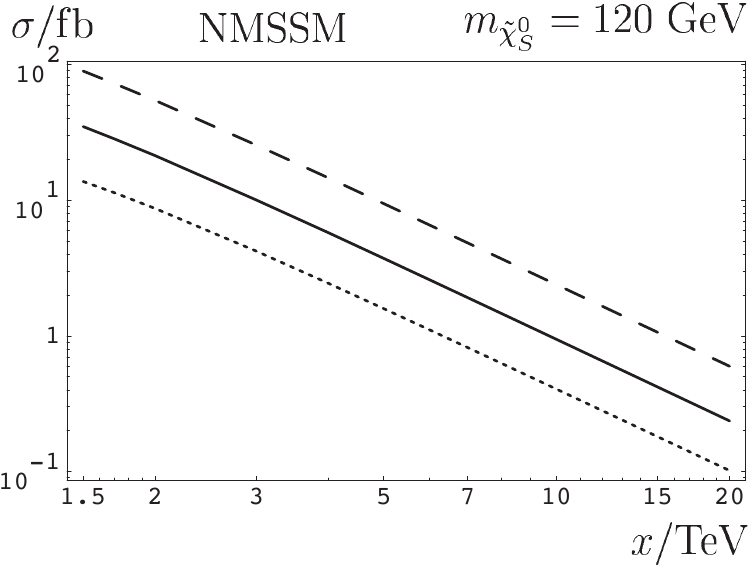}
\caption{\label{figFranke}
\it Total cross section for singlino-like neutralino production 
$e^+e^- \rightarrow \tilde{\chi}_1^0 \tilde{\chi}_S^0$ depending
on the vacuum-expectation-value $v_s$, here denoted by $x$ as presented
by Hesselbach and Franke~\citep{Hesselbach:2002nv}. 
The electron--positron center-of-mass energy is fixed to 500~GeV.
The model parameters
chosen are inspired by the SPS1a scenario in the MSSM with
$M_1=99$~GeV, $M_2=193$~GeV, $\tan(\beta)=10$, $\mu= \lambda v_s= 352$~GeV.
The full line refers to unpolarized beams, the dashed line to beam polarizations
$P_-=+0.8$, $P_+=-0.6$ and the dotted line to beam polarizations
$P_-=-0.8$, $P_+=+0.6$. The mass of the neutralino $\tilde{\chi}_S^0$ 
is fixed at 120~GeV by the $\kappa$-parameter.}
\end{figure}
As shown in Fig.~\ref{figFranke} the total cross section
drops with increasing singlet vacuum-expectation-value $v_s$ 
(denoted by $x$ in this work),
as is evident from the increasing singlino-component in $\tilde{\chi}_S^0$, that
is, smaller couplings for rising $v_s$. The effect of beam polarization is
also shown in this figure and we see that we get total
cross sections of the order of 10~femtobarn for
not too large $v_s$, depending also on the beam polarizations.
The chosen parameters in this example are given in the figure caption.\\

In case that four light neutralinos are detected, the distinction of
the minimal supersymmetric model from further extension is
in general difficult. As noted already, a heavy singlino-like neutralino
would be omitted in cascade decays of even heavier supersymmetric
partner particles. 
In case there is substantial mixing in
the neutralino sector there is a method discussed 
by Choi et al. \citep{Choi:2001ww} to discriminate
the NMSSM from the MSSM at a future electron--positron collider:
the total cross section,
summed over the four light neutralinos and normalized
to its asymptotic form in the limit of infinite
center-of-mass energies, is sensitive to the
total number of neutralinos.
\begin{figure}[h!] 
\centering
\includegraphics[width=0.4\linewidth, angle=0,clip]{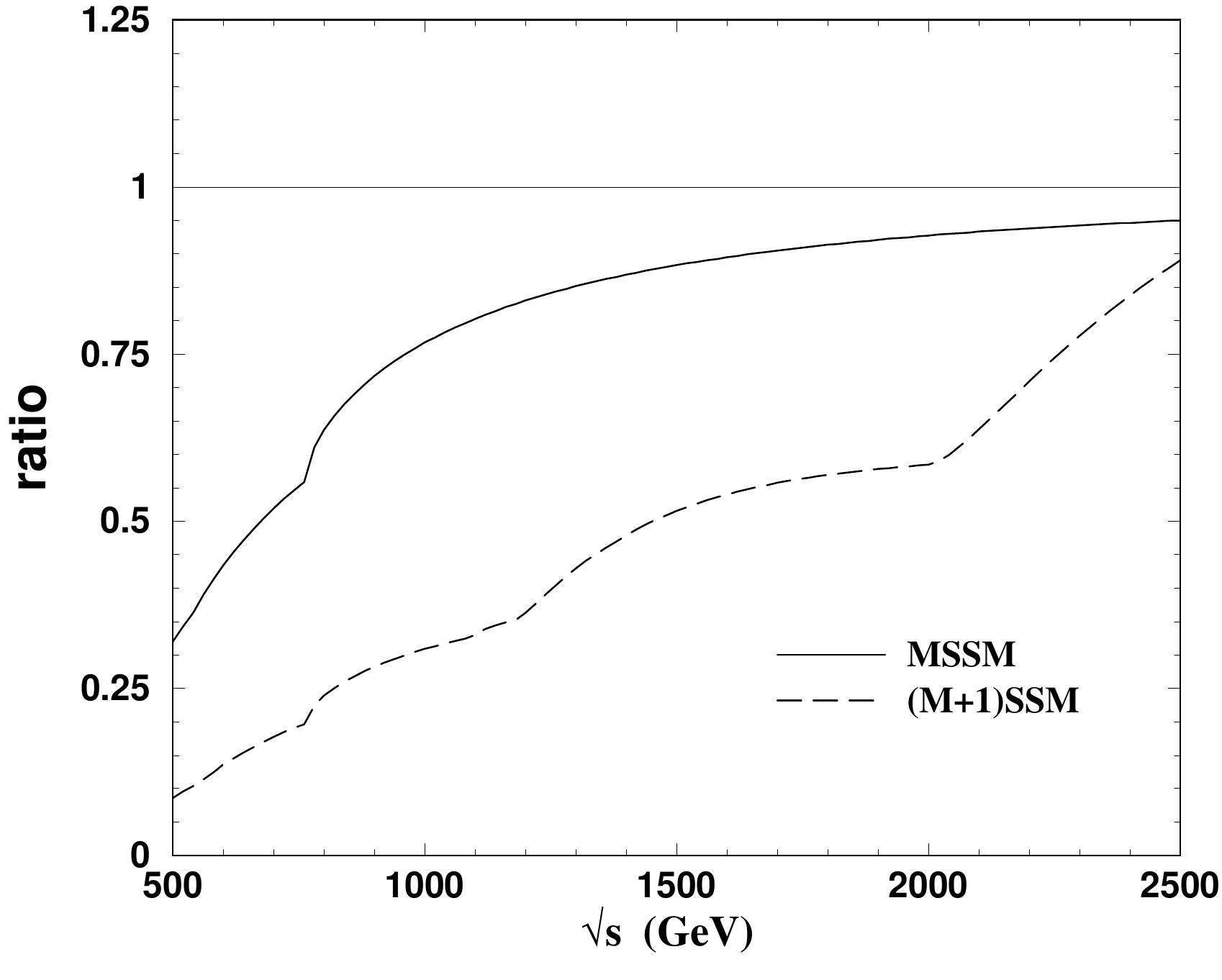}
\caption{\label{figZerwas} \it 
Energy dependence of the four light neutralino production cross
sections summed up
and normalized to its asymptotic form for infinite
center-of-mass energies. The full line corresponds
to the MSSM, whereas the dashed line corresponds to
the NMSSM. The assumed parameters are
$\tan(\beta)=10$, $|M_1|=100.5$~GeV, $M_2=190.8$~GeV,
$|\mu|=|\lambda v_s|=365.1$~GeV where CP-invariance
in the Higgs-boson sector is assumed. The selectron masses
are fixed to $m_{\tilde{e}_L}=208.7$~GeV and
$m_{\tilde{e}_R}=144.1$~GeV.
Figure taken from~\citep{Choi:2001ww}.
}
\end{figure}
The energy dependence of the ratios for the MSSM and the NMSSM is shown
in Fig.~\ref{figZerwas}.
The model parameters chosen in this plot are given in the figure caption.\\

There remains the possibility that the singlino-like neutralino
is the LSP as discussed in more detail in Sect.~\ref{sect-scan}.
Parameter scans in the NMSSM show that there is indeed large parameter
space with this possibility passing all nowadays known theoretical
and experimental constraints. Due to conserved matter parity
this LSP singlino-like neutralino would escape detection. However,
due to its small couplings, the next-to-lightest supersymmetric
particle (NLSP) would decay very slowly into the singlino-like LSP.
This opens the possibility to 
observe {\em displaced vertices} \citep{Ellwanger:1997jj,Ellwanger:1998vi,Hesselbach:2000qw}.
The partial decay width of a sfermion $\tilde{f}$
decaying into the LSP neutralino $\tilde{\chi}_1^0$ and a fermion $f$
is~\citep{Bartl:1996wt,Kraml:2005nx,Djouadi:2008uj}
\begin{equation}
\Gamma(\tilde{f} \rightarrow \tilde{\chi}_1^0 f) =
\frac{\rho^{1/2}(m_{\tilde{f}},m_{\tilde{\chi}_1^0}, m_f)}{16 \pi m_{\tilde{f}}^3}
\left(
(a^2 + b^2)(m_{\tilde{f}}-m_{\tilde{\chi}_1^0}- m_f) - 4\; a\; b\; m_{\tilde{f}} m_{\tilde{\chi}_1^0} 
\right)
\end{equation}
with kinematic function $\rho(x,y,z)=x^2+y^2+z^2-2xy-2xz-2yz$ and $a$ and $b$ 
the left- and right couplings in the Lagrangian
\begin{equation}
{\cal L}_{\tilde{f} f \tilde{\chi}_1^0} = \bar{f} (a P_R + b P_L)\tilde{\chi}_1^0 \tilde{f} + c.c.
\end{equation}
The length of flight of the sfermion is simply
$l_{\tilde{f}}= \hbar c/\Gamma(\tilde{f} \rightarrow \tilde{\chi}_1^0 f)$.
Scenarios of the NMSSM are studied for which a NLSP $\tilde{\chi}_2^0$ decays
into a singlino-like LSP $\tilde{\chi}_1^0$. As is pointed out, rather large
vacuum-expectation-values $v_s$ are required in order to get
a very pure singlino-like neutralino and suppressed couplings accompanied
by observable displaced vertices.\\

In the approach of Kraml, Raklev and White \citep{Kraml:2008zr}
a special scenario is studied, motivated by the 
parameter set SPS1a 
in the MSSM and extended to the NMSSM. In this scenario,
the LSP is a singlino-like neutralino $\tilde{\chi}_1^0$ and the NLSP a
bino-like neutralino $\tilde{\chi}_2^0$ with a small mass difference
$\Delta m = m_{\tilde{\chi}_2^0}-m_{\tilde{\chi}_1^0}$. 
Their analysis is based on
cascade decays which eventually end up with the decay
\begin{equation}
\tilde{\chi}_2^0 \rightarrow \tilde{\chi}_1^0\, l^+\, l^- \,.
\end{equation}
Due to the small mass difference, the final state leptons
are soft and could escape detection. As the authors emphasize,
with a typical kinematical cut on the minimal transversal momentum of
leptons at the LHC this could lead to a wrong interpretation
of a $\tilde{\chi}_2^0$ LSP, since non of the true final state particles
would be seen in this case. Accepting low momenta
of the final state leptons, the measurement
of the invariant di-lepton mass squared distribution
$m_{ll}^2 \equiv (p_l^+ + p_l^-)^2$, with the momenta of the leptons denoted 
with $p_l^\pm$, would allow to determine the mass difference  
$\Delta m = m_{\tilde{\chi}_2^0}-m_{\tilde{\chi}_1^0}$ at the edge of a
measured $m_{ll}^2$ distribution.

\newpage
%%%%%%%%%%%%%%%%%%%%%%%%%%%%%%%%%%%%%%%%%%%%%%%%%%%%%%%%%%%%%%%%%%5
%%%%%%%%%%%%%%%%%%%%%%%%%%%%%%%%%%%%%%%%%%%%%%%%%%%%%%%%%%%%%%%%%%5
% Parameter constraints
% revised 30.04.09
%%%%%%%%%%%%%%%%%%%%%%%%%%%%%%%%%%%%%%%%%%%%%%%%%%%%%%%%%%%%%%%%%%5
%%%%%%%%%%%%%%%%%%%%%%%%%%%%%%%%%%%%%%%%%%%%%%%%%%%%%%%%%%%%%%%%%%5
\section{Parameter constraints}
\label{sec-par}

In this section we will discuss some parameter constraint studies,
performed in the NMSSM. We start with theoretical constraints.
Constraints originating from a viable global minimum of
the Higgs potential, that is, a global minimum which has
the correct electroweak symmetry breaking behavior,
are discussed in detail in Sect.~\ref{sec-global}.
We mention the requirement of
perturbativity of couplings in considering
renormalization group equations.
Then we turn to fine-tuning. In a series of recent
papers some attention is payed to fine-tuning in
extensions of the SM; 
for instance in~\citep{Dermisek:2005ar,Dermisek:2005gg,Dermisek:2008uu},
where fine-tuning is quantified in terms of a simple function.
We proceed with constraints coming from the requirement of
a vacuum which does not break color- and electric charge. Also
an approximative constraint can be gained from the condition
to have a non-vanishing vacuum-expectation-value for
the singlet $\langle S \rangle=v_s$. 
On the experimental side we first consider the
firm exclusion limits from collider experiments. 
Further constraints with respect to 
the anomalous magnetic moment of the muon, the b-meson decay, 
cosmological constraints from indirect as well as direct 
detection of cold dark matter and strong 
first order electroweak phase transitions are considered.
Finally, we review some generic parameter scans over large 
parameter regions, revealing the viable parameter space.

%%%%%%%%%%%%%%%%%%%%%%%%%%%%%%%%%%%%%%%%%%%%%%%%%%%%%%%%%%%%%%%%%%5
% Theoretical constraints
% revised 29.4.09
%%%%%%%%%%%%%%%%%%%%%%%%%%%%%%%%%%%%%%%%%%%%%%%%%%%%%%%%%%%%%%%%%%5

\subsection{Theoretical constraints}
\label{sub-thconstraints}

Let us start with the Landau pole exclusion constraint.
The renormalization group equations for the NMSSM have been determined
to two-loop order.
Here we present the one-loop results of the
gauge couplings and the superpotential parameters $\lambda$ and $\kappa$
as well as $y_t$ in the case of a CP-conserving Higgs sector~\citep{Miller:2003ay}. 
\begin{equation}
\label{eq-betafunc}
\begin{split}
\frac{\D}{\D t} \bar{g}_a   =& \frac{1}{16 \pi^2}\;
\frac{b_a}{2} \bar{g}_a^3, \qquad (a=1,2,3)\;,\\
\frac{\D}{\D t} y_t^2 =& \frac{1}{16 \pi^2}\; y_t^2 
\left( \lambda^2 +6 y_t^2-\frac{16}{3} \bar{g}_3^2-\frac{13}{15} \bar{g}_1^2 \right)\;,\\
\frac{\D}{\D t} \lambda^2 =& \frac{1}{16 \pi^2}\; \lambda^2 
\left( 4 \lambda^2 +2 \kappa^2+3 y_t^2- 3 \bar{g}_2^2-\frac{3}{5} \bar{g}_1^2 \right)\;,\\
\frac{\D}{\D t} \kappa^2 =& \frac{1}{16 \pi^2}\; 6 \kappa^2 
\left( \lambda^2 + \kappa^2 \right)\;.
\end{split}
\end{equation}
The coefficient for the couplings are $b_1=33/5$, $b_2=1$, $b_3=-3$. The 
context of the gauge couplings $\bar{g}_1$, $\bar{g}_2$ with the conventional couplings is
$\bar{g}_1=\sqrt{5/3}g_1$, $\bar{g}_2=g_2$ with, as usual, 
$e=g_2 \sin (\theta_W)=g_1 \cos (\theta_W)$. 
The scale factor is defined as $t=\ln(Q^2/m_{\text{GUT}})$ with unification 
scale $m_{\text{GUT}} \approx 3\cdot 10^{16}$~GeV and $Q$
the scale under consideration.
Note that the unification
of the gauge couplings 
at the GUT scale occurs also in the NMSSM, like in the MSSM.
This is due to the fact, that the additional superfield $\hat{S}$ is
a gauge-singlet.

\begin{figure}[h] 
\centering
\includegraphics[width=0.4\linewidth, angle=0,clip]{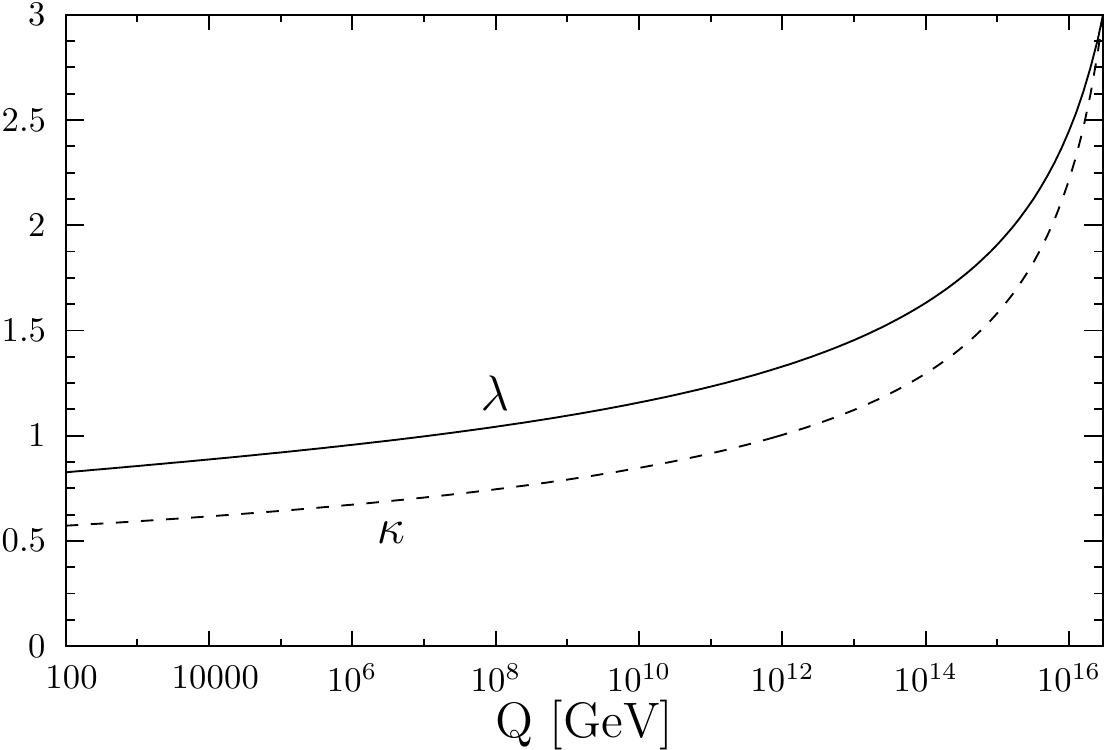}
\caption{\label{fig-RGE}
\it Evolution of the dimensionless NMSSM parameters $\lambda$ and $\kappa$ 
supposed to be real~\citep{Miller:2003ay}. 
The Yukawa coupling $h_t$ is set to $0.8$ at the grand unification scale,
$Q_{\text{GUT}}=3 \cdot 10^{16}$~GeV. For the evolution of $\lambda$ we
start with $\lambda(Q_{\text{GUT}})=3$ and $\kappa(Q_{\text{GUT}})=1$.
The evolution of $\kappa$ is given for vice versa values, that is
$\kappa(Q_{\text{GUT}})=3$ and $\lambda(Q_{\text{GUT}})=1$.
See~\citep{Miller:2003ay} for more details.}
\end{figure}

The requirement of {\em perturbativity} of the couplings means
that quantum corrections are assumed not to become too large 
at all scales
from the electroweak scale of about $100$~GeV up to the gand unification
scale of about $3\cdot 10^{16}$~GeV
(typically, the couplings are required not to exceed $2 \pi$).
At least there should not be any Landau pole for the couplings.
Due to the $\beta$ functions given
in~(\ref{eq-betafunc}) we see that $\lambda$ and $\kappa$ drop 
with a decreasing energy scale $t$. That is, perturbativity
up to the GUT scale gives a strong constraint of these coupling
parameters at the electroweak scale. As an example, in Fig.~\ref{fig-RGE} the
running of the couplings $\lambda$ and $\kappa$ with the assumption of
real parameter values is shown.
It is worthwhile to note, that the constraint of perturbativity up
to the GUT or Planck scale is based on the assumption, that
the NMSSM is valid up to the very high GUT or Planck scale.
In spite of the unification of gauge couplings, which indicate
that this could be true, this is not granted even if
the NMSSM may be correct as an effective theory at
the electroweak scale.\\

% Fine-tuning quantified
One of the main motivations for the
introduction of supersymmetry is to avoid the
fine-tuning problem we encounter in the SM.
that is, we expect the 
model under consideration not to reintroduce
fine-tuning again. In order to quantify
fine-tuning an appropriate quantity is introduces~\citep{Barbieri:1987fn}, 
\begin{equation}
\label{eq-fine}
F = \mathop{\max}_{a_i} \left| \frac{\D \ln (m_Z(a_i))}{\D \ln ( a_i)} \right| \,,
\end{equation} 
with $a_i$ denoting all soft supersymmetry breaking parameters.
Note, that in the original work the $Z$-boson mass occurs squared in
this expression.
With help of this quantity~$F$, parameter sets within a certain model 
may be compared on a
quantitative basis and the
parameters giving lower values of $F$ correspond to a more {\em natural}
parameter choice. Moreover this allows also for a comparison
of different models with respect to fine-tuning. 
This approach is employed 
in some studies on the Higgs-boson spectrum in the NMSSM; 
see Sect.~\ref{sect-scan}.\\

A further restriction for the parameters in the potential comes from the requirement
of a color and electric charge-invariant vacuum~\citep{Frere:1983ag,Gunion:1987qv}.
This is evident if we consider for instance the scalar part
of the slepton--slepton--Higgs superpotential term
$-{\tilde{e}_R^*} y_e (\tilde{L}^\trans \epsilon H_d)$;
see~\eqref{eq-Wscalar2}.
With view on Tab.~\ref{NMSSM-cont} we see that the supermultiplets have
hypercharges $Y(\tilde{e}_R)=1$, $Y(\tilde{L})= -1/2$ and $Y(H_d)=-1/2$ and
thus, the superpotential term is invariant under $U_Y(1)$ transformations. However
a non-zero vacuum-expectation-value of the scalar field
$\tilde{e}_R$ corresponds to a electric charge breaking minimum.
In a analogous way also color breaking minima may arise from
the potential.
The undesired global minima of the
scalar fields can be translated into 
charge- and color-breaking bounds (CCB). Let us
sketch the bounds found on the $A$-parameter, where we 
follow closely~\citep{Gunion:1987qv}.
We start with a generic trilinear superpotential term 
$W_\phi= \lambda \hat{\phi}_1 \hat{\phi}_2 \hat{\phi}_3$ with a corresponding
scalar part
$W_\phi= \lambda \phi_1 \phi_2 \phi_3$. As shown in App.~\ref{app-A} we get
from this superpotential term  a {\em physical} potential by collecting
the F-terms, D-terms as well as the corresponding soft supersymmetry
breaking terms, yielding mass terms and trilinear $A$-parameter terms. The physical
potential thus reads
\begin{equation}
\begin{split}
V_\phi(\phi_1, \phi_2, \phi_3) =& 
|\lambda|^2 \left( |\phi_1|^2 |\phi_2|^2 + |\phi_2|^2 |\phi_3|^2 +|\phi_1|^2 |\phi_3|^2 \right)\\
&+ g_a^2 \left( Y_{\phi_1}^a |\phi_1|^2 + Y_{\phi_2}^a |\phi_2|^2 + Y_{\phi_3}^a |\phi_3|^2 \right)^2\\
&+ m_{\phi_1}^2 |\phi_1|^2 + m_{\phi_2}^2 |\phi_2|^2 + m_{\phi_3}^2 |\phi_3|^2\\
&+ \left ( A \lambda \phi_1 \phi_2 \phi_3 + c.c. \right) \,.
\end{split}
\end{equation}
Here we denote by $Y_{\phi_i}^a$, $i=1,2,3$ the eigenvalues of the gauge group generators with adjoint
index $a$ and corresponding couplings $g_a$, originating from the D-terms. Since the initial 
superpotential term (along with its derived Lagrangian terms) is supposed
to be gauge invariant we have to have $Y_{\phi_3}^a = -(Y_{\phi_1}^a+Y_{\phi_2}^a)$. Now we are looking
for the global minimum of the potential $V_\phi$. To this end we examine the directions in
field space with $\phi\equiv\phi_1=\phi_2=\phi_3$. In this direction in field space,
the so-called {\em D-flat direction}, the
quartic D-terms of the potential vanish, 
and do not protect the potential from a stationary
solution for non-vanishing fields. 
Moreover, in the D-flat direction the potential becomes very simple, that is
\begin{equation}
V_\phi(\phi) = 3 |\lambda|^2 |\phi|^4 
+ \left(m_{\phi_1}^2+m_{\phi_2}^2+m_{\phi_3}^2\right) |\phi|^2- 2 A \lambda \phi^3\,.
\end{equation}
Of course we have the desired stationary solution for $\phi=0$ with $V_\phi(0)=0$. 
In order to avoid a vacuum for non-vanishing fields and thus to avoid a charge breaking minimum,
there has {\em not} to be a stationary solution with a negative potential value.
This eventually restricts the $A$-parameter not to be too large, that is
we find the constraint
\begin{equation}
\label{eq-Aconstraint}
A^2 < 3 \left(m_{\phi_1}^2+m_{\phi_2}^2+m_{\phi_3}^2\right)\,.
\end{equation}\\

A further approximative constraint arises from the Higgs potential with respect to the 
Higgs singlet~\citep{Frere:1983ag,Gunion:1987qv,Djouadi:2008uj}.
The dominant singlet-dependent part of the Higgs potential~\eqref{eq-V}
reads
\begin{equation}
\label{eq-VSapprox}
V_S(S) = \kappa^2 S^4 + \frac{2}{3} \kappa A_\kappa S^3 + m_S^2 S^2 +...\;,
\end{equation}
where the ellipsis denote terms which have a lower
dependence on $S$.
Generally, we want to have a non-vanishing $\mu$-term, with $\mu= \lambda v_s$, requiring
a non-vanishing vacuum-expectation-value $v_s$.
That is, in this case we have to have a minimum with a lower potential value compared to
the symmetric minimum at $V_S(0)\approx 0$. This immediately translates into
the approximate parameter condition
\begin{equation}
\label{eq-msconstraint}
m_S^2 \lesssim \frac{1}{9} A_\kappa^2\,.
\end{equation}
Note, that this relation is only approximatively valid, since
terms in the potential~\eqref{eq-VSapprox} are neglected.\\

\begin{figure}[h] 
\centering
\includegraphics[width=0.45\linewidth, angle=0,clip]{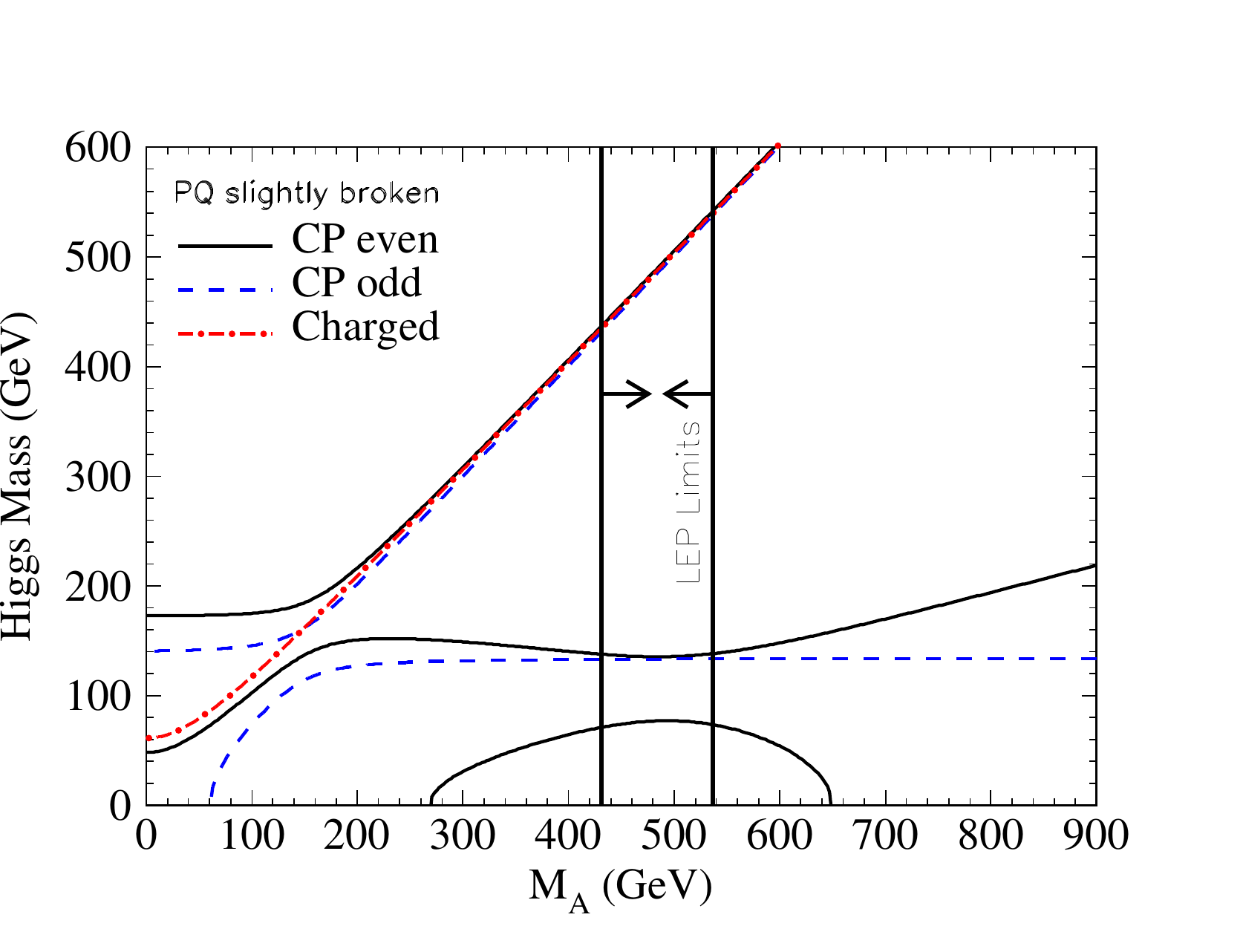}
\caption{\label{slightlykappaZerwas} \it 
The one-loop Higgs-boson mass spectrum as a function of $M_A$, the upper left
entry in the CP-odd mixing matrix. The remaining parameters are fixed to
$\kappa=0.1$, depicted as {\em slightly broken PQ-symmetry}, $v_s=3v$,
$\tan(\beta)=3$ and $A_\kappa=-100$~GeV. The arrows denote the region
allowed by LEP searches with 95\% confidence. Figure taken from~\citep{Miller:2003ay}.
}
\end{figure}

In~\citep{Miller:2003ay} the impact 
of the vacuum stability constraints as well
as the experimental constraints on the Higgs-boson sector parameter space
is studied.
Simple analytic expressions for the physical Higgs masses in
the CP-conserving case are derived, taking the one-loop contributions
to the Higgs-boson masses from top and stop loops into account.
Using a power expansion in both $1/\tan(\beta)$ and $1/M_A$, where
$M_A$ is the upper left entry in the pseudoscalar mass 
squared matrix \eqref{eq-MPMA}, analytic expressions are found,
approximative valid for moderate or large values of
$\tan (\beta)$ and a larger scale $M_A$. Three distinct
regions are considered with respect to
the PQ-symmetry breaking parameter $\kappa$.
The region with
vanishing $\kappa$ corresponds to 
the PQ-symmetric NMSSM, the region 
with large values of $\kappa$, denoted
as the NMSSM with {\em strongly broken} PQ-symmetry
and the region with small values of $\kappa$, that is, $\kappa \ll 1$,
denoted as {\em slightly broken} PQ-symmetry.
Of course, the negative searches for a light axion
on the one hand and the requirement of absence
of a Landau-pole through renormalization group 
equations for $\kappa$ up to the GUT scale
favor the slightly broken PQ-symmetry scenario (see Sect.~\ref{sub-thconstraints}).
As an example, the Higgs-boson masses  
are plotted in a 
slightly broken
PQ-symmetry scenario, namely with $\kappa=0.1$
in Fig.~\ref{slightlykappaZerwas}. The other 
parameters choices are given in the figure caption.
Also the strong experimental constraint on the
$m_A$ parameter is indicated in this figure. Note,
that $m_A$ in general is not a CP-odd Higgs boson mass in the NMSSM,
but the upper left entry in pseudoscalar mixing matrix squared;
see~\eqref{eq-mAparameter}.
The authors point out, that the spectrum, based on their assumptions,
in the NMSSM
is quite different from what is to
expect from the MSSM. So, even if some of the Higgs bosons are too heavy to be detected, the
discovery of the lighter Higgs-bosons may allow to distinguish the NMSSM from the MSSM.

%%%%%%%%%%%%%%%%%%%%%%%%%%%%%%%%%%%%%%%%%%%%%%%%%%%%%%%%%%%%%%%%%%5
%%%%%%%%%%%%%%%%%%%%%%%%%%%%%%%%%%%%%%%%%%%%%%%%%%%%%%%%%%%%%%%%%%5
% Experimental constraints
%%%%%%%%%%%%%%%%%%%%%%%%%%%%%%%%%%%%%%%%%%%%%%%%%%%%%%%%%%%%%%%%%%5
%%%%%%%%%%%%%%%%%%%%%%%%%%%%%%%%%%%%%%%%%%%%%%%%%%%%%%%%%%%%%%%%%%5

\subsection{Experimental constraints}

%%%%%%%%%%%%%%%%%%%%%%%%%%%%%%%%%%%%%%%%%%%%%%%%%%%%%%%%%%%%%%%%%%5
% Collider constraints.
% revised 29.4.09
%%%%%%%%%%%%%%%%%%%%%%%%%%%%%%%%%%%%%%%%%%%%%%%%%%%%%%%%%%%%%%%%%%5
\subsubsection{Collider constraints}

The collider constraints give very clear bounds originating from
very different observations like the Z-boson width as well
as the direct searches of supersymmetric partner particles 
in $e^+e^-$ collisions at LEP. Here we discuss the
bounds which are applied in the NMHDECAY program~\citep{Ellwanger:2004xm,Ellwanger:2005dv};
see App.~\ref{app-tools} for an overview of some currently
available program tools with respect to the NMSSM.

% neutralino constraint

A strong constraint arises form the Z-boson precision measurements
at LEP~\citep{Amsler:2008zzb}. 
A possible contribution for the decay into the lightest and stable
neutralino pairs $Z \rightarrow \tilde{\chi}^0_1 \tilde{\chi}^0_1$ with mass
$m_{\tilde{\chi}^0_1} < m_Z/2$ may spoil the measurement of the
{\em invisible} decay width
$\Gamma_Z^{\text{invisible}} \equiv \Gamma_Z^{\text{tot}}-\Gamma_Z^{\text{visible}}$.
From a resonance scan the total width $\Gamma_Z^{\text{tot}}=2.4952 \pm 0.0023$~GeV
(along with the Z-boson mass, $m_Z=91.1876\pm 0.0021$~GeV) is known.
The visible decay width consists of decays into charged leptons 
($\Gamma_Z^{l^+ l^-}=83.984\pm0.086$~MeV, with $l=e,\mu,\tau$)
and hadrons 
($\Gamma_Z^{\text{had}}=1744.4\pm 2.0$~MeV). From this the invisible
decay width is measured as $\Gamma_Z^{\text{invisible}}=499.0 \pm 1.5$~MeV.
This is about what we expect from the SM, which predicts an
invisible part, consisting solely of neutrinos as
$\Gamma_Z^{\text{invisible, SM}}=501.3 \pm 0.6$~MeV. This means
a possible contribution of decays into neutralinos should not exceed the 
measurement too much,
\begin{equation}
\Gamma (Z \rightarrow \tilde{\chi}^0_1 \tilde{\chi}^0_1) \lesssim 2~\text{MeV}\,.
\end{equation}
Neutralinos may be produced in pairs at LEP via s-channel Z-boson or Z-boson $\gamma$
interference or via t-channel exchange of a selectron. 
A further constraint arises from searches for
pair production of non-LSP neutralinos
with subsequent decay. These processes 
were searched for at LEP up to center-of-mass
energies of 208~GeV~\citep{Abdallah:2003xe}. 
From this negative search results the limits on the cross sections
\begin{equation}
\begin{split}
\sigma (e^+ e^- \rightarrow \tilde{\chi}_1^0 \tilde{\chi}_i^0) &< 0.01~\text{pb}\,,\\
\sigma (e^+ e^- \rightarrow \tilde{\chi}_i^0 \tilde{\chi}_j^0) &< 0.1~\text{pb}
\end{split}
\end{equation}
with $i,j = 2,...,5$ can be deduced, viable for a sum of neutralino 
masses in the final state not exceeding the center-of-mass energy
of the colliding electrons.\\

Charginos could have been produced in pairs at LEP via s-channel exchange or
via t-channel exchange of a sneutrino. The negative search can be translated
into a minimal chargino mass limit of~\citep{lepsusywg}
\begin{equation}
\label{eq-charginoconstraint}
m_{\tilde{\chi}^\pm} > 103.5~\text{GeV}\,.
\end{equation}

%charged Higgs boson constraint
The production of charged Higgs bosons in pairs $e^+ e^- \rightarrow H^+ H^-$
was investigated at LEP~\citep{:2001xy} for general SM extensions with
two Higgs doublets, like the MSSM or the NMSSM. Assuming main decay channels
of the charged Higgs bosons $H^+ \rightarrow c \bar{s}$ and
$H^+ \rightarrow \tau^+ \nu_\tau$ and charged conjugated decay products for the $H^-$,
a combination of all four LEP experiments ALEPH, DELPHI, L3 and OPAL with
center-of-mass energies up to 209~GeV, give a lower mass limit of
\begin{equation}
m_{H^\pm} > 78.6~\text{GeV}\,.
\end{equation}

% neutral Higgs-boson constraints

There are constraints from the negative searches of the lightest 
CP-even Higgs boson $H_1$ based on CP-conserving
NMSSM Higgs sector studies. The dominant production at LEP 
proceeds via Higgs-strahlung off a s-channel $Z$-boson 
($e^+ e^- \rightarrow Z^* \rightarrow Z H$). 
This 
Higgs-boson production channel was searched for in various subsequent decay 
channels. The bounds derived in this way depend on the
Higgs-boson mass. Here we mention the subsequent decays
$H \rightarrow b\bar{b}$ and
$H \rightarrow \tau^+ \tau^-$~\citep{Barate:2003sz},
$H \rightarrow j j$, with $j$ denoting a jet~\citep{:2001yb,Abbiendi:2003gd},
$H \rightarrow \gamma \gamma$~\citep{lephwg},
$H \rightarrow$~invisible, that is, into LSP neutralinos~\citep{Buskulic:1993gi,:2001xz},
$H \rightarrow X$, that is, independent of the decay product $X$ in $e^+ e^- \rightarrow Z X$. 
The latter channel is accessible by studying the recoil mass spectrum in $Z \rightarrow e^+ e^-$
and $Z \rightarrow \mu^+ \mu^-$ events and by searching for 
$X \rightarrow e^+ e^-$ or $X \rightarrow \gamma \gamma$
and $Z \rightarrow \nu \bar{\nu}$~\citep{Buskulic:1993gi,Abbiendi:2002qp},
$H \rightarrow A A$ with subsequent decay 
$A A \rightarrow 4j$,
$A A \rightarrow 2j\, c \bar{c}$,
$A A \rightarrow 2j\, \tau^+ \tau^-$,
$A A \rightarrow \tau^+ \tau^- \tau^+ \tau^-$,
$A A \rightarrow c \bar{c} c \bar{c}$,
$A A \rightarrow \tau^+ \tau^- c \bar{c}$~\citep{Abbiendi:2002in}.
Also it was looked for the associated production of two Higgs bosons,
$e^+ e^- \rightarrow H A$, with subsequent decay 
$H A \rightarrow b \bar{b} b \bar{b}$,
$H A \rightarrow \tau^+ \tau^- \tau^+ \tau^-$, and
$H A \rightarrow A A A \rightarrow 3 (b \bar{b})$~\citep{Abdallah:2004wy}.

\newpage
%%%%%%%%%%%%%%%%%%%%%%%%%%%%%%%%%%%%%%%%%%%%%%%%%%%%%%%%%%%%%%%%%%5
% Muon anomalous magnetic moment
% revised 29.4.09
%%%%%%%%%%%%%%%%%%%%%%%%%%%%%%%%%%%%%%%%%%%%%%%%%%%%%%%%%%%%%%%%%%5
\subsubsection{Muon anomalous magnetic moment}

The muon anomalous magnetic moment is a quantum effect which 
is very valuable in studying the possible deviations from the SM.
For a recent review we refer the reader to St\"{o}ckinger~\citep{Stockinger:2006zn}.

The magnetic moment of an electron or muon is
\begin{equation}
\tvec{\mu} = g \bigg( \frac{e}{2 m} \bigg) \tvec{s}
\end{equation}
with $m$ mass, $e$ charge, and $\tvec{s}$ spin of the electron or muon and $g$ 
the gyromagnetic ratio, which is $g=2$ following from
the Dirac equation at the classical level. Quantum effects lead
to a deviation from this value, called the anomalous magnetic moment
\begin{equation}
\label{eq-defa}
a \equiv \frac{1}{2} (g-2)\,.
\end{equation}
The one-loop QED correction was first calculated by Schwinger and is
$a=\frac{\alpha}{2 \pi}$~\citep{Schwinger:1948iu}. 
\begin{figure}[h] 
\centering
\includegraphics[width=0.6\linewidth, angle=0,clip]{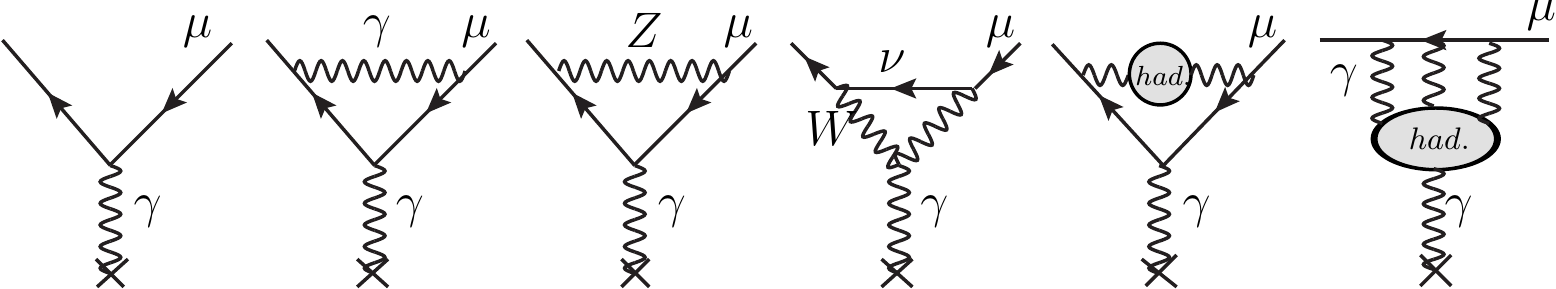}
\caption{\label{fig-muonan}
Contributions to the muon magnetic moment. From left to right the first diagram gives the
contribution to the {\em normal} gyromagnetic ratio $g=2$. The deviations to $g=2$, that is
the {\em anomalous magnetic moment} come from the quantum corrections shown
in the five right diagrams.
The second diagram gives the leading QED Schwinger contributing 
$\frac{\alpha}{2 \pi}$ to $a$ in
\eqref{eq-defa}, the third and forth diagrams are the leading electroweak contributions.
The fifth diagram shows
the leading hadronic contribution via vacuum polarization and the last one
the hadronic light-by-light contribution; see the review~\citep{Stockinger:2006zn}. 
\it }
\end{figure}
The contributions to the muon magnetic moment are shown in Fig~\ref{fig-muonan}. 
Generally, loop contribution from heavy particles with mass $M$ are suppressed by
a factor $m^2/M^2$~\citep{Stockinger:2006zn}. This makes clear that the anomalous magnetic
moment of the muon, $a_\mu$, is enhanced by a factor $(m_\mu/m_e)^2 \approx 40,000$ compared
to the electron with respect to these contributions.

For a muon anomalous magnetic moment calculation in the
SM we refer to the more recent publications mentioned by the Particle Data Group,
\citep{Davier:2002dy,Davier:2003pw,deTroconiz:2004tr,Hagiwara:2006jt,Davier:2007ua,Jegerlehner:2007xe}.
For the corresponding calculation of the
muon anomalous magnetic moment in SUSY models see
\citep{Lopez:1993vi,Chattopadhyay:1995ae,Moroi:1995yh,Carena:1996qa,Stockinger:2006zn,Hertzog:2007hz,Marchetti:2008hw}.

\begin{figure}[h] 
\centering
\includegraphics[width=0.4\linewidth, angle=0,clip]{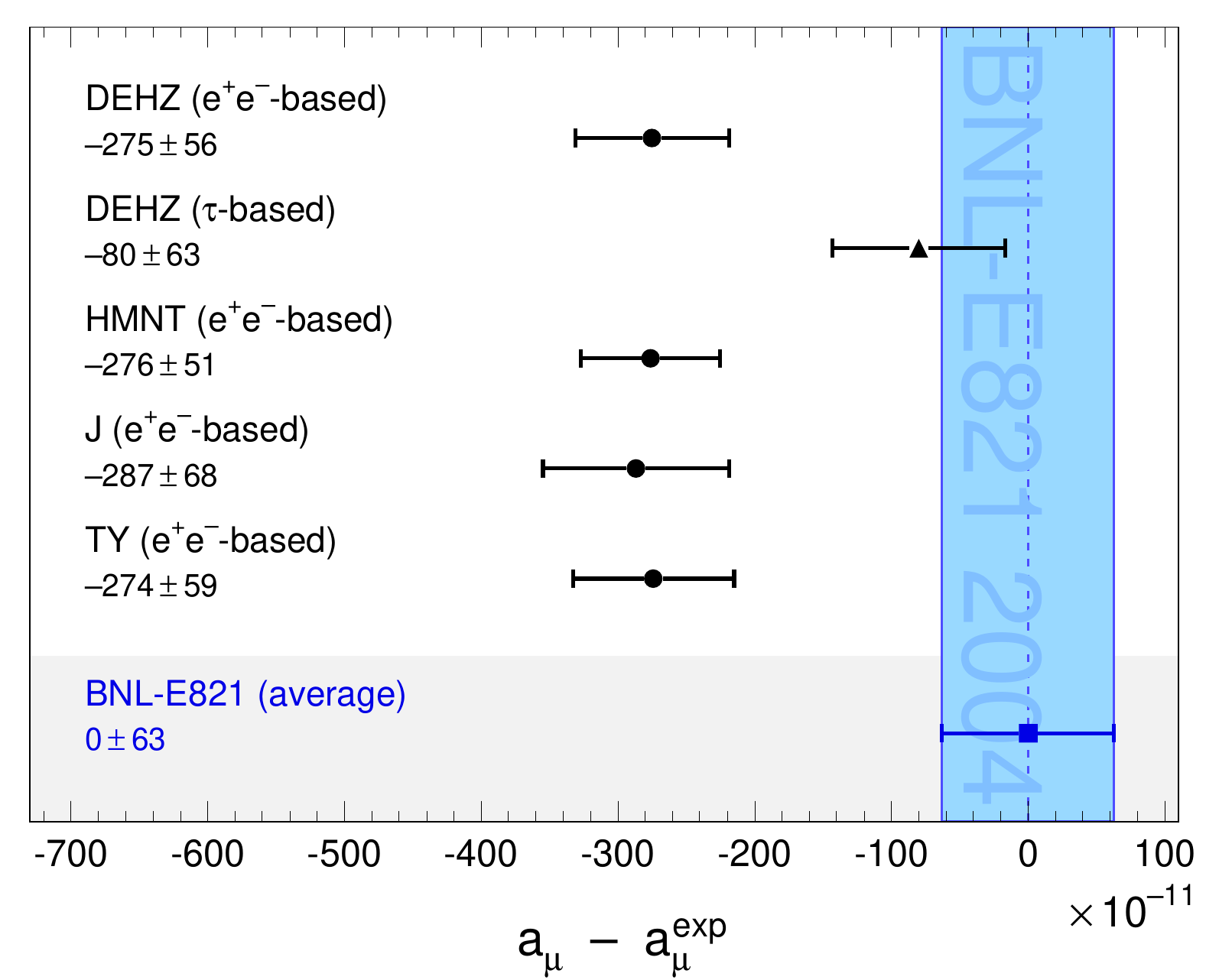}
\caption{\label{fig-amures}
Recently published predictions of the SM anomalous magnetic moment of the muon $a_\mu$, subtracted from the
mean experimental value from Brookhaven E821, as presented by the 
Particle Data Group~\citep{Amsler:2008zzb}.
The dots are predictions based on $e^+e^-$ data, the small triangle the
prediction based on $\tau$ data and the experimental measurement
is given by a little square.
The predictions are taken from
\citep{Davier:2002dy,Davier:2003pw,deTroconiz:2004tr,Hagiwara:2006jt,Davier:2007ua,Jegerlehner:2007xe}.
\it }
\end{figure}

Let us briefly present the current results.
The Brookhaven $g-2$ experiment E821~\citep{Bennett:2004pv,Bennett:2006fi} measures 
an anomalous magnetic moment of the muon of
\begin{equation}
\label{eq-amuex}
a^{\text{exp}}_\mu = (116,592,080 \pm 63) \times 10^{-11}
\end{equation}
compared to the prediction of the SM, see the review \citep{Miller:2007kk},
\begin{equation}
a^{SM}_\mu = (116,591,785 \pm 61) \times 10^{-11}\,.
\end{equation}
Thus we get a deviation of
\begin{equation}
\label{eq-anmudev}
\delta a_\mu = a^{\text{exp}}_\mu - a^{SM}_\mu = (295 \pm88) \times 10^{-11}\,,
\end{equation}
where we added the errors quadratically. If we trust the measurement as well as the
SM prediction, this is a more than 3--$\sigma$ deviation. 
The situation is presented by the Particle Data Group \citep{Amsler:2008zzb}
and displayed in Fig.~\ref{fig-amures}. In this figure the deviations of the
predictions from the BNL measurement are shown. Note that the prediction based
on the $\tau$ data which enter 
the hadronic vacuum polarization contribution do not deviate very much from the measurement,
whereas all other predictions 
based on $e^+ e^-$ data
show a large deviation. 
With an E821 upgrade proposal, called
E969~\citep{LeeRoberts:2005uy},
the error is aimed to be at least halved and with the same improvement of the
theoretical predictions~\citep{Miller:2007kk}, 
the 5--$\sigma$ discovery limit may in principle be reached.\\

\begin{figure}[h] 
\centering
\includegraphics[width=0.35\linewidth, angle=0,clip]{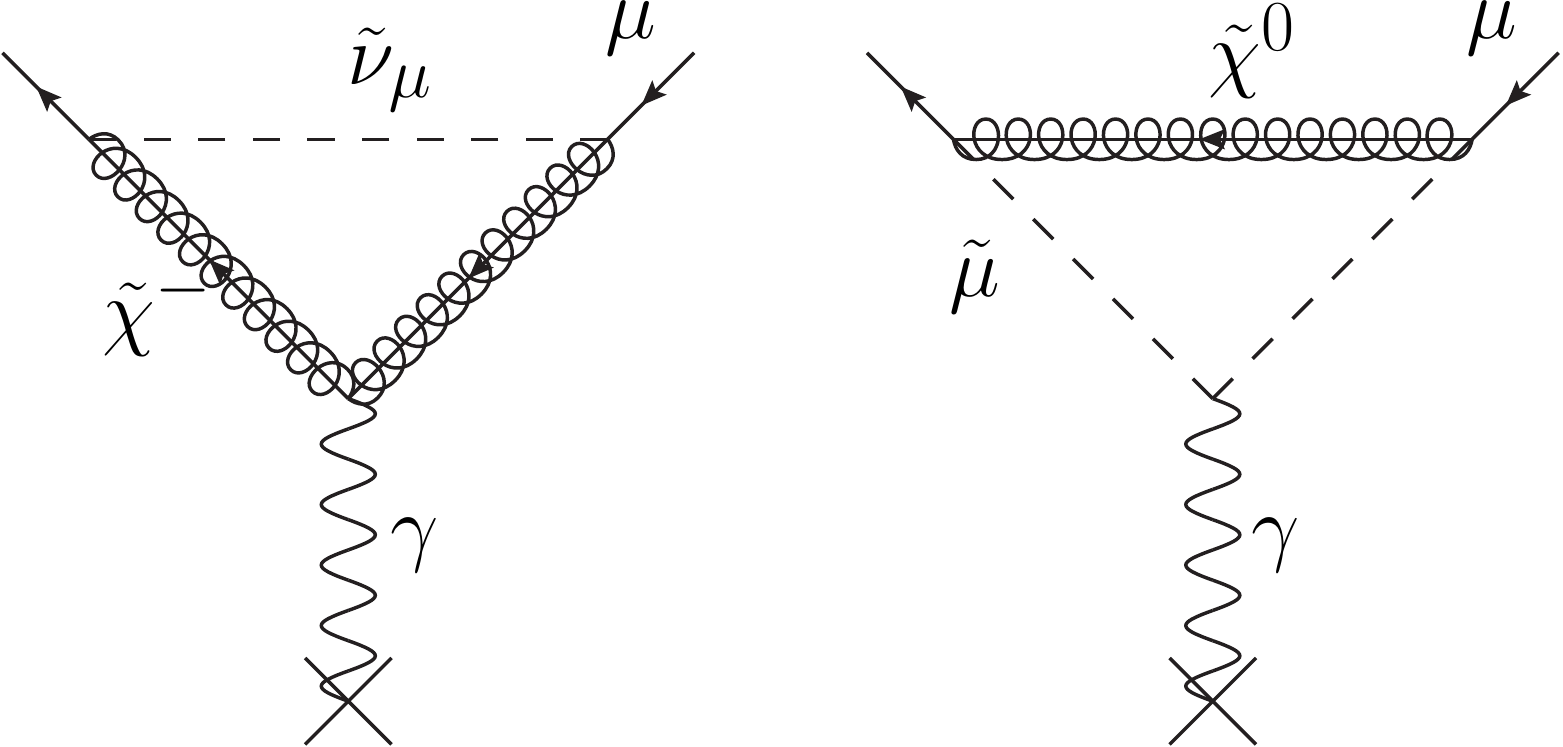}
\caption{\label{fig-amuresNMSSM}
Additional Feynman diagrams contributing to the anomalous magnetic moment
of the muon $a_\mu$ in the MSSM and the NMSSM.
\it }
\end{figure}

Of course supersymmetry changes the predictions of $a_\mu$, since we get also
quantum correction contributions from the superpartners. The outcome depends on the masses and couplings,
not to mention the supersymmetric model under consideration itself.
In any case, the experimental result \eqref{eq-amuex} 
gives a strong constraint on the model.
Additional Feynman diagrams to lowest order
contributing to the anomalous magnetic moment $a_\mu$ in the MSSM and
the NMSSM are shown in Fig.~\ref{fig-amuresNMSSM}.
For large $\tan(\beta)$ the dominant supersymmetric contribution comes
from the chargino--sneutrino loop diagram and is approximately given by~\citep{Czarnecki:2001pv}
\begin{equation}
|\delta a_\mu^{\text{SUSY}}| \approx
\frac{\alpha}{8 \pi s_W^2}
\frac{m_\mu^2}{\tilde{M}^2} \tan(\beta)\;
\left(1 - \frac{4 \alpha}{\pi} \ln\left( \frac{\tilde{M}}{m_\mu}\right) \right)\;,
\end{equation}
with $\alpha$ the fine structure constant and $\tilde{M}$ the heavier
of the masses in the loop, that is, either the
mass of the chargino or the sneutrino.

\begin{figure}[h] 
\centering
\includegraphics[width=0.45\linewidth, angle=0,clip]{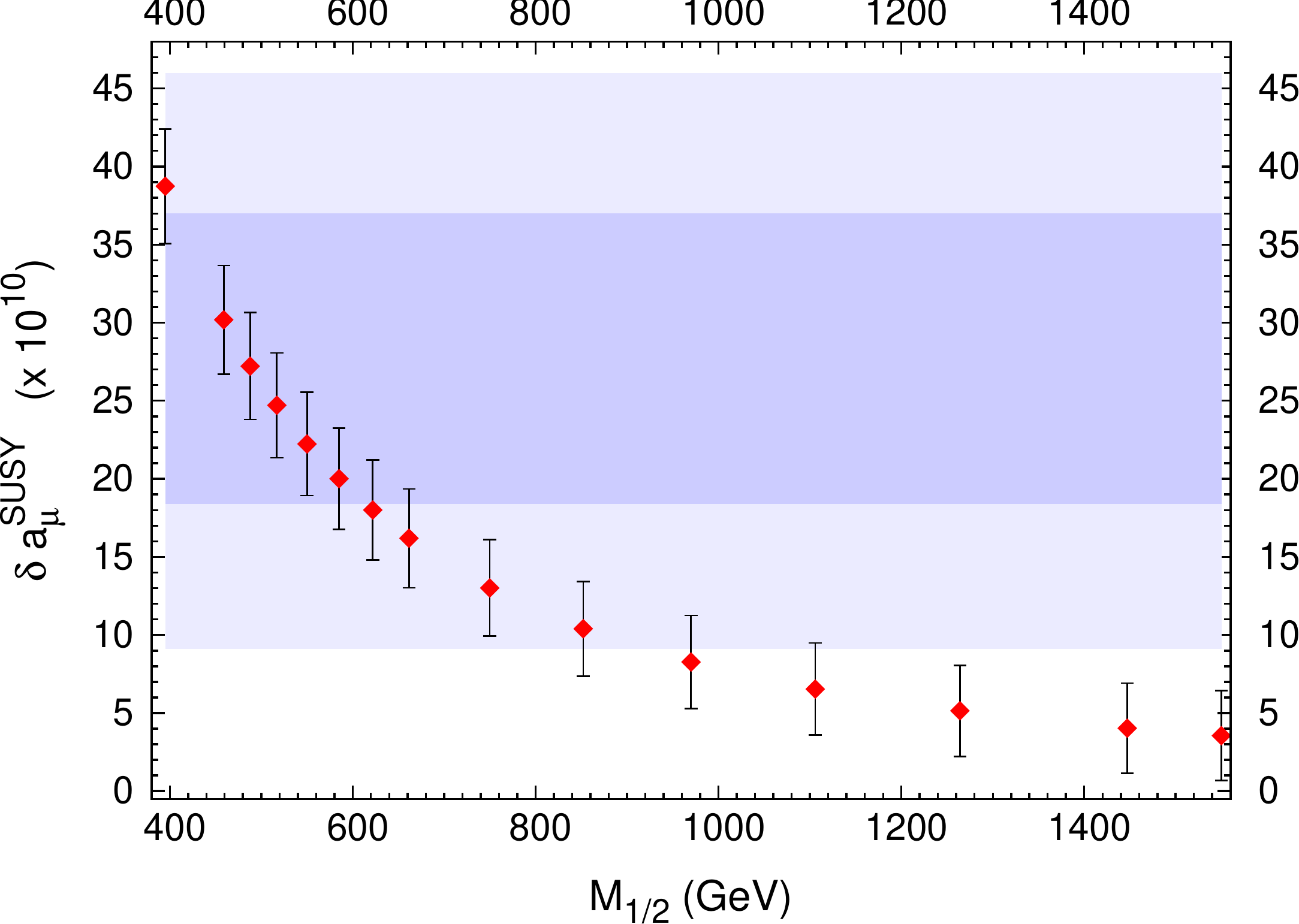}
\caption{\label{fig-g-2ellwanger} \it
$\delta a_\mu^{\text{SUSY}}$ in the cNMSSM as a function
of the unified gaugino mass $M_{1/2}$.
The darker and brighter shaded areas show the 1--$\sigma$
respectively 2--$\sigma$ deviations from
$\delta a_\mu = a^{\text{exp}}_\mu - a^{SM}_\mu$.
Figure taken from \citep{Djouadi:2008uj}. }
\end{figure}

The additional contribution to $a_\mu$ in the constrained NMSSM (cNMSSM)
(The cNMSSM is introduced in App.~\ref{sec-varNMSSM})
is shown in Fig.~\ref{fig-g-2ellwanger} in dependence
on the unified gaugino mass $M_{1/2}$, where the unified scalar mass $m_0$ is
fixed to zero~\citep{Djouadi:2008uj}. We see in this plot that
the cNMSSM gives for $M_{1/2} \lesssim 1$~TeV a contribution in
agreement with the observation at BNL.\\

It is worthwhile to add some critical remarks: first of all the
experimental value with this high precision of ($0.54$ ppm) is
not confirmed by an alternative laboratory.
Secondly, the uncertainty of the predictions is obvious from the
deviating results based on the one hand on $e^+e^-$ data
and on the other hand on $\tau$ data.
Moreover, even if the theoretical predictions of the hadronic leading
contribution may be under control, 
the hadronic light by light scattering contribution (HLLS) is
more or less estimated with a rather large contribution 
of $a_\mu^{\text{HLLS}}=110 \pm 40$ ~\citep{Miller:2007kk,Bijnens:2007pz}.
Some effort is currently done to make this prediction more 
reliable; see~\citep{Stockinger:2006zn}. 

\newpage
%%%%%%%%%%%%%%%%%%%%%%%%%%%%%%%%%%%%%%%%%%%%%%%%%%%%%%%%%%%%%%%%%%5
% B-meson decay
%%%%%%%%%%%%%%%%%%%%%%%%%%%%%%%%%%%%%%%%%%%%%%%%%%%%%%%%%%%%%%%%%%5
\subsubsection{B-meson decay}

\begin{figure}[h] 
\centering
\includegraphics[width=0.4\linewidth, angle=0,clip]{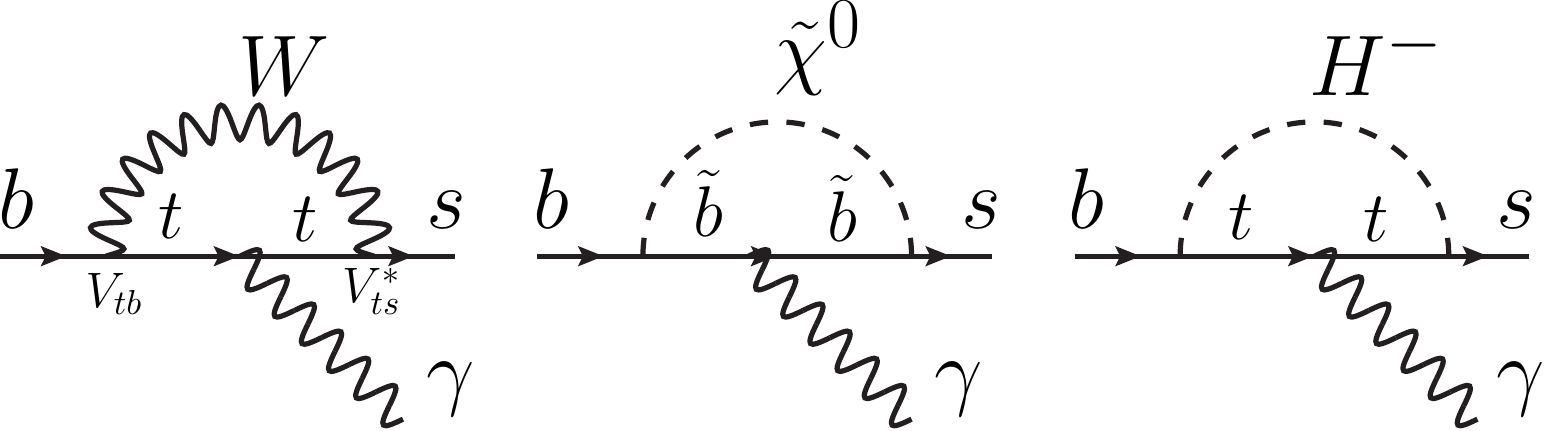}
\caption{\label{fig-bsg}
Example diagrams contributing to the $b \rightarrow s \gamma$ decay.
The left diagram is a SM contribution, where also the
CKM-matrix entries $V_{tb}$ and $V_{ts}^*$ are depicted. The mid- and right-diagrams
show contributions occurring in supersymmetry via a neutralino--sbottom
loop and a Higgs-boson--top loop, respectively.
\it }
\end{figure}

%Cerdeno S.6
The flavor changing neutral current decay $\Gamma(b \rightarrow s \gamma)$
is very sensitive to physics beyond the SM. The reason is, that
this decay can only proceed via loops in the SM, thus is
suppressed.
Hypothetically postulated new particles may contribute to the decay.
In supersymmetry there are lots of contributions, two examples are
shown in Fig.~\ref{fig-bsg}, that is, a neutralino-sbino loop and
a Higgs-boson--top loop.
However a quite good agreement of experimental observation and
SM prediction of the $b \rightarrow s \gamma$ decay 
is found not allowing for a large additional total contribution.
But, on the other hand, stringent
constraints are put on every new model which postulates new particles
which may contribute to this decay channel.

The experimental world average for the branching ratio (for a minimal
photon energy of $E_{\text{min}}=1.6$~GeV) is presented
by the Heavy Flavor Averaging Group \citep{Barberio:2008fa}
based on the measurements of 
BABAR~\citep{Aubert:2007my},
Belle~\citep{Abe:2001hk} and 
CLEO~\citep{Chen:2001fja}:
\begin{equation}
\BR^{\text{exp}}(b \rightarrow s \gamma) = (3.52\pm0.25) \times 10^{-4}\,.
\end{equation}

In contrast, the corresponding SM prediction~\citep{Gambino:2005dp} is, including
the charm loop~\citep{Gambino:2001ew} as well as two-loop contributions
\citep{Adel:1993ah,Misiak:1994zw,Ali:1995bi,Pott:1995if,Greub:1996tg,Chetyrkin:1996vx,Greub:1997hf,Buras:1997xf,Kagan:1998ym,Gambino:2001ew,Buras:2002tp,Hiller:2004ii}
\begin{equation}
\BR^{\text{SM}}(b \rightarrow s \gamma) = (3.73\pm0.30) \times 10^{-4}\,.
\end{equation}
That is, within the errors we have good agreement of the experimental data
with the SM prediction. Thus, additional contributions from extensions
of the SM are very restricted in case they contribute to
the $b \rightarrow s \gamma$ decay.

\newpage
%%%%%%%%%%%%%%%%%%%%%%%%%%%%%%%%%%%%%%%%%%%%%%%%%%%%%%%%%%%%%%%%%%5
%%%%%%%%%%%%%%%%%%%%%%%%%%%%%%%%%%%%%%%%%%%%%%%%%%%%%%%%%%%%%%%%%%5
% Cosmological constraints
%%%%%%%%%%%%%%%%%%%%%%%%%%%%%%%%%%%%%%%%%%%%%%%%%%%%%%%%%%%%%%%%%%5
%%%%%%%%%%%%%%%%%%%%%%%%%%%%%%%%%%%%%%%%%%%%%%%%%%%%%%%%%%%%%%%%%%5
\subsection{Cosmological constraints}
\label{sub-coscon}

Supersymmetric extensions of the SM predict a stable lightest
supersymmetric particle (LSP), as long as matter parity (or R-parity)
is conserved. This LSP may account
for the observed cold dark matter in the Universe. In the early Universe the LSP
would be present in large numbers in thermal equilibrium. As the Universe
has cooled the LSP's only may reduce their density by 
pair annihilation and co-annihilation~\citep{Hut:1977zn,Vysotsky:1977pe,Lee:1977ua}. 
As their density decreases, the chance to find another particle to
annihilate decreases and at a certain point the comoving density 
can become constant (``freeze out''). The remaining {\em relic density} can
be calculated depending on the considered model. In case the {\em relic density} of
the LSP is identified with the cold dark matter candidate, a comparison
with the cold dark matter observations constrain the model.
We distinguish between {\em direct}
detection of the relic density LSP particle in a detector and the {\em indirect} detection
via astrophysical observations.\\

%%%%%%%%%%%%%%%%%%%%%%%%%%%%%%%%%%%%%%%%%%%%%%%%%%%%%%%%%%%%%%%%%%5
% Indirect dark matter detection
%%%%%%%%%%%%%%%%%%%%%%%%%%%%%%%%%%%%%%%%%%%%%%%%%%%%%%%%%%%%%%%%%%5
\subsubsection{Indirect dark matter detection}

The most stringent bound comes from the 
cosmic microwave background (CMB) observation performed by WMAP.
The observed CMB is compared to the predictions of the standard model
of cosmology, the so-called 
$\Lambda$-cold-dark-matter model ($\Lambda$CDM). This model depends
on six parameters,
the Hubble parameter $H_0$, the baryon density $\Omega_b h_0^2$,
the cold dark matter density $\Omega_c h_0^2$, 
the dark energy density $\Omega_{\Lambda}$,
the scalar spectral index $n_s$,
the optical depth $\tau$,
as well as the variation of the spectral index $n_s$ and
the curvature perturbation $\Delta_{\cal R}^2$.
%the scalar fluctuation amplitude $A_s$, as well as .
By adjusting the parameters of the $\lambda$CDM, the
prediction of the CMB power-spectrum can
be fitted to the observation. 
Taking into account the recent five year 
WMAP data, combined with measurements of type Ia supernovae and 
baryon acoustic oscillations in the distribution of galaxies, the
current 1--$\sigma$ limits on the $\Lambda$CDM parameters are given as~\citep{Hinshaw:2008kr}
\begin{equation}
\begin{split}
\label{eq-WMAP}
\Omega_b h_0^2 &= 0.02267^{+ 0.00058}_{-0.00059}\;,\\
\Omega_c h_0^2 &= 0.1131 \pm 0.0034\;,\\
\Omega_{\Lambda} &= 0.726 \pm 0.015\;,\\
n_s &= 0.960 \pm 0.013\;,\\
\tau &=0.084 \pm0.016\;,\\
\Delta_{\cal R}^2 &= (2.445 \pm 0.096) \times 10^{-9} \text{ at } k=0.002~\text{Mpc}^{-1}\,,
\end{split}
\end{equation}
where we are in particular interested on the cold dark matter density $\Omega_c h_0^2$
constraint. Note that the Hubble constant is defined in terms of $h_0$, 
that is, $H_0=100\; h_0$~km/s/Mpc.\\

Now let us briefly sketch the calculation of the cold dark matter density in
supersymmetric models. 
This relic density of the LSP as cold dark matter candidate, supposed to be a neutralino, can be computed
by the solution of the continuity equation \citep{Lee:1977ua,Vysotsky:1977pe}
\begin{equation}
\label{eq-cont}
\frac{d}{dt} (n_{\tilde{\chi}} R^3) = - \langle \sigma_{\text{ann}} v\rangle
\left( n_{\tilde{\chi}}^2 - (n_{\tilde{\chi}}^{\text{eq}})^2 \right) R^3\,.
\end{equation}
This continuity equation states that the number of neutralinos,
with density $n_{\tilde{\chi}}$, in a comoving volume with radius~$R$, 
is governed by the competition
between annihilations and creations. The creations depend on the
density of neutralinos in equilibrium $n_{\tilde{\chi}}^{\text{eq}}$. 
It is assumed that the
annihilating particles have non-relativistic relative velocity $v$.
Further, it is assumed that the annihilating particles
as well as the annihilation products are maintained in kinetic equilibrium
with the background thermal plasma through rapid scattering processes
\citep{Bernstein:1985th,Gondolo:1990dk}.
The model-dependent input is given by the thermally averaged
annihilation cross section $\langle \sigma_{\text{ann}} v\rangle= a + b v^2$ with
parameters $a$ and $b$.
With this annihilation cross section the 
cosmic neutralino density can be deduced \citep{Griest:1989zh,Kolb:1990vq,Drees:1992am}
\begin{equation}
\Omega_{\tilde{\chi}^0} h_0^2= 
\frac{1.07 \times 10^9\, x_{fr}}{\sqrt{g_*} M_{P} (a+\frac{3 b}{x_{fr}})}
\frac{1}{\text{GeV}}\;,
\end{equation}
with $M_{P}$ the Planck mass and $g_* \approx 81$ the effective
number of degrees of freedom at the freeze-out temperature $T_{fr}$ and
$x_{fr}=m_{\tilde{\chi}}/T_{fr}$, where $m_{\tilde{\chi}}$ is the
neutralino mass. 
In models like the MSSM and the NMSSM the annihilation cross section 
can be computed from s-channel Higgs and Z-boson exchange
into fermions and gauge bosons, and from
heavier neutralino or chargino 
t-channel exchange with gauge bosons in the final state.
In particular for s-channel resonances, rapid
annihilation can take place, reducing the relic density strongly.
In the NMSSM, in contrast to the MSSM we have additional Higgs-boson
resonances, contributing to the annihilation.
In the case of light Higgs bosons in the NMSSM also new
decay channels of neutralinos decaying into a $Z$ boson and a Higgs boson or
two light Higgs bosons are available 
\citep{Greene:1986th,Olive:1990aj,Abel:1992ts,Stephan:1997ds,Belanger:2005kh,Gunion:2005rw}.
Contrary, a large singlino component of the LSP reduces the annihilation
cross section, since the singlino does not couple to
gauge bosons at tree level.\\

\begin{figure}[htbp] 
\centering
\includegraphics[width=0.4\linewidth, angle=0,clip]{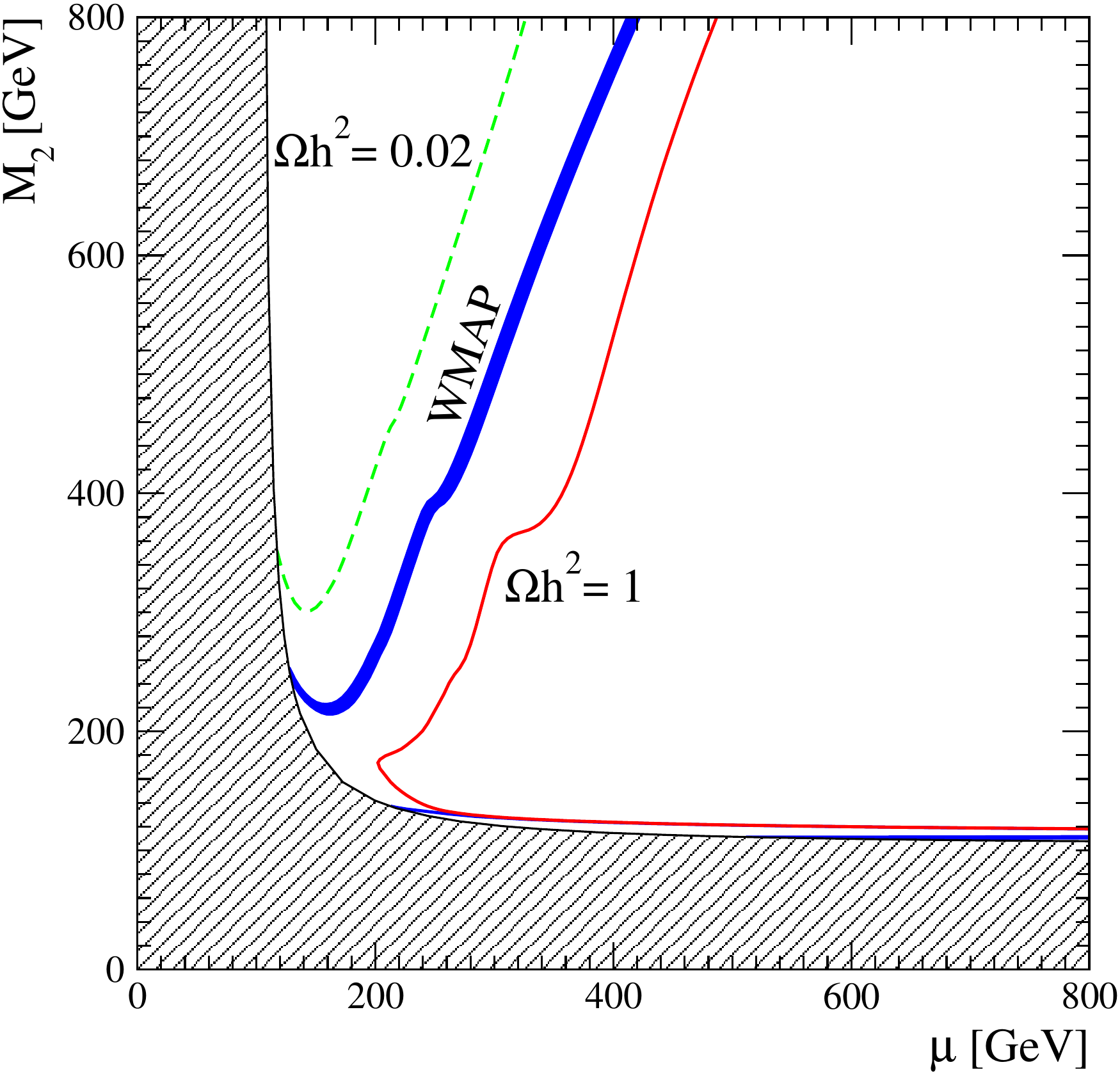}
\caption{\label{fig-belanger} \it 
Contours for $\Omega_{\tilde{\chi}^0} h_0^2 = 0.02$, $0.0945$,
$0.1287$, $1$ in the $\mu$--$M_2$ plane, as 
presented by~\citep{Belanger:2005kh}. The remaining parameters
are fixed to $\lambda=0.1$, $\kappa=0.1$, $\tan(\beta)=5$,
$A_\lambda=500$~GeV, $A_\kappa=0$. The gaugino masses are
assumed to unify at the GUT scale, corresponding at
the electroweak scale to $M_2= 2 M_1$ and
$M_{\tilde{g}}=3.3 M_2$. The hatched region is
parameter space which is excluded by LEP due to searches for chargino
pairs~\eqref{eq-charginoconstraint};
see~\citep{Belanger:2005kh} for further details.
}
\end{figure}

In the investigation of Belanger et al.~\citep{Belanger:2005kh}
the parameters of the NMSSM are
scanned in particular with view on the dark matter constraint in~\eqref{eq-WMAP}.
In this publication the program package micrOMEGAs~\citep{Belanger:2001fz}
is employed, which, starting from the particle spectrum and couplings, computes
the relevant annihilation and co-annihilation neutralino cross sections
and eventually the relic density. The particle and coupling spectrum 
is provided by the NMHDECAY program~\citep{Ellwanger:2005dv}.
Assuming gaugino mass unification at the GUT scale ($M_{1/2}$),
the independent parameters are $\lambda$, $\kappa$, $\tan(\beta)$,
$\mu=\lambda v_s$, $A_\lambda$, $A_\kappa$.
The authors find parameter space
corresponding to a large singlino component of the neutralino LSP.
This parameter space passes also the current collider constraints.
Some representative examples are given in Tab.~\ref{tab-belanger}~\citep{Belanger:2001fz}.
\begin{table}[htbp]
 \centering
 \begin{tabular} {l|l|l|l|l|l|l}
 \hline
 parameters & 1 & 2 & 3 & 4 & 5 & 6\\
 \hline
 $\lambda$       & 0.6    & 0.24   & 0.4    & 0.23   & 0.31   & 0.0348 \\
 $\kappa$        & 0.12   & 0.096  & 0.028  & 0.0037 & 0.006  & 0.0124 \\
 $\tan(\beta)$   & 2      & 5      & 3      & 3.1    & 2.7    & 5 \\
 $\lambda v_s$~[GeV]     & 265    & 200    & 180    & 215    & 210    & 285 \\
 $A_{\lambda}$~[GeV]      &  550   & 690    & 580    & 725    & 600    & 50 \\
 $A_{\kappa}$ ~[GeV]     & -40    & -10    & -60    & -24    & -6     & -150 \\
 $M_2$~[GeV]         & 1000   & 690    & 660    & 200    & 540    & 470 \\
 \hline
 \multicolumn{7}{l}{spectrum}\\
 \hline
 $m_{\chi^0_1}$~[GeV]      & 122    & 148    & 35     & 10     & 15     & 203 \\
 $U_{13}^2+U_{14}^2$ & 0.12   & 0.29   & 0.12   & 0.03   & 0.06   & 0.02 \\
 $U_{15}^2$          & 0.88   & 0.69   & 0.87   & 0.95   & 0.94   & 0.96 \\
 $m_{\chi^0_2}$~[GeV]      & 259    & 199    & 169    & 87     & 182    & 214 \\
 $m_{\chi^\pm_1}$~[GeV]      & 258    & 193    & 171    & 139    & 196    & 266 \\
 $m_{H_1}$~[GeV]     & 117    & 116    & 36     & 22     & 34     & 115 \\
 $R_{13}^2$          & 0.88   & 0.04   & 0.98   & 1.00   & 1.00   & 0.04 \\
 $m_{H_2}$~[GeV]     & 128    & 158    & 117    & 114    & 113    & 163 \\
 $R_{23}^2$          & 0.11   & 0.96   & 0.01   & 0.00   & 0.00   & 0.96 \\
 $m_{A_1}$~[GeV]     & 114    & 59     & 56     & 18     & 15     & 214 \\
 $R_{45}^{2}$       & 0.99   & 1.00   & 0.99   & 1.00   & 0.99   & 1.00 \\
 \hline
 \multicolumn{7}{l}{dark matter study}\\
 \hline 
 $\Omega_{\tilde{\chi}^0} h_0^2$              & 0.1092 & 0.1179 & 0.1155 & 0.1068 & 0.1124 & 0.1023 \\
 \hline
 channels & $HA$ (73\%) & $VV$ (75\%) & $qq$ (65\%) & $qq$ (93\%) & $AA$ (92\%) &
 $\tilde{\chi}^0_2\tilde{\chi}^0_2\to X$ (81\%) \\
 & $VV$ (13\%) & $HA$ (17\%) & $ll$ (35\%) & $ll$ (7\%) & $qq$ (7\%) &
 $\tilde{\chi}^0_1\tilde{\chi}^0_2 \to X$ (15\%) \\
 & $ZH$ (8\%) & $ HH$ (5\%) & & & $ll$ (1\%) & $\tilde{\chi}^0_1\tilde{\chi}^\pm_1 \to X$ (2\%) \\
 & $HH$ (3\%) & $ ZH$ (2\%) & & & & $qq$ (2\%)\\
 & $qq$ (2\%) & & & & & \\
 & $ll$ (1\%) & & & & & \\
 \hline
 \end{tabular}
\caption{Parameters with a LSP neutralino with a large singlino component ($U_{15}^2$)
respecting the current collider as well as the cosmological dark matter
constraints as presented in~\citep{Belanger:2005kh}. The neutralino mass mixing 
matrix $U$ is defined in~\eqref{eq-chirot}, the Higgs-boson mass mixing matrix $R$
is defined in~\eqref{eq-Hrot}. Here a CP-conserving Higgs sector is assumed. }
\label{tab-belanger}
\end{table}
In the upper part the initial NMSSM parameters are given, followed in the mid part
by the particle spectrum and mixings and in the lower part by the dark matter 
study. Presented are the annihilation and co-annihilation channels which lead to 
the required relic density $\Omega_{\tilde{\chi}^0} h_0^2$ in the 3-$\sigma$ range
of $\Omega_c h_0^2$; with view on~\eqref{eq-WMAP}.
For the parameter points 3--5 there are very light LSP neutralinos which have,
following from the neutralino mass mixing parameter $U_{15}^2$ (the matrix $U$
is defined in~\eqref{eq-chirot}), a large singlino component. 
In particular in these
cases the dominating
annihilation proceeds via Z- and Higgs-boson resonances. The parameter points
4 and 5 correspond to a light scalar Higgs boson which is dominantly a Higgs-boson singlet,
as can be seen from the Higgs-boson mass mixing parameter $R_{13}^2 \approx 1$
(the matrix $R$ is defined in~\eqref{eq-Hrot}). 
In this case the lightest
non-singlet SM-like Higgs boson has to pass the LEP bound. This in turn drives $\tan(\beta)$ close
to 3, where the mass $m_{H_2}$ is maximized. 
In Fig.~\ref{fig-belanger} the impact of the WMAP constraint on the
parameter space is shown. The contours for different
values of $\Omega_{\tilde{\chi}^0} h_0^2$  are given, where
the black region corresponds to the
WMAP constraint, that is, $0.0945<\Omega_{\tilde{\chi}^0} h_0^2<0.1287$.
The chosen independent parameters are described in the figure caption.
For further details see~\citep{Belanger:2005kh}.
\\

The PAMELA satellite experiment finds an excess of positrons
in cosmic rays~\cite{Adriani:2008zr}, whereas there is no excess
of anti--protons found~\cite{Adriani:2008zq}. This could be explained 
in the NMSSM in a rather specific scenario~\cite{Bai:2009ka}.
In this scenario the dark matter neutralinos can annihilate into
the heavier pseudoscalar $A_2$ which subsequently decays into
the lightest pseudoscalar $A_1$ and the CP-even $H_1$.
In order to get a large annihilation cross section a resonant s-channel
decay is assumed, forcing to have
$m_{A_2} \approx 2 m_{\tilde{\chi}^0}$. Further the
neutralinos have to be lighter than the top quark to avoid a decay into
a pair of top quarks which would yield an anti--proton access, 
which is not observed.
For the same reason it is assumed that $m_{A_1}<1$~GeV such that
the lightest pseudoscalar $A_1$ mainly decays into muons giving the 
observed positron excess
and suppressing the decay of $A_1$ into mesons. 
In a parameter scan employing NMHDECAY (see App.~\ref{app-tools})
parameter space is found which could explain both
PAMELA observations with no contradiction to other constraints.

%%%%%%%%%%%%%%%%%%%%%%%%%%%%%%%%%%%%%%%%%%%%%%%%%%%%%%%%%%%%%%%%%%5
% Direct dark matter detection
%%%%%%%%%%%%%%%%%%%%%%%%%%%%%%%%%%%%%%%%%%%%%%%%%%%%%%%%%%%%%%%%%%5

\subsubsection{Direct dark matter detection}

\begin{figure}[th] 
\centering
\includegraphics[width=0.5\linewidth, angle=0,clip]{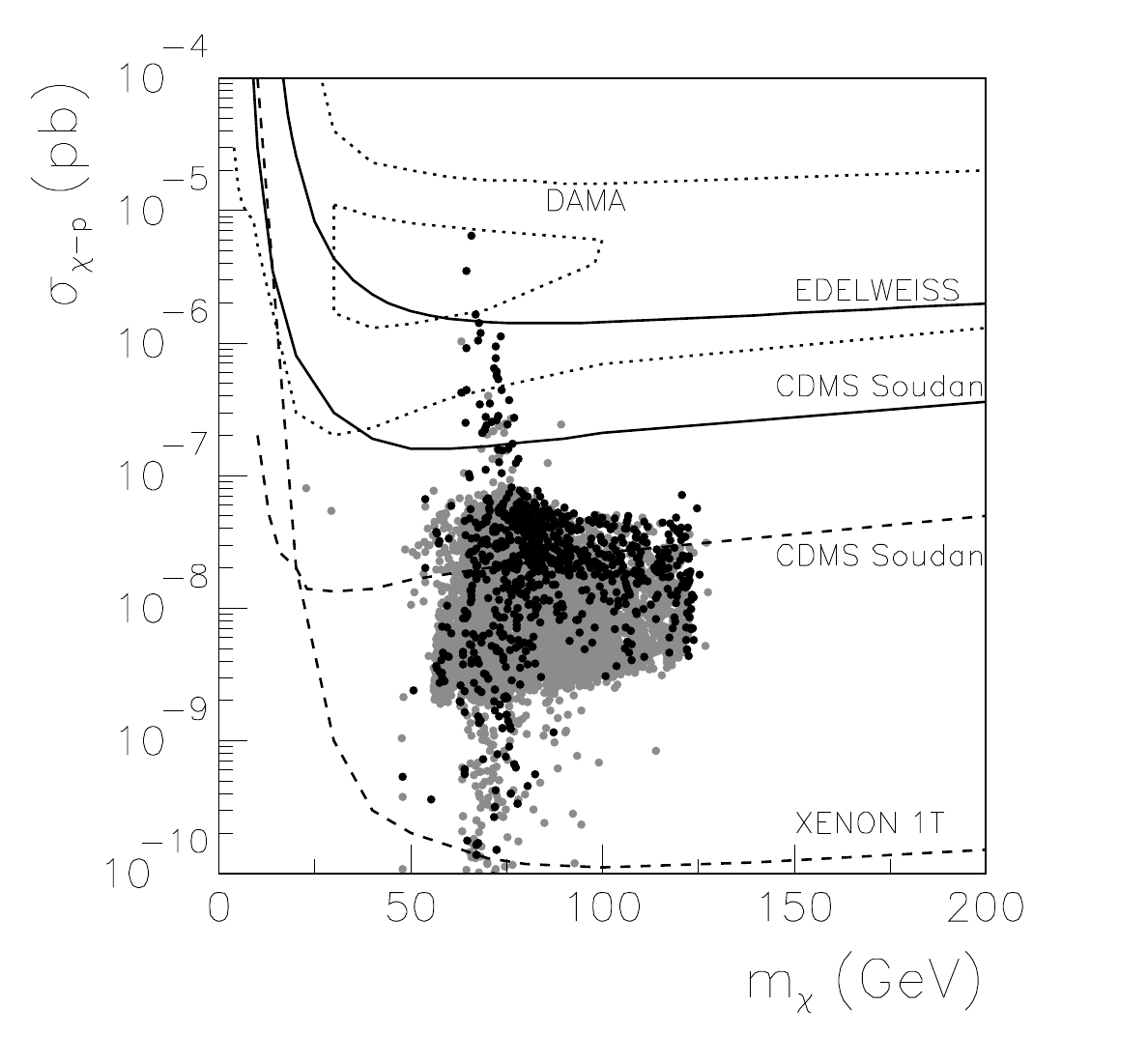}
\caption{\label{fig-cerdeno}
\it Scatter plot of the neutralino--nucleon cross section as a function of the neutralino
mass, here denoted by $m_\chi$ for $\tan(\beta)=5$, and the remaining parameters varied in the ranges 
$0.01 \le \lambda$, $\kappa \le 0.7$, 
$110~\text{GeV}< M_2 < 430~\text{GeV}$,
$-300~\text{GeV}< A_\kappa < 300~\text{GeV}$,
$-800~\text{GeV}< A_\lambda < 800~\text{GeV}$,
$110~\text{GeV}< v_s \lambda < 300~\text{GeV}$. All points pass
the LEP/TEVATRON collider, muon anomalous magnetic moment, 
and $BR(b \rightarrow s \gamma)$
constraints and have a relic density in the range
$0.1 < \Omega_c h_0^2 < 0.3$ (grey dots) or
$0.095 < \Omega_c h_0^2 < 0.112$ (black dots)
reflecting the three year WMAP data. 
The sensitivities of experiments 
are given by full lines, whereas the dashed lines give the
sensitivities expected in the future. Figure taken from~\citep{Cerdeno:2007sn}.}
\end{figure}

If the dark matter consists of unidentified particles, the earth should
be passing through a flux of these particles which constitute
the dark halo of our Milky Way.
Direct detection of dark matter may then proceed via elastic scattering
from a target nucleus. The nuclear recoil could then be detected in 
an appropriate detector.
The corresponding event rate depends on the dark matter density
in the solar vicinity and the model-dependent
neutralino--nucleus elastic scattering cross section, under the assumption
that neutralinos make up dark matter.
For a fixed dark matter density
(which is, following cosmological arguments about
$\rho_{\text{DM}} \approx 300$~MeV/cm$^3$) the
flux of dark matter particles through a detector is
obviously inverse proportional to its masses.\\
For investigations of direct neutralino dark matter detection 
in context with the NMSSM we mention the approaches 
\citep{Flores:1991rx,Bednyakov:1998is,Cerdeno:2004xw,Gunion:2005rw,Cerdeno:2007sn}.
On the experimental side we draw the attention to the experiments of the
DAMA collaboration~\citep{Bernabei:2008yi}, %new
CDMS~\citep{Ahmed:2008eu}, %new
EDELWEISS~\citep{Lubashevskiy:2008zz}, %new
ZEPLIN~I~\citep{Alner:2005wq}, %ok
as well as the 
XENON10 experiment~\citep{Angle:2007uj}. Up to now there %new
are no confirmed detections of any weakly interacting massive
particle (WIMP). The sensitivities of some direct detection 
experiments are shown in Fig.~\ref{fig-cerdeno} depending
on the mass of the neutralino~\citep{Cerdeno:2007sn}. 
The dots give predictions of the NMSSM varied in certain parameter ranges, as
indicated in the figure caption. These points pass the LEP/TEVATRON,
muon anomalous magnetic moment,
and $BR(b \rightarrow s \gamma)$ constraints as discussed
above in this section. The grey dots have passed
the relic density constraint
$0.1 < \Omega_{\tilde{\chi}^0} h_0^2 < 0.3$ and the black dots
even the much stronger constraint from the three-year WMAP data
$0.095 < \Omega_{\tilde{\chi}^0} h_0^2 < 0.112$.
Obviously there is parameter space left which passes all
the experimental and indirect/direct cosmological constrains in the NMSSM.
As indicated in the figure, future direct dark matter detection experiments may give
strong constraints of viable parameter space. 

%%%%%%%%%%%%%%%%%%%%%%%%%%%%%%%%%%%%%%%%%%%%%%%%%%%%%%%%%%%%%%%%%%5
% Baryogenesis
%%%%%%%%%%%%%%%%%%%%%%%%%%%%%%%%%%%%%%%%%%%%%%%%%%%%%%%%%%%%%%%%%%5
\subsubsection{Baryogenesis}
The absence of antimatter, that is, the observed baryon asymmetry
in our Universe may be generated by strong {\em electroweak phase
transitions} (EWPT) of first order~\citep{Kuzmin:1985mm,Shaposhnikov:1987tw,Rubakov:1996vz,Shaposhnikov:1996th}. 
Alternatives to EWPT like GUT baryogenesis~\citep{Buchmuller:1996pa}, 
baryogenesis via leptogenesis~\citep{Fukugita:1986hr} 
and the Affleck--Dine mechanism~\citep{Affleck:1984fy} are not discussed here. For an overview, see
for instance~\citep{Riotto:1999yt}. One of the attractive properties of 
EWPT is that this mechanism is understandable in terms of physics at
the electroweak scale, accessible at current colliders.
Our focus lies on the discussion of EWPT with respect to the NMSSM. We start
with a brief sketch of EWPT and show how parameter constraints
are deduced in the SM, the MSSM and in the NMSSM.\\

Phase transitions are generally characterized by an order parameter
which suddenly changes at a {\em critical temperature} $T_c$. In case the
order parameter has a discontinuity at $T_c$, the transition is
called a first-order phase transition. In contrast, if the order
parameter changes continuously at $T_c$, it is called a 
second-order phase transition.
In the electroweak phase transition the order parameter is
identified with the Higgs vacuum-expectation-value. Assuming a first order electroweak
phase transition we have with decreasing temperature a transition
from the symmetric phase with a global minimum at the origin
for $T>T_c$ to a symmetry breaking phase with a global minimum
with a vacuum-expectation-value at $v \neq 0$ for $T<T_c$. For $T=T_c$ we have degenerate
minima at $v=0$ and $v=v_c$.

It has been shown that a {\em strong} first order phase transition
is needed \citep{Dine:1992vs,Dine:1992wr}, that is
\begin{equation}
\label{eq-strong}
\frac{v_c}{T_c} > \xi \approx 1
\end{equation}
in order to avoid a wash-out of the generated baryon asymmetry after it
is generated.\\

Following closely the argumentation found in \citep{Cline:2000fh,BasteroGil:2000bw} 
let us consider a generic
Higgs potential
\begin{equation}
\label{eq-genV}
V_{\text{eff}}= m^2 \phi^2 - \alpha \phi^3 + \beta \phi^4
\end{equation}
with a generic scalar Higgs field $\phi$. From this potential we see that 
we get a first order phase
transition, that is, degenerate minima with a symmetry breaking 
minimum at $v_c$, for
\begin{equation}
\label{eq-minV}
v_c = \frac{\alpha}{2 \beta}.
\end{equation}
Obviously we need a non-vanishing parameter $\alpha$, that is, a cubic term in the
potential in order to get
the wanted symmetry breaking minimum.\\

In the SM and in the MSSM we do not have a cubic term at all. 
Nevertheless, an effective cubic
term is generated by thermal loops~\citep{Dolan:1973qd,Weinberg:1974hy}
\begin{equation}
V_{\text{eff}}=V + V^{T}_{\text{1-loop}}
\end{equation}
with
\begin{equation}
V^{T}_{\text{1-loop}}= T \sum_i 
\mp \int \frac{d^3 p}{(2 \pi)^3}
\ln \bigg(1 \pm e^{-\sqrt{p^2+m_i^2(\phi)/T}} \bigg)
\end{equation}
with $i$ running over all fermions (upper sign) and bosons (lower sign), which
couple to the Higgs boson. The expressions~$m_i(\phi)$ are the corresponding 
field-dependent masses, that is,
the masses depending on the Higgs field, before electroweak symmetry is broken.
For high temperature $T$ the effective potential can be expanded as \citep{Cline:2000fh}
\begin{equation}
\label{eq-thermic}
V^{T}_{\text{1-loop}}= \sum_i 
\frac{1}{48} m_i^2(\phi)\; T^2 \cdot 
\left\{ \begin{matrix} 2\\4\end{matrix} \right\}
-\frac{1}{12 \pi} m_i^3(\phi)\; T \cdot 
\left\{ \begin{matrix} 1\\0\end{matrix} \right\}
+\frac{1}{64 \pi^2} m_i^4(\phi)\; \bigg(\ln \frac{m_i^2(\phi)}{T^2} - c_i \bigg) \cdot 
\left\{ \begin{matrix} -1\\4\end{matrix} \right\}
+{\cal O}\frac{m_i^5(\phi)}{T}\,,
\end{equation}
where the upper line is valid for bosons and the lower one for fermions. 
The constants are
\begin{equation}
c_i = 
\begin{cases}
\frac{3}{2} + 2 \ln (4 \pi) - 2 \gamma_E\,, &{\text{for bosons}}\\
c_{\text{boson}} - 2 \ln(4)\,, &{\text{for fermions}}
\end{cases}\,.
\end{equation}
Obviously, we get for each boson in \eqref{eq-thermic} an effective cubic term.
For instance, for the field-dependent charged $W^\pm$-boson masses
$m_{W^\pm}(\phi)= g_2 \phi/\sqrt{2}$ a cubic
$\alpha$-term in \eqref{eq-genV} is generated 
and first order phase transitions are possible.\\

In the SM however, these loop effects are too small, thus give only a too small $\alpha$-term.
From the condition \eqref{eq-strong} one can deduce an upper bound on the Higgs-boson mass
and gets~\citep{Cline:1998hy}
\begin{equation}
m_H^{\text{SM}} < 32~\text{GeV},
\end{equation}
which is excluded from the lower bound on the Higgs-boson mass from the LEP experiment,
$m_H^{\text{SM, exp}} > 114$~GeV at 95\% CL~\citep{Barate:2003sz}.
The conclusion is, that this mechanism of strong first order EWPT is {\em not} possible
in the SM.\\

In the MSSM the situation is quite different. Let us sketch briefly
the situation in this model.
In Sect.~\ref{sub-Higgspheno} we have already seen that the 
tree-level mass of
the lightest Higgs boson,
$(m_{H_1}^{\text{MSSM}})^2 < m_Z^2 \cos^2 (2 \beta)$,
is ruled out experimentally by LEP
from the lower bound of $m_{H}^{\text{MSSM, exp}} > 92$~GeV~\citep{Schael:2006cr}. 
However, as we have discussed, there are quite large quantum corrections to this tree level mass. 
The largest contribution to the quantum corrections arise from top and stop loops, which
couple strongly to the Higgs-bosons. 
We recall the one-loop radiative contribution to the lightest
Higgs-boson mass~\eqref{eq-shift1},
\begin{equation}
\label{eq-shift1again}
(\Delta m_{H_1}^{\text{MSSM}})^2 = c \frac{m_t^4}{v^2} 
\ln\left( \frac{m_{\tilde{t}_L} m_{\tilde{t}_R}}{m_t^2} \right)\,.
\end{equation}
This correction allows
for larger Higgs-boson masses passing the current LEP limits.
From the logarithmic term we see that large 
radiative corrections require at least one of
the masses
$m_{\tilde{t}_L}$, $m_{\tilde{t}_R}$ to be very large.
From precision electroweak constraints we know that
$\tilde{t}_L$ has to be the heavy scalar. This follows
from the fact that $\tilde{t}_L$ couples stronger
to the electroweak gauge bosons than $\tilde{t}_R$ and a
suppression of this loop contributions requires that
$\tilde{t}_L$ is the heavier scalar. However, from
the logarithmic dependence on $m_{\tilde{t}_L}$
of the mass shift~\eqref{eq-shift1again} it
follows that a substantial shift
of the Higgs-boson mass requires
a very large mass $m_{\tilde{t}_L}$.

\begin{figure}[t] 
\centering
\includegraphics[width=0.4\linewidth, angle=0,clip]{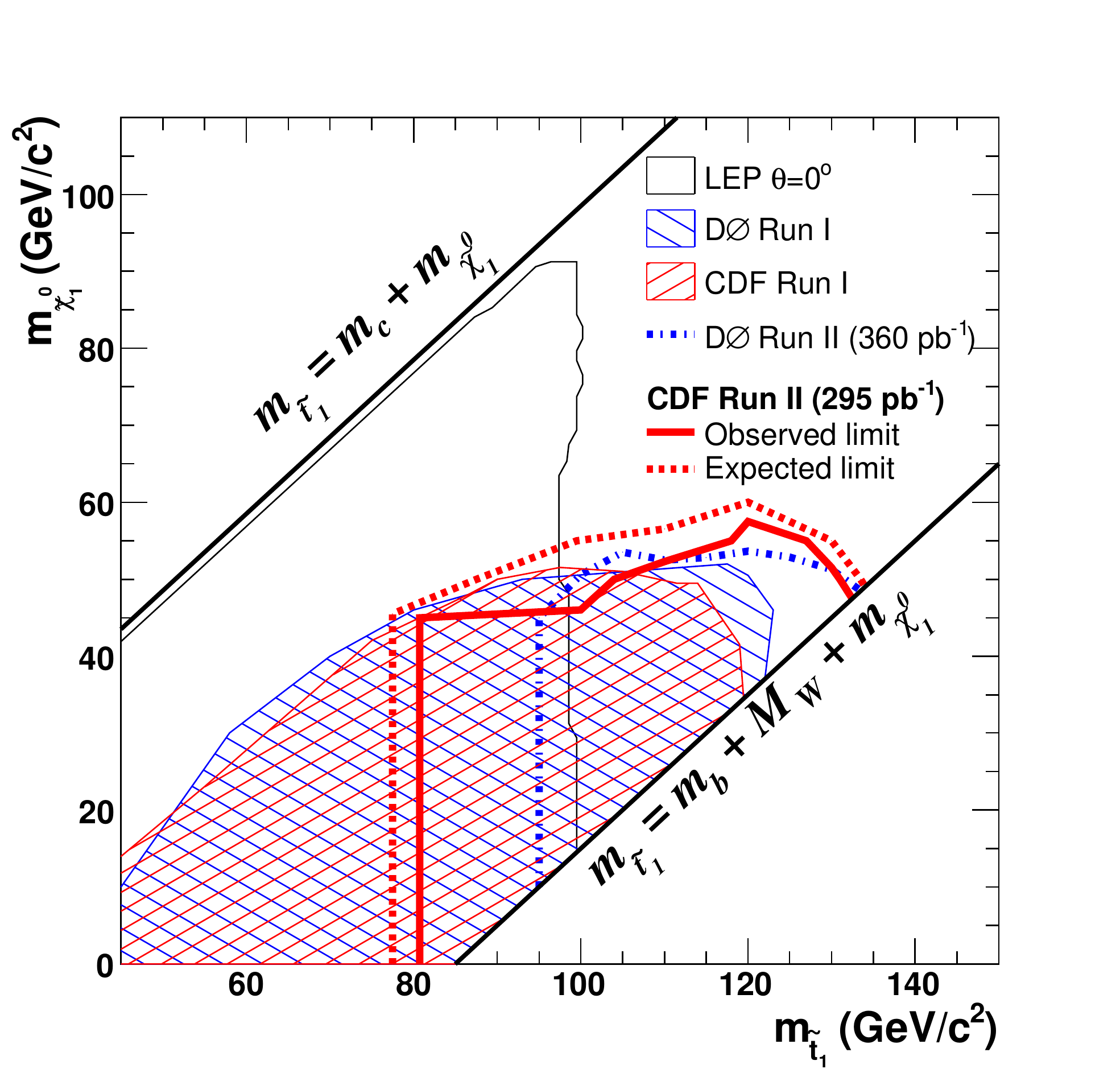}
\caption{\label{fig-stop} \it 
95\% exclusion regions in the stop--neutralino mass plane for the MSSM. 
Figure taken from \citep{Aaltonen:2007sw}.
}
\end{figure}

Now we come to the thermic loops which potentially give a
cubic term in \eqref{eq-thermic}
in the potential required for a {\em strong} first order EWPT
\citep{Carena:1996wj,Espinosa:1996qw,Bodeker:1996pc,Cline:1998hy}.
The field-dependent stop mass $m_{\tilde{t}_R}$ reads, taking also
thermic corrections of the stop mass itself into account,
\begin{equation}
m_{\tilde{t}_R}^2 = m_{\tilde{u}}^2 + |y_t|^2 |H_u^0|^2 + c_R T^2\,,
\end{equation}
with $c_R = 4/9 g_3^2+1/6 |y_t|^2(1+\sin^2(\beta))$.
A {\em strong} EWPT requires a large cubic term 
($m_i^3(\phi) \sim \phi^3$ in~\eqref{eq-thermic}) 
and this in
turn requires $m_{\tilde{u}}^2 \approx -c_R T^2$. In the limit of
zero temperature
this means that we have to have a rather light $\tilde{t}_R$-mass. 
Altogether we find
a mass hierarchy in the MSSM $m_{\tilde{t}_R}<m_t<<m_{\tilde{t}_L}$, which
is considered to be unnatural, since it requires fine-tuning.
In Fig.~\ref{fig-stop} the experimental 95\% exclusion regions in the stop--neutralino mass plane
are shown as presented by~\citep{Aaltonen:2007sw}. We see that for a light neutralino,
$m_{\tilde{\chi}}<45$~GeV, stop masses below 120~GeV are excluded.
Thus, with respect to \eqref{eq-shift1again} we certainly need corrections
of the order of 30~GeV in the MSSM, that is, rather large heavy scalar top
mass $m_{\tilde{t}_L}$ compared to $m_{\tilde{t}_R}$, that is, large fine-tuning.

In the NMSSM the {\em strong} electroweak phase transition requirement \eqref{eq-strong}
is generically much easier fulfilled since already at tree level
we have cubic terms in form of additional soft-breaking terms~\citep{Pietroni:1992in},
\begin{equation}
V_{\text{soft, trilinear}}= 
  \lambda A_{\lambda} (H_u^\trans  \epsilon H_d) S
+ \frac{\kappa}{3} A_{\kappa} S^3 + c.c.
\end{equation}
which are absent in the SM and the MSSM. 
Note that a cubic term in the
generic Higgs potential can not be constructed in a gauge invariant
way from Higgs doublets, since gauge invariance requires the Higgs
doublets to occur in a bilinear or quartic form.
Thus, an additional
singlet Higgs field $S$, as introduced in the NMSSM, opens the possibility of cubic terms
without relying on quantum corrections.
Moreover, the
tree-level mass limit on the lightest Higgs-boson mass
in relaxed in the NMSSM.\\

There was some effort spent in systematical numerical parameter scans
in order to find
parameter space fulfilling the
{\em strong} EWPT constraint 
(see for instance~\citep{Pietroni:1992in,Davies:1996qn,Huber:1998ck,BasteroGil:2000bw}). 
Let us start with the study of Pietroni~\citep{Pietroni:1992in}.
The Coleman-Weinberg one-loop corrections
\eqref{eq-colwei} as well as the non-vanishing temperature contributions to
the potential were taken into account~\citep{Dolan:1973qd,Weinberg:1974hy}.
The following assumptions were imposed.
\begin{itemize}
\item Common soft-breaking masses $m_Q^2=M_1^2=M_2^2 \equiv 1$~TeV.
\item Fixed values for $\lambda^2(m_Z)=0.274$, $y_t(m_Z)=0.97$ and
$\kappa^2=1/2 \lambda^2$.
\item A minimal value for the singlet VEV $\lambda v_s \le 45$~GeV, following
approximatively from the experimental lower bound on the chargino mass.
\item The ratio $\tan(\beta)$ is fixed to the values 2 and 10 and
for the Higgs-boson mass parameters it is assumed that $m_{H_d}^2>0$, $m_{H_u}^2<0$
and $m_{S}^2>0$. 
\end{itemize}
From a parameter scan it is found that the allowed parameter space leads to an upper bound of
$m_H^{\text{NMSSM}} < 170$~GeV for the lightest Higgs boson, far beyond the LEP limits.\\

In the work of Bastero-Gil et al.~\citep{BasteroGil:2000bw}
also the parameter space is examined targeting on the requirement
of strong first order EWPT.
With different assumptions for the soft-breaking parameters
as well as for the experimental LEP bounds it is shown that there
is parameter space available in the NMSSM which also respects the LEP
Higgs-boson mass bound.\\

Let us also mention the paper of Davies et al.~\citep{Davies:1996qn}. The authors report
that in a parameter scan passing appropriate constraints about one half of parameter points
pass the strong first order EWPT condition. Under certain assumptions, not given here, they find
masses of the lightest Higgs boson not exceeding 120~GeV. In this study
also a generalization of the superpotential $W$ in \eqref{eq-Wscalar}
is introduced with an explicit $\mu$-term
and a linear term:
\begin{equation}
\label{eq-davies}
W' = W + \mu (H_{d}^T \epsilon H_{u}) - r S\,.
\end{equation}
Of course, by setting the dimensional parameters $\mu$ and $r$ to zero
the original NMSSM is restored.\\

Eventually in the publication of Huber and Schmidt~\citep{Huber:1998ck} emphasis
is placed on a subtle point
in context with first order EWPT. As is pointed out, for an EWPT it is required,
that phase transition in $v_d$, $v_u$ and $v_s$ have to occur {\em simultaneously}.
This means that all three vacuum-expectation-values should be of the same order of magnitude. If one has for
instance a much larger value $v_s$ compared to $v_d$ and $v_u$, a cascade
of phase transitions would occur with decreasing temperature. 
This would not correspond to a first order phase transition which needs a 
bump in the effective potential, requiring a trilinear term in the Higgs fields.
The authors refer to the parameter scans of Ellwanger et al. 
(see the discussion in Sect.~\ref{sect-scan}), which predict
a generic larger value for $v_s$ compared to the doublet VEVs. 
In order to circumvent this problem it is proposed that the
supplemented superpotential~\eqref{eq-davies} should be taken into account. In this
way a compatible vacuum-expectation-value $v_s$ of the electroweak order arises.
However, as is pointed out by the authors, the generic larger values
of $v_s$ arise in studies, where the assumption of unification
of soft-breaking parameters at the GUT scale or other additional constraints
are considered. 
As we will see in 
the next subsection, dropping the unification condition meets the
criterion of a VEV $v_s$ of the electroweak order also without
a modification of the NMSSM superpotential.

\newpage
%%%%%%%%%%%%%%%%%%%%%%%%%%%%%%%%%%%%%%%%%%%%%%%%%%%%%%%%%%%%%%%%%%5
% Parameter scans
%%%%%%%%%%%%%%%%%%%%%%%%%%%%%%%%%%%%%%%%%%%%%%%%%%%%%%%%%%%%%%%%%%5

\subsection{Parameter scans}
\label{sect-scan}

A quite generic approach in order to constrain the parameter space 
of a model is
to scan over the parameter ranges of interest.
For each parameter set in this scan the particle mass and coupling spectrum
is derived. Then, the viability of this spectrum is checked, for instance
with respect to the corresponding minimum structure: only a parameter set
corresponding to a global minimum in the Higgs potential with the
observed electroweak symmetry breaking is to be accepted in this scan.
Moreover, the different experimental and cosmological constraints on the masses and couplings
have to be passed. From the eventually gained {\em allowed} parameter space, which
passes all constraints,
restrictions can be read off. 
In this section we want to present only some of the various scans
which have been performed in the 
NMSSM~\citep{Ellwanger:1991bq,Ellwanger:1995ru,Ellwanger:1996gw,Stephan:1997rv,Djouadi:2008uj}.
Typically, they are based on 
computer tools which calculate the spectrum and apply various theoretical
and experimental constraints.
Due to the large number of parameters in the NMSSM
typically not the full parameter space is scanned over for practical reasons. 
This is in particular true for 
the rich part of parameters originating from the soft supersymmetry breaking terms.
Note that in the discussion of
Higgs-boson phenomenology in Sect.~\ref{sub-Higgspheno} already 
some studies based on parameter scans
were mentioned.
Assumptions like the unification of parameters at a
high scale like the GUT scale lead to an
enormous reduction of available parameter space and thus simplify
the investigation.
Therefore, most of the parameter scan studies are applied to the
constrained NMSSM (cNMSSM); see App.~\ref{sec-varNMSSM}. We start
with briefly discussion some of these studies in the cNMSSM followed
by a parameter scan in the general NMSSM without unification assumption.\\

%Ellwanger cNMSSM
Let us first mention a series of investigations of the cNMSSM by
Ellwanger et al.~\citep{Ellwanger:1991bq,Ellwanger:1995ru,Ellwanger:1996gw}.
For each of about one million points in the five-dimensional 
parameter space, the particle and coupling spectrum of the model at the electroweak scale
is generated by the renormalization group equations. Then, the
following constraints are applied:
\begin{itemize}
\item Large trilinear soft $A$-parameters may induce electrically charged or
colored vacua; see \eqref{eq-Aconstraint}. Since this is phenomenologically unacceptable,
the corresponding parameter sets are discarded. 
\item The global minimum is searched for numerically and the global
minimum is required to have non-vanishing Higgs VEVs $v_d$, $v_u$ and $v_s$.
In this, the effective one-loop Higgs potential is considered, taking the Coleman--Weinberg
contributions~\eqref{eq-colwei} into account.
The ratio of the VEVs $v_d$ and
$v_u$ is forced to comply with $\tan (\beta)<30$. 
\item The top-quark mass, as derived from the particle mass spectrum,
is forced to fulfill the
experimental constraints, that is,
168~GeV~$< m_t <$~192~GeV. Also it is checked that
the negative search for 
charginos $m_{\tilde{\chi}^{\pm}}$ is not violated; see~\eqref{eq-charginoconstraint}.
\end{itemize}
It is found that low values of the parameters
$\lambda$ and $\kappa$ are favored as well as a large value of the
singlet VEV $v_s$. This means that the singlet is decoupled and the
NMSSM mimics the MSSM under the unification assumption.
Nevertheless, there remains parameter space where one neutralino
has a large singlino component
and thus substantial differences may arise compared to the MSSM.
It is pointed out (see for instance \citep{Djouadi:2008uj}) 
that for small parameters $\lambda$ and $\kappa$, the two parameters
$m_S^2$ and $A_\kappa$ change not very much between the electroweak and the GUT
scale by the renormalization group equations, that is, from \eqref{eq-msconstraint} 
the unification parameter constraints $m_0^2 \lesssim 1/9 A_0^2$
is evident.
Note that there is a subtle difference between the cMSSM and the cNMSSM:
In the cMSSM small values of the $m_0$ parameter are disfavored since
they lead to a charged slepton LSP, unacceptable as a dark matter candidate. 
Contrary, in the cNMSSM
a small parameter $m_0$ is favored due to a viable global minimum for non-vanishing $v_s$.
The slepton LSP in the cNMSSM is avoided since the possibility of
a singlino-like LSP arises in parameter space \citep{Hugonie:2007vd}.\\

\begin{figure}[h] 
\centering
\includegraphics[width=0.5\linewidth, angle=0,clip]{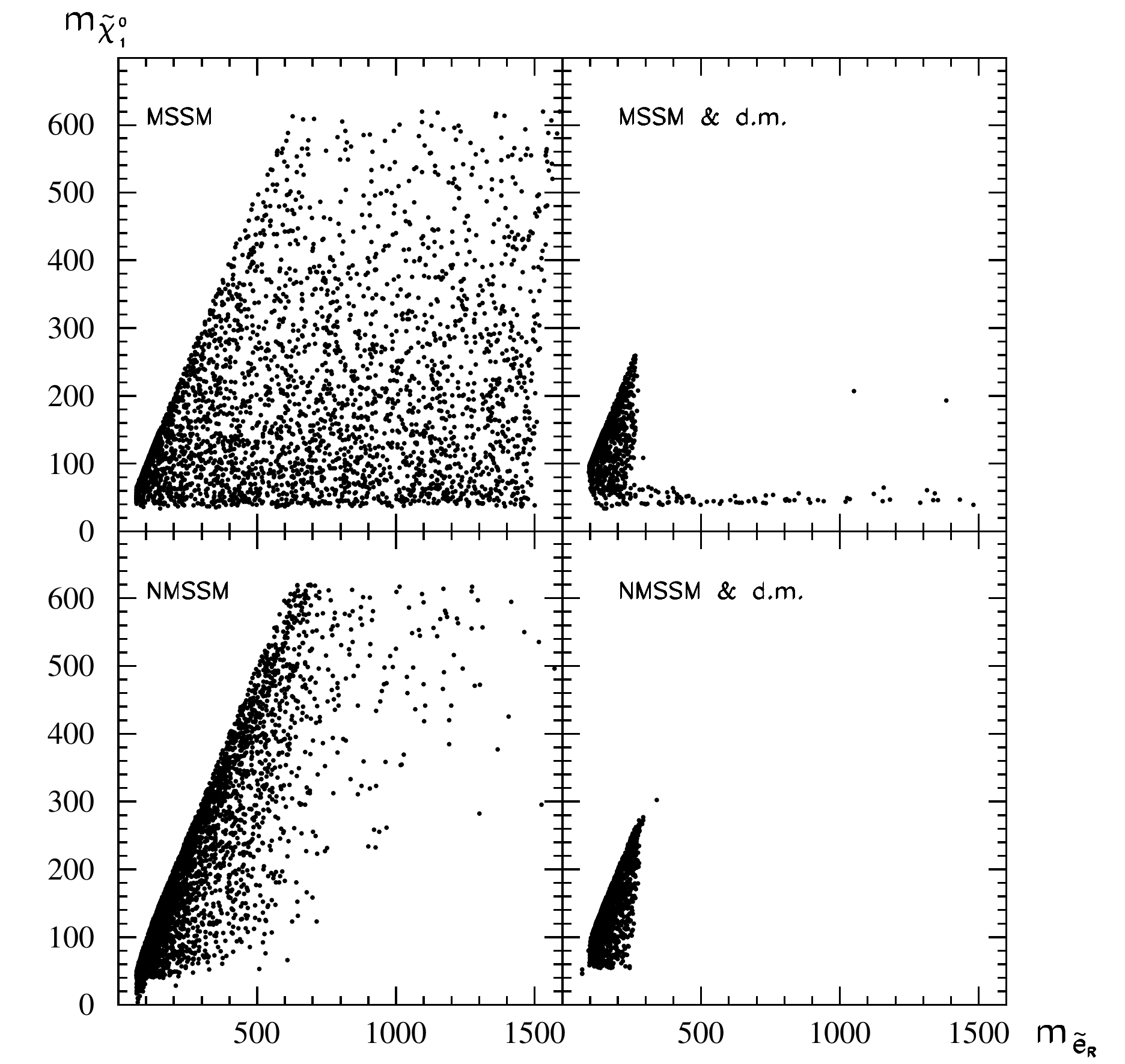}
\caption{\label{fig-stephan} \it 
Scatter plots in the constraint cMSSM (upper part) compared to the constrained cNMSSM (lower part) in the
$m_{\tilde{e}_R}$--$m_{\tilde{\chi}_1^0}$ plane. The right part shows the scatter plots 
passing also the WMAP dark matter (denoted by d.m.) constraint. Figure taken from~\citep{Stephan:1997rv}.
}
\end{figure}

%Stephan
We would also like to mention the approach of Stephan~\citep{Stephan:1997rv}.
This study is also performed in the constrained NMSSM.
The main difference to the previously discussed approaches is that
in this work in addition the dark matter constraint, that is, the relic abundance of the LSP,
supposed to be a neutralino, is taken into account. 
There are also some subtle points concerning some
deviating theoretical and experimental constraints, which are applied. Starting 
with $5.5 \cdot 10^8$ points in the five-dimensional parameter space, about 4900 points
pass the theoretical and experimental constraints but only 2000 points
of these pass the cosmological constraint given by the relic neutralino abundance bound.
Quite restrict explicit bounds are derived in this way and given already in the abstract of
this work:
\begin{gather*}
m_{H_1} < 140 \text{ GeV}\,,\\ 
m_{\tilde{\chi}^0} < 300 \text{ GeV}\,,\\
m_{\tilde{e}_R} < 300 \text{ GeV}\,,\\
300 \text{ GeV}~< m_{\tilde{u}_R} < 1900 \text{ GeV}\,,\\
200 \text{ GeV}~< m_{\tilde{t}_1} < 1500 \text{ GeV}\,,\\
350 \text{ GeV}~< m_{\tilde{g}} < 2100 \text{ GeV}\,.\\
\end{gather*}
Scatter plots in the $m_{\tilde{e}_R}$--$m_{\tilde{\chi}_1^0}$
plane are shown in Fig.~\ref{fig-stephan} for the cMSSM as well as
for the cNMSSM. Note that in the cNMSSM the selectron mass
$m_{\tilde{e}_R}$ is stronger constrained than in the cMSSM. As mentioned
above this arises from favored lower values of $m_0$, whereas in the cMSSM
small values of $m_0$ are disfavored. 
An upper bound of $m_0 \approx m_{\tilde{e}_R}$ can be deduced based
on the relic density constraint which requires to have nearly
degenerate masses of the LSP and the NLSP~\citep{Djouadi:2008uj}. 
From an approximation
for the RGE running of the NLSP mass it is found that $m_0^2 \lesssim 1/15 M_{1/2}^2$,
favoring low values of $m_0 \approx m_{\tilde{e}_R}$ in the cNMSSM in contrast
to the cMSSM.
Also the strong effect of the application of the relic abundance dark matter constraint
is demonstrated in this figure.

\begin{figure}[h] 
\centering
\includegraphics[width=0.45\linewidth, angle=0,clip]{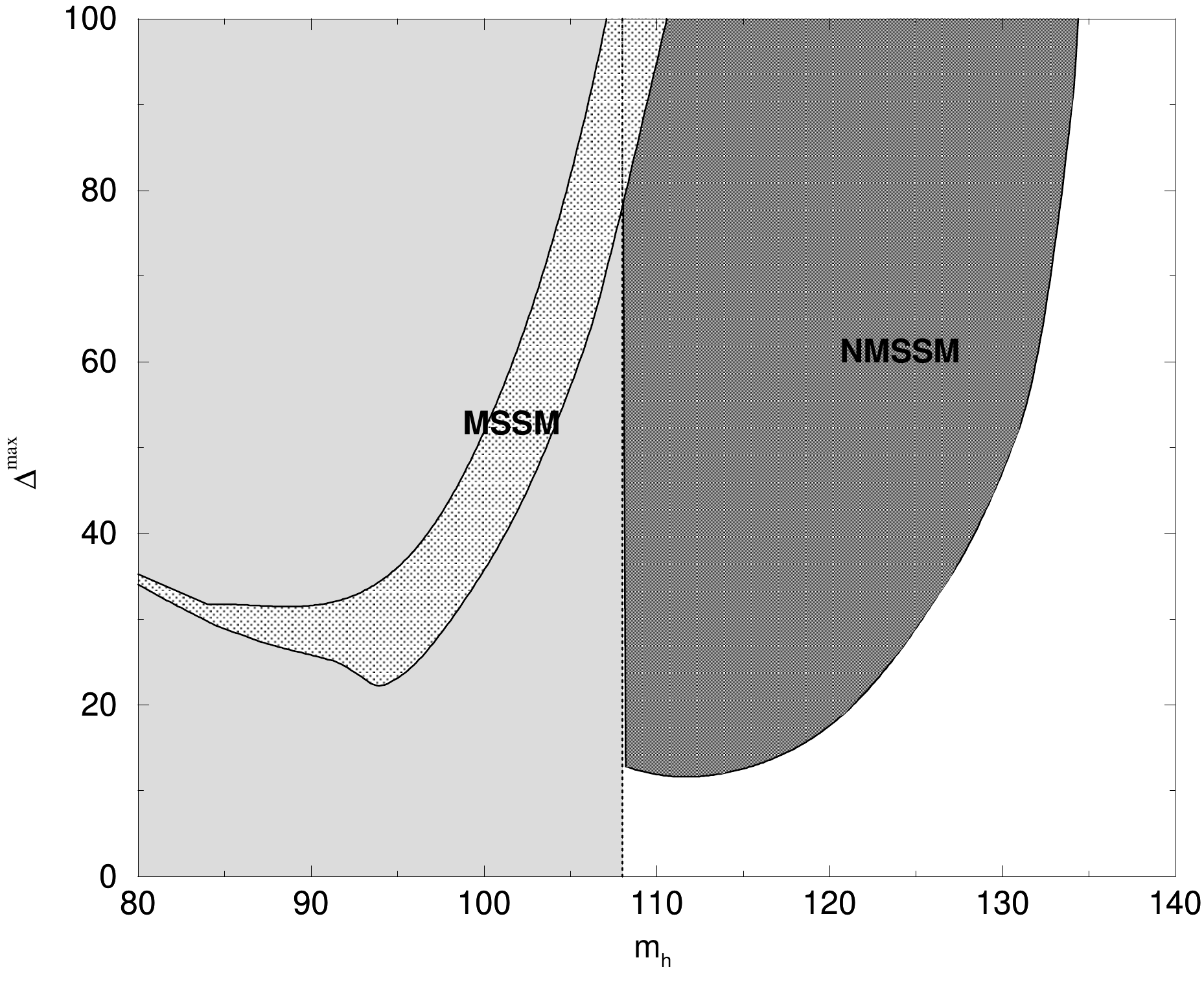}
\includegraphics[width=0.45\linewidth, angle=0,clip]{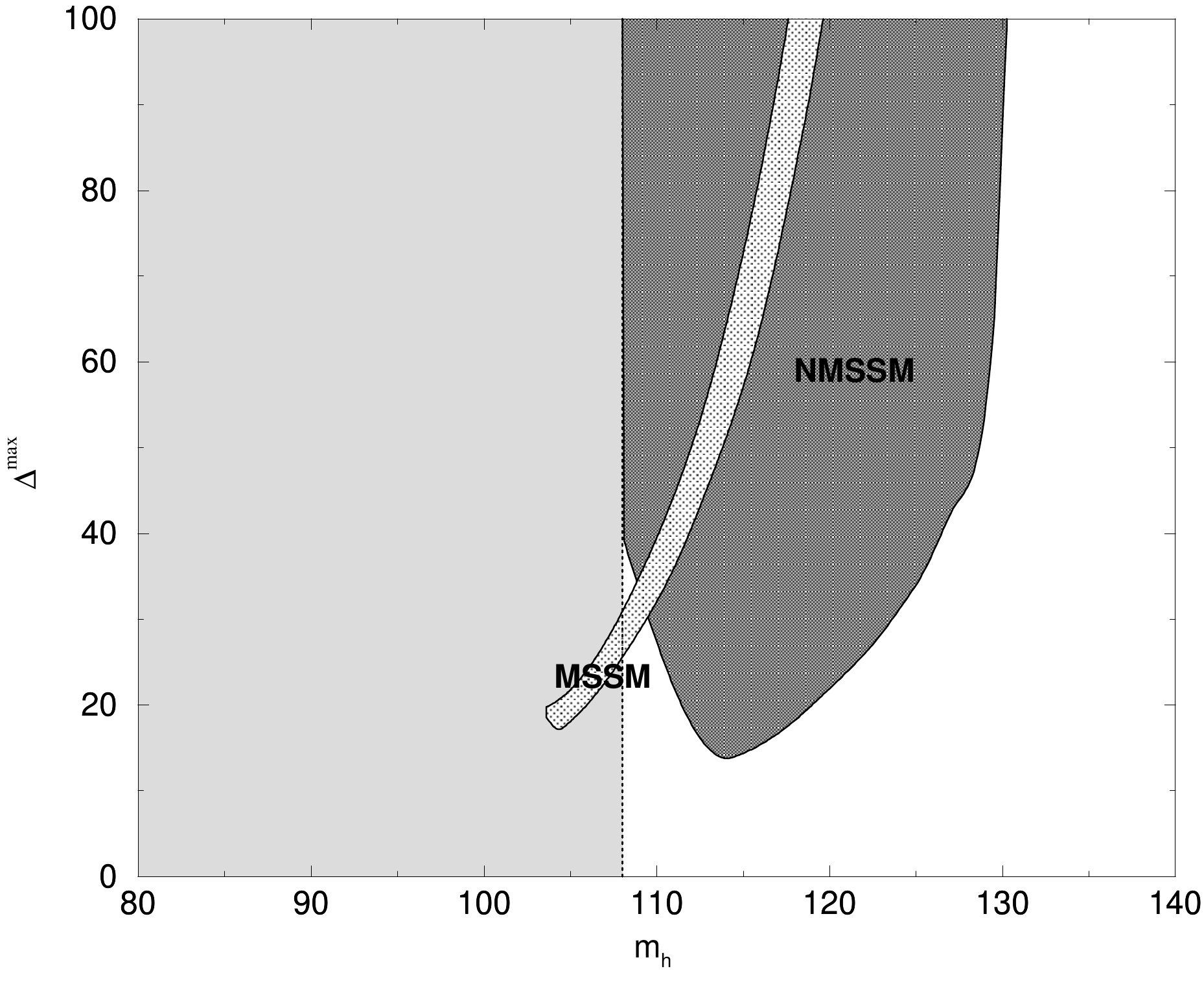}
\caption{\label{fig-Deltafine} \it 
Scatter plot of fine-tuning as defined in \eqref{eq-Deltafine} as a function of
the lightest physical Higgs-boson mass for $\tan(\beta)=3$, left, 
and $\tan(\beta)=5$, right. 
In this scan there are fixed values of $m_0=100$~GeV, $A_t(\Lambda_G)=0$, 
$M_1(\Lambda_G)=M_2(\Lambda_G)=500$~GeV,
$\mu=\lambda v_s<0$ and varied parameters
$M_{\tilde{g}}(\Lambda_G)$, $m_{H_d}(\Lambda_G)$ and $m_{H_u}(\Lambda_G)$.
The dark shaded regions shows the fine-tuning values corresponding to the NMSSM
and the brighter shaded regions corresponding to the MSSM. The LEP limit on
the SM Higgs-boson with $m_H^{\text{SM}}<108$ mass is also shaded. 
Figure taken from~\citep{BasteroGil:2000bw}. 
}
\end{figure}

%Roy et al
In the work of Bastero-Gil et al.~\citep{BasteroGil:2000bw}, 
the MSSM is compared 
to the NMSSM. 
In this study emphasis is placed on the fine-tuning required in both models in order
to comply with the theoretical and experimental constraints. Fine-tuning is quantified 
quite similar to \eqref{eq-fine}, in the form
\begin{equation}
\label{eq-Deltafine}
\Delta^{\text{max}} = 
\mathop{\max}_{a_i}  \left| \frac{a_i}{m_Z^2} \frac{\D m_Z^2}{\D a_i} \right|\;,
\end{equation}
where $a_i$ denotes all the soft supersymmetry breaking parameters.
In a parameter scan the unified scalar mass is fixed, $m_0=100$~GeV, as well
as the $A$-parameters, $A_\kappa(m_Z)=0$, $A_t(\Lambda_G)=0$ and the gaugino masses
$M_1(\Lambda_G)=M_2(\Lambda_G)=500$~GeV, where $\Lambda_G$ denotes the GUT scale. 
The gaugino mass $M_{\tilde{g}}$ is varied in the range
100 GeV~$<M_{\tilde{g}}(\Lambda_G)< 600$~GeV, and the Higgs-boson mass
parameter $m_{H_d}$ is devoted to  $0 < m_{H_d}(\Lambda_G)<1$~TeV, with
$\mu=\lambda v_s<0$. In the NMSSM also the additional parameters $\lambda(\Lambda_G)=1$
and $\kappa(\Lambda_G)=0.1$ are fixed. Requiring further for the chargino masses
$m_{\tilde{\chi}^\pm}>90$~GeV, the values for the fine-tuning function $\Delta^{\text{max}}$ are given
in Fig.~\ref{fig-Deltafine} for $\tan(\beta)=3$ and $\tan(\beta)=5$, respectively.
Let us cite the authors: {\em The plots are a striking demonstration that the physical Higgs boson
can be heavier and involve less fine-tuning in the NMSSM compared to the MSSM at low
values of $\tan \beta$}.
The large fine-tuning in the MSSM for low $\tan(\beta)$ originates from the LEP bounds on the
minimal Higgs-boson mass confronting the tree-level prediction of a mass below
the $Z$-boson mass $m_Z$. Thus, large quantum corrections to the Higgs-boson mass
in turn require a very large stop mass, that is, large fine-tuning.
On the other hand it is pointed out that low values of $\tan(\beta)$ are favored in order to
have a strong electroweak phase transition. The authors conclude that in
the NMSSM the situation is much better with respect to the LEP bound on the
minimal Higgs-boson mass, fine-tuning and baryogenesis via first order EWPT.\\

% Djouadi, Ellwanger and Teixeira
\begin{figure}[h] 
\centering
\includegraphics[width=0.45\linewidth, angle=0,clip]{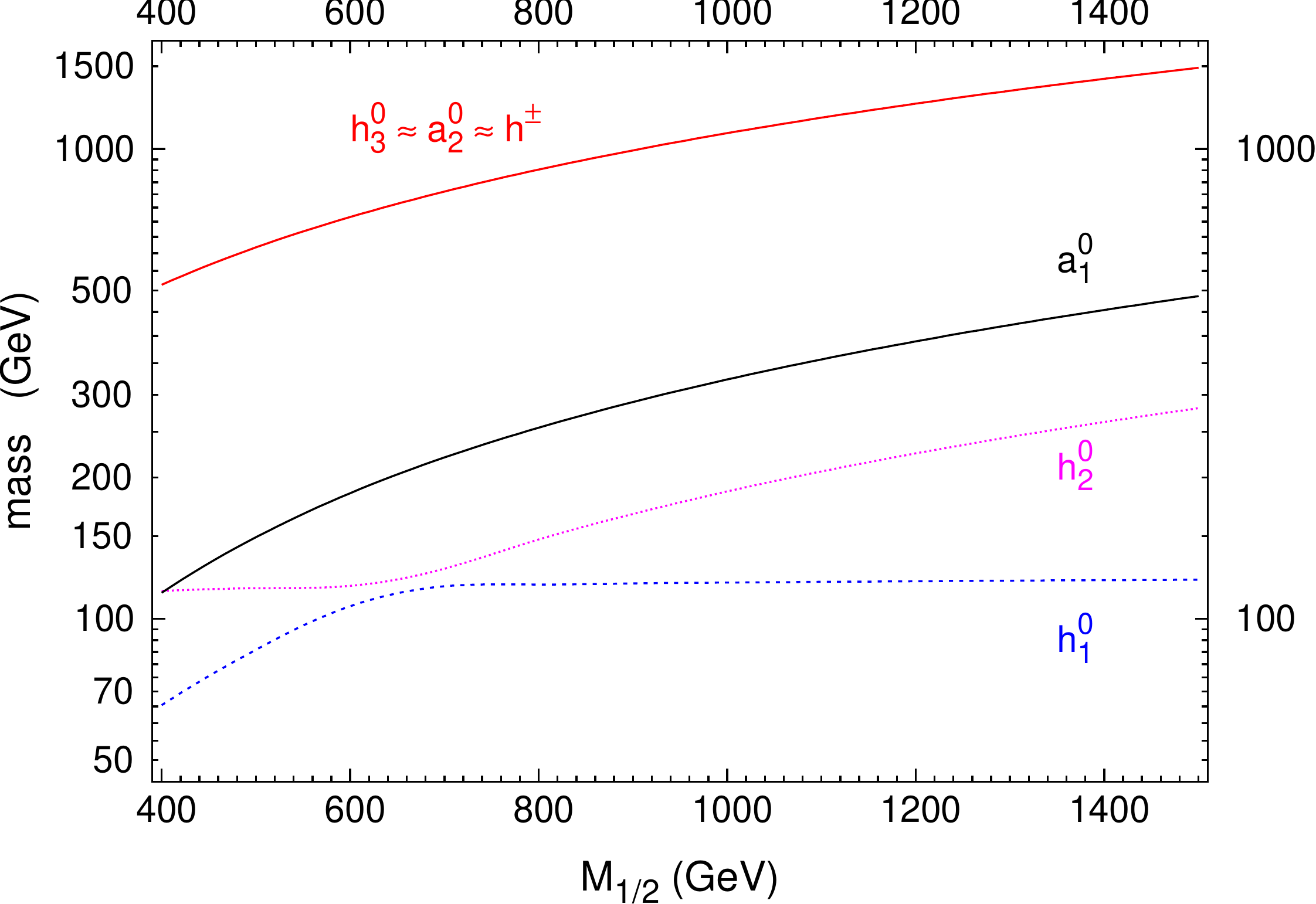}\\
\includegraphics[width=0.45\linewidth, angle=0,clip]{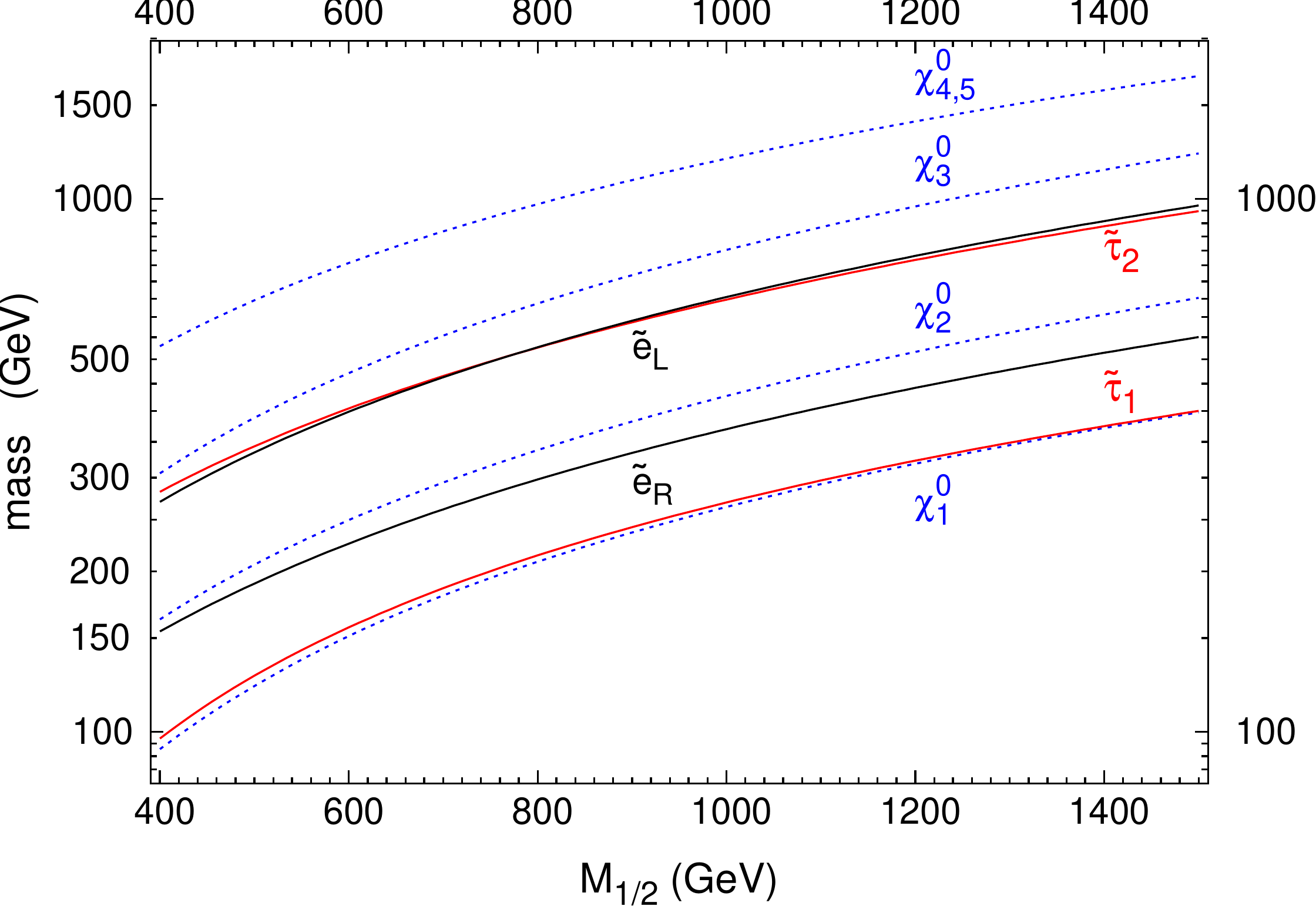}
\includegraphics[width=0.45\linewidth, angle=0,clip]{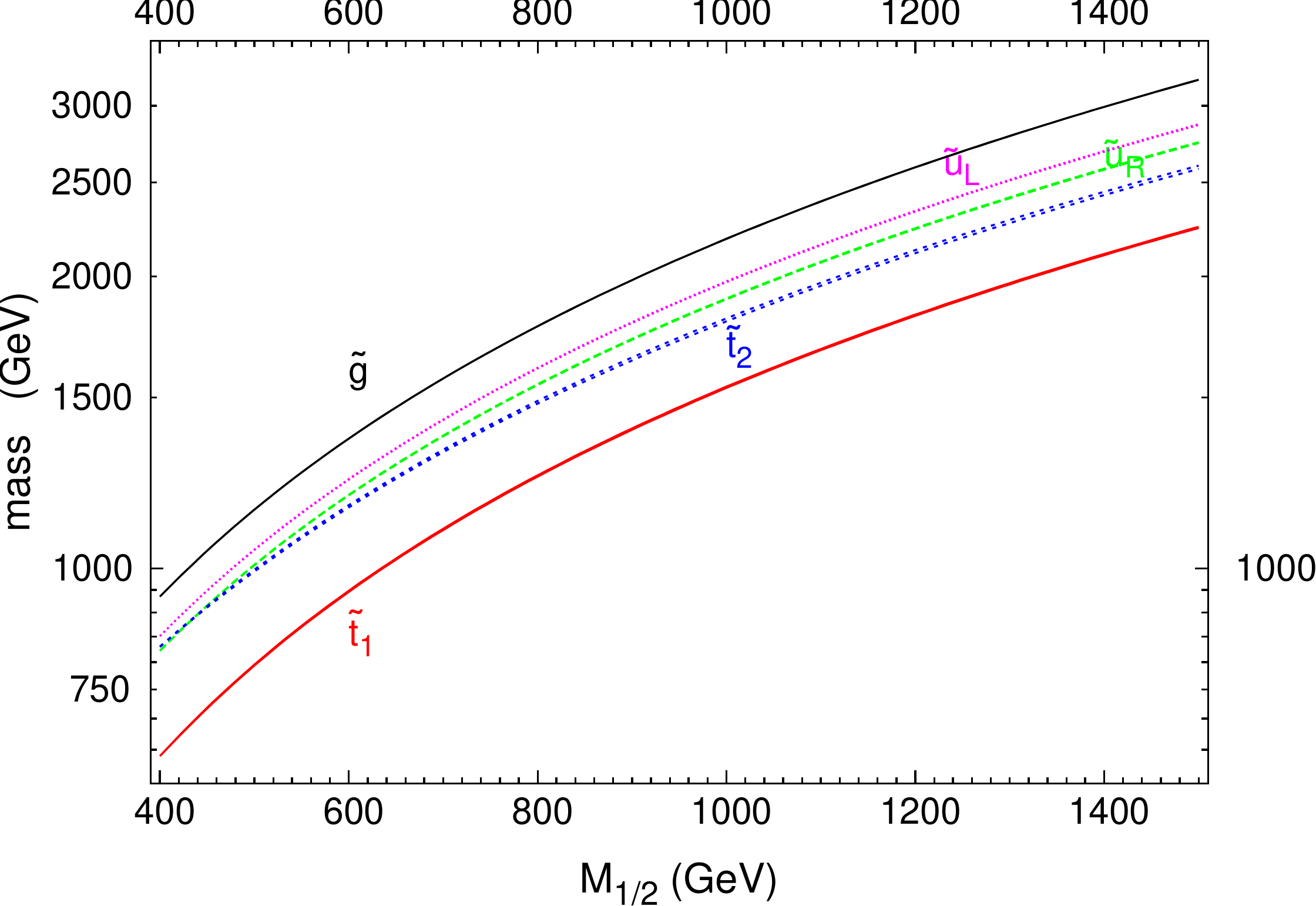}
\caption{\label{fig-spectrumEllwanger} \it
Spectrum of Higgs bosons in the upper plot (here denoted with small letters), neutralino-, selectron- and stau-mass spectrum
in the left plot and
squark- and gluino-mass spectrum in the right plot. The unified scalar mass is 
set to $m_0=0$. Degeneracy of the masses $m_{\tilde{\chi}^\pm_1}=m_{\tilde{\chi}_3}$
and
$m_{\tilde{\chi}^\pm_2}=m_{\tilde{\chi}_{4/5}}$ is found. Plots 
taken from~\citep{Djouadi:2008uj}.
 }
\end{figure}
In a recent study by Djouadi, Ellwanger and Teixeira~\citep{Djouadi:2008uj}, 
focusing also on the cNMSSM, very severe constraints are found.
After applying the theoretical as well as current experimental constraints, including the new
WMAP data on the relic density, there remains a quite narrow window of allowed parameter space.
It is reported that in the cNMSSM,
the gluino $\tilde{g}$ turns out to be generically heavier than all squarks $\tilde{q}$. 
The NLSP is found to be generically stau-like and nearly
degenerate with  the LSP mass which is a neutralino, reflecting the
constraints coming from the relic density.
The supersymmetric
partner particles eventually decay via the stau-like NLSP into the LSP neutralino.
 This opens the possibility of the observation of displaced vertices as discussed
in Sect.~\ref{sec-nc}.
The particle spectra of the neutralinos, charginos, selectron and stau
as well as the spectra of squarks and gluino are shown in Fig.~\ref{fig-spectrumEllwanger}.
In this figures, the unified scalar mass is fixed to zero, $m_0=0$ (also another
choice of the unified scalar mass is discussed in the paper). 
We see that the spectra of supersymmetric partner particles are highly
restricted by the theoretical and experimental bounds, where also the
current WMAP constraint for the relic density is applied; see~\eqref{eq-WMAP}. 
As the authors of this investigation have stressed, the measurement of
one sparticle mass or mass difference would allow to predict quite
accurately the complete sparticle spectrum in the cNMSSM.\\

%Bednyakov
Let us also mention the approach of Bednyakov and 
Klapdor-Kleingrothaus~\citep{Bednyakov:1998is},
targeting mainly at the {\em direct detection} 
of a LSP neutralino dark matter candidate. 
This direct detection refers to elastic scattering of a dark matter
neutralino from a nucleus producing a nuclear recoil detected in an
appropriate detector; see Sect.~\ref{sub-coscon}.
In this work no unification at the GUT scale of soft parameters is assumed, that is, there is no need to apply the RGE's from the GUT scale starting point in this case. 
The scan is done over the eleven-dimensional parameter space
\begin{equation}
M_1, M_2, \tan(\beta), v_s, \lambda, \kappa, m_{Q_{1}}, m_{Q_3}, A_{u_3}, A_\lambda, A_\kappa,
\end{equation}
with $m_{Q_1}=m_{Q_2}$, $m_{Q_3}$ the soft mass parameters with
generation index. For simplicity reasons it is assumed that
$m_{Q_2}^2 \equiv m_{\tilde{u}_{1,2}}^2=m_{\tilde{d}_{1,2}}^2=m_{L_{1,2}}^2=m_{\tilde{e}_{1,2}}^2$
and
$m_{Q_3}^2 \equiv m_{\tilde{u}_{3}}=m_{\tilde{d}_{3}}=m_{L_{3}}=m_{\tilde{e}_{3}}$. 
Moreover it is set $A_u=A_d=A_e \equiv 0$ for all generations except $A_{u_3}$.
The soft supersymmetry breaking parameters are defined in~\eqref{eq-Vsoft}.
Negative results for the search for supersymmetric particles at LEP and at TEVATRON
are included in the analysis. 
The resulting allowed parameter space is compared 
once without any cosmological constraint on the co-annihilation rate of the LSP and
once taken the constraint
$0.025 < \Omega_{\tilde{\chi}^0} h_0^2 < 1$
into account. This cosmological constraint alone gives, as is reported, a
20\% cut on the allowed parameter space. As is pointed out in their conclusion, 
under their assumptions, but without unification of soft supersymmetry
breaking parameters at the GUT scale, there
remain domains in parameter space where the lightest neutralino has a quite 
small mass, even as small as 3~GeV. In this work it is stressed that in a proposed germanium
$^{73}Ge$ detector there are large event rates expected in the NMSSM. 
Let us remark, that with view on the current WMAP data, 
it would be very interesting to see the effect of the current much 
tighter bounds in this study; see~\eqref{eq-WMAP}.\\

Finally, it is clear that the results of the
here discussed parameter scans depend 
strongly on the assumptions taken into account, like the unification of
parameters. On the other hand are the results also strongly dependent on
the constraints applied to the derived particle and coupling
spectra. 
To summarize, we have seen that at least for the general NMSSM,
without any unification assumptions, there is a lot of viable parameter
space, even if the cosmological LSP cold dark matter constraint is taken into account.
Nevertheless, even that the cNMSSM seems to be very restricted, it
remains a viable model with respect to all applied theoretical and experimental bounds.

\newpage
%%%%%%%%%%%%%%%%%%%%%%%%%%%%%%%%%%%%%%%%%%%%%%%%%%%%%%%%%%%%%%%%%%5
%%%%%%%%%%%%%%%%%%%%%%%%%%%%%%%%%%%%%%%%%%%%%%%%%%%%%%%%%%%%%%%%%%5
% Determining the global minimum of the Higgs potential
%%%%%%%%%%%%%%%%%%%%%%%%%%%%%%%%%%%%%%%%%%%%%%%%%%%%%%%%%%%%%%%%%%5
%%%%%%%%%%%%%%%%%%%%%%%%%%%%%%%%%%%%%%%%%%%%%%%%%%%%%%%%%%%%%%%%%%5

\section{Determining the global minimum of the Higgs potential}
\label{sec-global}

It is a non-trivial task to find the global minimum of Higgs potentials
with a large number of Higgs-boson fields. For instance in the NMSSM
the Higgs-boson sector consists of
two complex electroweak doublets and one complex electroweak singlet,
that is, 8 real fields from the doublets plus 2 real fields from the singlet.
The conventional approach based on the unitary gauge requires
the global minimum to be found in a 7-dimensional field space.
Typically, numerically methods are applied in such involved cases in 
order to determine the global minimum;
in contrast, here we want to discuss an algebraic approach~\citep{Maniatis:2006jd}.
In this approach all stationary points of 
the tree-level Higgs potential are found and
supposed, the potential is bounded from below,
the global minimum is identified from
the corresponding lowest value of the potential.
The method is applied to the NMSSM, revealing  
a quite surprising structure of stationary points, that is
minima, maxima, and saddle points with different behavior with respect to
the symmetry breaking of the \eweakgroup electroweak gauge group.

The global minimum of the Higgs potential gives the expectation
values of the Higgs fields at the stable vacuum.
Parameter values for the Higgs potential are thus considered to be acceptable only,
if the global minimum of the Higgs potential occurs for 
Higgs-field vacuum-expectation-values, which induce the spontaneous breakdown
of \eweakgroup to the electromagnetic \emgroup at the electroweak
scale $v\approx 246$~GeV.

Firstly, the tree level Higgs potential for general models with two
Higgs doublets and an arbitrary number of additional Higgs singlets is considered.
The first step is to recognize that the potential is restricted by renormalizability
and gauge invariance. Renormalizability restricts the potential
to at most quartic terms in the Higgs fields.
Electroweak gauge invariance restricts the possible doublet terms
in the potential, since
only gauge invariant scalar products of doublets can occur. 
Substituting the doublet fields by appropriate functions of
these invariant scalar products, all gauge degrees of freedom can be eliminated.
The method to base the analysis on bilinear gauge-invariant functions
was introduced already in context with the general 
two-Higgs doublet model~\citep{Nagel:2004sw, Maniatis:2006fs}.

The global minimum is among the stationary points of the potential.
The stationarity conditions of the Higgs potential form non-linear, multivariate,
inhomogeneous polynomial systems of equations of third order.
A systematic approach to solve these
-- in general quite involved -- systems of polynomial equations is
possible by a Groebner basis computation, well established
in ideal theory~\citep{Buchberger, Bose, Weispfenning}.
The Groebner basis was originally introduced to solve the {\em ideal membership problem}.
Constructing this Groebner basis with an appropriate ordering
of the {\em monomials} (see~App.~\ref{ap-buchberger} for
an illustration of the Groebner basis approach and further details),
for instance the {\em lexicographical ordering},
and subsequent triangularization allows to solve the
initial system of equations algorithmically for any finite number of complex solutions.
The introduction of gauge invariant functions 
just avoids continuous gauge symmetries in the potential and the
finiteness of the set of complex solutions
can be easily checked within this algorithmic approach.
Moreover, this approach guarantees that all stationary points
are found.

In~\cite{Maniatis:2006jd} the method is applied to the NMSSM.
For the computation of Groebner bases as well as the subsequent steps to solve
the systems of equations the freely available open-source
algebra program \mbox{SINGULAR~\citep{Singular}} is employed.
It is found that large parts of the parameter space of the NMSSM
Higgs potential can be excluded
by requiring the global minimum to have the electroweak symmetry
breaking observed in Nature.
This is illustrated by determining the allowed and forbidden ranges
for some generic parameters of the model.

%%%%%%%%%%%%%%%%%%%%%%%%%%%%%%%%%%%%%%%%%%%%%%%%%%%%%%%%%%%%%%%%%
%%%%%%%%%%%%%%%%%%%%%%%%%%%%%%%%%%%%%%%%%%%%%%%%%%%%%%%%%%%%%%%%%

\subsection{Stationary points in the NMSSM}
\label{NMSSM}

The NMSSM Higgs potential is given in Sect.~\ref{sec-Hpot}
with parameters 
\begin{equation}
\label{eq-par}
\lambda, \kappa, m_{H_u}^2, m_{H_d}^2, m_{S}^2, A_{\lambda}, A_{\kappa}.
\end{equation}
The quartic terms of the potential~(\ref{eq-V})
are positive for any non-trivial field configuration,
if both $\lambda$ and $\kappa$ are non-vanishing.
The potential is therefore bounded from below for all cases considered here,
and stability needs not to be checked any further.

The NMSSM Higgs potential is translated to the formalism described
in App.~\ref{quadratics}, where 
all Higgs-doublet scalar products are replaced
by real {\em gauge-invariant functions}, $K_0$, $K_1$,
$K_2$, $K_3$ and
the complex singlet field is decomposed into
two real fields according to \mbox{$S=S_{re} + i S_{im}$}.
In this notation the Higgs potential $V=V_{\text{F}}+V_{\text{D}}+V_{\text{soft}}$ 
is given by
\begin{align}
\begin{split}
\label{eq-NMSSMpot2}
V_{\text{F}} =&\;
  \frac{1}{4} |\lambda|^2 \left( K^2_1+K^2_2 + 4 K_0 (S^2_{re}+ S^2_{im}) \right)\\
&+ |\kappa|^2 ( S^2_{re}+S^2_{im} )^2\\
&- \operatorname{Re}(\lambda \kappa^*) 
	\left(K_1 (S_{re}^2-S_{im}^2) + 2 K_2 S_{re}S_{im}\right)\\
&+ \operatorname{Im}(\lambda \kappa^*) 
	\left(K_2 (S^2_{re}-S^2_{im} )- 2 K_1 S_{re}S_{im}\right),\\
V_{\text{D}} =&\;
   \frac{1}{8} (g_1^2+g_2^2) K_3^2 
+ \frac{1}{8} g_2^2 \; \left( K^2_0-K^2_1-K^2_2-K^2_3 \right),\\
V_{\text{soft}} =&\;
  \frac{1}{2} m_{H_u}^2 \; (K_0-K_3) 
+ \frac{1}{2} m_{H_d}^2 \; (K_0+K_3)\\ 
&+ m_{S}^2 \; (S^2_{re} + S^2_{im})\\ 
&- \operatorname{Re}(\lambda A_{\lambda})
	 \left( K_1 S_{re} - K_2 S_{im} \right)\\
&+ \operatorname{Im}(\lambda A_{\lambda})
	 \left( K_2 S_{re} + K_1 S_{im} \right)\\
&+ \frac{2}{3} \operatorname{Re}(\kappa A_{\kappa})
	 \left( S_{re}^3 - 3 S_{re} S_{im}^2 \right)\\
&+ \frac{2}{3} \operatorname{Im}(\kappa A_{\kappa})
	 \left( S_{im}^3 - 3 S_{re}^2 S_{im} \right).
\end{split}
\end{align}
For given values of the potential parameters~\eqref{eq-par}
all stationary points of the NMSSM can be found by 
solving the systems 
of equations~\eqref{eq-stationarityU}, \eqref{eq-stationarityF}, \eqref{eq-stationarityP}
as described in App.~\ref{quadratics}.

The initial parameters in the Higgs potential \eqref{eq-par}
are translated to the set of parameters as described in Sect.~\ref{sub-Higgspara}.
This enables for
instance to fix the vacuum expectation values $v$, $\tan(\beta)$ and $v_s$.
As a numerical example the parameter are chosen as
\begin{equation}
\label{eq-centralvals}
\begin{gathered}
  \lambda = 0.4, \quad
  \kappa = 0.3, \quad
  \abs{A_\kappa}  = 200\text{~GeV},\\% \quad
  \tan\beta = 3,\quad
  m_{H^\pm} = 2 v,\\% \quad
  \sign{R_\kappa}=-, \quad
  \dedm = 0, \quad
  \dcp = 0
\end{gathered}
\end{equation}
and the parameter $v_s$ is varied.
The roots of the univariate polynomials are found numerically,
with a precision of $100$ digits.
The approach allows to use arbitrary precision.
The errors of the approximate statements
described in the following are checked to be under control.
For generic values of the parameters $52$ complex solutions are found:
$7$ corresponding to the unbroken, $38$ to the partially broken, 
and $7$ to the fully broken cases.
The number of real and therefore relevant solutions depends on the
specific values of the parameters.

As expected from the $\Zthree$ symmetry of the potential, either 1 or 3
solutions sharing the same value of the potential are found within
the accuracy of the numerical roots.
From the computed stationary points only those may be
accepted as global minima 
which correspond to the initial vacuum expectation values 
(up to the complex phases), that is, which fulfill
\begin{equation}
\label{eq-vac}
\sqrt{2 K_0} \approx v,\;\;
\sqrt{\frac{K_0-K_3}{K_0+K_3}} \approx \tan\beta,\;\;
\sqrt{2 ({S_{re}^2+S_{im}^2})} \approx v_s.
\end{equation}
Since for non-vanishing parameters $\lambda, \kappa$ the potential is bounded from below,
the stationary point with the lowest value of the potential is
the global minimum.

\begin{figure*}[t!]
\begin{center}
\includegraphics[width=0.95\linewidth]{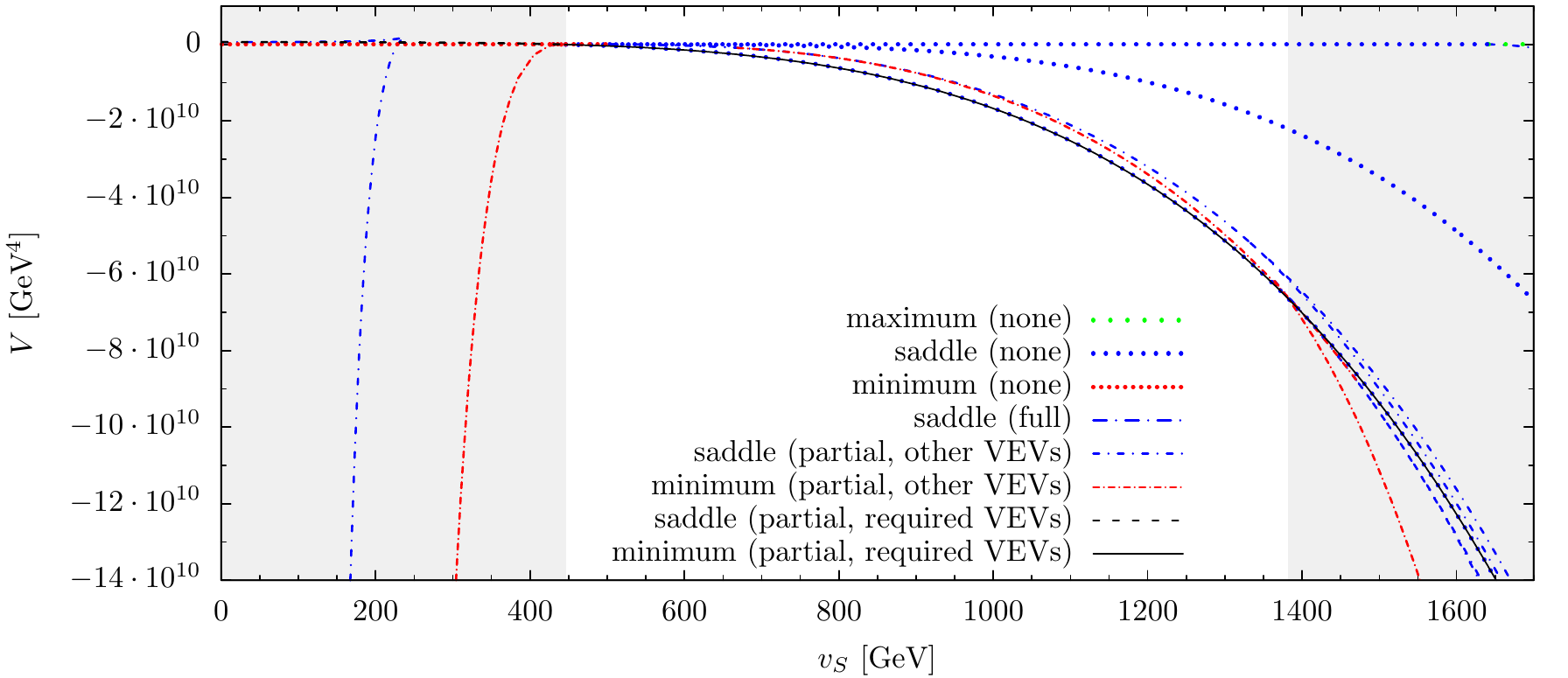}
\end{center}
\vspace*{-2mm}
\caption{\label{fig-ManiatisGlobal} \it
Values of the NMSSM Higgs potential at its stationary points in dependence
on $v_s$, as presented in~\citep{Maniatis:2006jd}. The parameters of the Higgs potential 
are chosen as
$\lambda=0.4, \kappa=0.3, \abs{A_\kappa}=200\text{ GeV},
\tan (\beta)=3, m_{H^\pm} = 2 v, \sign{R_\kappa} = -,
\dedm=\dcp=0$.
Each line corresponds to 1 or 3 stationary points sharing the same
value of the potential.
The different line styles denote saddle points, maxima, and minima.
The labels
'none', 'full', and 'partial' denote 
solutions of the classes with
unbroken~(\ref{eq-stationarityU}), 
fully broken~(\ref{eq-stationarityF}), and
partially broken~(\ref{eq-stationarityP})
\mbox{$SU(2)_L \times U(1)_Y$}, respectively.
For solutions of the partially broken class, it is also denoted
whether they correspond to the 'required VEVs' $v_u,v_d,v_s$ or
to 'other VEVs'.
Excluded parameter values, where the global minimum does not exhibit
the required vacuum expectation values, are shown by shaded area.
}
\end{figure*}

In Fig.~\ref{fig-ManiatisGlobal} the values of the potential
at all stationary points for the parameter values~(\ref{eq-centralvals}) 
with varied $v_s$ is shown~\citep{Maniatis:2006jd}.
%%%%%%%%%%%%%%%%%%%%%%%%%%%%%%%%%%%%%%%%%%%%%%%%%%%%%%%%%%%%%%%%%
%%%%%%%%%%%%%%%%%%%%%%%%%%%%%%%%%%%%%%%%%%%%%%%%%%%%%%%%%%%%%%%%%
%%%%%%%%%%%%%%%%%%%%%%%%%%%%%%%%%%%%%%%%%%%%%%%%%%%%%%%%%%%%%%%%%
%%%%%%%%%%%%%%%%%%%%%%%%%%%%%%%%%%%%%%%%%%%%%%%%%%%%%%%%%%%%%%%%%
Each curve in the figure represents 1- or 3-fold degenerate stationary potential
values, where the gauge symmetry breaking behavior of the solutions is denoted by
different line styles.
Excluded parameter regions, where the global minimum does not exhibit the
required expectation values~(\ref{eq-vac}) are shown shaded.
As is illustrated in the figure, substantial regions
of the NMSSM parameter space are excluded.
For some excluded parameter regions, the partially breaking solutions
with the required vacuum expectation values~(\ref{eq-vac})
are saddle points.
This means they can be discarded as global minima without
calculation of the other stationary points.
However, this is not always the case.
Obviously from Fig.~\ref{fig-ManiatisGlobal}, an upper bound
for $v_s$ is found.
For the plotted $v_s$ larger than this
upper exclusion bound the solutions fulfilling~(\ref{eq-vac})
are still pronounced minima, i.e. the mass
matrices have positive eigenvalues, but they are no
longer the global minima.
Also there
are non-breaking saddle points
with potential values slightly above those of the 
{\em wanted} global minimum. 
It is found that this effect is not coincidental for the
initial parameters~(\ref{eq-centralvals}) chosen, but rather a generic feature of the NMSSM.
Within the CP conserving parameter range
\begin{equation}\label{eq-randomrange}
\begin{gathered}
\lambda \in \,]0,1],\quad \kappa \in \,]0,1],\quad A_\kappa \in \pm\,]0,2500]\text{ GeV},\\
\tan\beta \in \,]0,50],\quad  v_s \in \,]0,5000]\text{ GeV},\\% \quad
m_{H^\pm} \in \,]0,2500]\text{ GeV}
\end{gathered}
\end{equation}
samples are selected, producing the wanted global minimum and
typically non-breaking saddle points are found,
where the relative separation of the potential values for the saddle points
and the global minimum is below the per-mill level, in many cases even far below.
No fully breaking global minima are found for scenarios in the range~\eqref{eq-randomrange}
where the solutions with the required vacuum expectation values~(\ref{eq-vac}) are local minima.
Eventually, there are examples, 
where CP conserving parameters with the ``wrong'' global minimum
produce the wanted global minimum 
if a non-vanishing phase $\dcp$ is introduced.

\newpage
%%%%%%%%%%%%%%%%%%%%%%%%%%%%%%%%%%%%%%%%%%%%%%%%%%%%%%%%%%%%%%%%%%5
% Summary
%%%%%%%%%%%%%%%%%%%%%%%%%%%%%%%%%%%%%%%%%%%%%%%%%%%%%%%%%%%%%%%%%%5
%%%%%%%%%%%%%%%%%%%%%%%%%%%%%%%%%%%%%%%%%%%%%%%%%%%%%%%%%%%%%%%%%
\section{Summary}
\label{sec-summary}

The MSSM suffers from the $\mu$-problem, that is, the dimensionful $\mu$-parameter
has to be adjusted by hand to the electroweak scale. In the NMSSM an
effective $\mu$-term is generated dynamically. To this purpose
an extra singlet superfield is introduced. Via spontaneous
symmetry breaking the Higgs-boson singlet
acquires a vacuum-expectation-value and generates the
required $\mu$-term dynamically.\\

However, as we have seen, absence of the $\mu$-term in 
the superpotential imposes
a continuous symmetry, the Peccei--Quinn-symmetry.
An additional cubic self-coupling term of the singlet is introduced in order
to break the Peccei--Quinn-symmetry which would lead to 
an unobserved axion. 
That is, the PQ-symmetry is promoted to a $\Zthree$-symmetry, which,
since spontaneously broken, leads to the formation
of dangerous domain walls. The loophole is to impose
higher order operators and additional symmetries,
breaking the $\Zthree$-symmetry but
not disturbing physics at the electroweak scale.\\

In the NMSSM we encounter two more Higgs bosons 
as well as a fifth neutralino; compared to the MSSM.
The modified mixing matrices were recalled in the Higgs sector 
as well as in the neutralino sector. Also the
physical potential was derived, yielding,
in particular, additional trilinear $A$-parameter terms.\\

The theoretical and experimental constraints 
on the model were discussed. On the theoretical
side restrictions arise from the symmetries
of the physical potential. Thus a global minimum
which is electric- or color-charge breaking is
forbidden, as well as a minimum with the wrong
electroweak symmetry breaking behavior.
Stability of the potential is of no concern,
since for a non-vanishing Peccei--Quinn-symmetry
breaking $\kappa$-parameter quartic terms
in the potential ensure a potential which
is bounded from below.

Further theoretical constraints come
from the requirement of perturbativity of the couplings up
to the GUT or Planck scale. Of course,
this condition relies on the viability of 
the model up to large scales.
In a series of recent publications 
the argument of fine-tuning got
a lot of attention.
On a quantitative basis, studies of fine-tuning
in different scenarios in the NMSSM as
well as compared to the MSSM were performed,
revealing that in general there is
much less fine-tuning in the NMSSM.
This fine-tuning arises mainly from
the stop-loops entering
logarithmically in the Higgs-boson selfenergy.
In particular, the LEP bounds require to have large 
radiative corrections accompanied by
very large stop masses in the MSSM.
Compared to the MSSM, the Higgs sector is much less constraint,
that is,
the NMSSM passes the LEP constraints with much less fine-tuning.\\

We proceeded with a discussion of the experimental constraints.
First of all the electroweak precision measurements
were discussed, which agree well with
the SM predictions. Especially the
invisible $Z$-boson decay puts severe
constraints on the NMSSM, but also
the constraint of limits on neutralino/chargino-
and charged Higgs-boson pair production at LEP
was discussed.\\

The impact of the muon anomalous magnetic moment, measured
at BNL, on the NMSSM was presented. 
From chargino--sneutrino and neutralino--smuon
loops we get additional contributions to the anomalous 
magnetic moment compared to the SM. 
Based on accurate SM predictions, which are 
not yet available, strong constraints on
the parameter space in the NMSSM could
be derived.\\

Also the $b \rightarrow s \gamma$ decay was 
considered, which, since loop-induced, is
very sensitive to new particles, which couple
to SM particles. Here the measurement agrees
well with the SM prediction, yielding an
additional constraint for new contributions
which arise in extensions of the SM.
Also in this respect the NMSSM does
not violate this constraint significantly
in large parts of parameter space.\\

The new five-year WMAP data give
accurate predictions of cold dark matter
in the Universe based on the $\Lambda$CDM. 
Since supersymmetric models,
respecting matter-parity, predict a new
stable particle, they provide
a natural cold dark matter candidate, the LSP. 
The WMAP constraint on the LSP turns out to
be very strong in the NMSSM and
restricts the available parameter space
enormously. Likewise, in the constraint NMSSM,
where unification of the scalar mass,
of the gaugino mass, and of the trilinear
$A$-parameter at the GUT scale
is assumed, the theoretical
and experimental constraints
are shown to be highly restrictive.\\

Strong first order electroweak phase transitions in order
to account for the observed
baryon--antibaryon asymmetry in our 
Universe require 
a cubic term in the physical potential.
With this respect the SM as well
as the MSSM rely on loop contributions
in the effective potential which are
generically small. In contrast, in the
NMSSM, the additional trilinear terms
allow to accomplish for this mechanism
of baryogenesis without large fine-tuning.\\

Some recent parameter scans were reviewed
with quite interesting results, constraining
the NMSSM parameter space. Some emphasis was placed on
a ``no-lose''-theorem, that is, the question whether
at least one Higgs-boson is detectable
at the LHC with high integrated luminosity.
The Higgs-to-Higgs decays are found to
be rather difficult to detect, which however
correspond to large parameter space.
New ideas to detect such signatures
were reviewed. However there seems to
be some parameter space left, where
all supersymmetric partner particles
could escape detection.
An electron--positron collider
could close this gap in detecting 
the recoil mass in signatures
of Higgs-boson production in invisible $Z$-strahlung. 
Some publications were reviewed discussing how to distinguish
the NMSSM from the MSSM.
The fifth neutralino may lead to a very different signature
in colliders. In case the LSP is a singlino-like neutralino, this
LSP has suppressed couplings to non-Higgs particles. Since eventually
all superpartner particles decay into the LSP this would cause
very different signatures, possibly displaced vertices.\\

Eventually, we draw attention to the determination of the global minimum
in the Higgs potential.
An algebraic method based on Groebner bases computation was introduced, showing a surprising
rich structure of stationary solutions. However, this approach is
yet limited to studies of the tree-level potential.\\

The NMSSM is an intriguing model, which deserves
a lot of attention, since it is a coherent
supersymmetric extension of the Standard Model.
Moreover the NMSSM complies with collider and cosmological
precision data and could
be discovered at the LHC.

\acknowledgments{I am very grateful to O.~Nachtmann for
encouraging me to write this review, for giving me uncountable
helpful advises and for many valuable discussions. A lot of thanks go
to P.M.~Zerwas for carefully studying this review and especially
for his critical remarks and fruitful suggestions. Last but not least let
me greatly acknowledge discussions with P.~Fayet and E.~Ma.}

\newpage
%%%%%%%%%%%%%%%%%%%%%%%%%%%%%%%%%%%%%%%%%%%%%%%%%%%%%%%%%%%%%%%%%
%%%%%%%%%%%%%%%%%%%%%%%%%%%%%%%%%%%%%%%%%%%%%%%%%%%%%%%%%%%%%%%%%

\appendix

%%%%%%%%%%%%%%%%%%%%%%%%%%%%%%%%%%%%%%%%%%%%%%%%%%%%%%%%%%%%%%%%%%5
% Conventions and Abbreviations
%%%%%%%%%%%%%%%%%%%%%%%%%%%%%%%%%%%%%%%%%%%%%%%%%%%%%%%%%%%%%%%%%%5
%%%%%%%%%%%%%%%%%%%%%%%%%%%%%%%%%%%%%%%%%%%%%%%%%%%%%%%%%%%%%%%%%
\section{Conventions and abbreviations}
\label{app-conv}

We use the space-time metric $(g_{\mu \nu})= \diag(1,-1,-1,-1)$.\\
The generalized Pauli matrices are
\begin{equation}
\sigma_0 = \bar{\sigma}_0 =
\begin{pmatrix}
1 & 0 \\ 0 & 1
\end{pmatrix},
\qquad
\sigma_1 = - \bar{\sigma}_1 =
\begin{pmatrix}
0 & 1 \\ 1 & 0
\end{pmatrix},
\qquad
\sigma_2 = - \bar{\sigma}_2 =
\begin{pmatrix}
0 & -i \\ i & 0
\end{pmatrix},
\qquad
\sigma_3 = - \bar{\sigma}_3 =
\begin{pmatrix}
1 & 0 \\ 0 & -1
\end{pmatrix}\;.
\end{equation}
The $\epsilon$ symbol matrix is 
\begin{equation}
\epsilon =
\begin{pmatrix}
\phantom{+}0 & 1\\
-1 & 0
\end{pmatrix}\;.
\end{equation}
We use the usual representation for the  Dirac matrices
\begin{equation}
\gamma_\mu =
\begin{pmatrix}
0 & \sigma_\mu \\ \bar{\sigma}_\mu & 0
\end{pmatrix},
\qquad 
\text{and}
\quad
\gamma_5=
\begin{pmatrix}
\unitmatrix_2 & 0 \\ 0 & -\unitmatrix_2
\end{pmatrix}\;.
\end{equation}
Further the projectors are
\begin{equation}
P_L \equiv \frac{\unitmatrix_4+\gamma_5}{2}=
\begin{pmatrix}
\unitmatrix_2 & 0 \\ 0 & 0
\end{pmatrix},
\qquad
P_R \equiv \frac{\unitmatrix_4-\gamma_5}{2}=
\begin{pmatrix}
0 & 0 \\ 0 & \unitmatrix_2
\end{pmatrix}\;.
\end{equation}
The convention for the vacuum-expectation-values include the square-root of two:
\begin{equation}
\langle H_d \rangle=
\begin{pmatrix}
v_d/\sqrt{2} \\ 0
\end{pmatrix},
\qquad
\langle H_u \rangle=
e^{i \phi_u}
\begin{pmatrix}
0 \\ v_u/\sqrt{2} 
\end{pmatrix},
\qquad
\langle S \rangle= e^{i \phi_s} v_s/\sqrt{2} \,,
\end{equation}
and we use the definition
\begin{equation}
\tan (\beta) = t_\beta = \frac{v_u}{v_d},
\qquad \text{with }\quad
v = \sqrt{v_u^2+v_d^2} \approx 246~\text{GeV}\;.
\end{equation}
We write $c_\beta = \cos(\beta)$, $s_\beta = \sin(\beta)$,
$\cot_\beta = \cot(\beta)$. 
After electroweak symmetry breaking we thus get
the tree level gauge-boson masses $m_W= g_2 v/2$ and $m_Z= \sqrt{g_1^2+g_2^2} v/2$.
As usual, $g_2$ and $g_1$ are the \eweakgroup
gauge couplings with the positron charge
$e= g_1 c_W = g_2 s_W$.
We use the abbreviations related to 
the Weinberg angle $c_W = \cos (\theta_W)$ as well
as $s_W = \sin (\theta_W)$.
For the partial derivative acting on fields, the
short notation
$(F \dlr{\mu} G) \equiv F (\partial_\mu G) - (\partial_\mu F) G$
is used.

\newpage
%%%%%%%%%%%%%%%%%%%%%%%%%%%%%%%%%%%%%%%%%%%%%%%%%%%%%%%%%%%%%%%%%%5
% Computer tools for the NMSSM
%%%%%%%%%%%%%%%%%%%%%%%%%%%%%%%%%%%%%%%%%%%%%%%%%%%%%%%%%%%%%%%%%%5
%%%%%%%%%%%%%%%%%%%%%%%%%%%%%%%%%%%%%%%%%%%%%%%%%%%%%%%%%%%%%%%%%
\section{Computer tools for the NMSSM}
\label{app-tools}

Here we list some frequently used computer tools for the calculations in
the NMSSM. In the parameter scans discussed in
Sec.~\ref{sect-scan} these tools appear in different contexts.

\begin{itemize}
\item NMHDECAY \citep{Ellwanger:2004xm,Ellwanger:2005dv}.
This Fortran code computes the masses of all sparticles and Higgs bosons.
The couplings and decay width of the Higgs bosons are also calculated.
The input parameters are $\lambda$, $\kappa$, $A_\lambda$, $A_\kappa$, $\tan(\beta)$,
$\mu \equiv \lambda v_s$ taken at the electroweak scale. The computation
of the Higgs-boson spectrum is done including leading electroweak corrections as
well as certain two loop terms. 
The decay width refers to HDECAY \citep{Djouadi:1997yw},
but without taking
into account three body decays.\\
Current experimental exclusion limits are taken into account.
The program code may be downloaded from the url:
\url{http://www.th.u-psud.fr/NMHDECAY/nmssmtools.html}

\item NMSPEC \citep{Ellwanger:2006rn}.
In contrast to NMHDECAY here the soft-breaking parameters have to
be specified at the GUT scale. Also the spectrum and couplings are
computed from this input.
Download at url: \url{http://www.th.u-psud.fr/NMHDECAY/nmssmtools.html}

\item MicrOMEGAS \citep{Belanger:2006is,Belanger:2008sj}.
This Tool calculates the relic density
of a stable massive particle together with the rates
for direct and indirect detection of dark matter.
MicrOMEGAS includes already a model file for the NMSSM in
addition to various other models. It may even be extended
to further models by the user.
Download available at url: \url{http://wwwlapp.in2p3.fr/lapth/micromegas}

\item CompHEP \citep{Boos:2004kh}.
This package calculates total and differential cross sections at
tree-level accuracy. Multi-particle final states in collisions as well
as decay processes can be computed in a completely automatic way. We refer
to the url: \url{http://comphep.sinp.msu.ru}

\item SuperIso \citep{Mahmoudi:2009zz}.
The SuperIso program evaluates different flavor physics observables in the
MSSM and the NMSSM. Examples are the branching ratios of
$B \to X_s \gamma$,
$B_s \to \mu^+ \mu^-$,
$B \to \tau \nu_\tau$,
$B \to D \tau \nu_\tau$,
$K \to \mu \nu_\mu$,
$D_s \to \tau \nu_\tau$,
$D_s \to \mu \nu_\mu$,
as well as the isospin asymmetry
$B \to K^* \gamma$.
The corresponding homepage can be found at the
url: \url{http://superiso.in2p3.fr}

\item LanHEP \citep{Semenov:1997qm}.
LanHEP computes the Feynman rules for a given Lagrangian. The initial
Lagrangian can be written in a compact form. The output Feynman rules
are given in terms of physical fields and independent parameters. It can
also be used to directly generate a model file as input for CompHEP. Download
at the url: \url{http://theory.sinp.msu.ru/~semenov/lanhep.html}
\end{itemize}

%%%%%%%%%%%%%%%%%%%%%%%%%%%%%%%%%%%%%%%%%%%%%%%%%%%%%%%%%%%%%%%%%%5
% Construction of a supersymmetric model
%%%%%%%%%%%%%%%%%%%%%%%%%%%%%%%%%%%%%%%%%%%%%%%%%%%%%%%%%%%%%%%%%%5
%%%%%%%%%%%%%%%%%%%%%%%%%%%%%%%%%%%%%%%%%%%%%%%%%%%%%%%%%%%%%%%%%
\section{Construction of a supersymmetric model}
\label{app-A}
Before we start to present the Lagrangian of the NMSSM we want
to sketch how an arbitrary supersymmetric model is constructed.
Based on this sketch it is easy to construct the specific NMSSM Lagrangian
in the next section. Here we follow closely the excellent introduction
given in~\citep{Martin:1997ns}. First we recall the meaning
of matter parity.

%%%%%%%%%%%%%%%%%%%%%%%%%%%%%%%%%%%%%%%%%%%%%%%%%%%%%%%%%%%%%%%%%%5
% Matter parity
%%%%%%%%%%%%%%%%%%%%%%%%%%%%%%%%%%%%%%%%%%%%%%%%%%%%%%%%%%%%%%%%%%5
\subsection{Matter parity}
\label{sub-matter}

Writing down the superpotential in the
MSSM \eqref{eq-Wscalar3} or the NMSSM \eqref{eq-Wscalar}, lepton and baryon number violating 
terms are omitted. Additional lepton number ($B$) and baryon number ($L$) 
terms in
the superpotential with dimensionless couplings would be
\begin{equation}
W_{\Delta B,\; \Delta L} =
\frac{1}{2} \lambda \hat{e} \left( \hat{L}^\trans \epsilon \hat{L} \right)
+ \lambda' \hat{d} \left( \hat{L}^\trans \epsilon \hat{Q} \right)
+ \frac{1}{2} \lambda'' \hat{d} \hat{d} \hat{u} \;.
\end{equation}
The non-appearance of such terms in the superpotential can
be gained by imposing an additional symmetry principle, called
{\em matter parity} \citep{Dimopoulos:1981zb,Weinberg:1981wj,Sakai:1981pk,Dimopoulos:1981dw} 
or {\em R-parity} \citep{Farrar:1978xj}.
Matter parity is defined as a multiplicative quantum number
\begin{equation}
P_M = (-1)^{3 (B-L)},
\end{equation}
such that only terms in the Lagrangian or the superpotential
are allowed with multiplicative matter parity $P_M= +1$.
In this way the forbidden terms are excluded from the model.
Equivalently one can impose also {\em R-parity}
instead of matter parity, defined as the multiplicative
quantum number
\begin{equation}
P_R = (-1)^{3 (B-L) +2 s}
\end{equation}
with $s$ the spin of the particle. Since the product of $(-1)^{2 s}$
is always $1$ in angular momentum conserving interaction vertices,
R-parity is indeed equivalent to matter parity but
has the advantage to give for the SM particles and Higgs bosons
$P_R=+1$ and for the superpartners $P_R=-1$.
R-parity conserving immediately translates to the fact that
the lightest supersymmetric particle (LSP) is stable.
Since the NMSSM like the MSSM is R-parity conserving
we expect to have a LSP which might be a candidate for
the up to now missing cold dark matter.

%%%%%%%%%%%%%%%%%%%%%%%%%%%%%%%%%%%%%%%%%%%%%%%%%%%%%%%%%%%%%%%%%%5
% Chiral supermultiplets
%%%%%%%%%%%%%%%%%%%%%%%%%%%%%%%%%%%%%%%%%%%%%%%%%%%%%%%%%%%%%%%%%%5
\subsection{Chiral supermultiplets}
The boson and fermion fields are cast into {\em chiral} supermultiplets.
Each of the $n$ boson fields of the theory, $\phi_i$ with $i=1,...,n$ is accompanied by a Weyl-fermion $\psi_i$
and an additional auxiliary field $F_i$ necessary to close the supersymmetry algebra
off-shell. Note that the auxiliary fields $F_i$ are no physical fields since
they do not propagate.
The supersymmetric Lagrangian, that is, the Lagrangian which is invariant under
supersymmetry transformations which turn bosons into fermions and vice versa, is
\begin{equation}
\label{eq-lagfree}
{\cal L}_{\text{chiral, free}}=
-(\partial_\mu \phi^{i})^\dagger (\partial^\mu \phi_i)
- i \psi^{i *} \bar{\sigma}^\mu \partial_\mu \psi_i
+ F^{i*} F_i \; .
\end{equation}
The generalized Pauli matrices are $\sigma_0=-\bar{\sigma}_0= \unitmatrix_2$, $\bar{\sigma}_1=-\sigma_1$,
$\bar{\sigma}_2=-\sigma_2$, $\bar{\sigma}_1=-\sigma_3$.
The most general set of renormalizable non-gauge interactions of these chiral
supermultiplets are
\begin{equation}
\label{eq-lagchir}
{\cal L}_{\text{chiral, int}}=
-\frac{1}{2} W^{ij} \psi_i \psi_j 
+ W^i F_i + c.c.\; ,
\end{equation}
where $W^i$ as well as $W^{ij}$ are determined from one function, 
the so-called {\em superpotential} $W$:
\begin{equation}
W = \frac{1}{2} \mu^{ij} \phi_i \phi_j
+ \frac{1}{6} \lambda^{ijk} \phi_i \phi_j \phi_k
\end{equation}
with
\begin{equation}
\label{eq-Wij}
\begin{split}
W^i &= \frac{\delta W}{\delta \phi_i} = \mu^{ij} \phi_j + \frac{1}{2} \lambda^{ijk} \phi_j \phi_k\;,\\
W^{ij} &= \frac{\delta^2 W}{\delta \phi_i \delta \phi_j} = \mu^{ij} + \lambda^{ijk} \phi_k \;.
\end{split}
\end{equation}
This general form of the interactions \eqref{eq-lagchir} is dictated by supersymmetry itself.
From the Lagrangians~(\ref{eq-lagfree}) and (\ref{eq-lagchir}) we find the
equation of motion 
\begin{equation}
\label{eq-F}
F_i=-W_i^* \;,
\end{equation}
thus the auxiliary fields $F_i$ can be expressed in terms of the scalar fields.\\

%%%%%%%%%%%%%%%%%%%%%%%%%%%%%%%%%%%%%%%%%%%%%%%%%%%%%%%%%%%%%%%%%%5
% Gauge supermultiplets
%%%%%%%%%%%%%%%%%%%%%%%%%%%%%%%%%%%%%%%%%%%%%%%%%%%%%%%%%%%%%%%%%%5
\subsection{Gauge supermultiplets}
The gauge boson fields $A_\mu^a$ are paired with Weyl fermions $\lambda^a$ into gauge
supermultiplets, where
auxiliary fields $D^a$ are needed in order to close the supersymmetry algebra 
off-shell. The adjoint representation of the gauge group is denoted by the index~$a$
here. The Lagrangian of the supersymmetric gauge supermultiplet fields is
\begin{equation}
{\cal L}_{\text{gauge}} =
- \frac{1}{4} F_{\mu \nu}^a F^{\mu \nu a} 
- i \lambda^\dagger \bar{\sigma}^\mu D_\mu \lambda^a
+ \frac{1}{2} D^a D^a \;,
\end{equation}
where, as usual, the Yang-Mills field strength is
$F_{\mu \nu}^a = \partial_\mu A_\nu^a - \partial_\nu A_\mu^a - g f^{abc} A_\mu^b A_\nu^c$
and the covariant derivative of the gaugino field reads
\begin{equation}
D_\mu \lambda^a = \partial_\mu \lambda^a - g f^{abc} A_\mu^b \lambda^c
\end{equation}
with structure
constants $f^{abc}$ and gauge coupling $g$.

%%%%%%%%%%%%%%%%%%%%%%%%%%%%%%%%%%%%%%%%%%%%%%%%%%%%%%%%%%%%%%%%%%5
% Interactions
%%%%%%%%%%%%%%%%%%%%%%%%%%%%%%%%%%%%%%%%%%%%%%%%%%%%%%%%%%%%%%%%%%5

\subsection{Interactions}
As usual the gauge interactions of the bosons and fermions are given by turning the partial derivatives in the
kinetic terms in the Lagrangian~(\ref{eq-lagfree}) into covariant derivatives.
Suppose the chiral supermultiplet fields transform under a gauge group in a representation
with hermitian matrices $T^a$ satisfying $[T^a, T^b]= i f^{abc} T^c$. 
Then the covariant derivatives of the multiplet fields are
\begin{equation}
\label{eq-cov}
\begin{split}
D_\mu \phi_i &= \partial_\mu \phi_i + i g A_\mu^a (T^a \phi)_i \;, \\
D_\mu \psi_i &= \partial_\mu \psi_i + i g A_\mu^a (T^a \psi)_i \; .
\end{split}
\end{equation}
Thus, the gauge invariant form of the Lagrangian~(\ref{eq-lagfree}) becomes
\begin{equation}
\label{eq-laggauge}
{\cal L}_{\text{chiral}}=
-(D_\mu \phi^{i})^\dagger (D^\mu \phi_i)
- i \psi^{i *} \bar{\sigma}^\mu D_\mu \psi_i
+ F^{i*} F_i \;.
\end{equation}

We have further to introduce all additional terms in the Lagrangian, which do not
violate any symmetry of the theory. The following Yukawa terms are neither
forbidden by gauge invariance nor by renormalizability. The couplings
in these Yukawa interaction terms are determined by supersymmetry. 
The Yukawa couplings are
\begin{equation}
\label{eq-yuk}
{\cal L}_{\text{Yukawa}}=-\sqrt{2} g \left[ (\phi^* T^a \psi) \lambda^a 
+ \lambda^{a \dagger} (\psi^\dagger T^a \phi) \right]
+ g ( \phi^* T^a \phi ) D^a \;.
\end{equation}
From ${\cal L}_{\text{gauge}}$ and ${\cal L}_{\text{Yukawa}}$ we get
the equation of motion for the auxiliary field
\begin{equation}
\label{eq-D}
D^a = -g (\phi^* T^a \phi)
\end{equation}
and we see that like for the auxiliary fields $F_i$ we can express $D^a$ 
in terms of scalar fields. We emphasize that the Yukawa coupling of a 
fermion with a scalar and a gaugino is determined by the gauge coupling,
one of the firm predictions of supersymmetry.

%%%%%%%%%%%%%%%%%%%%%%%%%%%%%%%%%%%%%%%%%%%%%%%%%%%%%%%%%%%%%%%%%%5
% Soft breaking terms
%%%%%%%%%%%%%%%%%%%%%%%%%%%%%%%%%%%%%%%%%%%%%%%%%%%%%%%%%%%%%%%%%%5
\subsection{Soft breaking terms}
We know from experiment that supersymmetry must be broken. 
This breaking is expected to be spontaneous, such that the Lagrangian
is invariant under supersymmetry transformations but the vacuum is not.
The breaking mechanism itself is up to date unknown. In order to keep
the theory as general as possible all explicit breaking terms
are introduced which do not lead to quadratical divergences.
A necessary condition for this is to introduce only
Lagrangian terms with couplings of positive mass dimension.
The most general 
soft breaking terms are~\citep{Girardello:1981wz}
\begin{equation}
\label{eq-soft}
{\cal L}_{\text{soft}}= 
-(m^2)^i_j \phi^{j \ast} \phi_i
-\frac{1}{2} \left( M_\lambda \lambda^a \lambda^a + c.c. \right)
-\left( \frac{1}{2} b^{ij} \mu^{ij} \phi_i \phi_j + \frac{1}{6} a^{ijk} \lambda^{ijk} \phi_i \phi_j \phi_k + c.c.\right)\,.
\end{equation}
The first term gives masses to the scalar superpartners and the second one masses to the gauginos.
In this way the degeneracy among the superpartners is removed. The bilinear ($b^{ij}$)
and trilinear ($a^{ijk}$) terms in the bracket are terms associated to
the superpotential bilinear and trilinear terms. Note that there is
implicit summation over the indices~$i,j$, respectively~$i,j,k$
running over all scalar fields.

%%%%%%%%%%%%%%%%%%%%%%%%%%%%%%%%%%%%%%%%%%%%%%%%%%%%%%%%%%%%%%%%%%5
% Complete supersymmetric Lagrangian
%%%%%%%%%%%%%%%%%%%%%%%%%%%%%%%%%%%%%%%%%%%%%%%%%%%%%%%%%%%%%%%%%%5
\subsection{Complete supersymmetric Lagrangian}
\label{sec-tot}
Eventually we have the general supersymmetric Lagrangian of a renormalizable
theory consisting of the chiral- and gauge-supermultiplets in gauge invariant form
as well as the Yukawa and the soft breaking part: 
\begin{equation}
{\cal L} = {\cal L}_{\text{chiral}} + {\cal L}_{\text{gauge}} 
+ {\cal L}_{\text{Yukawa}} + {\cal L}_{\text{soft}} \;.
\end{equation}

We may also isolate the potential part of this Lagrangian. 
We have the $F$ term of ${\cal L}_{\text{chiral}}$, the $D$ term of
${\cal L}_{\text{gauge}}$ and the scalar part of ${\cal L}_{\text{soft}}$,
altogether
\begin{equation}
\label{eq-pot}
\begin{split}
V &=
V_F+V_D+V_{\text{soft}}\\
&=
-F^{i*} F_i
-\frac{1}{2} D^a D^a 
+(m^2)^i_j \phi^{j \ast} \phi_i
+\left( \frac{1}{2} b^{ij} \mu^{ij} \phi_i \phi_j + \frac{1}{6} a^{ijk} \lambda^{ijk} \phi_i \phi_j \phi_k + c.c.\right)\,,
\end{split}
\end{equation}
where the equation of motion for the auxiliary fields~(\ref{eq-F}) and (\ref{eq-D}) have to be inserted.

\newpage
%%%%%%%%%%%%%%%%%%%%%%%%%%%%%%%%%%%%%%%%%%%%%%%%%%%%%%%%%%%%%%%%%%5
% Feynman rules of the NMSSM
%%%%%%%%%%%%%%%%%%%%%%%%%%%%%%%%%%%%%%%%%%%%%%%%%%%%%%%%%%%%%%%%%%5
\section{Feynman rules of the NMSSM}
\label{app-feyn}
In this appendix some of the essential Feynman rules of the NMSSM 
are derived.
With the previous
work in appendix~\ref{app-A} this is straightforward.
We focus on the Higgs sector and
restrict ourselves to the trilinear couplings.
In Ref.~\citep{Gunion:1984yn} the Feynman rules for a 
generic supersymmetric two-Higgs-doublet model are derived, taking
also an additional singlet into account. But this work is restricted
to the case that the singlet does not mix with the other neutral Higgs bosons -
in contrast to the case of the NMSSM. 
The Feynman rules for the NMSSM can also be found in
Refs.~\citep{Franke:1995tc,Ellwanger:2004xm}. Here we want to present
the Feynman rules for the general case, 
including CP violating
complex phases in the NMSSM parameters~\cite{graf}. In this case, the
scalar Higgs bosons mix with the pseudoscalar ones, as described 
in section~\ref{sub-massmatrices}. 
Note that we perform the $\beta$ rotation of the Higgs-boson
doublets explicitely.

In the case where the Feynman rules depend on 
any field momentum, the direction of this momentum
is denoted by an extra arrow. Further, we use
the abbreviations for the Weinberg angles, $s_W \equiv \sin (\theta_W)$, 
$c_W \equiv \cos (\theta_W)$, $\beta$ is the mixing angle
of the Higgs doublets with
$\tan (\beta) \equiv t_\beta = v_u/v_d$ and $v^2=v_u^2+v_d^2$. The mixing of
the Higgs bosons is determined by the mixing matrix $(R_{ij})$ 
and originates from the diagonalization~\eqref{eq-Hrot} with indices
$i,j=1,...,5$.
Since
the resulting entries in this matrix are very involved it is advantageous to perform the
diagonalization in a numerical way.
We use the short notation
$(F \dlr{\mu} G) \equiv F (\partial_\mu G) - (\partial_\mu F) G$.

%%%%%%%%%%%%%%%%%%%%%%%%%%%%%%%%%%%%%%%%%%%%%%%%%%%%%%%%%%%%%%%%%%5
% Higgs-boson gauge-boson interaction
%%%%%%%%%%%%%%%%%%%%%%%%%%%%%%%%%%%%%%%%%%%%%%%%%%%%%%%%%%%%%%%%%%5
\subsection{Higgs-boson gauge-boson interaction}
\label{sub-Hgauge}
The interaction of the Higgs bosons with the gauge bosons is determined
completely by the gauge invariant kinetic terms of the Higgs bosons (\ref{eq-laggauge})
\begin{equation}
\label{eq-higgskin}
{\cal L}_{\text{Higgs, kin}}=
-( D_\mu H_d)^\dagger ( D^\mu H_d)
-( D_\mu H_u)^\dagger ( D^\mu H_u)
-( \partial_\mu S^*) ( \partial^\mu S)\;.
\end{equation}
Note that the Higgs singlet $S$ is supposed not to have any gauge interactions,
that is, the covariant derivative in this case is just the usual partial derivative.
The covariant derivative for a field with electroweak interactions
reads (\ref{eq-cov})
\begin{equation}
\label{eq-kov}
D_\mu= \partial_\mu 
+ i g_1 B_\mu(x) {\mathbf Y}_W
+ i g_2 W_\mu^a(x)\; \opw
\end{equation}
with hypercharge operator ${\mathbf Y}_W$ and isoweak operators $\opw$ and
the corresponding gauge fields $B_\mu(x)$ respectively $W_\mu^a(x)$ .
The Higgs doublets have weak isospin and hypercharges
$\opw H_d=\sigma^a/2 H_d $ and  $\opw H_u=\sigma^a/2 H_u$ with $a=1,2,3$ and
${\mathbf Y}_W H_d = -1/2 H_d$ and  ${\mathbf Y}_W H_u = +1/2 H_u$ (see Tab.~\ref{NMSSM-cont}).
The Higgs bosons are singlets with respect to the $SU(3)_C$ gauge group. 
Inserting the covariant derivative~\eqref{eq-kov} into (\ref{eq-higgskin}) 
and using the parameterization 
(\ref{eq-higgsespara}) for the Higgs-boson doublets in the unitary gauge 
we find the following Feynman rules. 
The $5 \times 5$ matrix $R$ is defined in~\eqref{eq-Hrot}.\\

% Feynman rules
\noindent
\begin{tabular}{m{0.25\linewidth}m{0.5\linewidth}}
\cline{1-2}\\
\includegraphics[width=.7\linewidth]{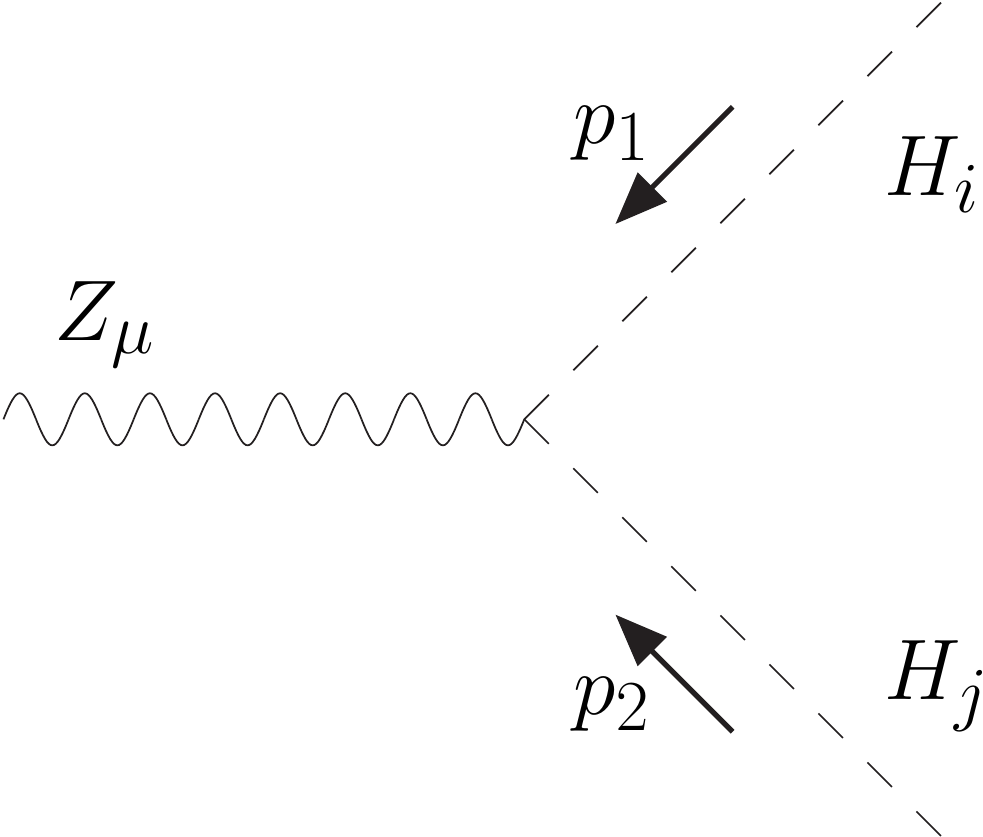} 
\hspace{0.1\linewidth}
&
\begin{multline} \nonumber
\frac{i \sqrt{g_1^2+g_2^2}}{2} (p_2^\mu - p_1^\mu)
\bigg(
	(R_{i 4} R_{j 1}-R_{i 1} R_{j 4}) s_\beta\\ + (R_{i 2} R_{j 4}-R_{i 4} R_{j 2}) c_\beta
\bigg)
\end{multline}
\\
\end{tabular}
\begin{tabular}{m{0.25\linewidth}m{0.5\linewidth}}
\includegraphics[width=.7\linewidth]{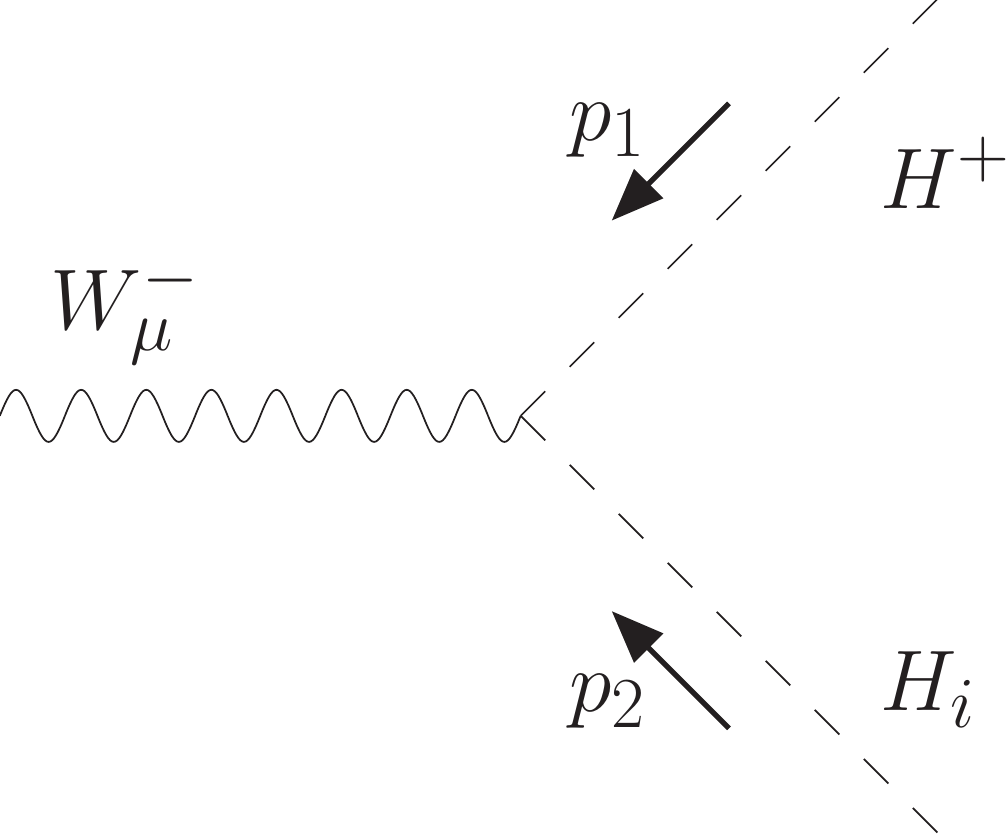} 
\hspace{0.1\linewidth}
&
$
\frac{g_2}{2} (p_2^\mu - p_1^\mu)
\bigg(
	R_{i 2} c_\beta - R_{i 1} s_\beta  - i R_{i 4}
\bigg)
$
\\
\end{tabular}
\begin{tabular}{m{0.25\linewidth}m{0.5\linewidth}}
%\cline{1-2}\\
\includegraphics[width=.7\linewidth]{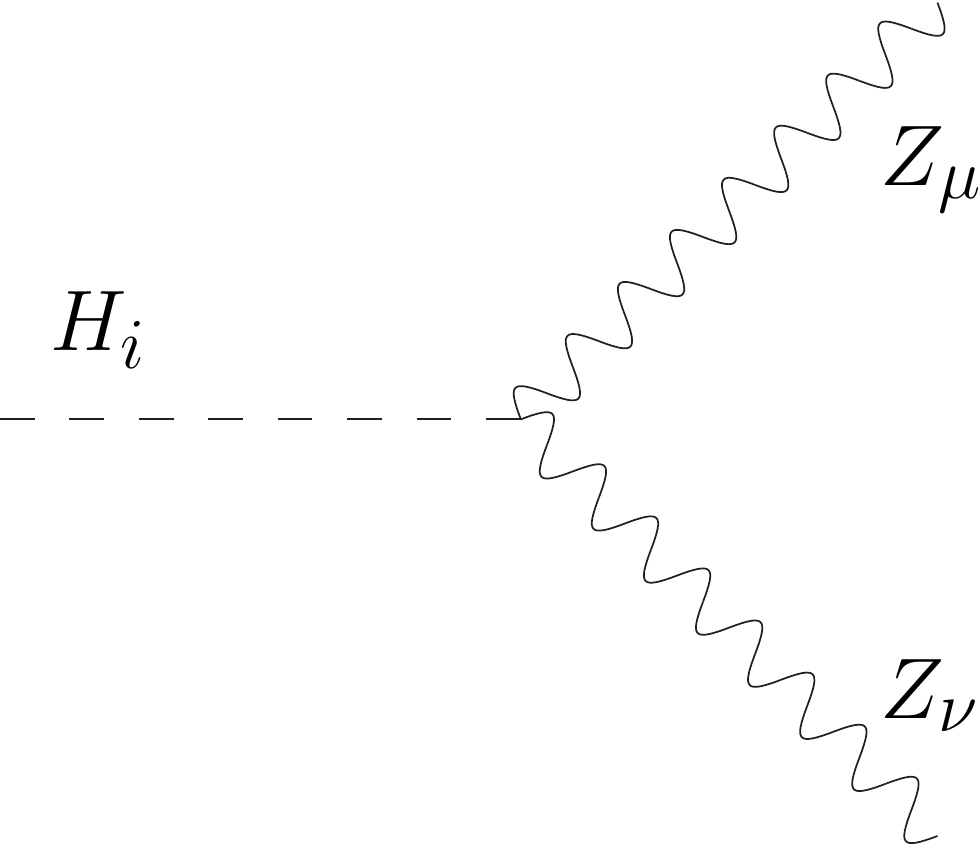} 
\hspace{0.1\linewidth}
&
$
\frac{g_1^2+g_2^2}{2}\; v\;
g^{\mu \nu}
\bigg(
	R_{i 1} c_\beta + R_{i 2} s_\beta
\bigg)
$\\
\end{tabular}
\begin{tabular}{m{0.25\linewidth}m{0.5\linewidth}}
\includegraphics[width=.7\linewidth]{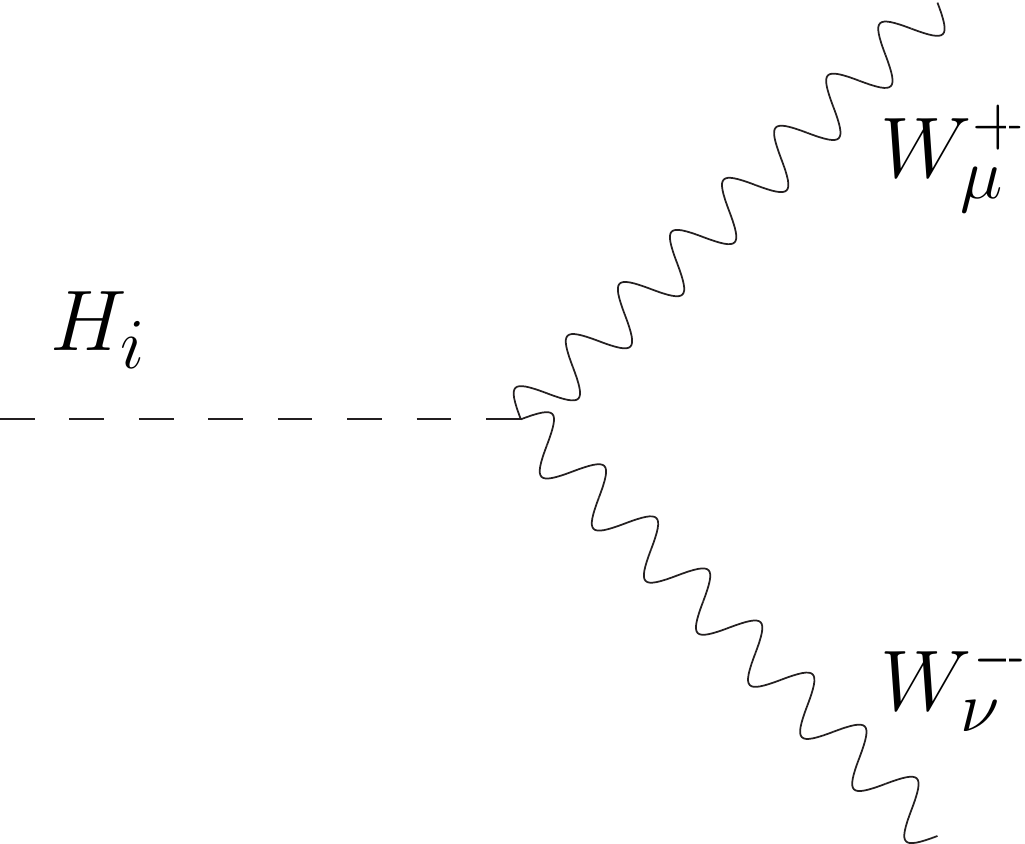} 
\hspace{0.1\linewidth}
&
$
\frac{g_2^2}{2}\;v\; g^{\mu \nu}
\bigg(
	R_{i 1} c_\beta + R_{i 2} s_\beta
\bigg)
$\\ 
%\cline{1-2}
\end{tabular}
\begin{tabular}{m{0.25\linewidth}m{0.5\linewidth}}
\includegraphics[width=.7\linewidth]{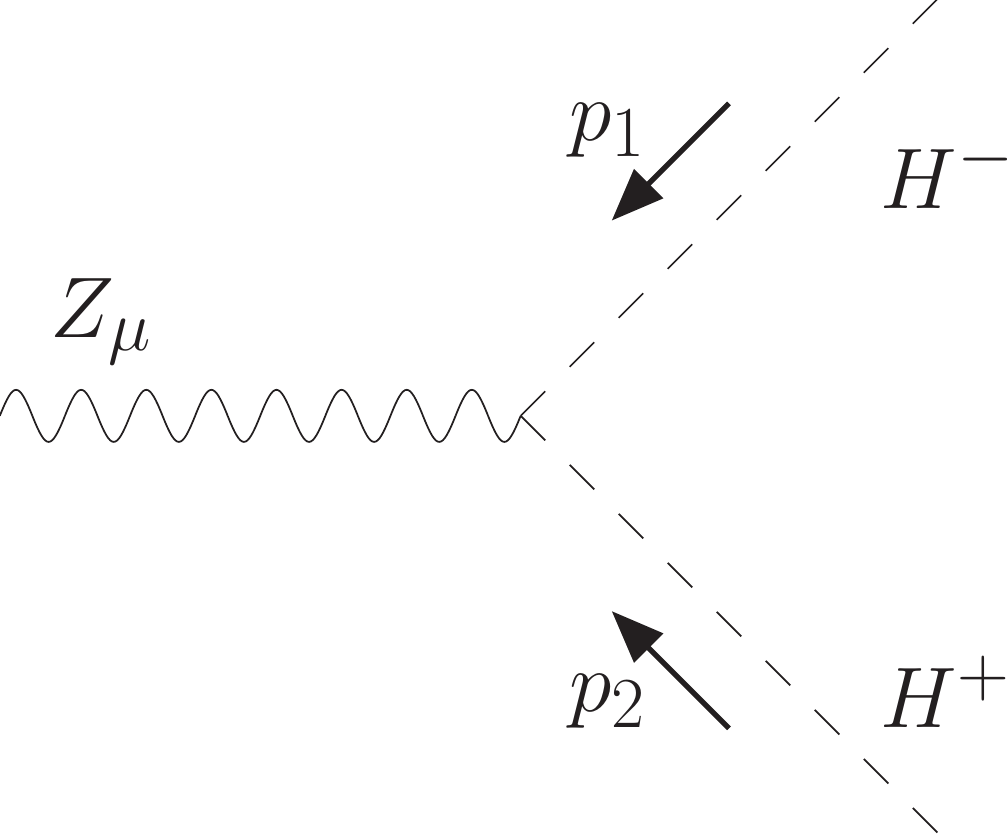} 
\hspace{0.1\linewidth}
&
$
\frac{g_1^2-g_2^2}{2 g_2}\;c_W (p_2^\mu - p_1^\mu)
$
\\
\cline{1-2}
\end{tabular}

%%%%%%%%%%%%%%%%%%%%%%%%%%%%%%%%%%%%%%%%%%%%%%%%%%%%%%%%%%%%%%%%%%5
% Higgs-boson fermion interaction
%%%%%%%%%%%%%%%%%%%%%%%%%%%%%%%%%%%%%%%%%%%%%%%%%%%%%%%%%%%%%%%%%%5
\subsection{Higgs-boson fermion interaction}
From the superpotential~\eqref{eq-Wscalar} we get via~\eqref{eq-lagchir} and \eqref{eq-Wij} the Higgs-boson 
fermion
interactions (note that there is no mass term proportional $\mu^{ij}$ in the NMSSM superpotential)
\begin{equation}
{\cal L}_{H \bar{f} f}= 
- u_R^\dagger y_u (Q^\trans \epsilon H_u) 
+ d_R^\dagger y_d (Q^\trans \epsilon H_d) 
+ c.c. 
\end{equation}
Inserting the doublets from Tab.~\ref{NMSSM-cont} this reads
\begin{equation}
{\cal L}_{H \bar{f} f}= 
-y_u u_R^\dagger u_L H_u^0
-y_d d_R^\dagger d_L H_d^0
+y_u u_R^\dagger d_L H_u^+
+y_d d_R^\dagger u_L H_d^-
+ c.c.
\end{equation}
Now we translate the Weyl two-component spinors into Dirac four-component spinors. For
arbitrary Dirac spinors decomposed into Weyl components $f=(f_{L}, f_{R})^\trans$ and
$g=(g_{L}, g_{R})^\trans$
we get
$\bar{f} g = f^\dagger \gamma_0 g = f^\dagger_L g_R + f^\dagger_R g_L$ and
$\bar{f} \gamma_5 g = - f^\dagger_{L} g_R + f^\dagger_R g_L$,
that is, we have
$f^\dagger_R g_L = \bar{f} P_L\; g $ and 
$f^\dagger_L g_R = \bar{f} P_R\; g $,
where $P_{L/R} \equiv (\unitmatrix_4 \pm \gamma_5)/2$.
Using
the parameterization~(\ref{eq-higgsespara}) and four-component Dirac spinors 
$u \equiv (u_L, u_R)^\trans$ and
$d \equiv (d_L, d_R)^\trans$ we get
\begin{equation}
{\cal L}_{H \bar{f} f}= 
- \frac{y_u}{\sqrt{2}} \bar{u} (v_u+h_u+ia_u \gamma_5) u
- \frac{y_d}{\sqrt{2}} \bar{d} (v_d+h_d+ia_d \gamma_5) d
+ (y_u \bar{u} P_L d H_u^+ + y_d \bar{u} P_R d H_d^+ + c.c.) \;.
\end{equation}
A complex phase in the parameterization of $H_u$ can be absorbed into the
parameter $y_u$.
The bilinear mass terms of this Lagrangian can be used
to relate the parameters $y_d$, $y_u$
to the fermion masses. We find
\begin{equation}
y_u= \frac{g_2 m_u}{\sqrt{2} m_W s_\beta}\;,
\qquad
y_d= \frac{g_2 m_d}{\sqrt{2} m_W c_\beta} \;.
\end{equation}
Employing the $\beta$ rotation~\eqref{eq-brot} and~\eqref{eq-chb} and the
mixing~\eqref{eq-Hrot}, the trilinear terms give the neutral Higgs--fermion--fermion interactions with Lagrangian
\begin{equation}
{\cal L}_{H^0 \bar{q} q} = 
- \frac{g_2 m_u}{2 m_W s_\beta} \;
H_i\; \bar{u}\;  \bigg (R_{i 2} + i c_\beta R_{i 4} \gamma_5 \bigg)\; u 
- \frac{g_2 m_d}{2 m_W c_\beta} \;
H_i\; \bar{d}\;  \bigg (R_{i 1} + i s_\beta R_{i 4} \gamma_5 \bigg)\; d
\end{equation}
as well as the charged Higgs-fermion-fermion interactions
\begin{equation}
{\cal L}_{H^\pm \bar{q} q} = 
+ \frac{g_2}{2 \sqrt{2} m_W}\; 
H^+ \; \bar{u}\; 
\bigg( m_u \ct_\beta + m_d t_\beta + (m_u \ct_\beta - m_d t_\beta)\gamma_5 \bigg)
\; d
+ c.c.
\end{equation}
The Lagrangian for the Higgs-lepton interaction may be derived in a quite analogous way.\\

% Feynman rules
\noindent
\begin{tabular}{m{0.25\linewidth}m{0.5\linewidth}}
\cline{1-2}\\
\includegraphics[width=.7\linewidth]{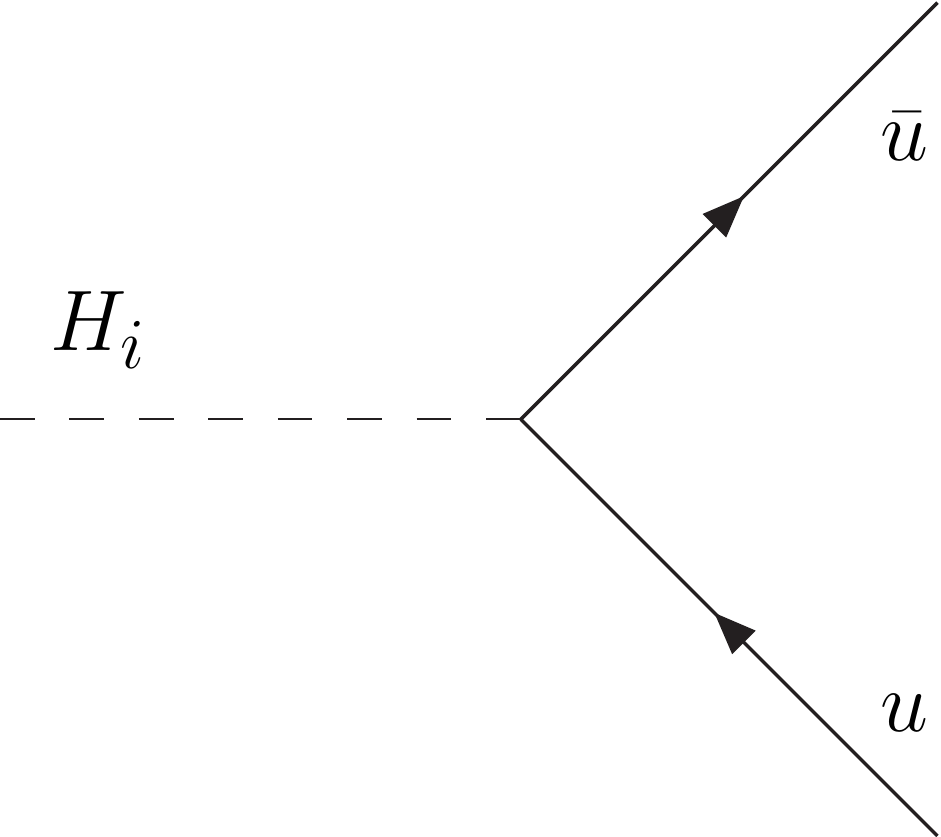} 
\hspace{0.1\linewidth}
&
$
 -\frac{g_2 m_u}{2 m_W s_\beta} \;
\bigg (R_{i 2} + i c_\beta R_{i 4} \gamma_5 \bigg)
$\\ 
\includegraphics[width=.7\linewidth]{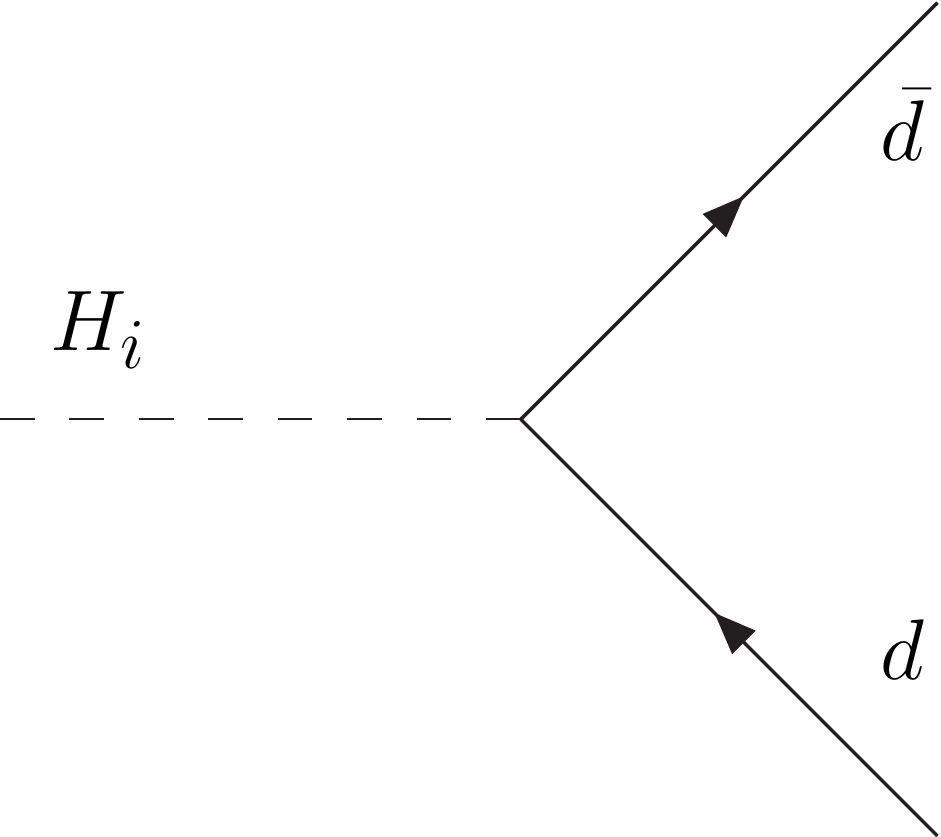} 
\hspace{0.1\linewidth}
&
$
-\frac{g_2 m_d}{2 m_W c_\beta} \;
\bigg (R_{i 1} + i s_\beta R_{i 4} \gamma_5 \bigg)
$\\ 
\includegraphics[width=.7\linewidth]{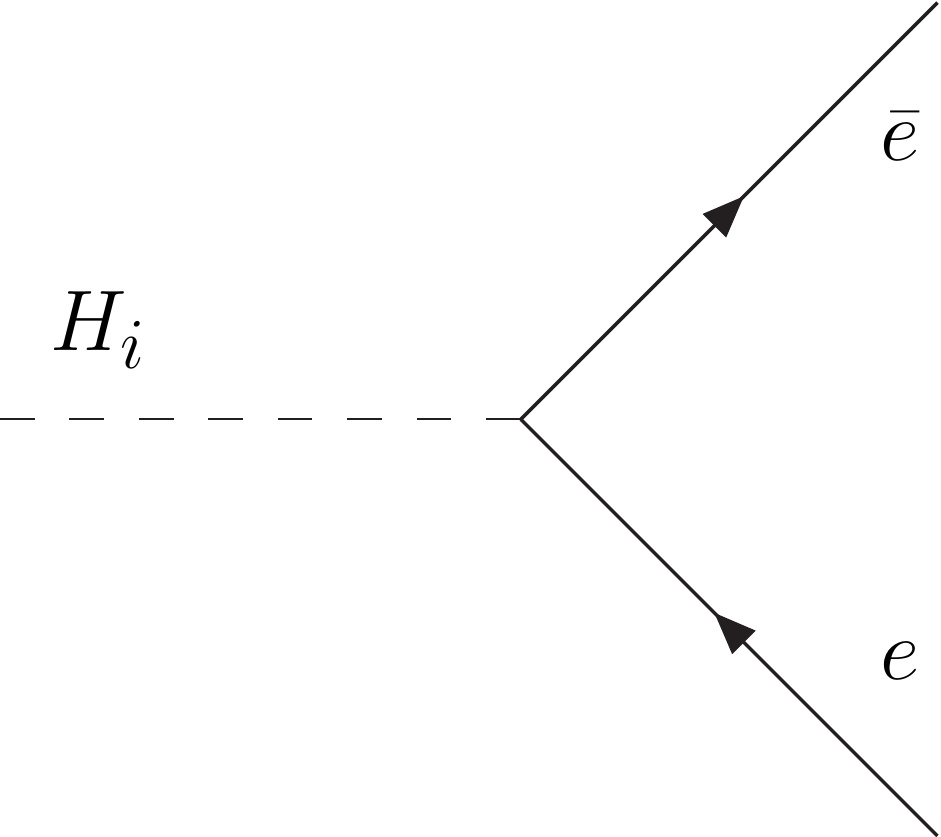} 
\hspace{0.1\linewidth}
&
$
-\frac{g_2 m_e}{2 m_W c_\beta} \;
\bigg (R_{i 1} + i s_\beta R_{i 4} \gamma_5 \bigg)
$\\ 
\includegraphics[width=.7\linewidth]{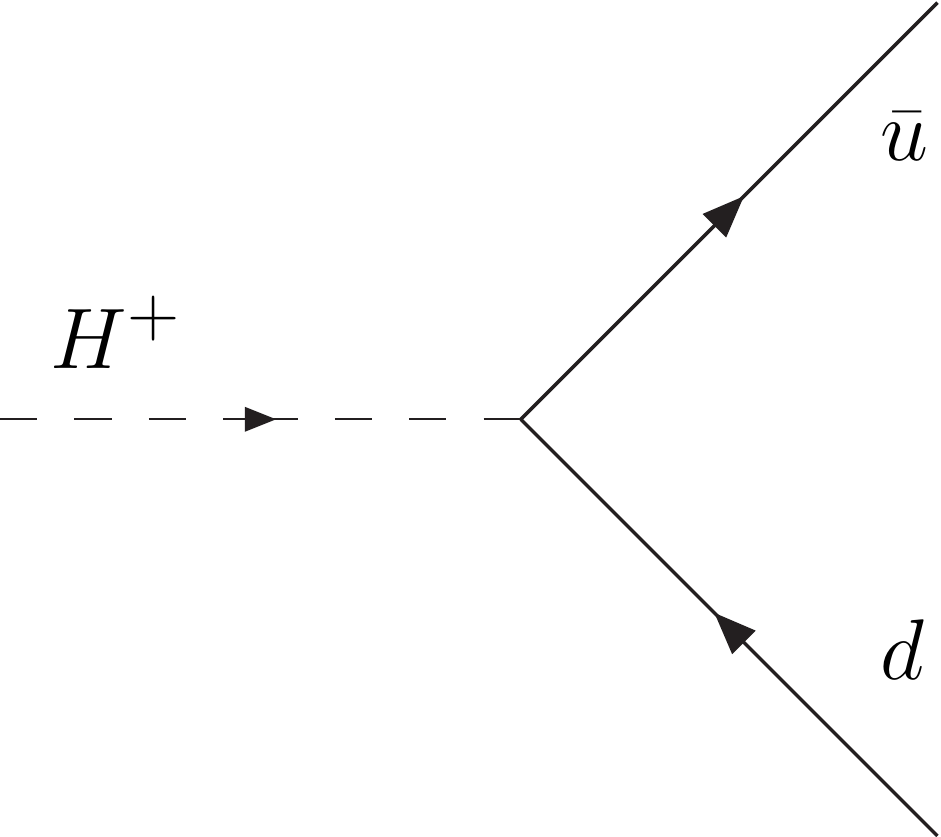} 
\hspace{0.1\linewidth}
&
$
 \frac{g_2}{2 \sqrt{2} m_W} \;
\bigg( m_u \ct_\beta + m_d t_\beta + (m_u \ct_\beta - m_d t_\beta)\gamma_5 \bigg)
$\\ 
\cline{1-2}
\end{tabular}\\

%%%%%%%%%%%%%%%%%%%%%%%%%%%%%%%%%%%%%%%%%%%%%%%%%%%%%%%%%%%%%%%%%%5
% The NMSSM potential
%%%%%%%%%%%%%%%%%%%%%%%%%%%%%%%%%%%%%%%%%%%%%%%%%%%%%%%%%%%%%%%%%%5
\subsection{The NMSSM potential}

In Sect.~\ref{app-A} the potential of 
a general supersymmetric theory is given. Here we want to derive
the full potential explicitly for the NMSSM. We get contributions
to the potential form the $F$-terms, $D$-terms, as well as from
the soft supersymmetry breaking terms ${\cal L}_{\text{soft}}$.
The $F$-terms read with~(\ref{eq-F})
\begin{equation}
V_F=F_i^* F^i = W_i^* W^i = 
\sum\limits_i \left| \frac{\delta W}{\delta \phi_i} \right|^2\; ,
\end{equation}
where $\phi_i$ are all scalar fields occurring in the chiral supermultiplets, that is
with view on Tab.~\ref{NMSSM-cont}, 
$\phi=(\tilde{u}_L$, $\tilde{d}_L$, $\tilde{u}_R^*$, 
$\tilde{d}_R^*$, $\tilde{\nu}_e$, $\tilde{e}_L$, $\tilde{e}_R^*$, 
$H_d^0$, $H_d^-$, $H_u^+$, $H_u^0$, $S)$.
In contrast to the MSSM the NMSSM has no dimensionful coupling
in the superpotential. As already mentioned this means that the function $W_i$ in
(\ref{eq-Wij}) has no term proportional to $\mu^{ij}$. 
After writing the isospin products out, the superpotential becomes
\begin{equation}
\label{eq-WscalarK}
W =
 \tilde{u}_R^* y_u (\tilde{u}_L H_u^0 - \tilde{d}_L H_u^+)
-\tilde{d}_R^* y_d (\tilde{u}_L H_d^- - \tilde{d}_L H_d^0)
-{\tilde{e}_R^*} y_e (\tilde{\nu}_e H_d^- - \tilde{e}_L H_d^0)
+ \lambda {S}  (H_d^- H_u^+ - H_d^0 H_u^0) 
+ \frac{1}{3} \kappa S^3
\end{equation}
and we can construct the functions $W_i$ and arrive at
\begin{equation}
\label{eq-VF2}
\begin{split}
V_F=&
\phantom{+}\left|\lambda (H_u \epsilon H_d) + \kappa S^2 \right |^2
+ \left| y_u (\tilde{Q} \epsilon H_u) \right|^2
+ \left| y_d (\tilde{Q} \epsilon H_d) \right|^2
+ \left| y_e (\tilde{L} \epsilon H_d) \right|^2
+ \left| y_u \tilde{u}_R^* H_u^0 - y_d \tilde{d}_R^* H_d^- \right|^2 \\
&+ \left| y_d \tilde{d}_R^* H_d^0  -y_u \tilde{u}_R^* H_u^+ \right|^2
+ \left| y_e \tilde{e}_R^* H_d^0 \right|^2
+ \left| y_e \tilde{e}_R^* H_d^- \right|^2
+ \left| y_u \tilde{u}_R^* \tilde{u}_L - \lambda S H_d^0  \right|^2\\
&+ \left| y_d \tilde{d}_R^* \tilde{d}_L + y_e \tilde{e}_R^* \tilde{e}_L - \lambda S H_u^0  \right|^2
+ \left| \lambda S H_u^+ - y_d \tilde{d}_R^* \tilde{u}_L - y_e \tilde{e}_R^* \tilde{\nu}_e   \right|^2
+ \left| \lambda S H_d^- - y_u \tilde{u}_R^* \tilde{d}_L   \right|^2\;.
\end{split}
\end{equation}

The next contribution to the potential comes from the $D$-terms
in (\ref{eq-pot}). With the equation of motion for the auxiliary field $D^a$
in (\ref{eq-D}) we get
\begin{equation}
\label{eq-VD}
V_D=-\frac{1}{2} D^a D^a = 
\frac{1}{2} \sum\limits_{i,j} g_a^2 
( \phi_i^\dagger \mathbf{T}^a \phi_i)( \phi_j^\dagger \mathbf{T}^a \phi_j),
\end{equation}
where $\phi_i$ denotes again all scalars of the
chiral supermultiplets and $g_a$ is the
gauge coupling corresponding to the gauge group generators $\mathbf{T}^a$ and
there is an implicit sum over the adjoint index $a$.
For the $SU(2)_L$ gauge group the gauge coupling is $g_2$ 
and the generators are half the Pauli matrices; $\mathbf{T}^a = \sigma^a/2$ with $a=1,2,3$. 
With view on Tab.~\ref{NMSSM-cont} we have the scalar weak isodoublets 
$\phi=(\tilde{Q}$, $\tilde{L}$, $H_u$, $H_d)$ available
and get for instance for the doublets $\tilde{Q}$ and $\tilde{L}$
a contribution to~\eqref{eq-VD}
\begin{equation}
\frac{1}{2} g_2^2 (\tilde{Q}^\dagger \frac{\sigma^a}{2} \tilde{Q})
(\tilde{L}^\dagger \frac{\sigma^a}{2} \tilde{L}) =
\frac{g_2^2}{8}  
(\tilde{Q}^\dagger_r \sigma^a_{rs} \tilde{Q}_s)
(\tilde{L}^\dagger_t \sigma^a_{tu} \tilde{L}_u) =
\frac{g_2^2}{8} \big( 2 |\tilde{Q}^\dagger \tilde{L} |^2 -
(\tilde{Q}^\dagger \tilde{Q})(\tilde{L}^\dagger \tilde{L}) \big),
\end{equation}
where $r,s,t,u = 1,2$ and we used the identity 
$\sigma^a_{rs} \sigma^a_{tu} = 2 \delta_{ru} \delta_{st} - 
\delta_{rs} \delta_{tu}$.
For the $U(1)_Y$ gauge group the gauge coupling is $g_1$ and
the generator is the hypercharge of the fields $\mathbf{T}^a=\mathbf{Y}_W$
given in Tab.~\ref{NMSSM-cont}.
We have the following scalar fields with non-vanishing
hypercharge $\phi=(\tilde{Q}$, $\tilde{u}_R^*$, $\tilde{d}_R^*$, 
$\tilde{L}$, $\tilde{e}_R^*$, $H_u$, $H_d)$.
Exemplarily, for the Higgs doublets $\tilde{Q}$ and $\tilde{L}$
with hypercharges $y_{\tilde{Q}}$ and $y_{\tilde{L}}$, respectively, we get 
a contribution to~\eqref{eq-VD} 
\begin{equation}
\frac{1}{2} g_1^2\; y_{\tilde{Q}}\; y_{\tilde{L}}\;
(\tilde{Q}^\dagger \tilde{Q})
(\tilde{L}^\dagger \tilde{L})\;.
\end{equation}
Collecting all D-term contributions we arrive at
\begin{equation}
\label{eq-VD1}
\begin{split}
V_D= &
\frac{g_2^2}{8}
\bigg(
(H_d^\dagger H_d)^2 
+ (H_u^\dagger H_u)^2 
+ (\tilde{Q}^\dagger \tilde{Q})^2 
+ (\tilde{L}^\dagger \tilde{L})^2 \\
&+ 4 |H_d^\dagger H_u|^2 - 2 (H_d^\dagger H_d)(H_u^\dagger H_u)
+ 4 |H_d^\dagger \tilde{Q}|^2 - 2 (H_d^\dagger H_d)(\tilde{Q}^\dagger \tilde{Q})
+ 4 |H_d^\dagger \tilde{L}|^2 - 2 (H_d^\dagger H_d)(\tilde{L}^\dagger \tilde{L})\\
&+ 4 |H_u^\dagger \tilde{Q}|^2 - 2 (H_u^\dagger H_u)(\tilde{Q}^\dagger \tilde{Q})
+ 4 |H_u^\dagger \tilde{L}|^2 - 2 (H_u^\dagger H_u)(\tilde{L}^\dagger \tilde{L})
+ 4 |\tilde{Q}^\dagger \tilde{L}|^2 - 2 (\tilde{Q}^\dagger \tilde{Q})(\tilde{L}^\dagger \tilde{L})
\bigg)\\
&+ \frac{g_1^2}{2}
\bigg( 
 \frac{1}{6} \tilde{Q}^\dagger \tilde{Q}
- \frac{2}{3} \tilde{u}_R^* \tilde{u}_R
+ \frac{1}{3} \tilde{d}_R^* \tilde{d}_R
- \frac{1}{2} \tilde{L}^\dagger \tilde{L}
+ \tilde{e}_R^* \tilde{e}_R
+ \frac{1}{2} H_u^\dagger H_u 
- \frac{1}{2} H_d^\dagger H_d
\bigg)^2\;.
\end{split}
\end{equation}

Now we come to the explicit soft supersymmetry breaking terms.
Writing down all terms as given in~(\ref{eq-soft})
we have
\begin{equation}
\label{eq-Vsoft}
\begin{split}
{\cal L}_{\text{soft}}=& 
-m_{H_d}^2 H_d^\dagger H_d
-m_{H_u}^2 H_u^\dagger H_u
-m_{S}^2 |S|^2
- \bigg( \lambda A_\lambda S (H_u^\trans \epsilon H_d) 
        + \frac{\kappa}{3} A_\kappa S^3 + c.c.
  \bigg)\\
&-  m_Q^2 \tilde{Q}^\dagger \tilde{Q}
-  m_L^2 \tilde{L}^\dagger \tilde{L}
- m_{\tilde{u}}^2 |\tilde{u}_R|^2
- m_{\tilde{d}}^2 |\tilde{d}_R|^2
- m_{\tilde{e}}^2 |\tilde{e}_R|^2\\
&- \bigg( \tilde{u}_R^* y_u A_u (\tilde{Q}^\trans \epsilon H_u)
	-\tilde{d}_R^* y_d A_d (\tilde{Q}^\trans \epsilon H_d)
	-\tilde{e}_R^* y_e A_e (\tilde{L}^\trans \epsilon H_d)
	+ c.c.
   \bigg)\\
&- \frac{1}{2} \bigg(
       	  M_{1} \tilde{B}^0 \tilde{B}^0
       	+ M_{2} \tilde{W} \tilde{W}
	+ M_{\tilde{g}} \tilde{g} \tilde{g}
       + c.c. 
       \bigg)
\end{split}
\end{equation}
with new soft parameters $m_{H_d}^2$, $m_{H_u}^2$, $m_S^2$,
$m_Q^2$, $m_L^2$, $m_{\tilde{u}}^2$, $m_{\tilde{d}}^2$, $m_{\tilde{e}}^2$,
$A_\lambda$, $A_\kappa$, $A_u$, $A_d$, $A_e$, $M_1$, $M_2$, $M_{\tilde{g}}$.
The idea is that these parameters one day emerge from
the supersymmetry breaking mechanism. As long as
this mechanism is unknown, this soft supersymmetry breaking
terms parameterize the generic model. Note that
due to missing bilinear terms in the NMSSM superpotential
there are no associated terms of this kind in the 
corresponding soft terms.

The scalar part of the soft supersymmetry breaking terms
contributes to the potential.
The soft breaking terms of the
first line on the right hand side of~\eqref{eq-Vsoft} contribute
to the Higgs potential.

%%%%%%%%%%%%%%%%%%%%%%%%%%%%%%%%%%%%%%%%%%%%%%%%%%%%%%%%%%%%%%%%%%5
% Higgs sfermion interaction
%%%%%%%%%%%%%%%%%%%%%%%%%%%%%%%%%%%%%%%%%%%%%%%%%%%%%%%%%%%%%%%%%%5
\subsection{Higgs sfermion interaction}

The Higgs sfermion interactions follow from the $F$-terms~(\ref{eq-VF2}),
the $D$-terms~(\ref{eq-VD1}) as well as from the soft terms~(\ref{eq-Vsoft}).
The corresponding Lagrangian is
\begin{equation}
\begin{split}
{\cal L}_{H \tilde{q} \tilde{q}} =&
- \left| y_u (\tilde{Q}^\trans \epsilon H_u) \right|^2
- \left| y_d (\tilde{Q}^\trans \epsilon H_d) \right|^2
- \left| y_u \tilde{u}_R^* H_u^0 - y_d \tilde{d}_R^* H_d^- \right|^2 
- \left| y_d \tilde{d}_R^* H_d^0 - y_u \tilde{u}_R^* H_u^+\right|^2\\
&
+y_u \big( \tilde{u}_R \tilde{u}_L^* \lambda S H_d^0 
+ \tilde{u}_R \tilde{d}_L^* \lambda S H_d^- + c.c \big)
+y_d \big( \tilde{d}_R \tilde{d}_L^* \lambda S H_u^0 
+ \tilde{d}_R \tilde{u}_L^* \lambda S H_u^+ + c.c \big)\\
&
-\frac{g_1^2}{12} \bigg(
(H_u^\dagger H_u)(\tilde{Q}^\dagger \tilde{Q})
-(H_d^\dagger H_d)(\tilde{Q}^\dagger \tilde{Q})
\bigg) \\
&
-\frac{g_2^2}{8} \bigg(
4 |H_d^\dagger \tilde{Q}|^2 - 2 (H_d^\dagger H_d)(\tilde{Q}^\dagger \tilde{Q})
+ 4 |H_u^\dagger \tilde{Q}|^2 - 2 (H_u^\dagger H_u)(\tilde{Q}^\dagger \tilde{Q})
\bigg) \\
&
- \bigg( \tilde{u}_R^* y_u A_u (\tilde{Q}^\trans \epsilon H_u)
	-\tilde{d}_R^* y_d A_d (\tilde{Q}^\trans \epsilon H_d)
	-\tilde{e}_R^* y_e A_e (\tilde{L}^\trans \epsilon H_d)
	+ c.c.
   \bigg)\;.
\end{split}
\end{equation}
The Higgs--slepton interaction is derived in a quite analogous way and not given
here explicitly.
The Feynman rules follow from the parameterization~(\ref{eq-higgsespara}),
the rotation of the Higgs gauge eigenstates into mass eigenstates~(\ref{eq-Hrot})
and the mixing of the $\tilde{q}_L$ and $\tilde{q}_R$ gauge eigenstates 
into mass eigenstates with mixing matrix $R_{\tilde{q}}$,
\begin{equation}
\label{eq-Rsq}
\begin{pmatrix}
\tilde{q}_1 \\
\tilde{q}_2
\end{pmatrix} 
=
\begin{pmatrix}
\phantom{+}\cos(\theta_{\tilde{q}}) & \sin(\theta_{\tilde{q}})\\
-\sin(\theta_{\tilde{q}}) & \cos(\theta_{\tilde{q}})
\end{pmatrix}
\begin{pmatrix}
\tilde{q}_L \\
\tilde{q}_R
\end{pmatrix}
=
R_{\tilde{q}}
\begin{pmatrix}
\tilde{q}_L \\
\tilde{q}_R
\end{pmatrix}\;,
\end{equation}
that is, by the diagonalization of the mass squared matrix $M$;
$\diag(m_{\tilde{q}_1}, m_{\tilde{q}_2}) \equiv R_{\tilde{q}} M R_{\tilde{q}}^\trans$.
%

%%%%%%%%%%%%%%%%%%%%%%%%%%%%%%%%%%%%%%%%%%%%%%%%%%%%%%%%%%%%%%%%%%5
% Higgs-boson neutralino/chargino interaction
%%%%%%%%%%%%%%%%%%%%%%%%%%%%%%%%%%%%%%%%%%%%%%%%%%%%%%%%%%%%%%%%%%5
\subsection{Higgs-boson neutralino/chargino interaction}
\label{sub-Hneu}

We start with collecting the bilinear neutralino/chargino interaction terms
as well as the Higgs-neutralino/chargino
interactions. 

From the soft breaking terms~(\ref{eq-Vsoft}) we get mass terms
for the {\em bino} and the {\em wino} with mass parameters $M_1$ and $M_2$, respectively.
\begin{equation}
\label{eq-BtWtsoft}
{\cal L}_{\text{soft}, \tilde{B}/\tilde{W}} =
-\frac{1}{2} M_1 \tilde{B}^0 \tilde{B}^0
-\frac{1}{2} M_2 \tilde{W}^a \tilde{W}^a
+ c.c. 
=
-\frac{1}{2} M_1 \tilde{B}^0 \tilde{B}^0
- M_2 \tilde{W}^- \tilde{W}^+  
-\frac{1}{2} M_2 \tilde{W}^3 \tilde{W}^3 + c.c.
\end{equation}
where we use the definition 
\begin{equation}
\label{eq-Wt}
\tilde{W}^\pm := \frac{1}{\sqrt{2}} \big(\tilde{W}^1 \mp i \tilde{W}^2 \big)
\end{equation}
in an analogous way to the vector bosons.

The chiral interactions~(\ref{eq-lagchir}) give us additional Higgs--Higgsino
terms.
\begin{equation}
{\cal L}_{\text{chiral}, H \tilde{H}} =
-\frac{1}{2} \frac{\delta^2 W}{ \delta \phi_i \delta \phi_j} \psi_i \psi_j + c.c. =
-\frac{1}{2} 
\bigg(
	  \lambda (H_u^\trans \epsilon \tilde{H}_d) \tilde{S}
	+ \lambda (\tilde{H}_u^\trans \epsilon H_d) \tilde{S}
	+ \lambda (\tilde{H}_u^\trans \epsilon \tilde{H}_d) S
	+ \kappa S \tilde{S} \tilde{S}
\bigg)
	+ c.c.
\end{equation}
Insertion of the Higgs-boson doublets and Higgsino doublets 
from Tab.~\ref{NMSSM-cont} gives
\begin{equation}
\label{eq-cHHt}
{\cal L}_{\text{chiral}, H \tilde{H}} =
-\frac{1}{2}
\bigg(
	\lambda H_u^+ \tilde{H}_d^- \tilde{S}
	-\lambda H_u^0 \tilde{H}_d^0 \tilde{S}
	-\lambda H_d^0 \tilde{H}_u^0 \tilde{S}
	+\lambda H_d^- \tilde{H}_u^+ \tilde{S}
	-\lambda S \tilde{H}_d^0 \tilde{H}_u^0
	+\lambda S \tilde{H}_d^- \tilde{H}_u^+
	+\kappa S \tilde{S} \tilde{S}
\bigg) + c.c.
\end{equation}

Further Higgs--Higgsino interactions come from the Yukawa interaction~(\ref{eq-yuk})
\begin{equation}
\begin{split}
{\cal L}_{\text{Yukawa}, H \tilde{H}} =&
- \sqrt{2} g ( \phi_i^\dagger T^a \psi^i) \lambda^a + c.c. =\\
&- \sqrt{2} g_2 \bigg( H_d^\dagger \frac{\sigma^a}{2} \tilde{H}_d + H_u^\dagger \frac{\sigma^a}{2} \tilde{H}_u \bigg)
\tilde{W}^a
- \sqrt{2} g_1 \bigg( -\frac{1}{2} H_d^\dagger \tilde{H}_d + \frac{1}{2} H_u^\dagger \tilde{H}_u \bigg)
\tilde{B}^0
+ c.c.
\end{split}
\end{equation}

We get, using \eqref{eq-Wt}
\begin{equation}
\label{eq-YHHt}
\begin{split}
{\cal L}_{\text{Yukawa}, H \tilde{H}} &=
- g_2 \bigg(
 (H_d^0)^* \tilde{H}_d^- \tilde{W}^+
+(H_d^-)^* \tilde{H}_d^0 \tilde{W}^-
-(H_u^+)^* \tilde{H}_u^0 \tilde{W}^+
-(H_u^0)^* \tilde{H}_u^+ \tilde{W}^-\\
&+ \frac{1}{\sqrt{2}} (H_d^0)^* \tilde{H}_d^0 \tilde{W}^3
- \frac{1}{\sqrt{2}} (H_d^-)^* \tilde{H}_d^- \tilde{W}^3 
+ \frac{1}{\sqrt{2}} (H_u^+)^* \tilde{H}_u^+ \tilde{W}^3
- \frac{1}{\sqrt{2}} (H_u^0)^* \tilde{H}_u^0 \tilde{W}^3 
	\bigg)\\
&+\frac{g_1}{\sqrt{2}} \bigg(
 (H_d^0)^* \tilde{H}_d^0 \tilde{B}^0
+(H_d^-)^* \tilde{H}_d^- \tilde{B}^0
-(H_u^+)^* \tilde{H}_u^+ \tilde{B}^0
-(H_u^0)^* \tilde{H}_u^0 \tilde{B}^0
\bigg) + c.c.
\end{split}
\end{equation}

Now, since we have collected all necessary Lagrangian terms we can easily give the
neutralino mass matrix, as presented already in \eqref{eq-neutralinomass}. We choose the basis
$\psi^0=(\tilde{B}^0, \tilde{W}^3, \tilde{H}_d^0, \tilde{H}_u^0, \tilde{S})$.
In this basis the neutralino mass part of the Lagrangian reads
\begin{equation}
{\cal L}_{\tilde{\chi}^0}= -\frac{1}{2} {\psi^0}^\trans M_{\tilde{\chi}^0} \psi^0 + c.c.
\end{equation}
with
\begin{equation}
M_{\tilde{\chi}^0}=
\begin{pmatrix}
M_1 & 0   & -\cb s_W m_Z &  \sb s_W m_Z  & 0\\
0   & M_2 &  \cb c_W m_Z & -\sb c_W m_Z & 0\\
-\cb s_W m_Z & \cb c_W m_Z & 0 & -\lambda v_s/\sqrt{2} & -\lambda v_u/\sqrt{2}\\
 \sb s_W m_Z &-\sb c_W m_Z & -\lambda v_s/\sqrt{2} & 0  & -\lambda v_d/\sqrt{2}\\
0 & 0 & -\lambda v_u/\sqrt{2} & -\lambda v_d/\sqrt{2} & \phantom{+}\sqrt{2} \kappa v_s
\end{pmatrix}.
\end{equation}
The upper left $2\times 2$ block arises from the soft breaking mass parameters of the 
bino $\tilde{B}^0$ respectively wino $\tilde{W}$~(\ref{eq-BtWtsoft}). 
The lower right $3\times 3$ block, mixing the Higgsinos among themselves, comes 
from~(\ref{eq-cHHt}) and the remaining part from the Yukawa
terms~(\ref{eq-YHHt}), where we get bilinear terms at the 
vacuum of the Higgs bosons; \eqref{higgsvev}.
The mass eigenstates follow from an unitary rotation, that is
\begin{equation}
\label{eq-chirot}
\chi_i^0= U_{ji} \psi^0_j
\qquad \text{ with }
\diag(m_{{\chi}_1^0}^2, m_{{\chi}_2^0}^2, m_{{\chi}_3^0}^2,
m_{{\chi}_4^0}^2, m_{{\chi}_5^0}^2)
= U^* M_{{\chi}^0} U^\dagger,
\end{equation}
with $i,j = 1,...,5$. Conventionally the neutralino masses are defined to be in ascending order,
that is, $\chi^0_1$ is the lightest neutralino. The result
of this diagonalization is rather involved and 
in practice performed numerically, whereas analytic approximations can be found
in~\citep{Pandita:1994ms,Pandita:1994vw,Choi:2004zx}.

From the chiral interactions~(\ref{eq-cHHt}) and the Yukawa interactions~(\ref{eq-YHHt})
we get the neutral Higgs boson neutralino interaction
\begin{equation}
\begin{split}
{\cal L}_{H \chi^0 \chi^0} =&
-\frac{1}{2}
\bigg(
	 -\lambda H_d^0 \tilde{H}_u^0 \tilde{S}
	-\lambda H_u^0 \tilde{H}_d^0 \tilde{S}
	-\lambda S \tilde{H}_d^0 \tilde{H}_u^0
	+\kappa S \tilde{S} \tilde{S}
\bigg) \\
&- \frac{g_2}{\sqrt{2}} \bigg(
 (H_d^0)^* \tilde{H}_d^0 \tilde{W}^3
- (H_u^0)^* \tilde{H}_u^0 \tilde{W}^3 
	\bigg)\\
&+\frac{g_1}{\sqrt{2}} \bigg(
 (H_d^0)^* \tilde{H}_d^0 \tilde{B}^0
- (H_u^0)^* \tilde{H}_u^0 \tilde{B}^0
\bigg) + c.c.
\end{split}
\end{equation}
Using the Higgs boson parameterization~(\ref{eq-higgsespara}), 
employing the Higgs-boson mixing~\eqref{eq-brot}, \eqref{eq-Hrot},
and the neutralino mixing~(\ref{eq-chirot}) we get
\begin{equation}
{\cal L}_{H \tilde{\chi}^0 \tilde{\chi}^0} =
 A^{ijk}  H_i \chi_j^0 \chi_k^0 + c.c.
\end{equation}
with
\begin{equation}
\begin{split}
 A^{ijk} =&
\frac{1}{2} 
\big[
  (R_{i1} - i R_{i4} \sb)(g_1 U_{j3}^* U_{k1}^* - g_2 U_{j3}^* U_{k2}^*)
- e^{-i \phi_u} (R_{i2} - i R_{i4} \cb)(g_1 U_{j4}^* U_{k1}^* - g_2 U_{j4}^* U_{k2}^*)\\
&+ \frac{\lambda}{\sqrt{2}}
\big(
(R_{i1} + i R_{i4} \sb)  U_{j4}^* U_{k5}^*
+ e^{i \phi_u} (R_{i2} + i R_{i4} \cb) U_{j3}^* U_{k5}^*
+ e^{i \phi_s} (R_{i3} + i R_{i5}) U_{j3}^* U_{k4}^* 
\big)\\
&- \frac{\kappa}{\sqrt{s}}
e^{i \phi_s} (R_{i3} + i R_{i5}) U_{j5}^* U_{k5}^* 
\big]\; .
\end{split}
\end{equation}
Now we write the
products of Weyl-fermions in terms of Dirac four component spinors
\begin{equation}
\label{eq-Dchi}
\tilde{\chi}_i^0 :=
\begin{pmatrix}
\chi_i^0\\
\bar{\chi}_i^0
\end{pmatrix}\;,
\quad 
\text{ with }
i=1,...,5\;.
\end{equation}
that is, we have for the products in terms of Dirac four-spinors
\begin{equation}
\label{eq-chi4}
\chi_j^0 \chi_k^0 = \bar{\tilde{\chi}}_j^0 P_L \tilde{\chi}_k^0
\qquad
\text{ and }\quad
(\chi_j^0 \chi_k^0)^* 
= \bar{\tilde{\chi}}_k^0 \gamma_0 P_L \gamma_0 \tilde{\chi}_j^0
= \bar{\tilde{\chi}}_k^0 P_R \tilde{\chi}_j^0
= \bar{\tilde{\chi}}_j^0 P_R \tilde{\chi}_k^0\;,
\end{equation}
where the fields denoted with a tilde symbol are the Dirac spinors and the last
equation holds only for Majorana spinors~\citep{Gunion:1984yn}. With \eqref{eq-chi4} we get
\begin{equation}
\begin{split}
{\cal L}_{H \tilde{\chi}^0 \tilde{\chi}^0} =&
A^{ijk} H_i \bar{\tilde{\chi}}^0_j P_L \tilde{\chi}_k^0 
+ (A^{ijk})^* H_i \bar{\tilde{\chi}}^0_j P_R \tilde{\chi}_k^0 \\
=& \bar{\tilde{\chi}}^0_j
\big[ A^{ijk} P_L + (A^{ijk})^* P_R \big]
\tilde{\chi}_k^0  H_i \;.
\end{split}
\end{equation}

% CHARGINO
From the soft Lagrangian terms~(\ref{eq-BtWtsoft}),
and the chiral Lagrangian terms~(\ref{eq-cHHt}) we can
immediately construct the chargino mass matrix.
With the chosen basis 
$\psi^\pm=(\tilde{W}^+, \tilde{H}_u^+, \tilde{W}^-, \tilde{H}_d^-)$
the chargino mass part of the Lagrangian reads
\begin{equation}
{\cal L}_{\text{chargino mass}}=
-\frac{1}{2} (\psi^\pm)^\trans M_{{\chi}^\pm} \psi^\pm +c.c.
\end{equation}
with
\begin{equation}
M_{{\chi}^\pm} =
\begin{pmatrix}
0 & X^\trans\\
X & 0
\end{pmatrix},
\qquad
X=
\begin{pmatrix}
M_2 & \sqrt{2} \sb m_W\\
\sqrt{2} \cb m_W & \lambda v_s/\sqrt{2}
\end{pmatrix}.
\end{equation}
The unitary rotations
\begin{equation}
\label{eq-Crot}
\begin{pmatrix}
\chi_1^+\\
\chi_2^+
\end{pmatrix}
= U^+
\begin{pmatrix}
\tilde{W}^+\\
\tilde{H}_u^+
\end{pmatrix},
\qquad
\begin{pmatrix}
\chi_1^-\\
\chi_2^-
\end{pmatrix}
= U^-
\begin{pmatrix}
\tilde{W}^-\\
\tilde{H}_d^-
\end{pmatrix}
\end{equation}
can be used to get the mass eigenstates
\begin{equation}
\diag( m_{\chi_1^\pm}, m_{\chi_2^\pm}) = (U^+)^* X (U^-)^\dagger\; .
\end{equation}
Explicit diagonalization yields the chargino masses
\begin{equation}
m_{\chi_1^\pm/\chi_2^\pm}^2 =
\frac{1}{2} 
\bigg(
	|M_2|^2 + |\lambda|^2 v_s^2 +2 m_W^2
	\mp
	\sqrt{(|M_2|^2 + |\lambda|^2 v_s^2 +2 m_W^2)^2
	    - 4|\lambda v_s M_2 - m_W^2 \sin (2 \beta) |^2}
\bigg)\; .
\end{equation}	    

From the chiral interactions~(\ref{eq-cHHt}) and the Yukawa interactions~(\ref{eq-YHHt})
we get the neutral Higgs boson interaction with two charginos.
\begin{equation}
{\cal L}_{H \chi^+ \chi^-} =
-\frac{\lambda}{2} S \tilde{H}_d^- \tilde{H}_u^+
- g_2 (
 (H_d^0)^* \tilde{H}_d^- \tilde{W}^+
-(H_u^0)^* \tilde{H}_u^+ \tilde{W}^-
	) + c.c.
\end{equation}
Using the Higgs-boson parameterization~(\ref{eq-higgsespara}),
the rotation of the gauge eigenstates into mass eigenstates
of the Higgs bosons~\eqref{eq-brot}, \eqref{eq-Hrot} and
of the charginos~(\ref{eq-Crot}), we get the trilinear terms
\begin{equation}
{\cal L}_{H \chi^+ \chi^-} =
B^{ijk} H_i \chi_j^- \chi_k^+ + c.c. 
\end{equation}
with
\begin{equation}
B^{ijk} = \frac{1}{\sqrt{2}}
\bigg\{
	- \frac{\lambda}{2} e^{i \phi_s} (R_{i3} + i R_{i5}) (U^-)_{j2}^* (U^+)_{k2}^*
	- g_2	(R_{i1} - i \sb R_{i4}) (U^-)_{j2}^* (U^+)_{k1}^*
	+ g_2	e^{-i \phi_u} (R_{i2} - i \cb R_{i4}) (U^-)_{j1}^* (U^+)_{k2}^*
\bigg\}
\end{equation}
Transferring to four-component Dirac spinor notation:
\begin{equation}
\tilde{\chi}_i^+ :=
\begin{pmatrix}
\chi_i^+\\
\bar{\chi}_i^-
\end{pmatrix},
\qquad
i=1,2 \;,
\end{equation}
where the charginos denoted with a tilde symbol are the Dirac four-spinors and the
charginos without a tilde symbol the two-component Weyl spinors.
This means we have the translation $\chi_i^+ = P_L \tilde{\chi}_i^+$ and
$\bar{\chi}_i^- = P_R \tilde{\chi}_i^+$, that is,
$\chi_i^-= \bar{\bar{\chi}}_i^- = \bar{\tilde{\chi}}_i^+ P_L$ and arrive at
\begin{equation}
{\cal L}_{H \chi^+ \chi^-} =
B^{ijk} H_i \bar{\tilde{\chi}}_j^0 P_L \tilde{\chi}_k^+ + c.c. 
=
\bar{\tilde{\chi}}_j^+ \; \big(
B^{ijk} P_L + (B^{ijk})^* P_R
	\big)\;
\tilde{\chi}_k^+ H_i. 
\end{equation}

Finally we have the charged Higgs interaction with one neutralino and one chargino. 
Collecting the appropriate terms in~(\ref{eq-cHHt}) and~(\ref{eq-YHHt}) we find
\begin{equation}
\begin{split}
{\cal L}_{H^\pm \chi^+ \chi^0} =\;&
-\frac{\lambda}{2}
\big(
	 H_d^- \tilde{H}_u^+ \tilde{S}
	+H_u^+ \tilde{H}_d^- \tilde{S}
\big)\\
&-g_2 \big(
 (H_d^-)^* \tilde{H}_d^0 \tilde{W}^-
+(H_u^+)^* \tilde{H}_u^0 \tilde{W}^+
+ \frac{1}{\sqrt{2}} (H_d^-)^* \tilde{H}_d^- \tilde{W}^3 
- \frac{1}{\sqrt{2}} (H_u^+)^* \tilde{H}_u^+ \tilde{W}^3
	\big)\\
&+\frac{g_1}{\sqrt{2}} \bigg(
 (H_d^-)^* \tilde{H}_d^- \tilde{B}^0
- (H_u^+)^* \tilde{H}_u^+ \tilde{B}^0
\bigg) + c.c.
\end{split}
\end{equation}
With the parameterization~(\ref{eq-higgsespara}), 
the $\beta$ rotation of the charged Higgs bosons~(\ref{eq-chb}),
the rotation of the neutralinos~(\ref{eq-chirot})
and the charginos~(\ref{eq-Crot}) we have
\begin{equation}
{\cal L}_{H^\pm \chi^\mp \chi^0} = C^{jk}  H^- \chi_j^0 \chi_k^+ + c.c.
\end{equation}
where we defined
\begin{equation}
\begin{split}
C^{jk} & = 
 -\frac{\lambda}{2} \big( \sb (U^+_{k2})^* U_{j5}^* 
	+ e^{-i \phi_u} \cb U^-_{k2} U_{j5} \big)\\
&+ g_2 \big(e^{-i \phi_u} \cb (U^+_{k1})^* U_{j4}^* 
	- \sb U^-_{k1} U_{j3} \big)\\
&-\frac{g_2}{\sqrt{2}} \big( e^{-i \phi_u} \cb (U^+_{k2})^* U_{j2}^* 
- \sb U^-_{k2} U_{j2} \big)\\
&-\frac{g_1}{\sqrt{2}} \big( e^{-i \phi_u} \cb (U^+_{k2})^* U_{j1}^* 
- \sb U^-_{k2} U_{j1} \big)\;.
\end{split}
\end{equation}
In Dirac four-component notation we find eventually
\begin{equation}
{\cal L}_{H^\pm \chi^\mp \chi^0} =
\bar{\tilde{\chi}}_j^0\;
\big( C^{jk} P_L + (C^{jk})^* P_R \big)\; \tilde{\chi}_k^+ H^- \;.
\end{equation}\\

% Feynman rules
\noindent
\begin{tabular}{m{0.25\linewidth}m{0.5\linewidth}}
\cline{1-2}\\
\includegraphics[width=.7\linewidth]{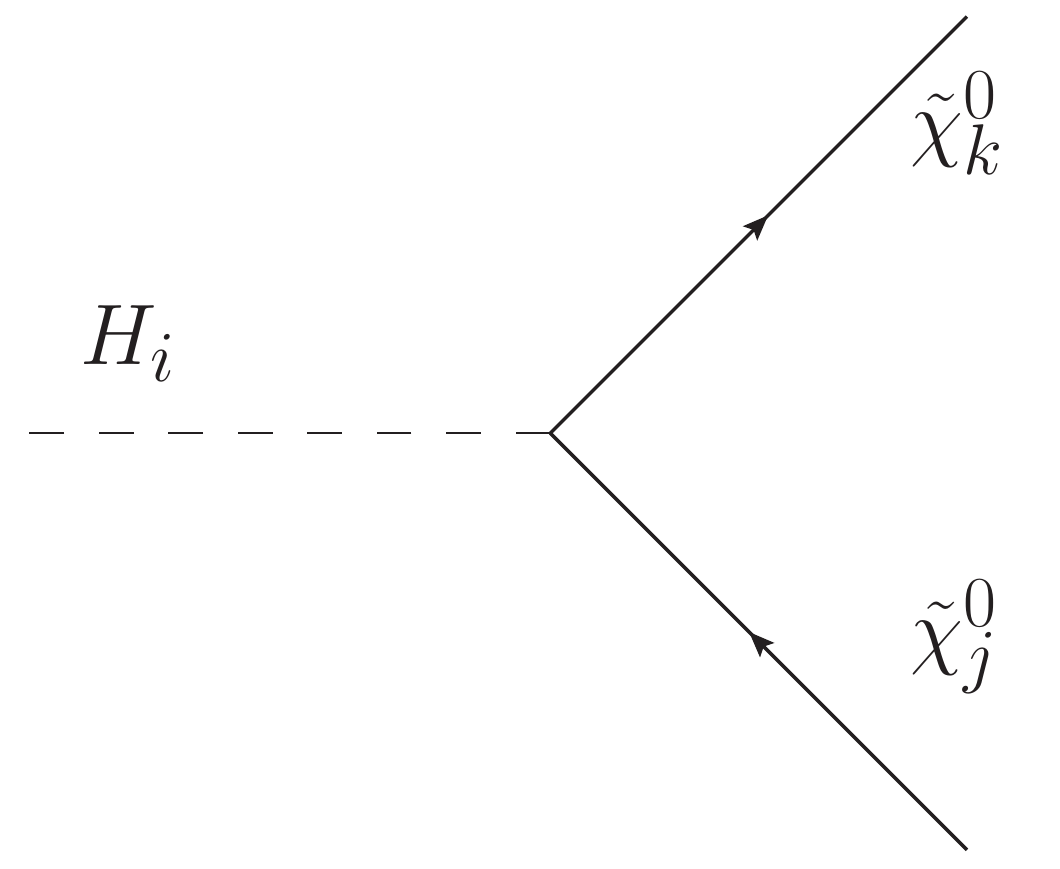} 
\hspace{0.1\linewidth}
&
$A^{ijk} P_L + (A^{ijk})^* P_R$
\\
\end{tabular}
\begin{tabular}{m{0.25\linewidth}m{0.5\linewidth}}
\includegraphics[width=.7\linewidth]{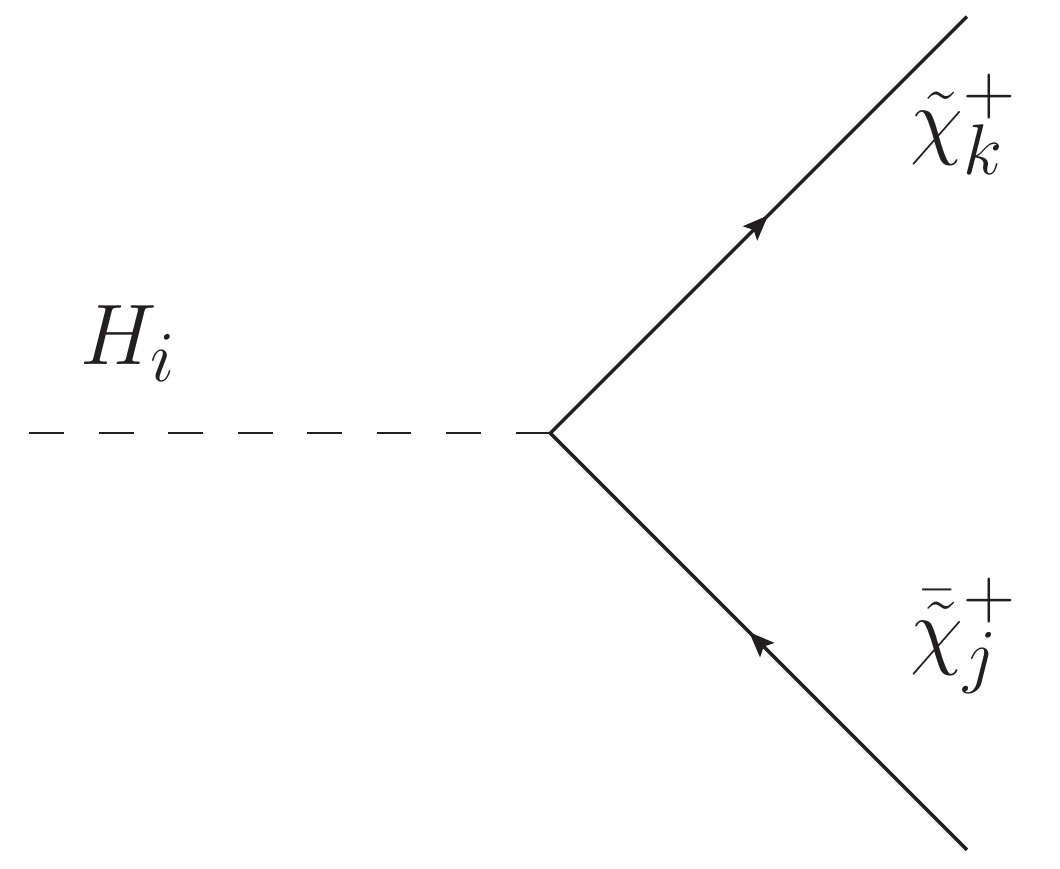} 
\hspace{0.1\linewidth}
&
$B^{ijk} P_L + (B^{ijk})^* P_R$
\\
\end{tabular}
\begin{tabular}{m{0.25\linewidth}m{0.5\linewidth}}
\includegraphics[width=.7\linewidth]{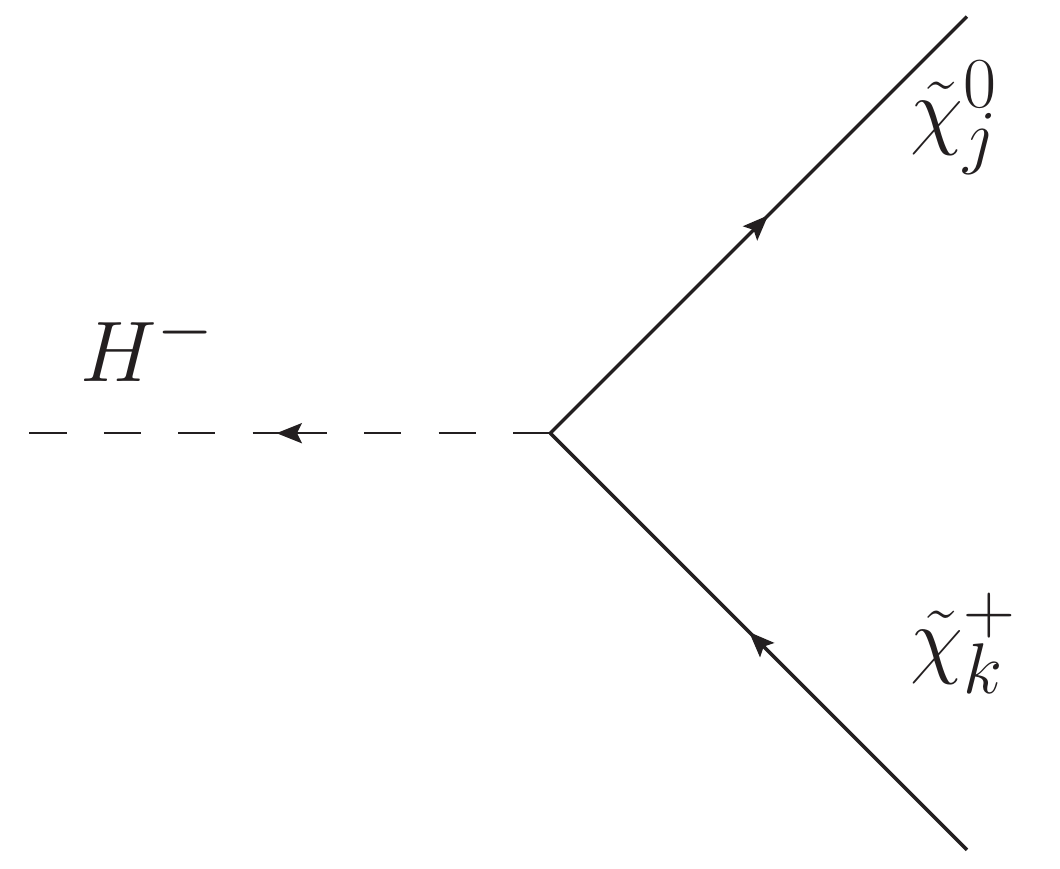} 
\hspace{0.1\linewidth}
&
$C^{jk} P_L + (C^{jk})^* P_R$
\\
\cline{1-2}\\
\end{tabular}
%

%%%%%%%%%%%%%%%%%%%%%%%%%%%%%%%%%%%%%%%%%%%%%%%%%%%%%%%%%%%%%%%5
% Higgs-boson self interaction
%%%%%%%%%%%%%%%%%%%%%%%%%%%%%%%%%%%%%%%%%%%%%%%%%%%%%%%%%%%%%%%5
\subsection{Higgs-boson self interaction}

The Higgs boson self interactions 
follow from the potential, that is, from
$F$-terms~(\ref{eq-VF2}),
$D$-terms~(\ref{eq-VD1}) as well as from the soft terms~(\ref{eq-Vsoft}).
Here we present the trilinear part of the corresponding Feynman 
rules~\citep{Ellwanger:2004xm}, where we set
\begin{equation}
P_{ijk}^{lmn} =
R_{il} R_{jm} R_{kn}+
R_{il} R_{km} R_{jn}+
R_{jl} R_{im} R_{kn}+
R_{jl} R_{km} R_{in}+
R_{kl} R_{im} R_{jn}+
R_{kl} R_{jm} R_{in}\;,
\end{equation}\\
with $R_{ij}$ defined in~\eqref{eq-Hrot} and
the phase dependent abbreviations
$R$, $I$, $R_\lambda$, $I_\lambda$,
$R_\kappa$, $I_\kappa$ defined in~\eqref{eq-phases}.\\

\noindent
\begin{tabular}{m{0.25\linewidth}m{0.5\linewidth}}
\cline{1-2}\\
\includegraphics[width=.7\linewidth]{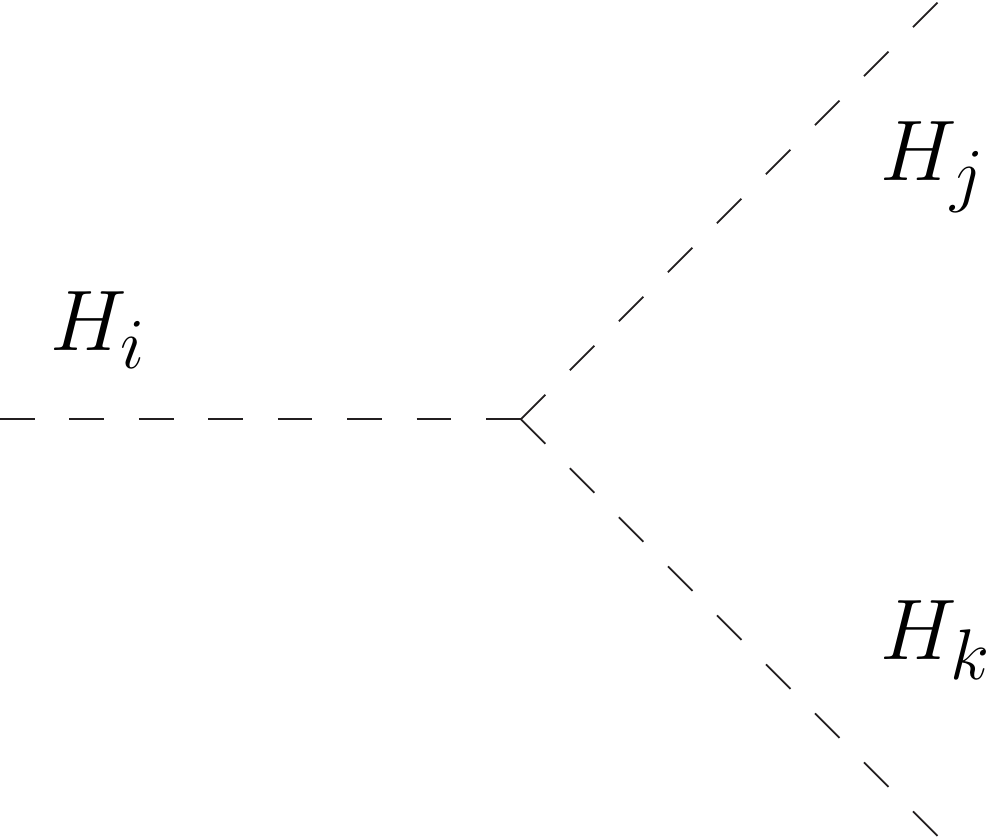} 
\hspace{0.1\linewidth}
&
\begin{multline*}
-\frac{1}{2} I v P^{334}_{ijk}
+I_\kappa P^{335}_{ijk}
+R v P^{345}_{ijk}
+\frac{1}{2} I v P^{455}_{ijk}\\
-\frac{1}{3} I_\kappa P^{555}_{ijk}
+Iv P^{235}_{ijk} c_{\beta }
+P^{111}_{ijk} \left(-\frac{1}{8} v c_{\beta } g_1^2-\frac{1}{8} v c_{\beta } g_2^2\right)\\
+P^{122}_{ijk} \left(-\frac{1}{2} v |\lambda|^2 c_{\beta }+\frac{1}{8} v c_{\beta } g_1^2+\frac{1}{8} v c_{\beta } g_2^2\right)\\
+P^{144}_{ijk} \left(|\lambda|^2 \left(-\frac{3 v c_{\beta}}{8}
	-\frac{1}{8} v c_{3 \beta }\right)
	+\left( v c_{\beta } + v c_{3 \beta }\right) \frac{g_1^2+g_2^2}{16}\right)\\
+I v P^{135}_{ijk} s_{\beta }
+P^{155}_{ijk} \left(-\frac{1}{2} v |\lambda|^2 c_{\beta }-\frac{1}{2}R v s_{\beta }\right)\\
+P^{133}_{ijk} \left(-\frac{1}{2} v |\lambda|^2 c_{\beta }+\frac{1}{2} R v s_{\beta }\right)
+P^{255}_{ijk} \left(-\frac{1}{2} R v c_{\beta }-\frac{1}{2} v |\lambda|^2 s_{\beta }\right)\\
+P^{233}_{ijk} \left(\frac{1}{2} R v c_{\beta }-\frac{1}{2} v |\lambda|^2 s_{\beta }\right)
+P^{222}_{ijk} \left(-\frac{1}{8} v g_1^2 s_{\beta }-\frac{1}{8} v g_2^2 s_{\beta }\right)\\
+P^{112}_{ijk} \left(-\frac{1}{2} v |\lambda|^2 s_{\beta }+\frac{1}{8} v g_1^2 s_{\beta }+\frac{1}{8} v g_2^2 s_{\beta }\right)\\
+P^{244}_{ijk} \left(
	 \frac{g_1^2+g_2^2}{16} \left(v s_{\beta }- v s_{3 \beta }\right)
	+|\lambda|^2 \left(-\frac{3 v s_{\beta }}{8}
	+\frac{1}{8} v s_{3 \beta }\right)\right)\\
-\frac{1}{2} |\lambda|^2 P^{113}_{ijk} v_s
-\frac{1}{2} |\lambda|^2 P^{223}_{ijk} v_s
+P^{125}_{ijk} \left(I v_s - I_\lambda\right)
+P^{123}_{ijk} \left(R_\lambda+R v_s\right)\\
-P^{333}_{ijk} \left(\frac{R_\kappa}{3}+|\kappa|^2 v_s\right)
+P^{355}_{ijk} \left(R_\kappa-|\kappa|^2 v_s\right)
-P^{134}_{ijk} c_{\beta } \left(I_\lambda +I v_s\right)\\
+P^{145}_{ijk} c_{\beta } \left(R v_s - R_\lambda  \right)
-P^{234}_{ijk} s_{\beta } \left(I_\lambda +I  v_s\right)
+P^{245}_{ijk} s_{\beta } \left(R v_s - R_\lambda \right)\\
+P^{445}_{ijk} \frac{s_{2 \beta }}{2} \left( I_\lambda - I v_s\right)
+P^{344}_{ijk} \left(-\frac{1}{2} R_\lambda s_{2 \beta }-\frac{1}{2} |\lambda|^2 v_s-\frac{1}{2} R s_{2 \beta } v_s\right)
\end{multline*}
   \\
\end{tabular}

\begin{tabular}{m{0.25\linewidth}m{0.5\linewidth}}
\includegraphics[width=.7\linewidth]{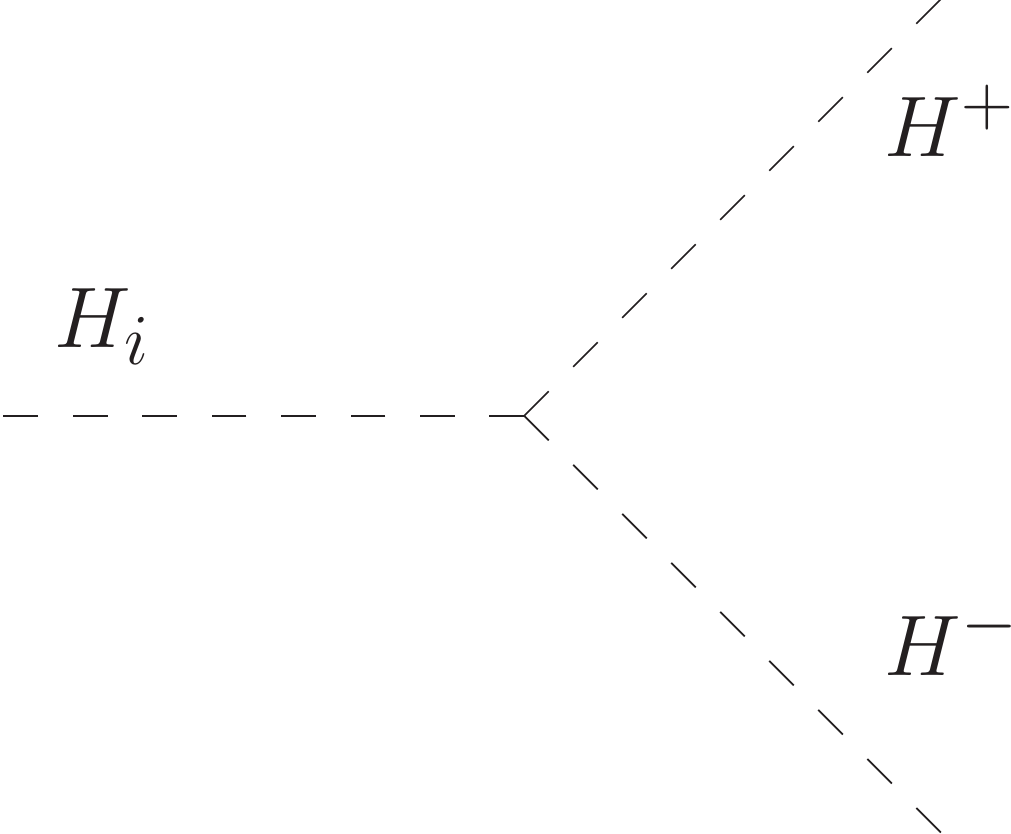} 
\hspace{0.1\linewidth}
&
\begin{multline*}
\frac{1}{4} 
\bigg( 
v R_{1i} c_{\beta } \left(c_{\beta }^2 
\left(g_1^2-g_2^2\right)+\left(4 |\lambda|^2-g_1^2-3 g_2^2\right) s_{\beta }^2\right) \\
-s_{\beta } \big( v R_{2i} \left(-4 |\lambda|^2 c_{\beta }^2+c_{\beta }^2 \left(g_1^2+3 g_2^2\right)+\left(g_2^2-g_1^2 \right) s_{\beta }^2\right) \\
-8 R_{5i} c_{\beta } \left( I_\lambda- I v_s\right) \big)\\
-4 R_{3i} \left(|\lambda|^2 c_{\beta }^2 v_s+|\lambda|^2 s_{\beta }^2 v_s+2 c_{\beta } s_{\beta } \left(R_\lambda + R v_s\right)\right)
\bigg)
\end{multline*}\\
\cline{1-2}\\
\end{tabular}

\newpage

%%%%%%%%%%%%%%%%%%%%%%%%%%%%%%%%%%%%%%%%%%%%%%%%%%%%%%%%%%%%%%%%%%5
%%%%%%%%%%%%%%%%%%%%%%%%%%%%%%%%%%%%%%%%%%%%%%%%%%%%%%%%%%%%%%%%%%5
% Non-minimal supersymmetric extensions
%%%%%%%%%%%%%%%%%%%%%%%%%%%%%%%%%%%%%%%%%%%%%%%%%%%%%%%%%%%%%%%%%%5
%%%%%%%%%%%%%%%%%%%%%%%%%%%%%%%%%%%%%%%%%%%%%%%%%%%%%%%%%%%%%%%%%%5
\section{Non-minimal supersymmetric extensions}
\label{sec-varNMSSM}

In this appendix we want to mention some 
of the alternative non-minimal supersymmetric
extensions. 
All here discussed models have in common to
solve the $\mu$ problem of the MSSM by the introduction of 
an additional singlet $\hat{S}$, generating
a $\mu$ term dynamically via a VEV
$\langle S \rangle$. We will
not discuss the phenomenology of these models
but refer the reader to the literature. 
For a comparison of the Higgs sector in
non-minimal supersymmetric extension models we refer to~\citep{Barger:2006dh}.\\

The {\em nearly minimal supersymmetric} SM (nMSSM) was 
introduced in~\citep{Fayet:1974pd, Cremmer:1982vy, Panagiotakopoulos:1999ah}. 
In this model also a singlet $\hat{S}$ is introduced in order
to generate the $\mu$-term dynamically in the superpotential,
that is, the $\mu$ term of the MSSM is replaced by
\begin{equation}
\label{eq-WHiggs}
W_\mu^{\text{nMSSM}}=\lambda \hat{S}  (\hat{H}_u^\trans \epsilon \hat{H}_d)\;.
\end{equation}
But in contrast to the NMSSM the trilinear $\kappa$ term
$\frac{1}{3} \kappa \hat{S}^3$ is abandoned. That is,
the nMSSM is in fact the truly next to minimal supersymmetric
extension of the SM.
Note that the $\kappa$ term fulfills in the
NMSSM the role to break the continuous Peccei--Quinn symmetry
and also contributes to the generation of a vacuum-expectation-value
of $S$ via the corresponding soft supersymmetry-breaking
terms (see the discussion in Sect.~\ref{sec-nmssm}).
In the NMSSM the $\kappa$ term leads to an elevation of
the PQ symmetry to a spontaneously broken discrete $\Zthree$ symmetry.
Non-renormalizable operators 
are introduced to avoid domain walls 
which are forced to obey a $\mathbbm{Z}_2^R$
symmetry, such that large tadpoles are
forbidden in order not to give
a much too large VEV for $S$ beyond the electroweak scale, 
which would destabilize the gauge hierarchy.
In the nMSSM in contrast
a discrete $\mathbbm{Z}_5^R$ is imposed~\citep{Panagiotakopoulos:1999ah}.
It is shown that a tadpole is generated
first appearing at six-loop order, generating
a contribution to the effective potential 
linear in S with small $\xi_n$,
\begin{equation}
\delta V =\xi_n m_s^3 S + c.c.
\end{equation}
That is, on the one hand by this term
the PQ symmetry is explicitely broken and on the
other hand $\xi_n$ is small enough not 
to destroy the gauge hierarchy.\\

Another approach is to start with a superpotential in
the form~\eqref{eq-WHiggs},
\begin{equation}
W_\mu^{\text{UMSSM}}=\lambda \hat{S}  (\hat{H}_u^\trans \epsilon \hat{H}_d)\;,
\end{equation}
but promote the PQ-symmetry to
a local, that is, gauge symmetry
$U(1)'$~\citep{Fayet:1976et,Fayet:1977yc,Farrar:1978xj,Fayet:1980ad,Kim:1983dt,Cvetic:1995rj,Cvetic:1996mf,Keith:1996fv,Keith:1996px,Keith:1997zb,Cvetic:1997ky,Cleaver:1997nj,Cvetic:1997wu,Langacker:1998tc},
giving an additional gauge boson~$Z'$ (UMSSM).
The charges
of the Higgs bosons with respect to this 
symmetry are denoted by $Q_d$, $Q_u$ and
$Q_s$ respectively and we have to have
$Q_d+Q_u+Q_s=0$ for reasons of gauge invariance.
Of course this gauge symmetry is, analogously to the
electroweak SM, spontaneously broken when the Higgs bosons get 
vacuum-expectation-values
and the Goldstone mode is absorbed by
the additional gauge boson $Z'$ which becomes massive.
Both gauge bosons $Z$ and $Z'$ mix, with mixing
matrix
\begin{equation}
M_{Z-Z'} =
\begin{pmatrix}
M_Z^2 & \Delta^2\\
\Delta^2 & M_Z'^2
\end{pmatrix}\;,
\end{equation}
where
\begin{equation}
\label{eq-Umix}
\begin{split}
M_Z^2    &= \frac{1}{4} G^2 (v_d^2 +v_u^2)\;,\\
M_Z'^2   &= g_1'^2 (v_d^2 Q_d^2 + v_u^2 Q_u^2 + v_s^2 Q_s^2)\;,\\
\Delta^2 &= \frac{1}{2} g_1'^2 G (v_d^2 Q_d - v_u^2 Q_u)
\end{split}
\end{equation}
with $G^2= g_2^2+ 3 g_1^2/5$ and $g_1'$ the gauge coupling
associated with $U(1)'$.
Obviously, the UMSSM introduces no PQ axion and
also no spontaneously broken discrete symmetry.
On the other hand the additional gauge symmetry
must have escaped detection so far.
From experiments the mixing of the $Z$ and $Z'$ bosons
is constraint to be below the per mill level
as well as the mass of the gauge boson has the lower limit 
$m_{Z'}>500$~GeV~\citep{Abe:1994ns,Abe:1994zg,Park:1995td,Abe:1997fd,Erler:2002pr}.
In deed, we see from~\eqref{eq-Umix}
that we may have a cancellation in the
off-diagonal terms, given by $\Delta^2$, that is, the 
mixing can be very small. Moreover, for
a sufficient large VEV $v_s$ 
we see that $m_{Z'}$ may be heavy enough
to be undetected up to now. 
Another possibility is that the new gauge coupling
$g_1'$ is small. In this case the new $Z'$ could be
light and even has escaped detection~\citep{Fayet:1980ad}.
There
may occur chiral anomalies in the UMSSM, but
these may be avoided in versions with family-dependent
$U(1)'$ charges~\citep{Demir:2005ti}.\\

The {\em secluded  $U(1)'$-extended minimal supersymmetric} SM 
(sMSSM) was introduced in~\citep{Erler:2002pr}. 
In the sMSSM four singlets 
are introduced, $\hat{S}$, $\hat{S}_1$,
$\hat{S}_2$, $\hat{S}_3$ with
a superpotential with the $\mu$ term of 
the MSSM replaced by
\begin{equation}
W_\mu^{\text{sMSSM}} = 
 \lambda \hat{S}  (\hat{H}_u^\trans \epsilon \hat{H}_d)
 + \lambda_s \hat{S}_1 \hat{S}_2 \hat{S}_3 \;.
\end{equation}
In this way only $\hat{S}$ contributes to
the $\mu$ term but all four singlets
are charged under $U(1)'$ and contribute
to $m_{Z'}$. For large values
of $\langle S_1 \rangle$, $\langle S_2 \rangle$, $\langle S_3 \rangle$
the additional singlets decouple and
give a natural explanation for
a very heavy $Z'$. In this limit
the Higgs-boson phenomenology of the sMSSM approaches
that of the nMSSM~\citep{Barger:2006dh}.\\

\begin{table}[t]
\begin{tabular}{r|c|c|l|l|c}
\hline
Model & Symmetry & Superpotential $W_\mu$ & P-even & CP-odd & Charged\\
\hline
MSSM   & --                             & $\mu \hat H_u \cdot \hat H_d$                                               & $H_1^0, H_2^0$        & $A_1^0$        & $H^\pm$ \\
NMSSM  & $\mathbb Z_3$                  & $h_s \hat S \hat H_u \cdot \hat H_d + \frac{\kappa}{3} \hat S^3$            & $H_1^0, H_2^0, H_3^0$ & $A_1^0, A_2^0$ & $H^\pm$\\
nMSSM  & $\mathbb Z^R_5$ 		& $h_s \hat S \hat H_u \cdot \hat H_d$          			      & $H_1^0, H_2^0, H_3^0$ & $A_1^0, A_2^0$ & $H^\pm$\\
UMSSM  & $U(1)'$                        & $h_s \hat S \hat H_u \cdot \hat H_d$                                        & $H_1^0, H_2^0, H_3^0$ & $A_1^0$        & $H^\pm$ \\
sMSSM  & $U(1)'$                        & $h_s \hat S \hat H_u \cdot \hat H_d + \lambda_s \hat S_1 \hat S_2 \hat S_3$ & $H_1^0, H_2^0, H_3^0, H_4^0, H_5^0, H_6^0$ & $A_1^0, A_2^0, A_3^0, A_4^0$ & $H^\pm$\\
\hline
\end{tabular}
\caption{\em \label{tab-barger} 
Overview over several non-minimal supersymmetric extensions, similar to the presentation in~\citep{Barger:2006dh}. Given
are the symmetries, the superpotential of the Higgs part $W_\mu$ and the types of Higgs bosons for
the case of conserved CP symmetry.}
\end{table}
In Tab.~\ref{tab-barger} a comparison of the various non-minimal supersymmetric extensions
is given, quite similar to the presentation in~\citep{Barger:2006dh}.\\

%Constrained NMSSM
Let us also give the definition of the
constrained NMSSM. Actually this is only a special
case of the NMSSM since only parameters are constraint.
In the NMSSM there is a very rich parameter space, mostly originating from
the soft-breaking terms. In contrast, the remaining part of the model is highly
restricted by supersymmetry. Essentially, the soft supersymmetry-breaking part parameterizes
the missing knowledge of the supersymmetry breaking mechanism. 
Motivated by the unification of the gauge couplings at the GUT scale, 
a rather generic approach is to assume unifications at this scale
of the soft-breaking parameters. 
%Let us start with subset of parameter space of the NMSSM, often considered in
%the literature; for instance in the recent work of Djouadi, Ellwanger and Teixeira 
%\citep{Djouadi:2008yj}.
The NMSSM with this unification 
assumption is called {\em constrained} NMSSM, or cNMSSM,
introduced analogously to the constraint MSSM, or cMSSM.
The initial set of parameters is reduced by imposing
soft parameter unification at the GUT scale of about
$\Lambda_{\text G} \equiv 10^{16}$~GeV. Explicitely, the parameters, taken
at the GUT scale are:\\
Trilinear $A$-parameter unification,
\begin{equation}
A_0 \equiv A_{\lambda}(\Lambda_{\text{G}})=A_{\kappa}(\Lambda_{\text{G}})=A_{u}(\Lambda_{\text{G}})
=A_{d}(\Lambda_{\text{G}})=A_{e}(\Lambda_{\text{G}})\;.
\end{equation}
Gaugino mass unification,
\begin{equation}
M_{1/2} \equiv M_1(\Lambda_{\text{G}}) = M_2(\Lambda_{\text{G}}) 
= M_{\tilde{g}}(\Lambda_{\text{G}})\;.
\end{equation}
Scalar mass unification,
\begin{equation}
m_{0}^2 \equiv {m_{H_d}^2}(\Lambda_{\text{G}}) = {m_{H_u}^2}(\Lambda_{\text{G}}) = {m_{S}^2}(\Lambda_{\text{G}})
= {m_Q^2}(\Lambda_{\text{G}}) = {m_L^2}(\Lambda_{\text{G}})
= {m_{\tilde{u}}^2}(\Lambda_{\text{G}})
= {m_{\tilde{d}}^2}(\Lambda_{\text{G}})
= {m_{\tilde{e}}^2}(\Lambda_{\text{G}})\;.
\end{equation}
With these unification assumptions there remain only
five undetermined parameters in the cNMSSM 
which may for instance be chosen as the
dimensionless parameters \citep{Djouadi:2008yj}:
\begin{equation}
y_{t}(\Lambda_G),\quad \lambda(\Lambda_G),\quad \kappa(\Lambda_G),\quad m_0/M_{1/2},\quad  A_0/M_{1/2}\;.
\end{equation}
Some studies of the cNMSSM are presented in Sect.~\ref{sect-scan}.\\

\newpage
%%%%%%%%%%%%%%%%%%%%%%%%%%%%%%%%%%%%%%%%%%%%%%%%%%%%%%%%%%%%%%%%%%5
% Field-dependent masses
%%%%%%%%%%%%%%%%%%%%%%%%%%%%%%%%%%%%%%%%%%%%%%%%%%%%%%%%%%%%%%%%%%5

\section{Field-dependent masses}
\label{ap-fielddependentmasses}

In this appendix the field-dependent masses of the top,
stop and of the gauge bosons are given,
taken from~\citep{Funakubo:2004ka}. This field-dependent masses
arise from the Lagrangian {\em before} spontaneous symmetry
breaking occurs. All field-dependent masses are denoted with
a bar sign in order to distinguish them from the physical masses.

The quark masses are expressed in terms of the
Higgs fields and vacuum expectation values,
\begin{align}
 &\bar{m}_b^2=|y_b|^2| H_d^0|^2
 =\frac{1}{2}|y_b|^2(v_d^2+2v_d h_d+h_d^2+a_d^2)
,\\
 &\bar{m}_t^2=|y_t|^2|H_u^0|^2
 =\frac{1}{2}|y_t|^2(v_u^2+2v_u h_u+h_u^2+a_u^2),
\end{align}
where, recalling the neutral part of~\eqref{eq-higgsespara},
\begin{equation}
H_d^0=(v_d+h_d+ia_d)/\sqrt2\,,\quad
H_u^0=e^{i\phi_u}(v_u+h_u+ia_u)/\sqrt2\,,\quad 
S=e^{i\phi_s}(v_s+h_s+ia_s)/\sqrt2\,.
\end{equation}
The physical masses of the particles arise, if
we take the vacuum-expectation-values of the Higgs bosons,
$\langle H_d^0 \rangle = v_d/\sqrt{2}$,
$\langle H_u^0 \rangle = e^{i\phi_u} v_u/\sqrt{2}$,
$\langle S \rangle = e^{i\phi_s} v_s/\sqrt{2}$,
and we get
\begin{equation}
\langle\bar{m}^2_b\rangle=m_b^2=\frac{1}{2}|y_b|^2 v_d^2,\quad
\langle\bar{m}^2_t\rangle=m_t^2=\frac{1}{2}|y_t|^2 v_u^2\,.
\end{equation}
The field dependent masses of the gauge bosons are
\begin{equation}
\bar{m}_Z^2=\frac{g_2^2+g_1^2}{2}(|H_d^0|^2+|H_u^0|^2)
,\quad
\bar{m}_W^2=\frac{g_2^2}{2}(|H_d^0|^2+|H_u^0|^2).
\end{equation}
The gauge bosons masses at the vacuum are
\begin{equation}
\langle\bar{m}_Z^2\rangle=m_Z^2=\frac{g_2^2+g_1^2}{4}(v_d^2+v_u^2)
,\quad
\langle\bar{m}_W^2\rangle=m_W^2=\frac{g_2^2}{4}(v_d^2+v_u^2).
\end{equation}
The field-dependent top- and bottom-squark masses are
\begin{multline}
 \bar{m}_{\tilde{t}_{1,2}}^2
 =\frac{1}{2}\left[m_{\tilde{q}}^2+m_{\tilde{t}_R}^2
 +\frac{g_2^2+g_1^2}{4}(|H_d^0|^2-|H_u^0|^2)
 +2|y_t|^2|H_u^0|^2\right.\\
\left.\pm\sqrt{(m_{\tilde{q}}^2-m_{\tilde{t}_R}^2
 +\frac{1}{4}\left(g_2^2-\frac{5}{3}g_1^2\right) (|H_d^0|^2-|H_u^0|^2))^2
 +4|y_t|^2|\lambda S H_d^0-A_t^*H_u^{0*}|^2}\:\right],
\end{multline}
\begin{multline}
 \bar{m}_{\tilde{b}_{1,2}}^2
 =\frac{1}{2}\left[m_{\tilde{q}}^2+m_{\tilde{b}_R}^2
 -\frac{g_2^2+g_1^2}{4}(|H_d^0|^2-|H_u^0|^2)
 +2|y_b|^2|H_d^0|^2\right.\\
\left.\pm\sqrt{(m_{\tilde{q}}^2-m_{\tilde{b}_R}^2
 -\frac{1}{4}\left(g_2^2-\frac{1}{3}g_1^2\right)(|H_d^0|^2-|H_u^0|^2))^2
 +4|y_b|^2|\lambda S H_u^0-A_b^*H_d^{0*}|^2}\:\right]\,.
\end{multline}

The masses of the squarks at the vacuum are
\begin{multline}
 \langle\bar{m}_{\tilde{t}_{1,2}}^2\rangle=m_{\tilde{t}_{1,2}}^2=
  \frac{1}{2}\left[m_{\tilde{q}}^2+m_{\tilde{t}_R}^2
  +\frac{g_2^2+g_1^2}{8}(v_d^2-v_u^2)+|y_t|^2v_u^2\right.\\
  \left.\pm\sqrt{m_{\tilde{q}}^2-m_{\tilde{t}_R}^2
  +\frac{g_2^2-\frac{5}{3}g_1^2}{8}(v_d^2-v_u^2)+2|y_t|^2((\frac{1}{2}|\lambda|^2v_s^2-R_tv_s\tan\beta) v_d^2+(|A_t|^2-R_tv_s\cot\beta)v_u^2)}\right],
\end{multline}
\begin{multline}
 \langle\bar{m}_{\tilde{b}_{1,2}}^2\rangle=m_{\tilde{b}_{1,2}}^2=
  \frac{1}{2}\left[m_{\tilde{q}}^2+m_{\tilde{b}_R}^2
  -\frac{g_2^2+g_1^2}{8}(v_d^2-v_u^2)+|y_b|^2v_d^2\right.\\
  \left.\pm\sqrt{m_{\tilde{q}}^2-m_{\tilde{b}_R}^2
  -\frac{g_2^2-\frac{1}{3}g_1^2}{8}(v_d^2-v_u^2)+2|y_b|^2((\frac{1}{2}|\lambda|^2v_s^2-R_bv_s\cot\beta)v_u^2+(|A_b|^2-R_bv_s\tan\beta)v_d^2)}\right],
\end{multline}
with the abbreviation
\begin{equation}
 R_q=\frac{1}{\sqrt{2}}\re(\lambda A_qe^{i(\phi_u+\phi_s)}),
 \qquad
 (q=t,b)\,.
\end{equation}
%

%%%%%%%%%%%%%%%%%%%%%%%%%%%%%%%%%%%%%%%%%%%%%%%%%%%%%%%%%%%%%%%%%
%%%%%%%%%%%%%%%%%%%%%%%%%%%%%%%%%%%%%%%%%%%%%%%%%%%%%%%%%%%%%%%%%
\section{Gauge invariant functions}
\label{quadratics}

In this appendix let us discuss the introduction of
gauge invariant functions which turn out to be very helpful
in studies of two-Higgs-doublet models and further extensions
of the SM. In this we follow closely~\citep{Maniatis:2006fs,Maniatis:2006jd}.
Let us consider the tree-level Higgs potential of models
having \eweakgroup (weak isospin and hypercharge) electroweak
gauge symmetry. We draw the attention to models with
two Higgs doublets and $n$ additional real Higgs isospin and hypercharge
singlets.
This includes in particular THDMs, where we have no additional Higgs 
singlets, and the NMSSM with one additional complex Higgs singlet
corresponding to two real singlets.
It is assumed that both doublets carry hypercharge $y=+1/2$ and the
complex doublet fields are denoted by
\begin{equation}
\label{eq-doubldef}
\varphi_i(x) = \begin{pmatrix} \varphi^+_i(x) \\  \varphi^0_i(x) \end{pmatrix},
\quad
i=1,2 .
\end{equation}
For the singlets real fields are assumed, denoted by
\begin{equation}
\phi_i(x), \quad i=1,\ldots,n .
\end{equation}

We remark that in supersymmetric models like the NMSSM the two Higgs doublets $H_d$, $H_u$
carry hypercharges \mbox{$y=-1/2$} and \mbox{$y=+1/2$}, respectively.
This can be translated to the convention used here by setting
\begin{equation}
\label{eq-thdmsusytrafo}
\begin{split}
\varphi^\alpha_{1} & = - \epsilon_{\alpha \beta} ( H_d^{\beta} )^{*},\\
\varphi^\alpha_{2} & = H_u^\alpha,
\end{split}
\end{equation}
where $\epsilon$ is defined in~\eqref{eq-eps}.
Complex singlet fields are embedded in the notation by treating
the real and imaginary parts of the complex singlets as two real singlet
degrees of freedom.

In the most general \eweakgroup gauge invariant Higgs potential
with the field content described above,
the doublet degrees of freedom enter solely via products of the
following form:
\begin{equation}
\label{eq-potterms}
\varphi_i^{\dagger}\varphi_j
\qquad \text{with } i,j \in \{1,2\}.
\end{equation}
It is convenient to discuss the properties of the potential such
as its stability and its stationary points in terms of
these gauge invariant quadratic expressions.
This was discussed in detail for THDMs and
also extended for the case of more than two doublets
in~\citep{Maniatis:2006fs}.
We recall the main steps here.

All possible \eweakgroup invariant
scalar products are arranged into the hermitian $2\by 2$~matrix
\begin{equation}
\label{eq-kmat}
\twomat{K} :=
\begin{pmatrix}
  \varphi_1^{\dagger}\varphi_1 & \varphi_2^{\dagger}\varphi_1 \\
  \varphi_1^{\dagger}\varphi_2 & \varphi_2^{\dagger}\varphi_2
\end{pmatrix}
\end{equation}
and its decomposition reads
\begin{equation}
\label{eq-kmatdecomp}
\twomat{K}_{i j} =
 \frac{1}{2}\,\left( K_0\,\delta_{i j} + K_a\,\sigma^a_{i j}\right),
\end{equation}
where $\sigma^a$ are the Pauli matrices.
The four real coefficients in this decomposition are
\begin{equation}
\label{eq-kdef}
K_0 = \varphi_{i}^{\dagger} \varphi_{i},
\quad
K_a = ( \varphi_{i}^{\dagger} \varphi_{j} )\, \sigma^a_{ij} ,
\quad a=1,2,3,
\end{equation}
where here and in the following summation over repeated indices is understood.
The matrix~(\ref{eq-kmat}) is positive semi-definite, which implies
\begin{equation}
\label{eq-kconditions}
K_0 \ge 0, \quad K_0^2-K_1^2-K_2^2-K_3^2 \ge 0.
\end{equation}
On the other hand, for every hermitian 
$2\by 2$~matrix $\twomat{K}_{i j}$~\eqref{eq-kmatdecomp}, 
where~\eqref{eq-kconditions} holds
there exist fields~\eqref{eq-doubldef}
satisfying~\eqref{eq-kmat}, see~\citep{Maniatis:2006fs}.
It was also shown in~\citep{Maniatis:2006fs} that the four quantities
$K_0,K_a$ satisfying~\eqref{eq-kconditions} parameterize the
gauge orbits of the Higgs doublets.
Using the inversion of~(\ref{eq-kdef}),
\begin{equation}
\label{eq-phik}
\begin{alignedat}{2}
\varphi_1^{\dagger}\varphi_1 &= (K_0 + K_3)/2, &\quad
\varphi_1^{\dagger}\varphi_2 &= (K_1 + i K_2)/2, \\ 
\varphi_2^{\dagger}\varphi_2 &= (K_0 - K_3)/2, &
\varphi_2^{\dagger}\varphi_1 &= (K_1 - i K_2)/2,
\end{alignedat}
\end{equation}
the doublet terms of the potential can be replaced
-- due to renormalizability --
by at most quadratic terms in the real functions $K_0$, $K_1$,
$K_2$, and $K_3$, which simplifies the potential and
eliminates all \eweakgroup~gauge degrees of freedom.
Eventually, the potential is written in the form
\mbox{$V(K_0, K_1, K_2, K_3, \phi_1,\ldots,\phi_n)$}.

To determine the stationary points of the Higgs potential
a potential of the form \mbox{$V(K_0, K_1, K_2, K_3, \phi_1,\ldots,\phi_n)$}
is considered
and constraint \eqref{eq-kconditions} is taken into account.
The possible cases of stationary points are distinguished with respect
to the \eweakgroup symmetry breaking behavior
which a vacuum of this type would have~\citep{Maniatis:2006fs}:
\begin{itemize}\label{eq-kcondnon}
\item{{\bf unbroken~\eweakgroup}: A stationary point with
\begin{equation}
K_0=K_1=K_2=K_3=0.
\end{equation}
A global minimum of this type implies vanishing vacuum expectation
value for the doublet fields~(\ref{eq-doubldef})
and therefore the trivial behavior with respect to the gauge group.
The stationary points of this type are found by setting 
all Higgs-doublet fields (or correspondingly the $K_0$ as well as
the $K_a$ fields) in the potential to zero and requiring a vanishing
gradient with respect to the remaining real fields:
\begin{equation}\label{eq-stationarityU}
\nabla \; V(\phi_1, \ldots, \phi_n) =0.
\end{equation}
}
\item{{\bf fully broken~\eweakgroup}: A stationary point with
\begin{equation}\label{eq-kcondfull}
\begin{gathered}
K_0>0,\\
K_0^2-K_1^2-K_2^2-K_3^2 > 0.
\end{gathered}
\end{equation}
A global minimum of this type has non-vanishing
vacuum expectation values for the
charged components of the doublets fields in~(\ref{eq-doubldef}),
thus leads to a fully broken~\eweakgroup.
The stationary points of this type are found by requiring a vanishing gradient
with respect to all singlet fields and all gauge invariant functions:
\begin{equation}
\label{eq-stationarityF}
\nabla \; V(K_0, K_1, K_2, K_3, \phi_1, \ldots, \phi_n) =0.
\end{equation}
The constraints~\eqref{eq-kcondfull} on the gauge invariant functions
must be checked explicitly for the real solutions found.
}
\item{{\bf partially broken~\eweakgroup}: A stationary point with
\begin{equation}\label{eq-kcondpartial}
\begin{gathered}
K_0>0,\\
K_0^2-K_1^2-K_2^2-K_3^2 = 0.
\end{gathered}
\end{equation}
For a global minimum of this type follows the desired partial breaking
of~\eweakgroup down to~\emgroup.
Using the Lagrange multiplier method, these stationary points are given
by the real solutions of the system of equations
\begin{equation}\label{eq-stationarityP}
\begin{split}
\nabla
\big[
V(K_0, K_1, K_2, K_3, \phi_1, \ldots, \phi_n)\quad& \\
       - u \cdot (K_0^2-K_1^2-K_2^2-K_3^2)\; \big] &= 0,
\\
K_0^2-K_1^2-K_2^2-K_3^2 &= 0,
\end{split}
\end{equation}
where $u$ is a Lagrange multiplier.
The inequality in~\eqref{eq-kcondpartial} must be
checked explicitly for the solutions found for~\eqref{eq-stationarityP}.
}
\end{itemize}
For a potential which is bounded from below,
the global minima will be among these stationary points.
Solving the systems of equations~\eqref{eq-stationarityU},
\eqref{eq-stationarityF}, and \eqref{eq-stationarityP}, and inserting
the solutions in the potential, therefore
the global minima can be identified as those solutions which have the lowest value of
the potential.
Note that in general there can be more than one global minimum point.

From the mathematical point of view
with~\eqref{eq-stationarityU},
\eqref{eq-stationarityF}, \eqref{eq-stationarityP} 
non-linear, multivariate, inhomogeneous systems 
of polynomial equations 
of third order have to be solved.
It is demonstrated in~\cite{Maniatis:2006jd} that this is possible, even if the number of
fields is large, like in the NMSSM. The most involved case is given
by~(\ref{eq-stationarityP}), which for the NMSSM consists of seven equations
in seven indeterminates, namely six real fields and one Lagrange
multiplier.
In the following the algorithmic method to solve
\eqref{eq-stationarityU}, \eqref{eq-stationarityF}, \eqref{eq-stationarityP} is described
for the case that the number of complex solutions is finite.
The latter is indeed fulfilled for the NMSSM with generic values for the parameters,
and it is automatically checked by the method.
Note that the gauge invariant functions avoid spurious continuous sets of
complex solutions, which are found to arise in the case of the
MSSM as well as the NMSSM if the stationarity conditions are
formulated with respect to the Higgs fields~\eqref{eq-doubldef}
in an unitary gauge.
This is not surprising given the fact, that the gauge invariant functions
express the contribution of the doublets to the
potential by four real degrees of freedom in contrast to the five
encountered for the doublet components in the unitary gauge.

The solution of multivariate polynomial systems of equations
is the subject of polynomial ideal theory and can
be obtained algorithmically in the Groebner basis approach~\citep{Buchberger}. 
See appendix~\ref{ap-buchberger} for a brief introduction to
this subject.
Within this approach the system of equations is transformed into
a unique standard form with respect to
a specified underlying {\em ordering} of the polynomial summands ({\em monomials}).
This unique standard form of the system of equations is given by the
corresponding {\em reduced Groebner basis}.
If the underlying ordering is the {\em lexicographical} one,
the unique standard form consists of equations with a partial separation
in the indeterminates.
A variant of the $F_4$ algorithm~\citep{Faugere-F4} is used to compute the Groebner bases.
A Groebner basis computation is generally much faster if the standard form is 
computed with respect to {\em total degree} ordering and 
then transformed into a lexicographical Groebner basis.
The transformation of bases from total degree to lexicographical ordering
is done with the help of the FGLM algorithm~\citep{FGLM}.
Finally, the system of equations represented by the lexicographical Groebner basis
has to be triangularized.
The decomposition of the system of equations with a
finite number of solutions into triangular sets is 
performed with the algorithm introduced in~\citep{Moeller, Hillebrand}.
Each triangular system consists of one univariate equation, 
one equation in $2$ indeterminates, one equation in $3$ indeterminates
and so forth. This means that the solutions are found
by subsequently solving just univariate equations by inserting the
solutions from the previous steps.

The construction of the Groebner basis as well as the 
triangularization are done algebraically, 
so no approximations are needed. 
However, the triangular system of equations 
contains in general polynomials of high order, where the zeros
cannot be obtained algebraically.
Here numerical methods are needed to find the in general complex roots
of the univariate polynomials. 

In more involved potentials, like the NMSSM, the algorithmic solution
is considerably simplified (or even made accessible), if the coefficients
of the polynomials are given in form of rational numbers.
Since rational numbers are
arbitrarily close to real numbers and moreover the physical parameters are
given only with a certain precision this does not limit the
general applicability of the method in practice.

All algorithms for the computation of the Groebner basis with respect 
to a given ordering of the monomials, 
the change of the underlying ordering, the 
triangularization, and the solution of the triangular systems 
are implemented in the SINGULAR program package~\citep{Singular}.
The solutions obtained can be easily checked by inserting them into the 
initial system of equations. Moreover, the number of complex solutions, that is, the
multiplicity of the system, is known,
so it can easily be checked that no stationary point is missing.

\newpage
%%%%%%%%%%%%%%%%%%%%%%%%%%%%%%%%%%%%%%%%%%%%%%%%%%%%%%%%%%%%%%%%%%5
% Buchberger Algorithm
%%%%%%%%%%%%%%%%%%%%%%%%%%%%%%%%%%%%%%%%%%%%%%%%%%%%%%%%%%%%%%%%%%5
\section{Buchberger algorithm}
\label{ap-buchberger}
In this appendix we want to sketch the construction of the Buchberger algorithm which
transforms a given set of polynomials~$F$ into a Groebner basis~$G$. 
Here we follow closely~Refs.~\citep{Maniatis:2006jd,Bose}.
The Groebner basis~$G$ has exactly the same simultaneous zeros
as the initial set of polynomials~$F$, but allows better access to the actual calculation
of these zeros.
The general idea is to {\em complete} the set~$F$ by adjoining differences of polynomials.
Before the Buchberger algorithm is presented, the two basic
ingredients have to be introduced, 
that is
{\em Reduction} and the
{\em S-polynomial}. 
For a more detailed discussion we refer the reader to
the literature~\citep{Buchberger,Weispfenning, Bose}. 
First of all we recall some definitions.
\begin{definition}{Polynomial Ring\\}
	A Polynomial Ring $K[x_1,\ldots,x_n]\equiv K[\mathbf{x}]$
 	is the set of all $n$-variate polynomials with variables $x_1,\ldots,x_n$
	and coefficients in the field $K$.
\end{definition}
\begin{definition}{Generated Ideal\\}
	Let $F=\{f_1,\ldots,f_n\} \subset K[\mathbf{x}]$ be finite,
	$F$ generates an ideal defined by\\
$
	I(F)\equiv\bigg\{ \sum\limits_{f_i \in F} r_i \cdot f_i
  \;\bigg|\;
  r_i \in K[\mathbf{x}],\;
  f_i \in F,\;
  i=1,\ldots,n\bigg\}.
$
%$1 \le i \le n \bigg\}$.
\end{definition}
In the following we want to consider an explicit example, 
that is, a set 
$F=\{f_1,f_2,f_3\} \subset \mathbbm{Q}[x,y]$
of polynomials with rational coefficients:
\begin{align}
\label{eq-poly}
	\begin{split}
	f_1 &= 3 x^2y+2xy+y+9x^2+5x-3,\\
	f_2 &= 2 x^3y-xy-y+6x^3-2x^2-3x+3,\\
	f_3 &= x^3y+x^2y+3x^3+2x^2.
	\end{split}
\end{align}
The set $F$ generates an ideal~$I(F)$, which is given
by the set of sums of $f_1$, $f_2$, and $f_3$, where each polynomial is 
multiplied with another arbitrary polynomial from the ring $\mathbbm{Q}[x,y]$.
The summands of the polynomial are denoted as {\em monomials}
and each monomial is the 
product of a coefficient and a {\em power product}.

Further, we introduce an {\em ordering} ($\tl$) of the monomials.
In the {\em lexicographical ordering}~($\tlex$) the monomials are ordered with respect
to the power of each variable subsequently.
The ring notation~$\mathbbm{Q}[x,y]$
defines $y \tlex x$, that is for the
lexicographical ordering of monomials powers of $y$ are considered first,
then powers of $x$.
Explicitly, this means $2 x^2 y^3 \tlex 5 x y^2  $ because
the power of $y$ is larger in the first monomial and 
$2 x y^2 \tlex 5 y^2$, because both monomials have the same power of~$y$,
but the first monomial has a larger power of~$x$. 
The monomials of the polynomials~(\ref{eq-poly})
from the ring $\mathbbm{Q}[x,y]$ 
are ordered with respect to lexicographical ordering.
In {\em total degree ordering}~($\tdeg$) the monomials are ordered with respect
to the sum of powers in each monomial. 
If two monomials have the same sum of powers,
they are ordered with respect to another ordering, for instance lexicographical.
For polynomials in $\mathbbm{Q}[x,y]$ we have
$x^2 y \tdeg 4 x y$ since the sum of powers of the left power product is $3$ compared to
$2$ for the right power product. 

The largest power product with respect to the underlying  
ordering ($\tl$) of a polynomial~$f$ is denoted as the
{\em leading power product}, $\lp(f)$, the corresponding coefficient as
{\em leading coefficient}, $\lc(f)$. 
With help of these preparations
we can define the two essential parts of the Buchberger algorithm, 
that is, {\em Reduction} and the {\em S-polynomial}.
\begin{definition}{Reduction}\\
Let $f, p \in \kx$. We call $f$ reducible modulo $p$,
if for a power product $t$ of $f$ there exists a power product $u$ with 
$\lp(p) \cdot u=t$. 
Then we say, $f$ reduces to $h$
modulo $p$, where
$h= f - \frac{\text{Coefficient}(f,t)}{\lc(p)} \cdot u \cdot p$.
\\
\end{definition}
In the example~(\ref{eq-poly}) 
the polynomial~$f_3$ is reducible modulo $f_1$, 
since for example the second monomial of~$f_3$, that is, $x^2 y$, is
a multiple of the $\lp(f_1)$, and $h=f_3 - 1/3 f_1 = x^3y -2/3 x y-1/3y+3x^3-x^2-5/3x+1$.

Reduction of a polynomial modulo a set $P \subset \kx$
is accordingly defined if there is a $p \in P$ such that
$f$ is reducible modulo $p$. Further, we say, a polynomial~$h$
is in {\em reduced form} or {\em normal form}
modulo~$F$, in short $\normf(h,F)$,
if there is no $h'$ such that $h$ reduces to $h'$ modulo~$F$.
A  set $P \subset \kx$ is called reduced,
if each $p\in P$ is in reduced form modulo $P \backslash \{ p \}$.
Note that reduction is defined with respect to
the underlying ordering of the monomials, since the
leading power product is defined with respect to the ordering.
In general, a normal form is not unique, neither for a polynomial
nor for a set.

Now we can present an algorithm, to compute 
a normal form $Q \subset \kx$ of a finite $F \subset \kx$. 
\begin{algorithm}{Normal form}\\
For a given finite set~\mbox{$F \subset \kx$} determine a
normal form~\mbox{$Q \subset \kx$}.\\[0pt]
\begin{center}
\colorbox{lgray}{\parbox{0.4\textwidth}
	{
\begin{align*}
	&Q:=F\\
& \text{{\bf while} exists } p \in Q\\
& \text{which is reducible modulo } Q \backslash \{p \} \;\;\text{\bf do}\\
&	\di Q:=Q \backslash \{ p \}\\
&	\di h:= \normf(p, Q)\\
&	\di \text{\bf if } h  \neq 0 \;\;\text{\bf then}\\
&	\di \di Q:=Q \cup \{ h \}\\
&	\text{\bf return } Q
\end{align*}
	}}
\end{center}
\end{algorithm}
\noindent
Clearly, the simultaneous zeros of all~$f_i \in F$ 
are also simultaneous zeros of all~$q_i \in Q$ and vice versa.
\begin{definition}{S-polynomial}\\
For $g_1, g_2 \in \kx$ the S-polynomial of $g_1$ and $g_2$ is defined as
\begin{align*}
\spol(g_1,g_2) &\equiv
\frac{\lcm \big( \lp(g_1), \lp(g_2) \big)}{\lp(g_1)} g_1\\
&\quad-\frac{\lc(g_1)}{\lc(g_2)} \frac{\lcm \big( \lp(g_1), \lp(g_2) \big)}{\lp(g_2)} g_2,
\end{align*}
where $\lcm$ denotes the least common multiple.
\end{definition}
\noindent
Clearly, a simultaneous zero of~$g_1$ and~$g_2$ is also a
zero of~$\spol(g_1,g_2)$.
In the example~(\ref{eq-poly}) 
we can build the S-polynomial for any two polynomials, for instance
\begin{align*}
\spol(f_1,f_2) &= 
\frac{x^3y}{x^2y}\; f_1 - \frac{3}{2}\; \frac{x^3y}{x^3y}\; f_2
	= x\; f_1 - 3/2 \;f_2\\
&=2x^2y+5/2xy+3/2y+8x^2+3/2x-9/2.
\end{align*}

Finally we define the Groebner basis.
\begin{definition}{Groebner basis}\\
$G \subset \kx$ is called Groebner Basis,
if for all $f_1, f_2 \in G$ $\normf(\spol(f_1, f_2), G ) =0$. 
\end{definition}

Now we are in a position to present the Buchberger algorithm.
\begin{algorithm}{Buchberger}\\
For a given finite set $F \subset \kx$ determine the Groebner basis $G \subset \kx$
with $I(F)=I(G)$.\\[0pt]
\begin{center}
\colorbox{lgray}{\parbox{0.4\textwidth}
	{
\begin{align*}
&G:=F\\
&B:= \{ \{g_1, g_2\} | g_1, g_2 \in G \text{ with } g_1 \neq g_2 \}\\
&\text{\bf while } B \neq \emptyset \;\;\text{\bf do}\\
&	\di \text{choose }\{ g_1, g_2 \} \text{ from } B\\
&	\di B:=B \backslash \{\{g_1,g_2\}\}\\
&	\di h:= \spol(g_1, g_2)\\
&	\di h':= \normf(h,G)\\
&	\di {\bf if} \;\;h' \neq 0 \;\;\text{\bf then}\\ 
&	\di \di B:=B \cup \{\{g,h'\}|g \in G\}\\
&	\di \di G:=G \cup \{h'\}\\	
&	\text{\bf return } G
\end{align*}
	}}
\end{center}
\end{algorithm}

Note, that since~$G$ just follows by adjoining reduced S-polynomials
to~$F$ both sets generate the same ideal, especially,
both sets have exactly the same simultaneous zeros.
It can be proven, that the Buchberger algorithm 
terminates.
The final step is to construct the reduced Groebner basis
by applying the normal form algorithm defined above to the 
Groebner basis~$G$. It can be shown that the
reduced Groebner basis 
is unique~\citep{Weispfenning}.
If we apply the Buchberger algorithm to 
the set~(\ref{eq-poly}) with subsequent reduction
we end up with the reduced Groebner basis (with
underlying lexicographical ordering)
\begin{align*}
g_1 &=y + x^2-3/2x-3,\\
g_2 &=x^3-5/2x^2-5/2x.
\end{align*}
The system of equations $g_1=g_2=0$ is equivalent to $f_1=f_2=f_3=0$, but the
former allows to directly calculate the solutions:
Since $g_2=0$ is univariate it can be solved immediately
and subsequently $g_1=0$ for each partial solution inserted.

Despite the correctness of the Buchberger algorithm tractability
of practical examples requires to improve this algorithm. 
In particular, the number of iterations in the algorithm
drastically grows with an increasing number 
of polynomials and with higher degrees of the 
polynomials.
In this respect much progress has been made with the improvement
of this original Buchberger algorithm 
from 1965; see \citep{Weispfenning,Bose,Faugere-F4}.

%%%%%%%%%%%%%%%%%%%%%%%%%%%%%%%%%%%%%%%%%%%%%%%%%%%%%%%%%%%%%%%%%
%%%%%%%%%%%%%%%%%%%%%%%%%%%%%%%%%%%%%%%%%%%%%%%%%%%%%%%%%%%%%%%%%
% Bibliography
%%%%%%%%%%%%%%%%%%%%%%%%%%%%%%%%%%%%%%%%%%%%%%%%%%%%%%%%%%%%%%%%%
\bibliographystyle{utphys}
\bibliography{nmssm}

%%%%%%%%%%%%%%%%%%%%%%%%%%%%%%%%%%%%%%%%%%%%%%%%%%%%%%%%%%%%%%%%%
\end{document}